\documentclass[12pt]{report}
\usepackage{palatino,cite,epsfig,epsf,amsmath,amssymb,color}
\usepackage{graphics,graphicx,rotating,caption2}
\usepackage[english]{babel}
\input axodraw.sty
\allowdisplaybreaks
\def\greaterthansquiggle{\raise.3ex\hbox{$>$\kern-.75em\lower1ex\hbox{$\sim$}}}
\def\lessthansquiggle{\raise.3ex\hbox{$<$\kern-.75em\lower1ex\hbox{$\sim$}}}

\newcommand{\la}{\label}
\newcommand{\re}{\ref}
\newcommand{\beqn}{\begin{eqnarray}}
\newcommand{\eeqn}{\end{eqnarray}}
\newcommand{\bequ}{\begin{equation}}
\newcommand{\eequ}{\end{equation}}
\newcommand{\bsl}{\begin{sloppypar}}
\newcommand{\esl}{\end{sloppypar}}

\definecolor{White}{named}{White}
\definecolor{Black}{named}{Black}
\definecolor{Blue}{named}{Blue}
\definecolor{Red}{named}{Red}
\definecolor{Green}{named}{ForestGreen}
\definecolor{Black}{named}{Black}
\definecolor{Olive}{named}{OliveGreen}
\definecolor{Royal}{named}{RoyalBlue}
\definecolor{Orange}{named}{YellowOrange}
\definecolor{Yellow}{named}{Goldenrod}
\definecolor{Cornblue}{named}{CornflowerBlue}
\definecolor{Lila}{named}{DarkOrchid}
\definecolor{Brown}{named}{Mahogany}

\def\abstract{\list{}{\listparindent 0pt
    \setlength{\leftmargin}{2cm}
    \itemindent\listparindent
    \rightmargin\leftmargin }\item[]{\bf Abstract}\\}

\newcommand{\sweff}{\sin^2 \theta_{\mathrm{eff}}}

\newcommand{\gev}{\,\, \mathrm{GeV}}
\newcommand{\lsim}
{\;\raisebox{-.3em}{$\stackrel{\displaystyle <}{\sim}$}\;}
\newcommand{\gsim}
{\;\raisebox{-.3em}{$\stackrel{\displaystyle >}{\sim}$}\;}
\renewcommand{\thefootnote}{\fnsymbol{footnote}}
\newcommand{\es}{$\mathrm{E_6}$}
\renewcommand{\thefootnote}{\fnsymbol{footnote}}

\newcommand{\rp}{\mbox{$\not\!R_P$}}
\def\lsim{\raise0.3ex\hbox{$\;<$\kern-0.75em\raise-1.1ex\hbox{$\sim\;$}}}
\def\gsim{\raise0.3ex\hbox{$\;>$\kern-0.75em\raise-1.1ex\hbox{$\sim\;$}}}
\newcommand{\slashed}[1]{\not\!#1}

  \newlength{\absize}
\newcommand{\dd}{\mbox{{\rm d}}}

\renewcommand{\Re}{\,\mbox{Re}}
\renewcommand{\Im}{\,\mbox{Im}}

\newcommand{\TeV}{{\rm TeV}}
\newcommand{\Lumint}{{\cal L}_{\rm int}}

\allowdisplaybreaks %!!!!

\catcode`@=11
\def\citer{\@ifnextchar [{\@tempswatrue\@citexr}{\@tempswafalse\@citexr[]}}

\def\@citexr[#1]#2{\if@filesw\immediate\write\@auxout{\string\citation{#2}}\fi
  \def\@citea{}\@cite{\@for\@citeb:=#2\do
    {\@citea\def\@citea{--\penalty\@m}\@ifundefined
       {b@\@citeb}{{\bf ?}\@warning
       {Citation `\@citeb' on page \thepage \space undefined}}%
\hbox{\csname b@\@citeb\endcsname}}}{#1}}
\catcode`@=12

\newcommand{\PE}    {\ensuremath{P}_{e^-}}   
\newcommand{\PP}    {\ensuremath{P}_{e^+}}
\newcommand{\Peff}  {\ensuremath{P}_{\text{eff}}}

\newcommand{\ALR}   {\ensuremath{A_\text{LR}}}

\newcommand{\ee}    {\ensuremath{\mathrm{e}^+\mathrm{e}^-}}

\allowdisplaybreaks %!!!!

\catcode`@=11
\def\citer{\@ifnextchar [{\@tempswatrue\@citexr}{\@tempswafalse\@citexr[]}}

\def\@citexr[#1]#2{\if@filesw\immediate\write\@auxout{\string\citation{#2}}\fi
  \def\@citea{}\@cite{\@for\@citeb:=#2\do
    {\@citea\def\@citea{--\penalty\@m}\@ifundefined
       {b@\@citeb}{{\bf ?}\@warning
       {Citation `\@citeb' on page \thepage \space undefined}}%
\hbox{\csname b@\@citeb\endcsname}}}{#1}}
\catcode`@=12
\newcommand{\Cdgg}{\ensuremath{\Delta g^{\gamma}_1}}
\newcommand{\Cdgz}{\ensuremath{\Delta g^{Z}_1}}
\newcommand{\Cdkg}{\ensuremath{\Delta \kappa_{\gamma}}}
\newcommand{\Cdkz}{\ensuremath{\Delta \kappa_{Z}}}
\newcommand{\Cgg}[1]{\ensuremath{g^{\gamma}_{#1}}}
\newcommand{\Cgv}[1]{\ensuremath{g^V_{#1}}}
\newcommand{\Cgz}[1]{\ensuremath{g^Z_{#1}}}
\newcommand{\Ckg}{\ensuremath{\kappa_{\gamma}}}
\newcommand{\Ckv}{\ensuremath{\kappa_{V}}}
\newcommand{\Ckvt}{\ensuremath{\tilde{\kappa}_V}}
\newcommand{\Ckz}{\ensuremath{\kappa_{Z}}}
\newcommand{\Ckzt}{\ensuremath{\tilde{\kappa}_Z}}
\newcommand{\Clg}{\ensuremath{\lambda_{\gamma}}}
\newcommand{\Clv}{\ensuremath{\lambda_{V}}}
\newcommand{\Clvt}{\ensuremath{\tilde{\lambda}_V}}
\newcommand{\Clz}{\ensuremath{\lambda_{Z}}}
\newcommand{\Clzt}{\ensuremath{\tilde{\lambda}_Z}}
\newcommand{\invfb}{\ensuremath{\mathrm{fb^{-1}}}}
\newcommand{\SUU}{\ensuremath{\SU{L} \times \U}}
\newcommand{\U}{\ensuremath{\textrm{U(1)}_Y}}
\newcommand{\SU}[1]{\ensuremath{\textrm{SU(2)}_{#1}}}

\newcommand{\pb}{\ensuremath{\mathrm{pb}}}

\newcommand\ZJ{\mbox{$Z \rightarrow$ 4 jets}}
\newcommand\ZTJ{\mbox{$Z \rightarrow$ 3 jets}}

\newcommand\ycut{\mbox{$y_\text{cut}$ }}

\newcommand\hb{\mbox{$\widehat{h}_{b}$}}
\newcommand\hbn{\mbox{$\widetilde{h}_{b}$}}
\newcommand{\nc}{\newcommand}
\nc{\jc}{\frac{1}{4}} 
\nc{\sll}{S_{LL}}     
\nc{\slr}{S_{LR}}
\nc{\srl}{S_{RL}}      
\nc{\srr}{S_{RR}}     
\nc{\vll}{V_{LL}}
\nc{\vlr}{V_{LR}}      
\nc{\vrl}{V_{RL}}     
\nc{\vrr}{V_{RR}}
\nc{\tll}{T_{LL}}      
\nc{\tlrs}{T_{LR}}    
\nc{\trl}{T_{RL}}
\nc{\trr}{T_{RR}}      
\nc{\slld}{S_{LL}^D}  
\nc{\slrd}{S_{LR}^D}
\nc{\srld}{S_{RL}^D}   
\nc{\srrd}{S_{RR}^D}  
\nc{\vlld}{V_{LL}^D}
\nc{\vlrd}{V_{LR}^D}   
\nc{\vrld}{V_{RL}^D}  
\nc{\vrrd}{V_{RR}^D}
\nc{\tlld}{T_{LL}^D}   
\nc{\tlrd}{T_{LR}^D}  
\nc{\trld}{T_{RL}^D}
\nc{\trrd}{T_{RR}^D}   
\nc{\aqde}{\alpha_{qde}}
\nc{\alq}{\alpha_{\ell q}}        
\nc{\alqp}{\alpha_{\ell q'}}
\nc{\alqt}{\alpha_{\ell q}^{(3)}} 
\nc{\alqtc}{\alpha_{\ell q}^{(3)*}}
\nc{\alqj}{\alpha_{\ell q}^{(1)}} 
\nc{\alqjc}{\alpha_{\ell q}^{(1)*}}
\nc{\aeu}{\alpha_{eu}}      
\nc{\alu}{\alpha_{\ell u}}
\nc{\aqe}{\alpha_{qe}}      
\nc{\ber}{\begin{eqnarray*}}
\nc{\enr}{\end{eqnarray*}}  
\nc{\jmpb}{(1-\beta)/(1+\beta)}
\nc{\wspR}{r}      
\nc{\varx}{x}      
\nc{\bt}{\beta}
\nc{\non}{\nonumber}
\nc{\lspace}{\;\;\;\;\;\;\;\;\;\;}  
\nc{\llspace}{\lspace \lspace}
\nc{\jnl}{\frac{1}{{\mit\Lambda}^2}}
\nc{\jd}{\frac{1}{2}}
\nc{\ra}{\rightarrow}
\nc{\g}{\gamma}
\def\dps{\displaystyle}

\def\TeV{\ifmmode {\,\mathrm{ Te\kern -0.1em V}}\else
                   \textrm{Te\kern -0.1em V}\fi}%
\def\GeV{\ifmmode {\,\mathrm{ Ge\kern -0.1em V}}\else
                   \textrm{Ge\kern -0.1em V}\fi}%
\def\MeV{\ifmmode {\,\mathrm{ Me\kern -0.1em V}}\else
                   \textrm{Me\kern -0.1em V}\fi}%
\def\keV{\ifmmode {\,\mathrm{ ke\kern -0.1em V}}\else
                   \textrm{ke\kern -0.1em V}\fi}%
\def\eV{\ifmmode  {\,\mathrm{ e\kern -0.1em V}}\else
                   \textrm{e\kern -0.1em V}\fi}%
\newcommand{\fbi}{\rm{fb}^{-1}}
\newcommand{\abi}{\rm{ab}^{-1}}
\newcommand{\mch}{\multicolumn {2}{|c|}}
\newcommand{\mth}{\multicolumn {3}{|c|}}

\def\pd{\phantom{-}}

\def\deg{\hbox{$^{\circ}$}}

\begin{document}

\vspace*{-1.3cm}

\hfill CERN-PH-TH/2005-036\\
\hspace*{.5cm}\hfill DCPT-04-100\\
\hspace*{.5cm}\hfill DESY 05--059 \\
\hspace*{.5cm}\hfill FERMILAB-PUB-05-060-T\\
\hspace*{.5cm}\hfill IPPP-04-50\\
\hspace*{.5cm}\hfill KEK Preprint 2005-16\\
\hspace*{.5cm}\hfill PRL-TH-05/01\\
\hspace*{.5cm}\hfill SHEP-05-03\\
\hspace*{.5cm}\hfill SLAC-PUB-11087\\[-.5em]

\begin{center}

{\Large\bf
The role of polarized positrons and electrons in revealing}

\vspace{0.4cm}

{\Large\bf fundamental interactions 
at the Linear Collider%
}

\end{center}

\vspace{1cm}
\begin{minipage}{17.5cm}
{\sc G.~Moortgat-Pick$^{1,2}$, T.~Abe$^{3}$, G.~Alexander$^{4}$, 
B.~Ananthanarayan$^5$, A.A.~Babich$^{6}$, V. Bharadwaj$^{7}$, 
D.~Barber$^{8}$, A.~Bartl$^{9}$, A.~Brachmann$^{7}$,
S.~Chen$^{3}$, J.~Clarke$^{10}$, J.E.~Clendenin$^{7}$, 
J.~Dainton$^{11}$, K. Desch$^{12}$, M.~Diehl$^{8}$, B.~Dobos$^{3}$, 
T.~Dorland$^{3}$, H.~Eberl$^{13}$, J.~Ellis$^{1}$,
K.~Fl\"ottmann$^{8}$, H.~Fraas$^{14}$,  
F. Franco-Sollova$^{8}$, F. Franke$^{14}$,
A.~Freitas$^{15}$, J.~Goodson$^{3}$,
J.~Gray$^{3}$, A.~Han$^{3}$, S.~Heinemeyer$^{1}$,
S.~Hesselbach$^{9,16}$, T.~Hirose$^{17}$, K.~Hohenwarter-Sodek$^{9}$,
J.~Kalinowski$^{18}$, T.~Kernreiter$^{9}$, O.~Kittel$^{19}$, S.~Kraml$^{1}$,
W.~Majerotto$^{13}$, 
A.~Martinez$^{3}$, H.-U.~Martyn$^{20}$, W.~Menges$^{8}$, 
A.~Mikhailichenko$^{21}$, 
K.~M\"onig$^{8}$,
K.~Moffeit$^{7}$, S.~Moretti$^{22}$, 
O.~Nachtmann$^{23}$, F.~Nagel$^{23}$, T.~Nakanishi$^{24}$,
U.~Nauenberg$^{3}$, T.~Omori$^{25}$, P.~Osland$^{1,26}$, 
A.A.~Pankov$^{6}$, N.~Paver$^{27}$, R.~Pitthan$^{7}$, 
R.~P\"oschl$^{8}$,
W.~Porod$^{28,29}$, J.~Proulx$^{3}$,
P.~Richardson$^{2}$, S.~Riemann$^{8}$, S.D.~Rindani$^{30}$, 
T.G.~Rizzo$^{7}$,
P.~Sch\"uler$^{8}$, C.~Schwanenberger$^{19}$, D.~Scott$^{10}$,
J.~Sheppard$^{7}$, R.K. Singh$^{5}$, H.~Spiesberger$^{31}$,
A.~Stahl$^{32}$, H.~Steiner$^{33}$, A.~Wagner$^{14}$,
G.~Weiglein$^{2}$, G.W.~Wilson$^{34}$, M.~Woods$^{7}$, P.~Zerwas$^{8}$, 
J.~Zhang$^{3}$, F.~Zomer$^{35}$}
\end{minipage}
\vspace*{1cm}

\noindent
{\sl 
$^{1}$ CERN TH Division, Dept. of Physics, CH-1211 Geneva 23, Switzerland\\ 
$^{2}$ IPPP, University of Durham, Durham DH1 3LE, U.K.\\
$^{3}$ University of Colorado, Boulder, CO 80309-390, USA\\
$^{4}$ University of Tel-Aviv, Tel-Aviv 69978, Israel\\
$^{5}$ Centre for High Energy Physics, Indian Institute of Science, 
       Bangalore, 560012, India\\
$^{6}$ Pavel Sukhoi Technical University, Gomel, 246746, Belarus\\
$^{7}$ Stanford Linear Accelerator Center, 2575 Sand Hill Rd, Menlo Park, 
       CA 94025, USA\\
$^{8}$ DESY, Deutsches Elektronen Synchrotron, D-22603 Hamburg, Germany\\
$^{9}$ Institut f\"ur Theoretische Physik, Universit\"at Wien, 
       A-1090 Wien, Austria\\
$^{10}$ CCLRC, ASTeC, Daresbury Laboratory, Warrington, Cheshire,
        WA4 4AD, UK\\
$^{11}$ Dept. of Physics, University of Liverpool, Liverpool L69 7ZE, UK\\
$^{12}$ Phys. Institut, Albert-Ludwigs Universit\"at Freiburg, 
        D-79104 Freiburg, Germany\\
$^{13}$ Inst. f. Hochenergiephysik, \"Osterr. Akademie d.
        Wissenschaften, A-1050 Wien, Austria\\
$^{14}$ Institut f\"ur Theoretische Physik, Universit\"at W\"urzburg, 
        D-97074~W\"urzburg, Germany\\
$^{15}$ Fermi National Accelerator Laboratory, Batavia, ILO 60510-0500, USA\\
$^{16}$ High Energy Physics, Uppsala University, Box 535, S-751 21 Uppsala,
        Sweden\\
$^{17}$ Advanced Research Institute for Science and Engineering, 
        Waseda University, 389-5 Shimooyamada-machi, Machida, Tokyo 194-0202,
Japan\\
$^{18}$ Instytut Fizyki Teoretycznej, Uniwersytet Warszawski, PL-00681 Warsaw, 
        Poland\\
$^{19}$ Phys. Institut, Universit\"at Bonn, D-53115 Bonn,
        Germany\\
$^{20}$ I.\ Phys.\ Institut, RWTH Aachen, D-52074 Aachen, Germany\\
$^{21}$ Cornell University, Ithaca, NY 14853, USA\\
$^{22}$ School of Physics and Astronomy, University of Southampton, 
        Highfield, Southampton SO17 1BJ, UK.\\
$^{23}$ Institut f\"ur Theoretische Physik, Philosophenweg 16, 
        69120 Heidelberg, Germany\\
$^{24}$ Dept. of Physics, Nagoya University, Nagoya, Japan.\\
$^{25}$ KEK, Tsukuba-shi, Ibaraki, Japan\\
$^{26}$ Dept. of Physics and Technology, University of Bergen,
        All\'{e}gaten 55, N-5007 Bergen, Norway\\
$^{27}$ Dipartimento di Fisica Teorica, Universit\`a di Trieste and\\
        Istituto Nazionale di Fisica Nucleare, Sezione di Trieste, Trieste,
        Italy\\
$^{28}$ Instituto de F\'{\i}sica Corpuscular, C.\ S.\ I.\ C.\ /Universitat 
        Val{\'e}ncia, E-46071 Val{\'e}ncia, Spain\\
$^{29}$ Institut f\"ur Theoretische Physik, Universit\"at Z\"urich, 
        Switzerland\\
$^{30}$ Theory, Physical Research Laboratory, Navrangpura, Ahmedabad, 
        380 009, India\\
$^{31}$ Institut f\"ur Physik, Johannes-Gutenberg-Universi\"at Mainz, 
        D-55099 Mainz, Germany\\
$^{32}$ III. Physikalisches Institut, RWTH Aachen, D-52074 Aachen, Germany\\
$^{33}$ Dept. of Physics and Lawrence Berkeley Laboratory, 
        University of California, Berkeley, CA 94720, USA\\
$^{34}$ Dept. of Physics and Astronomy, University of Kansas, 
        Lawrence, KS-66045, USA\\
$^{35}$ LAL, Laboratoire de l'Accelerateur Lineaire,
Universite de Paris-Sud, F-91405 Cedex Orsay, France
}\\

\begin{abstract}
\begin{sloppypar}
The proposed International Linear Collider (ILC) 
is well-suited for discovering physics
beyond the Standard Model and for precisely
unraveling the structure of the
underlying physics. 
The physics return can be maximized by
the use of polarized beams.  
This report shows the paramount role of polarized beams and
summarizes the benefits obtained from
polarizing the positron beam, as well as the electron beam.  The
physics case for this option is illustrated explicitly by
analyzing reference reactions in different physics scenarios.  The
results show that positron polarization, combined with the clean
experimental environment provided by the linear collider, allows to
improve strongly the potential of searches for new particles and
the identification of their dynamics, which opens the road 
to resolve shortcomings of the Standard Model.
The report also presents an overview of
possible designs for polarizing both beams at the ILC, as well
as for measuring their polarization.
\end{sloppypar}
\end{abstract}

%%%%%%%%%%%%%%%%%%%%%%%%%%%%%%%%%%%%%%%%%%%%%%%%%%%

%%%%%%%%%%%%%%%%%%
\tableofcontents
\listoffigures
\listoftables
%%%%%%%%%%%%%%%%%%

%%%%%%%%%%%%%%%%%%%%%%
%%%%%%%%%%%%%%%%%%%%%%%%%%%%
\chapter{Introduction}
\section{Overview}
%%%%%%%%%%%%%%%%%%%%%%%%%%%%

The first exploration of the TeV energy scale will be made with the
proton--proton Large Hadron Collider (LHC) now under construction at CERN,
which is scheduled to start operation in the year 2007. Its discovery 
potential would
be complemented by the 
electron--positron International Linear Collider (ILC) now being
designed. The clean signatures and the precise measurements made possible by 
a  high-luminosity linear collider at a known and tunable beam energy could
bring revolutionary new insights into our understanding of the fundamental
interactions of nature and the structure of matter, space and time \cite{TDR,LC-general}.

The physics return from the investment in a linear collider would be maximized by the
possibility of providing polarized beams, particularly because a high
degree of polarization can be realized without a significant loss in
luminosity. A polarized electron beam would already provide a valuable
tool for stringent tests of the Standard Model and for diagnosing new physics. 
{\it The purpose of this report is to demonstrate that the full potential
of the linear collider could be realized only with a polarized positron beam as well}.
In addition to detailed studies of directly accessible new particles, it
would also make possible indirect searches for new physics with high
sensitivity in a largely model-independent approach.

In the hunt for physics beyond the Standard Model, only small signs
may be visible, and a linear collider (LC) provides optimal conditions
for searching for the unexpected. We recall that, in the recent past,
the availability of a polarized beam at the SLC (SLAC Linear
Collider), the prototype for the ILC, enabled it to compensate in some
respects for the fact that it had a lower luminosity than LEP (Large
Electron Positron collider).  In parallel, polarized lepton scattering
has been providing surprising revelations in hadronic structure, and
polarized beams play a crucial role in the experimental programmes of
RHIC (Relativistic Heavy Ion Collider) as well as HERA
(Hadron-Elektron-Ring-Anlage). It is recognized that beam polarization
can play an important role in the ILC programme, and polarization of
the electron beam is already foreseen for the baseline
design~\cite{scope}. A high degree of at least 80\% polarization is
already envisaged, and new results indicate that 90\% should be
achievable. The importance of polarizing the positron beam has already
been studied in \cite{Baltay,Moortgat-Pick:2001kg,TDR,LC-general} and
is discussed as an upgrade option for the ILC~\cite{scope}. This
report focuses on the physics case for choosing this option, as well
as reviewing its current technical status. Most of the studies are
explicitly evaluated at $\sqrt{s}=500$~GeV with an integrated
luminosity of ${\cal L}_{\rm int}=500$~fb$^{-1}$, which matches the
energy reach and annual luminosity goal for the first stage of the
ILC.  The qualitative features of the results with regard to beam
polarization, however, are more generally valid also for future energy
upgrades from 1 TeV up to multi-TeV
energies~\cite{TDR,LC-general,clic}.

The dominant processes in $e^+e^-$ experiments are 
annihilation ($s$-channel) and
scattering ($t$- and $u$-channel) processes. 
In annihilation processes, the helicities of the
electron and positron are correlated by the spin of the particle(s)
exchanged in the direct channel. Suitable combinations of the electron and
positron beam polarizations may therefore be used to enhance considerably
signal rates and also to  efficiently suppress unwanted background
processes. These capabilities are particularly welcome in planning
searches for new physics, where in many cases only very small rates are
predicted.  An increased signal/background ratio combined with large
luminosity gives additional opportunities for possible discoveries.

On the other hand, in scattering processes, the helicities of the electron
and positron can be related directly to the properties of any produced
(new) particles. The ability to adjust independently the polarizations of
both beams simultaneously will provide unique possibilities for directly
probing the properties of the new particles. In particular, it becomes
possible to gain direct access to their quantum numbers and chiral
couplings with a minimal number of assumptions.

The detailed advantages of polarized beams for many examples of physics
both within and beyond the Standard Model are discussed in later sections
of this report. Many models of physics beyond the Standard Model have a
large number of free parameters.%, and many of these violate CP. 
For
example, the Minimal Supersymmetric extension of the Standard Model (MSSM)
contains more than one hundred new physical parameters,
whose complete determination would require many
independent experimental observables. Having both beams polarized would
increase significantly the number of measurable observables, providing more
powerful diagnostic tools, which could be crucial for determining or
constraining the many free parameters. 
The examples show that the combination of
two polarized beams may not only be important for the discovery of new
particles, but may also be indispensable for revealing the structure of
the underlying new physics.

Another prominent example of potential new physics is provided by CP
violation. The measured baryon asymmetry of the Universe cannot be
explained by the small amount of CP violation present in the Standard
Model. Scenarios for new physics beyond the Standard Model, such as supersymmetry, 
predict numerous
additional sources of CP violation. However,
there are tight experimental bounds on CP-violating parameters beyond the
Standard Model, and ongoing experiments will strengthen these bounds,
constraining severely non-standard sources of CP violation, or perhaps
revealing them.  In the quest for CP violation beyond the Standard Model,
the simultaneous availability of two polarized beams would offer unique
access to powerful CP-odd observables. In particular, transverse beam
polarizations would give access to azimuthal asymmetries that can be
defined directly in terms of products of final particle momenta, without the
need to measure final-state polarizations. For $(V,A)$ interactions, due
to the negligible electron mass, observables involving
transversely-polarized beams are only available if both beams are
polarized.

In addition to direct searches for new physics, the simultaneous
polarization of both beams offers new prospects for model-independent
indirect searches. Some scales relevant to new physics, such as those
characterizing gravity in models with extra dimensions or the
compositeness scale of quarks and leptons, could be too large to be
directly accessible at the energies of future as well as present
accelerators. Optimal strategies for indirect searches, as sensitive as
possible to small deviations of cross sections from Standard Model
predictions due to new physics, may therefore be decisive in these cases.
Such searches will undoubtedly take advantage of the clean experimental
environment and the high luminosity available at a linear collider. In general, the
results may depend strongly on the particular models chosen for analyzing
the data. However, the larger class of observables available if both beams
are polarized will make it possible to minimize such model dependence.

Even within the Standard Model, having both beams polarized will make
it possible to perform tests of unprecedented precision, either at the
Z pole, the WW-threshold or at higher energies. In the GigaZ option
the precision on the weak mixing angle can be improved by two orders
of magnitude using the left-right asymmetry. This method was pioneered
by the SLD experiment at the SLC.  Such tests require knowledge of the
polarization degree at the per mille-level, which is not possible with
conventional polarimetry alone, but may be achieved with the
polarization of both beams by applying the Blondel scheme originally
proposed for LEP.

In the following, the physics case for positron
polarization is presented in detail. Section~\ref{sect:1} gives an overview of the 
possible contributing terms in polarized $e^+e^-$
processes. It provides---for the more concerned reader---in the beginning 
a formalism~\cite{Haber:1994pe,renard} 
for including longitudinally- as well as transversely-polarized 
beams. It continues with presenting useful definitions and possible statistical
gains in effective polarization, left-right asymmetry measurements and background suppression in 
searches for new physics are presented. Section~\ref{slc-results}
refers some experimental details from the SLD experiment at the SLC 
where already highly polarized $e^-$ beams have been used.
In chapter~\ref{smphysics}
the impact of the polarization of both beams for Standard Model physics 
at high energy and at the GigaZ option is discussed. Chapter~\ref{npphysics}
addresses the advantages of polarized beams in direct 
as well as indirect searches for different kinds of new physics models.
The physics results are summarised and the quantitative gain factors are 
listed in chapter~\ref{chap-sum}. 
One may  conclude as follows: in the search for unknown new physics that
we expect at a linear collider, a polarized
positron beam provides important new observables  so that
{\it two polarized beams could provide decisive tools} { and that one always gains  
when using both beams 
polarized, independent of the direction of physics beyond the Standard 
Model}.

In chapter~\ref{machine} we also give an overview of 
the ILC machine considerations for polarized beams, i.e. of sources, spin transport and 
polarization measurement. The status of designs for both undulator-based
and laser-based polarized positron sources are reviewed and compared to that of a conventional
unpolarized positron source.
It is technically feasible to provide colliding polarized beams with minimal
loss in luminosity, and without severe commissioning problems. The precise
technical details of these designs will be given in forthcoming technical
design reports.

%%%%%%%%%%%%%%%%%%%%%%%%%%%%%%%%%%%%%%%%%%%%%%%%%%%
\section{Polarized cross sections at an $e^+e^-$ collider \label{sect:1}}
\subsection{Formalism \label{subsec:1}}
\setcounter{equation}{0}
%%%%%%%%%%%%%%%%%%%%%%%%%%%%%%%%%%%%%%%%%%%%%%%%%%%

The helicity amplitudes for the process
\begin{equation}
e^-(p_{e^-},\lambda_{e^-}) e^+ (p_{e^+},\lambda_{e^+})\to X
\label{eq_intro0}
\end{equation}
can be expressed\footnote{A Fierz transformation may be required.}
as a sum of products of the electron/positron currents ${\mathcal{J}}_k$,
\begin{equation}
{\mathcal{J}}_k(\lambda_{e^-},\lambda_{e^+})
=\bar{v}(p_{e^+},\lambda_{e^+}) \Gamma_k u(p_{e^-},\lambda_{e^-}),
\label{eq_def1}
\end{equation}
and the contributions ${\mathcal{A}}_k$ of the transition amplitude
which depend on the respective final state.
In eq.~(\ref{eq_def1}) $\Gamma_k=(\gamma_{\mu}, \gamma_{\mu}\gamma_5, 1, 
\gamma_5,\sigma_{\mu\nu})$ are the basis elements (V, A, S, P, T) of the Dirac algebra. 
The four-component spinors $u$, $v$ are denoted with the four-momenta
$p_{e^-}$, $p_{e^+}$ in the $e^-e^+$-c.m.s. and the helicities $\lambda_{e^-}$, $\lambda_{e^+}$.
In the high-energy  limit, the vector and axial-vector (V, A) couple opposite-sign helicities,
the scalars (S, P) and tensors (T) equal-sign helicities. 

The helicity
amplitude can be written in compact form
\begin{equation}
F_{\lambda_{e^-}\lambda_{e^+}}=\bar{v}(p_{e^+},\lambda_{e^+}) \Gamma
u(p_{e^-},\lambda_{e^-}),
\label{eq_def3}
\end{equation}
where for simplification the matrix $\Gamma = \sum \Gamma_k {\mathcal{A}}_k$ is introduced.

The polarized electron/positron beam is described by the $2\times 2$ spin-density
matrix $\rho_{\lambda_{e^{\pm}}\lambda'_{e^{\pm}}}$ so that
the transition probability is given by:
\begin{eqnarray}
|{\cal M} |^2 &=& \sum_{\lambda_{e^-}\lambda_{e^+}\lambda'_{e^-}\lambda'_{e^+}}
\rho_{\lambda_{e^-}\lambda'_{e^-}} \rho_{\lambda_{e^+}\lambda'_{e^+}}
F_{\lambda_{e^-}\lambda_{e^+}}
F^*_{\lambda'_{e^-}\lambda'_{e^+}}\label{eq_intro2a}\\ &=&
\sum_{\lambda_{e^-}\lambda_{e^+}\lambda'_{e^-}\lambda'_{e^+}} \rho_{\lambda_{e^-}\lambda'_{e^-}}
\rho_{\lambda_{e^+}\lambda'_{e^+}} \bar{v}(p_{e^+},\lambda_{e^+}) \Gamma
u(p_{e^-},\lambda_{e^-}) \bar{u}(p_{e^-},\lambda'_{e^-}) \bar{\Gamma}
v(p_{e^+},\lambda'_{e^+}),
\label{eq_intro2b}
\end{eqnarray}
where $\bar{\Gamma}=\sum \bar{\Gamma}_{\ell} {\mathcal{A}}^{*}_{\ell}$ and
$\bar{\Gamma}_{\ell}=\gamma^0 \Gamma^{\dagger}_{\ell} \gamma^0$. 
To evaluate eq.~(\ref{eq_intro2b}) 
helicity projection operators are applied:
\begin{eqnarray}
u(p,\lambda^{'})\bar{u}(p,\lambda)&=&
\frac{1}{2}[\delta_{\lambda \lambda^{'}} + \gamma_5 {\slashed s}^a
\sigma^a_{\lambda \lambda^{'}}](\slashed p+m)\label{eq_bouch1}\\
v(p,\lambda^{'})\bar{v}(p,\lambda)&=&
\frac{1}{2}[\delta_{\lambda^{'} \lambda} + \gamma_5 {\slashed s}^a
\sigma^a_{\lambda^{'}\lambda}](\slashed p-m),
\label{eq_bouch2}
\end{eqnarray}
where the Pauli matrices are denoted by $\sigma^a$, $a=1,2,3$. The
three four-component spin vectors $s^a$ and the momentum $p/m$ form an
orthonormal system and can be chosen as
\begin{eqnarray}
s^{3\mu}&=&\frac{1}{m}(|{\bf p}|, E \hat{\bf p}),\label{basis_s3} \\
s^{2\mu}&=&\left(0,\frac{{\bf s}^3\times {\bf p}_{\rm ref}}{|{\bf s}^3\times
{\bf p}_{\rm ref}|}\right),\label{basis_s2}\\
s^{1\mu}&=&\left(0,\frac{{\bf s}^2\times{\bf s}^3}{|{\bf s}^2
\times{\bf s}^3|}\right),
\label{basis_s1}
\end{eqnarray}
where $\hat{\bf p}$ is the unit 3-vector in the direction of the
respective particle momentum and ${\bf p}_{\rm ref}$ denotes a
reference momentum, e.g. of the final state which defines with
${\bf p}_{e^-}$ the scattering plane. The three 4-vectors
$({ s}^1,{ s}^2,{ s}^3)$ build a 
right-handed-system and form together with
$p/m$ an orthonormal system (for
further details see \cite{Haber:1994pe}).  

In the 
high-energy limit, $(E_e\gg m_e)$,
eqn.~(\ref{eq_bouch1})-(\ref{eq_bouch2}), can be written as:
\begin{eqnarray}
\mbox{\hspace{-.7cm}}
u(p_{e^{\pm}}, \lambda_{e^{\pm}}^{'})
\bar{u}(p_{e^{\pm}},\lambda_{e^{\pm}})&=& \frac{1}{2}\{(1+2
\lambda_{e^{\pm}} \gamma_5) \delta_{\lambda_{e^{\pm}}
\lambda^{'}_{e^{\pm}}}\hspace{-.2cm}+\gamma_5 [\slashed
s^{1}_{e^{\pm}}\sigma^{1}_{\lambda_{e^{\pm}} \lambda^{'}_{e^{\pm}}}
\hspace{-.1cm}+\slashed s^2_{e^{\pm}}\sigma^2_{\lambda_{e^{\pm}}\lambda^{'}_{e^{\pm}}}
]\}\hspace{-.1cm}\slashed p_{e^{\pm}},
\label{eq_bouch1-he}\\
\mbox{\hspace{-.7cm}}
v(p_{e^{\pm}}, \lambda_{e^{\pm}}^{'})
\bar{v}(p_{e^{\pm}},\lambda_{e^{\pm}})&=& \frac{1}{2}\{ (1-2
\lambda_{e^{\pm}} \gamma_5)
\delta_{\lambda^{'}_{e^{\pm}}\lambda_{e^{\pm}}}\hspace{-.2cm}+ \gamma_5 [\slashed
s^1_{e^{\pm}}\sigma^1_{\lambda^{'}_{e^{\pm}} \lambda_{e^{\pm}}}\hspace{-.1cm}+\slashed s^2_{e^{\pm}}
\sigma^2_{\lambda^{'}_{e^{\pm}}\lambda_{e^{\pm}}}]\}\hspace{-.1cm}\slashed p_{e^{\pm}}
\label{eq_bouch2-he},
\end{eqnarray}

The $2\times 2$ spin-density matrix 
of the electron (positron) in  eq.(\ref{eq_intro2b}) 
can be expanded in terms of Pauli matrices
$\sigma^{1,2,3}$:
\begin{equation}
\rho_{\lambda_{e^{\pm}}\lambda'_{e^{\pm}}}=\frac{1}{2} 
(\delta_{\lambda_{e^{\pm}}\lambda'_{e^{\pm}}} 
+P_{e^{\pm}}^1  \sigma^1_{\lambda_{e^{\pm}}\lambda'_{e^{\pm}}}
+P_{e^{\pm}}^2  \sigma^2_{\lambda_{e^{\pm}}\lambda'_{e^{\pm}}}
+P_{e^{\pm}}^3 \sigma^3_{\lambda_{e^{\pm}}\lambda'_{e^{\pm}}} 
).
\label{eq_intro5}
\end{equation}
Here $P_{e^{\pm}}^3$ (in the following denoted by $P_{e^{\pm}}$) is
the longitudinal degree of polarization with $P_{e^\pm}>0$
(right-handed polarization) if it is parallel to the respective beam
direction and $P_{e^{\pm}}<0$ (left-handed polarization) if it is
antiparallel to the beam direction.  $P_{e^{\pm}}^2$ is the degree of
polarization perpendicular to the scattering plane (spanned by
${\bf p}_{e^-}$ and the reference momentum ${\bf p}_{\rm ref}$).
$P_{e^{\pm}}^{1}$ is the degree of transverse polarization in the
scattering plane. The signs of the transverse polarizations are chosen
with respect to the basis system,
eqn.(\ref{basis_s3})-(\ref{basis_s1}).

\begin{sloppypar}
For arbitrarily oriented transverse polarization
the components of the transverse polarizations
are 
$P_{e^\pm}^1=\mp P_{e^\pm}^{\rm T} \cos(\phi_{\pm}-\phi)$
and 
$P_{e^\pm}^2=P_{e^\pm}^{\rm T} \sin(\phi_{\pm}-\phi)$ where $\phi$ is the azimuthal angle 
in the c.m.s., $\phi_+$ and $\phi_-$ are the azimuthal angles of the $e^+$ and $e^-$
polarizations with respect to a 
fixed coordinate system, e.g.\ the lab system,
and 
$P_{e^{\pm}}^{\rm T}=\sqrt{(P_{e^{\pm}}^1)^2+(P_{e^{\pm}}^2)^2}$
is the degree of the transverse polarization, respectively.
\end{sloppypar}

The matrix element squared for a process
at an $e^+e^-$ collider with polarized beams can then be written as 
(cf. also~\cite{renard,Hikasa}):
\begin{eqnarray}
|{\cal M}|^2 &=&\frac{1}{4}
\Big\{
 (1-P_{e^-})(1+P_{e^+}) |F_{\rm LR}|^2
+(1+P_{e^-})(1-P_{e^+}) |F_{\rm RL}|^2\nonumber\\
&&+(1-P_{e^-})(1-P_{e^+}) |F_{\rm LL}|^2
+(1+P_{e^-})(1+P_{e^+}) |F_{\rm RR}|^2\nonumber\\
&&-2 P_{e^-}^{\rm T} P_{e^+}^{\rm T} \{
 [\cos(\phi_{-}-\phi_{+})\Re(F_{\rm RR}F_{\rm LL}^*)
+\cos(\phi_{-}+\phi_{+}-2 \phi)  \Re(F_{\rm LR}F_{\rm RL}^*)]\nonumber\\
&&\phantom{+P_{e^-}^{\rm T} P_{e^+}^{\rm T}}
+ [\sin(\phi_{-}+\phi_{+}-2 \phi) \Im(F_{\rm LR} F_{\rm RL}^*)
 +\sin(\phi_{-}-\phi_{+}) \Im(F_{\rm RR}^* F_{\rm LL})] \}\nonumber\\
&&+ 2 P_{e^-}^{\rm T}\{ \cos(\phi_{-} - \phi) 
[(1-P_{e^+})\Re(F_{\rm RL}F_{\rm LL}^*)
                                    +(1+P_{e^+})\Re(F_{\rm RR}F_{\rm LR}^*)]
\nonumber\\
&&\phantom{ +P_{e^-}^{\rm T}}-\sin(\phi_{-} - \phi) 
[(1-P_{e^+})\Im(F_{\rm RL}^* F_{\rm LL})
-(1+P_{e^+})\Im(F_{\rm RR}^* F_{\rm LR})]\}\nonumber\\
&&- 2 P_{e^+}^{\rm T}\{ \cos(\phi_{+} - \phi)
[(1-P_{e^-})\Re(F_{\rm LR}F_{\rm LL}^*)
+(1+P_{e^-})\Re(F_{\rm RR}F_{\rm RL}^*)]\nonumber\\
&&\phantom{+P_{e^+}^{\rm T}}+\sin(\phi_{+} - \phi)
[(1-P_{e^-})\Im(F_{\rm LR}^* F_{\rm LL})
-(1+P_{e^-})\Im(F_{\rm RR}^* F_{\rm RL})] \} \Big\},
\label{eq_general}
\end{eqnarray}
where $F_{\rm LL}$, etc. denote the helicity amlitudes, eq.~(\ref{eq_def3}), with 
L (R)$\equiv \lambda=-\frac{1}{2}(+\frac{1}{2})$. 

In the case of circular accelerators~\cite{Blondel:1993dv} one gets 
(due to the Sokolov-Ternov effect
\cite{Sokolov:1963zn})
$\phi_{+}=\phi_{-}+\pi$, whereas at a linear collider 
$\phi_{-}$ and $\phi_{+}$ are given by the experimental set-up and
can be changed independently.

A few comments may be added to interpret and illustrate the formula
(\ref{eq_general}):
\begin{itemize}
\item In this expression, the $F$'s contain the dependence of the differential
cross section on the polar angle $\theta$ and on $\sqrt{s}$.
\item The contributions of different helicity configurations add up 
incoherently for longi\-tu\-dinal\-ly-polarized beams. 
\item transversely-polarized beams generate interference terms between 
left- and right-helicity amplitudes.
\end{itemize}

In the limit $m_e \to 0$ and within the Standard Model (SM), $F_{\rm
RR}=F_{\rm LL}=0$, so that if the transverse polarization of the final state
particles is not measured, the effects of transverse polarizations are absent
in the $\phi$ averaged cross section~\cite{Hikasa}.

In table~\ref{tab_gencoup} the dependence of the cross section on beam polarization
for scalar, pseudo-scalar,  vector,
axial-vector and tensor interactions, $\Gamma=$(S,P,V,A,T), are listed for 
$m_e \to 0$ (see also~\cite{Dass:1975mj}).  In the
general case, $m_e\neq 0$, all combinations: bilinear, linear, and 
interference
of longitudinal with transverse polarization exist (with the exception of pure
S or P interactions where no linear polarization dependences occur).

\begin{table}[htb]
\renewcommand{\arraystretch}{1.3}
\begin{tabular}{|c|c||c|c|c|c|c|}
\hline
\multicolumn{2}{|c|}{Interaction structure} & \multicolumn{2}{c|}{Longitudinal} & 
\multicolumn{2}{c|}{Transverse} & Longitudinal/Transverse\\  
$\Gamma_{k}$ & $\bar{\Gamma}_{\ell}$ & Bilinear & Linear & Bilinear & Linear & Interference\\
\hline
S & S & {$\sim P_{e^-}P_{e^+}$} & --  & {$\sim P_{e^-}^T P_{e^+}^T$} & -- &
-- \\
S & P &  -- & 
{$\sim P_{e^\pm}$} & {$\sim P_{e^-}^T P_{e^+}^T$} & -- &
--  \\
S & V,A & -- 
&  -- & -- &
{$\sim P_{e^{\pm}}^T$} & {$\sim P_{e^{\pm}} P_{e^{\mp}}^T$} 
\\
S & T & {$\sim P_{e^-}P_{e^+}$} & 
{$\sim P_{e^\pm}$} &  {$\sim P_{e^-}^T P_{e^+}^T$} &  &
-- \\ \hline
P & P & {$\sim P_{e^-}P_{e^+}$} & -- &  {$\sim P_{e^-}^T P_{e^+}^T$} & -- & 
-- \\
P &  V,A & {$\sim P_{e^-}P_{e^+}$} & 
{$\sim P_{e^\pm}$} &  {$\sim P_{e^-}^T P_{e^+}^T$} & {$\sim P_{e^\pm}^T$} & 
{$\sim P_{e^{\pm}} P_{e^{\mp}}^T$}\\
P & T & {$\sim P_{e^-}P_{e^+}$} & 
{$\sim P_{e^\pm}$} & {$\sim P_{e^-}^T P_{e^+}^T$} &  &
--  \\ \hline
V,A & V,A & {$\sim P_{e^-}P_{e^+}$} & 
{$\sim P_{e^\pm}$} & {$\sim P_{e^-}^T P_{e^+}^T$} & -- &
-- \\
V,A & T & -- 
&  --  & -- & {$\sim P_{e^{\pm}}^T$} & 
{$\sim P_{e^{\pm}} P_{e^{\mp}}^T$} \\ \hline
T & T & {$\sim P_{e^-}P_{e^+}$} & {$\sim P_{e^\pm}$} 
& {$\sim P_{e^-}^T P_{e^+}^T$} & -- &
-- \\ 
\hline
\end{tabular} 
\caption[Beam polarization and Lorentz structure]
{Dependence on beam polarization of the transition probability,
for (pseudo)scalar-, (axial)vector- and
tensor-interactions in the limit $m_e\to 0$. \label{tab_gencoup}}
\end{table}

As can be seen from table~\ref{tab_gencoup}, 
in pure V,A-interactions for $m_e \to 0$ the
effects from transverse beam polarization occur only if both beams
are polarized. Effects from transverse beam polarization are
particularly interesting in searches for new sources of 
CP violation~\cite{Budny:1976tk}, see
later sections. A key issue for exploiting this option is to use specific
differential cross sections and define new CP-odd as well as CP-even
asymmetries.

%%%%%%%%%%%%%%%%%%%%%%%%%%%%%%%%%%%%%%%%%%%%
\subsection{Longitudinally-polarized beams \label{subsec:long}}
\begin{figure}[htb]
\setlength{\unitlength}{1cm}
\begin{picture}(12,5)
\put(1.5,0.0)
{\mbox{\epsfysize=6.0cm\epsffile{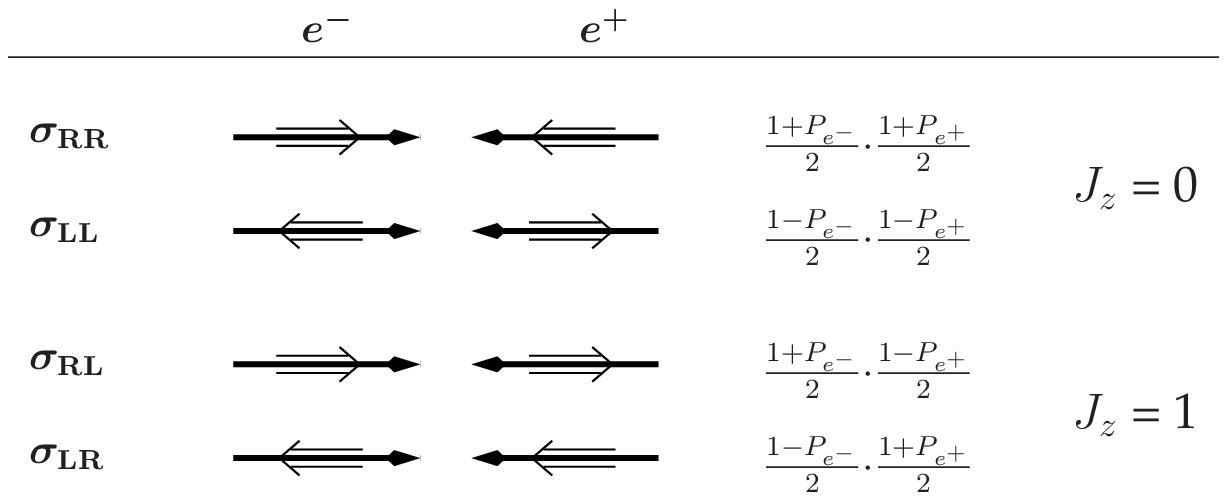}}}
\end{picture}
\caption[Possible different spin configurations in $e^+e^-$ 
annihilation]{The various longitudinal spin configurations in $e^+e^-$
collisions. The thick arrow represents the direction of motion of the particle
and the double arrow its spin direction.  The first column indicates the
corresponding cross section, the fourth column the fraction of this
configuration and the last column the total spin projection onto the $e^+e^-$
direction.}
\label{tab:SpinStates}
\end{figure}

%%%%%%%%%%%%%%%%%%%%%%%%%%%%%%%%%%%%%%%%%%%%
With longitudinally-polarized beams,
cross sections at an $e^+e^-$ collider can be subdivided (see
fig.~\ref{tab:SpinStates}) in \cite{Hikasa}:
\begin{eqnarray}
\sigma_{P_{e^-}P_{e^+}} &=& \frac{1}{4}\bigl\{
(1+P_{e^-})(1+P_{e^+}) \sigma_{\rm RR} 
+ (1-P_{e^-})(1-P_{e^+}) \sigma_{\rm LL} \nonumber \\
&& + (1+P_{e^-})(1-P_{e^+}) \sigma_{\rm RL} +
(1-P_{e^-})(1+P_{e^+})\sigma_{\rm LR} \bigr\},
\label{eq_intro1}
\end{eqnarray}
where $\sigma_{\rm RL}$ stands for the cross section 
if the $e^-$-beam is completely right-handed polarized ($P_{e^-}=+1$) and the
$e^+$-beam is completely left-handed polarized ($P_{e^+}=-1$). 
the cross sections $\sigma_{\rm LR}$, $\sigma_{\rm RR}$
and $\sigma_{\rm LL}$ are defined analogously. 
For partially polarized beams the corresponding
measurable cross sections will be denoted as $\sigma_{++}$, $\sigma_{+-}$,
$\sigma_{-+}$, $\sigma_{--}$, where the indices give 
the signs of the absolute electron/positron polarizations $|P_{e^-}|$ and $|P_{e^+}|$.

\begin{figure}[htb]
\setlength{\unitlength}{1cm}
\begin{picture}(12,3)
\put(2,0.0)
{\mbox{\epsfysize=3.5cm\epsffile{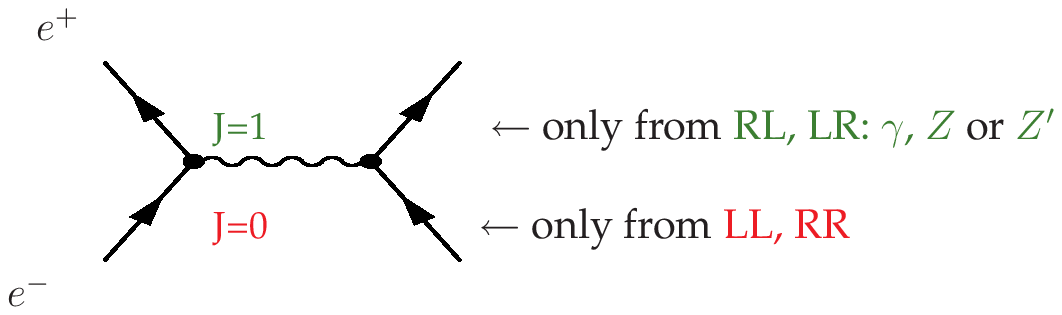}}}
\end{picture}
\caption[Polarization configurations in $s$-channel
diagrams]{Possible configurations in $s$-channel diagrams: the
helicities of the incoming $e^+e^-$ beams are directly coupled.
Within the Standard Model (SM) only the recombination into a
vector particle with $J=1$ is possible, which is given by the LR and
RL configurations. New physics (NP) models might contribute to $J=1$ but also 
to $J=0$, hence
the LL or RR configurations.\label{fig-s-channel}}
\end{figure}

\begin{figure}[htb]
\setlength{\unitlength}{1cm}
\begin{picture}(12,3)
\put(1.5,0.0)
{\mbox{\epsfysize=3.5cm\epsffile{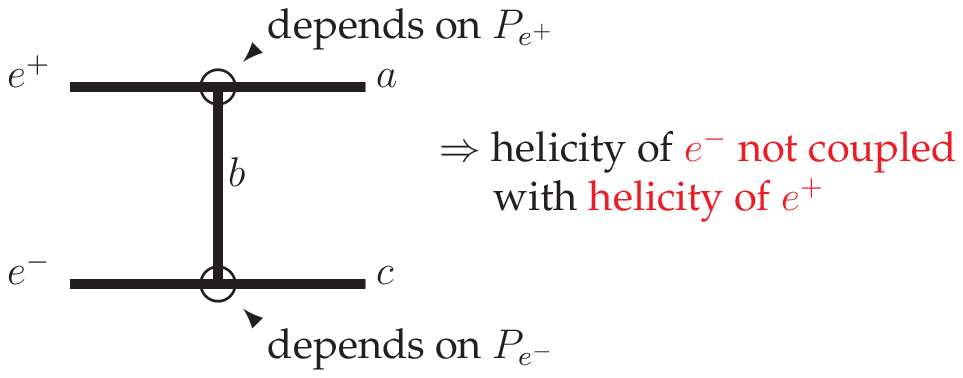}}}
\end{picture}
\caption[Polarization configurations in crossed channels]{Possible
configurations in $t$- and $u$-channel diagrams: the helicity of the incoming
beam is directly coupled to the helicity of the final particle and is
completely independent of the helicity of the second incoming particle.
\label{fig-t-channel}}
\end{figure}

One has to distinguish two cases:
\begin{itemize}
\item[a)]
in annihilation diagrams, see fig.~\ref{fig-s-channel}, the helicities of the incoming beams
are coupled to each other;
\item[b)]
in exchange diagrams, see fig.~\ref{fig-t-channel}, the helicities of the incoming beams are
directly coupled to the helicities of the final particles.
\end{itemize}

In case a) within the SM only the recombination into a vector particle with
total angular momentum $J=1$ is possible, i.e.,  the beams have to carry
opposite helicities, $F_{\rm LL}=F_{\rm RR}=0$. New physics (NP)
models can contribute to $J=1$ but 
might also allow to produce scalar particles, so that also $J=0$ would
be allowed, which would result in same-sign helicities of the incoming beams, see
fig.~\ref{fig-s-channel}.

In case b) the exchanged particle could be vector,
fermion or scalar; the helicity of the incoming
particle is directly coupled to the vertex and is independent of
the helicity of the second incoming particle. Therefore all
possible helicity configurations are in principle possible, see
fig.~\ref{fig-t-channel}.

\begin{figure}[htb]
\setlength{\unitlength}{1cm}
\begin{picture}(12,3)
\put(1.5,0.0)
{\mbox{\epsfysize=3.5cm\epsffile{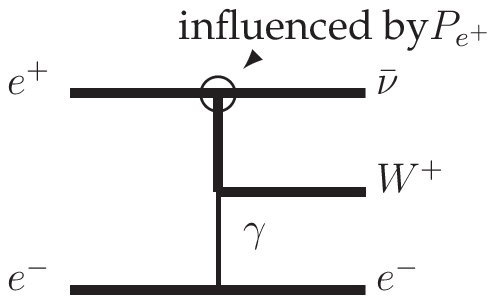}}}
\end{picture}
\caption[Single $W$ production]{Single $W^+$ production: 
the vertex $e^+ W^+ \bar{\nu}$ depends only on $P_{e^+}$. 
\label{fig-single-w} }
\end{figure}

SM candidates for case b) are single $W$ production, see
fig.~\ref{fig-single-w}, where the $e^+ W^+ \bar{\nu}$ coupling is only
influenced by $P_{e^+}$, and Bhabha scattering, where the $\gamma$, $Z$
exchange in the $t$-channel leads to an enhancement of the LL configuration so
that the cross sections for the configurations $(P_{e^-},P_{e^+})=(-80\%,0)$,
$(-80\%,+60\%)$ and $(-80\%,-60\%)$ can be of the same order of magnitude, see
table~\ref{tab-bhabha}.

\begin{table}[htb]
\begin{center}
\begin{tabular}{|l|c|c|c|c|}
\hline
$(P_{e^-},P_{e^+})$ & unpolarized &$(-80\%,0)$ &$(-80\%,-60\%)$ &$(-80\%, +60\%)$ \\ \hline
$\sigma(e^+e^-\to e^+e^-)$ & 4.50 pb & 4.63 pb & 4.69 pb & 4.58 pb\\
\hline
\end{tabular}
\end{center}
\caption[Bhabha scattering: cross sections with both beams
polarized]{Bhabha scattering at $\sqrt{s}=500$~GeV for
$45^\circ<\theta<135^\circ$. Due to the
$\gamma$, $Z$ exchange in the $t$-channel all possible
helicity configurations are allowed, e.g. the configuration LL
leads to higher cross sections than LR.\label{tab-bhabha}}
\end{table}

%%%%%%%%%%%%%%%%%%%%%%%%%%%%%%%%%%%%%%%%%%%%%%%%%%%
\subsection{Use of effective polarization and left-right asymmetry 
\label{subsec:2}}
%%%%%%%%%%%%%%%%%%%%%%%%%%%%%%%%%%%%%%%%%%%%%%%%%%%
 In the case of $e^+e^-$ annihilation into a vector particle (in the SM this
would be {$e^+e^- \rightarrow \gamma/Z^0$}) only the two $J=1$ configurations
$\sigma_{\rm RL}$ and $\sigma_{\rm LR}$ contribute, as already mentioned in
sect.~\ref{subsec:long}, and the cross section for arbitrary beam polarizations is
given by
\begin{eqnarray}
  \sigma_{P_{e^-}P_{e^+}} &=&
             \frac{1+P_{e^-}}{2}\:\frac{1-P_{e^+}}{2}\:\sigma_{\rm RL}
             \: + \:
             \frac{1-P_{e^-}}{2}\:\frac{1+P_{e^+}}{2}\:\sigma_{\rm LR}
             \nonumber \\ &=& \left(1-P_{e^-}P_{e^+}\right) \:
             \frac{\sigma_{\rm RL} + \sigma_{\rm LR}}{4} \: \left[1 \:
             - \: \frac{P_{e^-} - P_{e^+}}{1 - P_{e^+}P_{e^-}}\:
             \frac{\sigma_{\rm LR} - \sigma_{\rm RL}}{\sigma_{\rm LR}
             + \sigma_{\rm RL}}\right] \nonumber \\ &=& \left(1 -
             P_{e^+}P_{e^-}\right) \: \sigma_0 \: \left[1 \: - \:
             P_{\rm eff}\:A_{\rm LR} \right], \label{eq-sig}
\end{eqnarray}
with 
\begin{alignat}{2}
&\mbox{the unpolarized cross section:} & \quad
\sigma_0 &= \frac{\sigma_{\rm RL} + \sigma_{\rm LR}}{4}  \\
&\mbox{the left-right asymmetry:}     & \quad
A_{\rm LR} 
&= \frac{\sigma_{\rm LR} - \sigma_{\rm RL}}{\sigma_{\rm LR} + \sigma_{\rm RL}}
\label{eq_intro2} \\
&\mbox{and the effective polarization:}  & \quad
P_{\rm eff} &= \frac{P_{e^-} - P_{e^+}}{1 - P_{e^+}P_{e^-}}
\label{eq_peff}
\end{alignat}
The collision cross sections can be enhanced if both beams are polarized and
if $P_{e^-}$ and $P_{e^+}$ have different signs, see eq.~(\ref{eq-sig}).
Introducing the effective luminosity (where the ratio ${\cal L}_{\rm eff}/{\cal L}$ 
reflects the fraction of interacting
particles) by
\begin{equation} 
{\cal L}_\text{eff}
=\frac{1}{2}(1 - P_{e^-} P_{e^+}){\cal L}, 
\end{equation} 
eq.~(\ref{eq-sig}) can be rewritten as:
\begin{equation}
\sigma_{P_{e^-}P_{e^+}}=2 \sigma_0 ({\cal L}_\text{eff}/{\cal L})
\left[1 - P_{\rm eff}A_{\rm LR} \right]. 
\end{equation}
Some values for the
effective polarization as well as for the ratio  ${\cal L}_{\rm eff}/{\cal L}$  are
given in table~\ref{tab_effl}.

\begin{table}[htb]
\tabcolsep5mm
\begin{center}
\begin{tabular}{|ll||c|c|}
\hline && $P_{\rm eff}$ & ${\cal L}_\text{eff}/{\cal L}$\\ \hline
$P_{e^-}=0$,& $P_{e^+}=0$ & $0\%$ & 0.50\\
$P_{e^-}=-100\%$,& $P_{e^+}=0$ & $-100\%$ & 0.50\\
$P_{e^-}=-80\%$,& $P_{e^+}=0$ & $-80\%$ & 0.50\\
$P_{e^-}=-80\%$,& $P_{e^+}=+60\%$ & $-95\%$ & 0.74\\
\hline
\end{tabular}
\end{center}
\caption[Effective polarization and effective luminosity]
{Effective polarization and effective luminosity, 
for maximal and realistic values of beam polarization. \label{tab_effl}}
\end{table}

\begin{figure}[htb]
\begin{center}
\setlength{\unitlength}{1cm}
\begin{picture}(7,6)
\put(-.9,5.2){\small $\Peff [\%]$}
\put(6,.2){\small $P_{e^+} [\%]$}
\put(1.2,4.8){\small\color{Blue} $P_{e^-}=-90\%$}
\put(1.8,3.7){\small\color{Olive} $P_{e^-}=-80\%$}
\put(3.5,3){\small\color{Red} $P_{e^-}=-70\%$}
\put(0,0.5){\epsfig{file=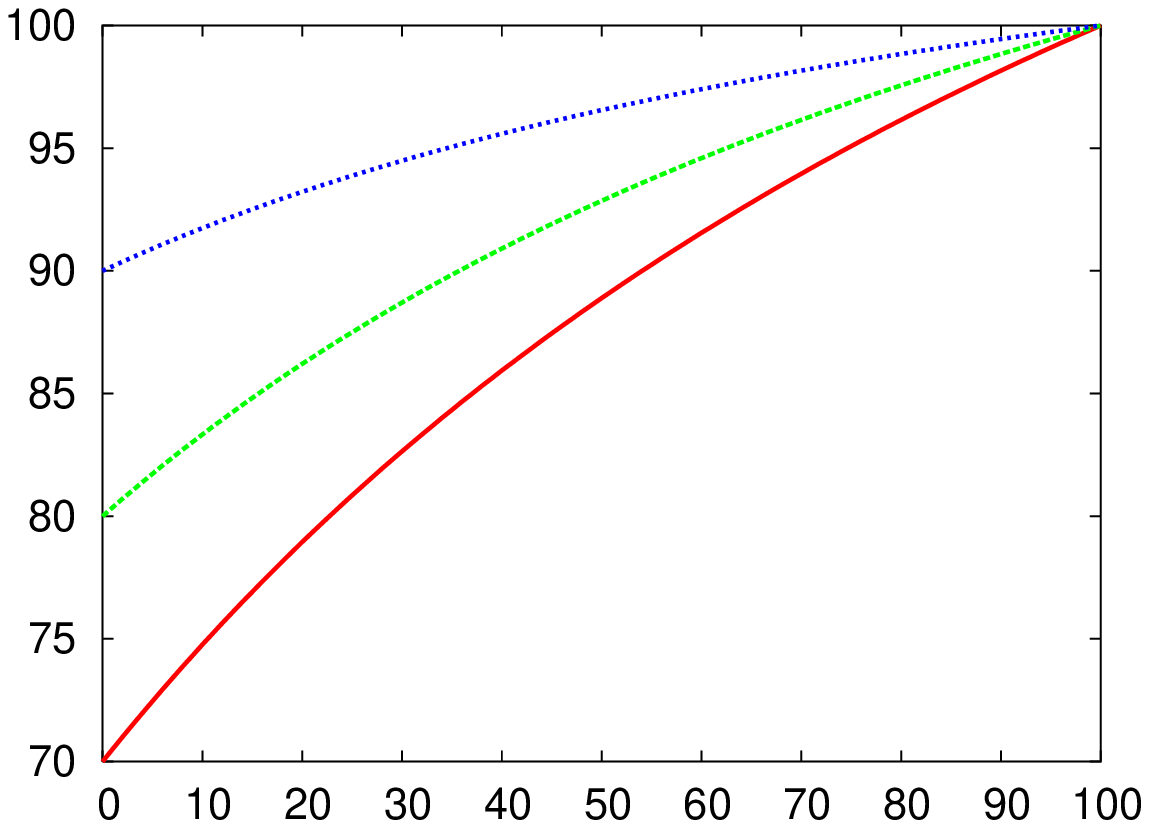,scale=.58}}
\end{picture}
\end{center}
\vspace*{-0.5cm}
\caption[Effective polarization vs.\ positron 
beam polarization]{Effective polarization vs.\ positron 
beam polarization.\label{fig:Peff}}
\end{figure}

The values of the effective polarization can be read off from
fig.~\ref{fig:Peff}. Notice that the effective polarization is closer to
{$100\%$} than either of the two beam polarizations. For further 
references, see \cite{renard,Omori:1999uz}.

\begin{figure}[htb]
\setlength{\unitlength}{1cm}
\begin{minipage}{10cm}
\begin{picture}(7,6)
\put(.1,5.2){\small $\frac{1}{x}\frac{\Delta \Peff}{|\Peff|}$}
%\put(.1,3.5){\small $\sim \frac{\Delta A_{\rm LR}}{A_{\rm LR}}$}
\put(7,0.2){\small $P_{e^+} [\%]$}
\put(4.,1.4){\small\color{Blue} $P_{e^-}=-90\%$}
\put(5.7,2.3){\small\color{Olive} $P_{e^-}=-80\%$}
\put(4.7,2.8){\small\color{Red} $P_{e^-}=-70\%$}
\put(2.2,5.5){\small errors completely independent}
\put(1.,0.5){\epsfig{file=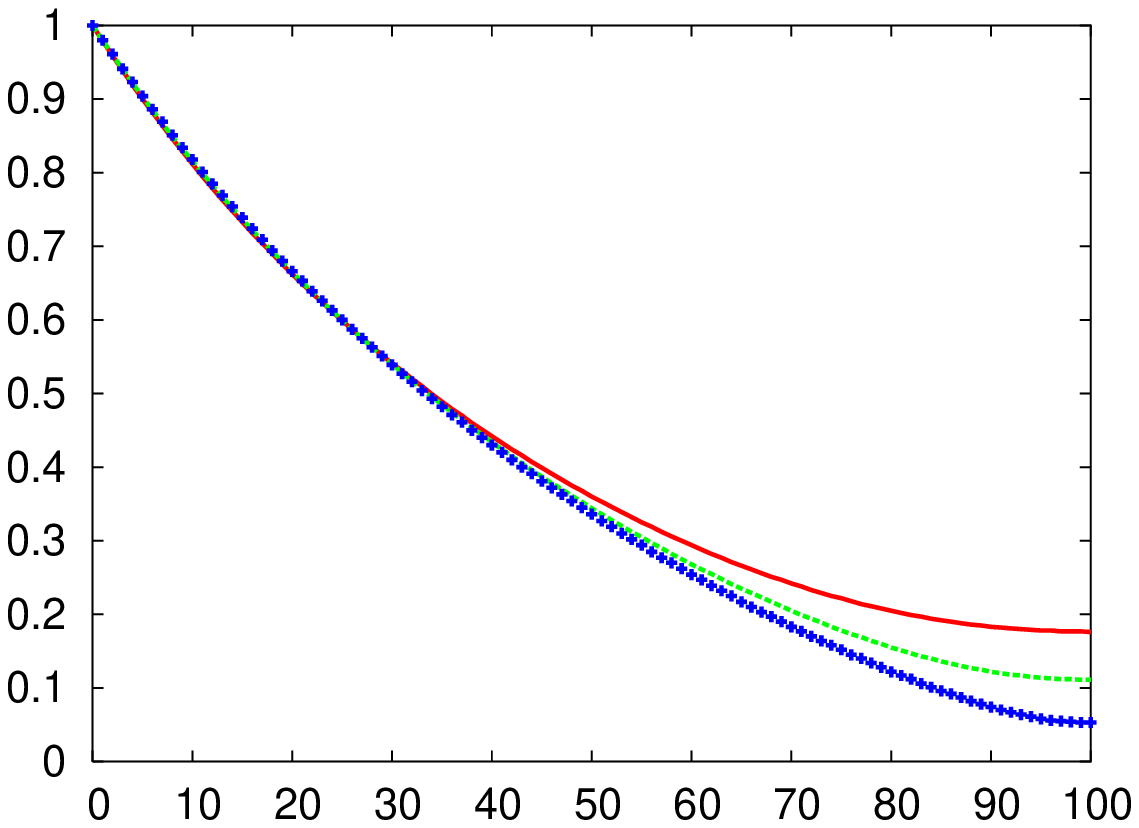,scale=.58}}
\end{picture}
\end{minipage}
    \hspace{-1.8cm}
\begin{minipage}{10cm}
\begin{picture}(7,6)
\put(.1,5.2){\small $\frac{1}{x}\frac{\Delta \Peff}{|\Peff|}$}
\put(7,0.2){\small $P_{e^+} [\%]$}
\put(4.,1.5){\small\color{Blue} $P_{e^-}=-90\%$}
\put(5.7,2.9){\small\color{Olive} $P_{e^-}=-80\%$}
\put(4.7,3.3){\small\color{Red} $P_{e^-}=-70\%$}
\put(2.8,5.5){\small errors fully correlated}
\put(1.,0.5){\epsfig{file=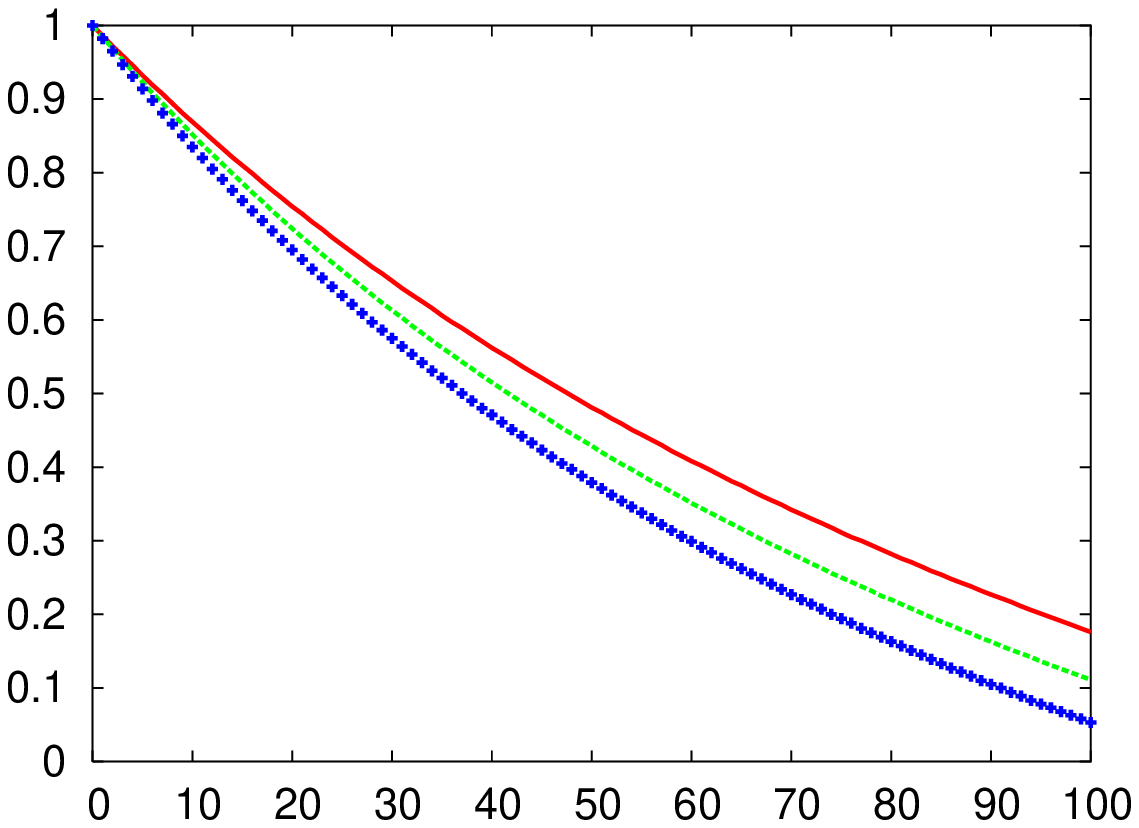,scale=.58}}
\end{picture}
\end{minipage}
\vspace*{-0.5cm}
\caption[Polarization uncertainty in $\Peff$] {Relative uncertainty on 
the effective polarization, 
$\Delta \Peff/|\Peff|\sim \Delta A_{\rm LR}/A_{\rm LR}$, 
normalized to the relative polarimeter
precision $x=\Delta P_{e^-}/P_{e^-}=\Delta P_{e^+}/P_{e^+}$ 
for independent and correlated errors on $P_{e^-}$ and $P_{e^+}$, see eqs.~(\ref{eq_pefferror}), 
(\ref{eq_pefferror-cor}).
\label{fig:EffPolErrors}}
\end{figure}

In an experiment one would like to extract, e.g., the two quantities
$\sigma_0$ and $A_{\rm LR}$, as determined by annihilation into a
vector particle, eq.~(\ref{eq-sig}).  
This can be done by running the experiment with two
different polarization configurations. One would choose one configuration with
the electron beam predominantly left-handed and the positron beam right-handed
and the second one with both spins reversed. The corresponding cross
sections, $\sigma_{-+}$ and $\sigma_{+-}$, can be expressed as
%%%%%%
\begin{align} 
{\sigma_{-+}}&= \frac{1}{4} \left\{(1+|P_{e^-}| |P_{e^+}| ) 
(\sigma_{\rm LR}+\sigma_{\rm RL})
+(|P_{e^-}|+|P_{e^+}|)(\sigma_{\rm LR}-\sigma_{\rm RL})\right\}, 
\label{eq_intro3}\\
{\sigma_{+-}}&= \frac{1}{4} \left\{(1+|P_{e^-}| |P_{e^+}| )
(\sigma_{\rm LR}+\sigma_{\rm RL}) -(|P_{e^-}|+|P_{e^+}|)
(\sigma_{\rm LR}-\sigma_{\rm RL})\right\}, 
\label{eq_intro4}
\end{align}
%%%%%%
where the subscripts denote the signs of $ P_{e^-}$ and $P_{e^+}$,
respectively.  It follows that
\begin{eqnarray}
\sigma_0 &=& \frac{{\sigma_{-+}} + {\sigma_{+-}}}{2\left(1 + |P_{e^+}||P_{e^-}|\right)} \nonumber \\
A_{\rm LR}   &=& \frac{1}{P_{\rm eff}}\: A^{\rm obs}_{LR}\: 
\:=\:\:\frac{1}{P_{\rm eff}} \:\:
             \frac{{\sigma_{-+}} - {\sigma_{+-}}}{{\sigma_{-+}} 
+ {\sigma_{+-}}}, \label{eq-def-alr}
\end{eqnarray}
where $A^{\rm obs}_{\rm LR}$ is the measured left-right asymmetry of
processes with partially polarized beams.

The contribution of the uncertainty of the
polarization measurement to the error  in $A_{\rm LR}$ is, under the assumption
that the errors are completely independent and added in quadrature:
\begin{align}
\frac{\Delta P_{\rm eff}}{P_{\rm eff}} 
&= \frac{x}
{\left(|P_{e^+}| + |P_{e^-}|\right)\:\left(1 + |P_{e^+}||P_{e^-}|\right)} \:
       \sqrt{\left(1 - |P_{e^-}|^2\right)^2 
P_{e^+}^2 + \left(1 - |P_{e^+}|^2\right)^2
       P_{e^-}^2}\quad \label{eq_pefferror}\\
\mbox{\hspace{-2cm}}\left|\frac{\Delta A_{\rm LR}}{A_{\rm LR}}\right|     
&= \left| \frac{\Delta P_{\rm eff}}{P_{\rm eff}} \right|. \label{eq_alrerror}
\end{align}
Equal relative precision $x\equiv\Delta P_{e^-}/P_{e^-}=\Delta
P_{e^+}/P_{e^+}$ of the two beam
polarizations is assumed.

In the case where the relative errors on $\PE$ and $\PP$ are fully correlated,
like for example by depolarization effects from Bremsstrahlung, the polarization contribution
to the uncertainty in $A_{\rm LR}$  is given by the linear sum:
\begin{eqnarray}
\frac{\Delta P_{\rm eff}}{P_{\rm eff}} 
&=& \frac{1-|P_{e^+}| |P_{e^-}|}{1+ |P_{e^+}| |P_{e^-}|} \: x.
\label{eq_pefferror-cor}
\end{eqnarray}
\begin{sloppypar}
It is immediately obvious from eqs.~(\ref{eq_pefferror}) and
(\ref{eq_pefferror-cor}), that in both cases $\Delta P_{\rm eff}/P_{\rm
eff}<\Delta P_{e^-}/P_{e^-}$. 
The improvement from positron polarization for the polarization
contribution to the error in $A_{\rm LR}$ is shown in fig.~\ref{fig:EffPolErrors}. 
The improvement due to positron beam polarization is substantial. For
a positron polarization of $60\%$ the error on $A_{\rm LR}$ is reduced by a
factor of about 3, see also~\cite{Flottmann:1995ga}.
\end{sloppypar}

%%%%%%%%%%%%%%%%%%%%%%%%%%%%%%%%%%%%%%%%%%%%%%%%%%%
\subsection{Suppression of background in new physics searches}
\label{sect:suppr-bck}
%%%%%%%%%%%%%%%%%%%%%%%%%%%%%%%%%%%%%%%%%%%%%%%%%%%
The use of both beams 
polarized compared with only electrons polarized 
can lead to an important gain in statistics and luminosity, i.e.\ 
it reduces the required running time.
Furthermore, in order to find even small traces of physics beyond the SM and detect
new signals, it is important to reduce possible background
processes as efficiently as possible. Beam polarization plays an
important role in this context by enhancing the signal and suppressing 
the background rates. It is parametrized by a scaling factor 
comparing the cross sections with two different polarization configurations a) and b): 
\begin{equation}
\mbox{\rm  scaling factor}=\sigma^{(P_{e^-},P_{e^+})_{b)}}/
\sigma^{(P_{e^-},P_{e^+})_{a)}}.
\label{def_scaling}
\end{equation}
Using both beams polarized in configuration b) instead of just the
electron beam polarized in configuration a) can lead to a scaling
factor between 0 and at most 2, cf.\ eq.~(\ref{eq_intro1}).

In cases where the background process depends on beam polarization in
a way different from the expected signal process, the gain in using
both beams polarized is obvious: suppressing the background and
enhancing the signal simultaneously with the suitable polarizaton
configuration.

However, also in cases where both processes, 
background as well as expected signal, show a similar 
dependence on beam polarization, it is advantageous to use this tool 
because of the immediate gain in statistical significance. 
In order to detect a signal, 
\begin{equation}
S=\sigma(e^+ e^-\to X_1 X_2, \ldots) \times \Lumint,
\label{eq_def_s}
\end{equation}
where $X_1$, $X_2$, $\ldots$ are any possible produced particles and
$\Lumint$ 
denotes the integrated luminosity, the signal (S) has to be separated from
the possible background ($B$) process(es). The background has  a 
statistical variation of $\sqrt{B}$ (Gaussian distribution assumed, which is 
suitable thanks to the high statistics at the LC).
In order to get a significance of ${\cal N}_{\sigma}$ 
standard deviations for the new signal,
it is required that 
\begin{equation}
S > {\cal N}_{\sigma} \times \sqrt{B}.
\label{eq_def_sig}
\end{equation}
Therefore, in order to evaluate the statistical gain correctly when
applying both beams polarized, one has to consider not only
$S/B$ but also $S/\sqrt{B}$. In table~\ref{tab_stat} the respective
values are listed for the two cases where the background and signal
processes have the same or an inverse scaling factor.
This clearly  shows that one gains even in the latter case, since the
significance is enhanced! 

\begin{table}
\begin{center} 
\begin{tabular}{|c||c|c||c|c|}
\hline
 & $\quad S\quad$ & $\quad B\quad$ & $\quad S/B\quad$ & 
$\quad S/\sqrt{B}\quad$ \\\hline
Example 1
& $\times 2$ & $\times 0.5$ & $\times 4$ & $\times 2 \sqrt{2}$ \\
Example 2 
& $\times 2$ & $\times 2$ & Unchanged & $\times \sqrt{2}$ \\ \hline
\end{tabular}
\caption[Statistical gain in significance]{ The gain in `signal over
background', $S/B$, and in significance, $S/\sqrt{B}$, when
both beams are polarized compared with the case of polarized
electrons only, for two examples where the background and signal
processes have the same or an inverse scaling factor.
 \label{tab_stat}}
\end{center}
\end{table}

%%%%%%%%%%%%%%%%%%%%%%%%%%%%%%%%%%%%%%%%%%%%%%%%%%%%%%%%%%%%%%%%%%%%%%%%%%
\section{Importance of electron polarization at the SLC \label{slc-results}}
%%%%%%%%%%%%%%%%%%%%%%%%%%%%%%%%%%%%%%%%%%%%%%%%%%%%%%%%%%%%%%%%%%%%%%%%%%
Before examining, in the next chapters, the physics motivations 
for polarizing
both beams, it is useful to highlight the importance the polarized
electron beam had for the SLD experiment at the SLAC Linear Collider (SLC)
\cite{slacpub8985}, and to review some of the technical aspects.  SLD made
detailed and precise measurements of parity violation in weak neutral current
interactions by studying $e^+e^-$ collisions at the $Z$ resonance, and
utilizing an electron beam with $75\%$ electron polarization.  This degree of
electron polarization provided an 
improvement in statistical power of approximately a factor 25 
for many $Z$-pole asymmetry observables, e.g. when 
using the left-right forward-backward 
asymmetry, $A_{\rm LR,FB}(\cos\theta)$ 
instead of the forward-backward asymmetry $A_{\rm FB}(\cos\theta)$.  In
particular, it allowed the SLD experiment to make the best individual measurement
of the weak mixing angle, which is a key ingredient for indirect predictions
of the SM Higgs mass.  The electron polarization at SLC also provided a
powerful tool for bottom quark studies providing a means for tagging $b$ or
$\overline{b}$ quark jets, by utilizing the large polarized forward-backward
asymmetry in this channel. Moreover, it allowed for studies of parity
violation in the $Zb\overline{b}$ vertex.

The flagship measurement for SLD was a high-precision measurement of
the left-right asymmetry,
and it is useful to recall this measurement in some detail.  With
the data obtained from 383500 $Z$ decays in 1996-98 (about three
quarters of the total SLD data sample), the pole value of the
asymmetry was measured with high precision, $A_{\rm LR}^0=0.15056\pm
0.00239$ \cite{Abe:2000dq}. The measurement required a precise
knowledge of the absolute beam polarization, but did not require
the knowledge of absolute luminosity, detector acceptance and
efficiency, provided the efficiency for detecting a fermion at some
polar angle was equal to the efficiency for detecting an antifermion at
the same polar angle.

The beam polarization was measured by a Compton-scattering polarimeter
that analyzed the Compton-scattered electrons in a magnetic
spectrometer. The accepted energy of the Compton electrons was in the
range between 17 and 30 GeV. The kinematical minimum was at 17.36 GeV.
Resolution and spectrometer effects were included and the derived analyzing power
of the detector
differed by typically about 1\% from the theoretical Compton
polarization asymmetry function \cite{gunst}.  Polarimeter data were
continuously acquired during the SLC operation~\cite{Woods:1996nz}.

Two additional detectors analyzed the Compton-scattered gammas and
assisted in the calibration of the primary spectrometer-based
polarimeter. Although they were not used during collision, both
provided a useful cross-check for the calibration procedure.

The systematic uncertainties affecting the polarization measurements
were a) laser polarization, b) detector linearity, c) analysing power
calibration and d) electric noise which led to an uncertainty of
$\Delta P_{e}/P_{e}\sim 0.5\%$, see table~\ref{tab_sldpol}.

Due to chromatic effects (finite beam energy spread, spin precession
in the SLC arcs and small chromatic aberrations in the final focus
optics) and the large angular divergence of the colliding beams, there
is an additional uncertainty in extracting the luminosity-weighted beam
polarization from the polarimeter measurements.  The depolarization
due to the $e^+e^-$ interaction itself turned out to be negligible.
These effects contribute to an additional 0.15$\%$
uncertainty, see table~\ref{tab_sldpol}.

There could have been small corrections to the left-right asymmetry
extracted by dividing the measured asymmetry by the
luminosity-weighted beam polarization.  These corrections can arise
from left-right asymmetries in luminosity, beam energy, beam
polarization, detector backgrounds and detector efficiency.  Another
potential source of correction is any positron polarization of
constant helicity, but this was measured to be negligible.  At the SLD
experiment these small corrections resulted in an additional 0.07$\%$
uncertainty.  The final relative uncertainty for $A_{\rm LR}$ was
0.52\%.

To obtain the pole asymmetry $\ALR^0$, the experimental asymmetry
$\ALR$ had to be corrected for effects arising from pure photon
exchange and $Z\gamma$ interference, where these electroweak
corrections sensitively depend on uncertainties in the
c.m.\ energy $\sqrt{s}$.  The relative systematic uncertainty for $\ALR^0$ was
finally around 0.64\%, see table~\ref{tab_sldpol}, and led to the
determination of $\sin^2\theta_W^\text{eff}=0.23107\pm 0.00030$ from the
1996-98 data sample.  Including all of the SLD data sample results in
$\sin^2\theta_W^\text{eff}=0.23097\pm 0.00027$.

The high-precision measurement of the left-right asymmetry at the $Z$-boson
shows the power of the polarized electron beam  
and also the numerous effects which must
be taken into account to achieve such high precision.  The SLD experiment has
demonstrated how successfully these effects can be controlled. 

\begin{table}
\begin{tabular}{|l|c|c|c|}
\hline
Uncertainty & $\Delta \PE/\PE$ & $\Delta \ALR/\ALR$ 
& $\Delta A^0_{\rm LR}/A^0_{\rm LR}$ \\
\hline
Laser polarization & 0.10\% &&\\
Detector linearity & 0.20\% &&\\
Analysing power calibration & 0.40\% && \\
Electronic noise & 0.20\% &&\\
Total polarimetry uncertainty & 0.50\% & 0.50\% & \\
Chromaticity and IP corrections & & 0.15\% & \\
Corrections between measured and theoretical $A_{\rm LR}$ & & 0.07\% &\\
$A_{\rm LR}$ systematic uncertainty & & 0.52\% & 0.52\% \\
Electroweak interference corrections & && 0.39\% \\
$A_{\rm LR}^0$ systematic uncertainty & && 0.64\% \\ \hline
\end{tabular}
\caption[Systematic uncertainties at SLD]{Systematic uncertainties that affect
the $A_{\rm LR}$ measurement. The uncertainty on the electroweak interference
correction is caused by the uncertainty on the SLC energy scale, see
\cite{Abe:2000dq}.\label{tab_sldpol} }
\end{table}

%%%%%%%%%%%%%%%%%%%%%%%%%%%%%%%%%%%%%%%%%%%%%%%%%%%
\chapter{Open questions of the Standard Model \label{smphysics}}
%%%%%%%%%%%%%%%%%%%%%%%%%%%%%%%%%%%%%%%%%%%%%%%%%%%
\section{Top couplings, influence of effective polarization
\label{sec:effpol}}
%%%%%%%%%%%%%%%%%%%%%%%%%%%%%%%%%%%%%%%%%%%%%%%%%%%
{\bf \boldmath
Top quark production occurs in the SM via $\gamma$,
$Z$ exchange and the influence of polarized beams can be described
by the effective polarization $\Peff$.  Availability of both beams
polarized allows for a substantial improvement in the $\ALR$
measurement and the determination of couplings and limits for non-standard top
physics.  Limits on flavour-changing neutral couplings or CP-violating
interactions are particularly improved.}
\smallskip

The top quark is by far the heaviest fermion observed, yet all the
experimental results obtained so far indicate that it behaves as would be
expected for a sequential third generation quark.  Its large mass, which is
close to the scale of electroweak symmetry breaking, makes the top quark a
unique object for studying the fundamental interactions in the attometer
regime. It is likely to play a key role in pinning down the origin of
electroweak symmetry breaking \cite{georg-top}.  High precision measurements
of the properties of the top quark will be an essential part of the ILC
research program.
%%%%%
\subsubsection*{a) Determination of static electroweak top properties 
}
%%%%%
A linear collider provides an ideal tool to probe the couplings of the
top quark to the electroweak gauge bosons. The neutral electroweak
couplings are accessible only at lepton colliders, because top quarks
at hadron colliders are pair-produced via gluon exchange.

The most general $(\gamma,Z)t \bar{t}$ couplings can be written 
as~\cite{Atwood,Hollik:1998vz,LC-general}
\begin{equation}
\Gamma^{\mu}_{t\bar{t}\gamma,Z}=i e
\left\{\gamma^{\mu}[F^{\gamma,Z}_{1V}
+F^{\gamma,Z}_{1A}\gamma^5]+\frac{(p_t-p_{\bar{t}})^{\mu}}{2
m_t}[F^{\gamma,Z}_{2V} +F^{\gamma,Z}_{2 A}\gamma^5]\right\},
\label{eq_topform}
\end{equation}
where the only form factors different from zero in the SM are
\begin{equation}
F^{\gamma}_{1V}=\frac{2}{3},\quad F^Z_{1V}=\frac{1}{2 \sin(2 \theta_W)}(1-\frac{8}{3}
\sin^2\theta_W),\quad F^Z_{1A}=-\frac{1}{2 \sin(2\theta_W)}.
\label{def-smtop}
\end{equation}
The form factor  $(e/m_t)F^{\gamma}_{2A}$ is the CP-violating electric dipole moment, 
$(e/m_t)F^{Z}_{2A}$ is the weak electric dipole moment.
The factors  $(e/m_t)F^{\gamma,Z}_{2V}$ are the electric and weak 
magnetic dipole moments. 

Polarization effects have been studied at the top threshold \cite{Kuehn}.
In the SM the main production process occurs via $\gamma$, $Z$ exchange and
the ratios between the polarized and unpolarized cross sections are given by
the scaling factors $(1-P_{e^-})(1+P_{e^+})$ and $(1+P_{e^-})(1-P_{e^+})$,
that can be parametrized with the effective polarization $\Peff$.
To determine the SM top vector coupling, $v_t=(1-\frac{8}{3}\sin^2\theta_W)$, 
one has to measure the left--right
asymmetry $\ALR$ with high accuracy. With an integrated luminosity of
${\cal L}_{\rm int}=300$~fb$^{-1}$ precisions in $\ALR$ and $v_t$ of about $0.4\%$
and $1\%$, respectively, can be achieved at the LC. The gain in using
simultaneously polarized $e^-$ and $e^+$ beams with
\begin{equation}
(\PE,\PP)=(\mp80\%,\pm60\%)
\end{equation}
is given by the higher effective polarization of $\Peff=95\%$ compared to the
case of only polarized electrons with $|P_{e^-}|=80\%$. This leads, 
according to fig.~\ref{fig:EffPolErrors} and eqs.~(\ref{eq_pefferror}) and
(\ref{eq_alrerror}) to a reduction of the relative uncertainty $\Delta A_{\rm
LR}/A_{\rm LR}\simeq \Delta\Peff/\Peff$ by about a factor of 3 \cite{LC-general}. 

Limits to all the above mentioned form factors have also been derived
in the continuum at $\sqrt{s}=500$~GeV for unpolarized beams and 
$(|P_{e^-}|,|P_{e^+}|)=(80\%,0)$~\cite{TDR,LC-general}.
It has been estimated that the polarization of both beams with
$(|P_{e^-}|,|P_{e^+}|)=(80\%,60\%)$ leads to an increase of the $t\bar{t}$ cross section 
by about a factor $\sim 1.5$ and improves again the bounds by about a 
factor 3~\cite{LC-general}.\\[.6em]

%%%%%
\subsubsection*{ b) Limits for CP-violating top dipole couplings }
\begin{figure}[htb]
\begin{center}
\setlength{\unitlength}{1cm}
\begin{picture}(12,3.7)
\put(1.7,0.)
{\mbox{\epsfysize=4.2cm\epsffile{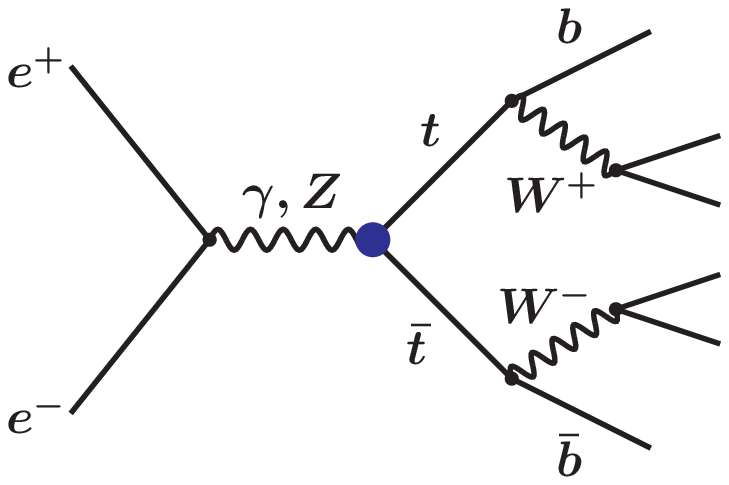}}}
\end{picture}
\caption[Anomalous top dipole interaction]{Anomalous top dipole interaction.
\label{fig:Feynman-ttbar-dipole}}
\end{center}
\end{figure}

%%%%%
Measurements aimed at testing top-quark couplings to gauge bosons (both
CP-conserving and CP-violating), see eq.~\ref{eq_topform}, 
certainly are an important issue to reveal
new fundamental interactions at the linear collider. Searches of
anomalous $t{\bar t}\gamma$ and $t{\bar t} Z$ couplings can be made by
studying the decay energy and angular distributions of $l^+$ ($l^-$) or 
$b$ ($\bar b$) in $e^+e^-\to t{\bar t}$ followed by the subsequent decays
$t\to l^+\nu_l b$ 
(${\bar t}\to l^- {\bar \nu}_l {\bar b}$),
as in 
fig.~\ref{fig:Feynman-ttbar-dipole}~\cite{Grzadkowski:2000nx,Rindani:2003av}.
 
Particularly interesting are the 
CP-violating couplings $F^{V}_{2A}$, $V=\gamma$, $Z$, that in the SM can be generated only 
at the
(extremely suppressed) two-loop level. Therefore, detection of the above
CP-violating couplings would be a clear manifestation of physics beyond the
SM.
 
In principle, also the $Wtb$ vertex can have an anomalous structure, viz.
\begin{equation}
\Gamma_{Wtb}^\mu=-\frac{g}{\sqrt 2}V_{tb}
\left[
  \gamma^\mu(f_1^{\rm L} P_{\rm L}+f_1^{\rm R}P_{\rm R})
 - \frac{i\sigma^{\mu\nu}(p_t-p_b)_\nu}{m_W} (f_2^{\rm L}P_{\rm L}
 + f_2^{\rm R}P_{\rm R})
\right],
\label{anomwtb}
\end{equation}
since in the SM only $f_1^{\rm L}=1$ while all
others vanish (in the limit $m_b\to 0$, which is used here). 
Also, it is assumed that the electron interactions with the
gauge bosons are described by the SM.
 
Focusing on CP violation, a suitable observable is represented by the
forward-backward charge asymmetry \cite{Grzadkowski:2000nx,Rindani:2003av}
\begin{equation}
{\cal A}_{\rm CP}^f(P_{e^-},P_{e^+})
=\frac{\int_{\theta_0}^{ \pi/2} d\cos\theta_f\frac{d\sigma^-}{d\cos\theta_f}
-\int_{\pi/2}^{\pi-\theta_0} d\cos\theta_f \frac{d\sigma^+}{d\cos\theta}}
{\int_{\theta_0}^{\pi /2} d\cos\theta_f\frac{d\sigma^-}{d\cos\theta_f}
+ \int_{\pi/2}^{\pi-\theta_0}d\cos\theta_f\frac{d\sigma^+}{d\cos\theta}},
\label{afbch}
\end{equation}
where $d\sigma^{\mp}$ refer to $f$ and $\bar f$ polar angle distribution in
the $e^+-e^-$ c.m. frame, respectively (with $f=l, b$), and $\theta_0$ is a
polar angle cut. The asymmetry (\ref{afbch}) is a genuine measure of
CP violation and, actually, for $f=l$ it is sensitive exclusively to the
CP violation at the $t{\bar t}$ production vertices.
Conversely, for $f=b$ the anomalous structure of $Wtb$ in eq.~(\ref{anomwtb}) appears.
 
Moreover, the asymmetry (\ref{afbch}) is defined, and proportional to
the CP-violating couplings 
for all values of beam
polarizations, although for $P_{e^-}\ne P_{e^+}$ the initial state is not a
CP eigenstate. This is due to the $s$-channel spin-one exchange character of
the production process, that forces only the ($e^-_{\rm L}e^+_{\rm R}$) or the
($e^-_{\rm R}e^+_{\rm L}$) components of the initial state to interact with
angular momentum conservation.
 
With $\sqrt s=500\hskip 3pt {\rm GeV}$,
${\cal L}_{\rm int}=500\hskip 3pt{\rm fb}^{-1}$, and 60\% reconstruction
efficiency for either lepton or $b$, the forward-backward charge asymmetry
could be measured at the 5.1-$\sigma$ (2.4-$\sigma$) level for $b$-quarks (leptons)
assuming CP-violating couplings of the order of $5\times 10^{-2}$ and
unpolarized beams. Having both beams 80\% polarized the reach on
${\cal A}_{\rm CP}^f$ would even increase up to 16-$\sigma$ (3.5-$\sigma$) and,
also, optimal observables and beam polarizations can be adjusted to minimize
the statistical error on the determination of the anomalous couplings
\cite{Grzadkowski:2000nx}.
 
Some more CP asymmetries, both in the polar and in the azimuthal angles, can be
combined and are studied in~\cite{Rindani:2003av}. The latter ones, however, require the
additional reconstruction of the $t$, $\bar t$ directions. At
$\sqrt s=500\hskip 3pt {\rm GeV}$,
${\cal L}_{\rm int}=500\hskip 3pt{\rm fb}^{-1}$, the simultaneous 90\% C.L.
limits on the CP-violating couplings with $\pm 80$\% polarized electrons and
unpolarized positrons are of the order of (1--2)$\times 10^{-1}$.
Including positron polarization, i.e.,
$(P_{e^-},P_{e^+})=(\pm 80\%,\mp 60\%)$, the sensitivity can be further
improved by 20--30\%.\footnote{Individual limits, assuming one non-zero
anomalous coupling at a time, are of course substantially more stringent.}
These sensitivities seem not so far from the predictions of some
models for new physics, that may be at the level of $10^{-2}-10^{-3}$. Also,
they compare favourably with the potential of probing anomalous $t{\bar t} V$
couplings at hadron colliders, see e.g. ref. \cite{Baur:2004uw}
for the LHC.
%%%%%
\subsubsection*{ c) Limits for flavour-changing neutral top couplings}
\begin{figure}[htb]
\begin{center}
\setlength{\unitlength}{1cm}
\begin{picture}(12,3.7)
\put(1.7,0.)
{\mbox{\epsfysize=4.2cm\epsffile{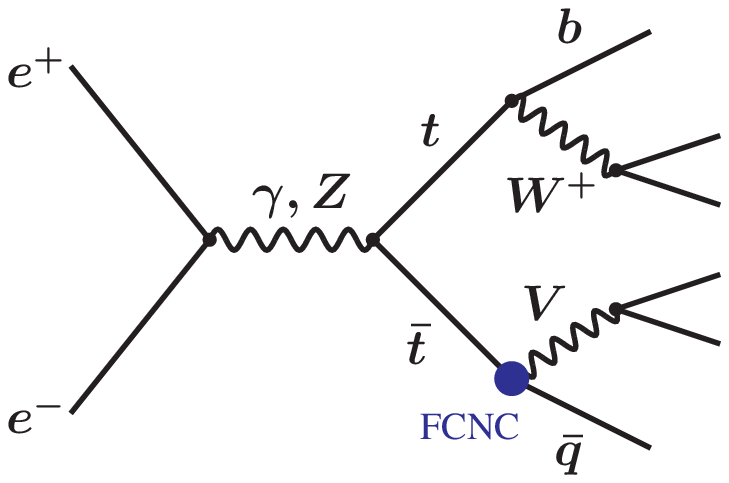}}}
\end{picture}
\caption[Flavour-changing neutral coupling in $t\bar t$ 
production]{Flavour-changing neutral coupling in $t\bar t$ production.
\label{fig:Feynman-ttbar}}
\end{center}
\end{figure}

\begin{figure}[htb]
\begin{center}
\setlength{\unitlength}{1cm}
\begin{picture}(12,3.7)
\put(2.2,-0.3)
{\mbox{\epsfysize=4.2cm\epsffile{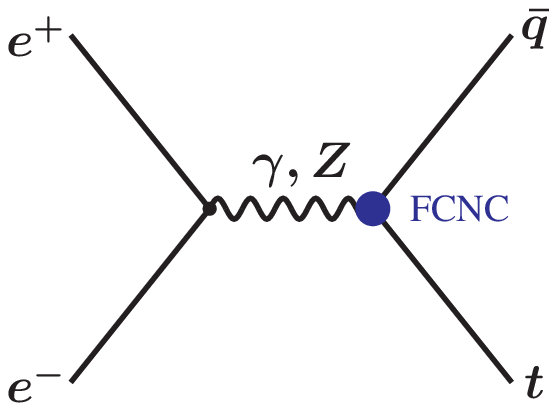}}}
\end{picture}
\caption[Feynman diagrams for $e^+e^-\to t\bar q$ via $Ztq$ or $\gamma tq$
FCNC]{Feynman diagrams for $e^+e^-\to t\bar q$ 
via $Ztq$ or $\gamma tq$
FCN couplings. The top quark is off-shell and has SM decays.
\label{fig:Feynman-tqbar}}
\end{center}
\end{figure}

%%%%%
Flavour-changing neutral (FCN) couplings of the top quark are relevant to
numerous extensions of the SM, and can represent an interesting field for
new-physics searches. Limits on top FCN decay branching ratios can be obtained
from top-pair production with subsequent $\bar t$ decay into $\gamma, Z$ plus
light quark governed by the FCN anomalous $tVq$ couplings
($V=\gamma, Z$ and $q=u,c$), $e^+e^-\to t{\bar t}\to W^+ b V {\bar q}$,
see fig.~\ref{fig:Feynman-ttbar},
or from single top production
$e^+e^-\to t{\bar q}\to W^+ b {\bar q}$ mediated by the anomalous
couplings at the production vertex as in fig.~\ref{fig:Feynman-tqbar}.
The general expression for the
effective-interaction describing the FCN $tVq$ couplings can be written 
%up to dimension-5 
as (effective top couplings to Higgs bosons are not considered in the 
following)~\cite{Aguilar-Saavedra:2001ab}:
\begin{align}
-{\cal L}
=&\frac{g_W}{2c_W}\,X_{tq}\bar t\gamma_\mu
(x_{tq}^{\rm L}\, P_{\rm L}+x_{tq}^{\rm R}\,P_{\rm R})qZ^\mu
+\frac{g_W}{2c_W}\,\kappa_{tq}\bar t(\kappa_{tq}^v-\kappa_{tq}^a\,\gamma_5)
\frac{i\sigma_{\mu\nu}q^\nu}{m_t}\,q\,Z^\mu \nonumber \\
&+e\lambda_{tq}\,\bar t(\lambda_{tq}^v-\lambda_{tq}^a\gamma_5)
\frac{i\sigma_{\mu\nu}\,q^\nu}{m_t}\,q\,A^\mu
\label{topfcn}
\end{align}
Here, colour indices are summed over, and the chirality-dependent couplings
are normalized as
$\left(x^{\rm L}_{tq}\right)^2+\left(x^{\rm R}_{tq}\right)^2=1$, 
$\left(\kappa^v_{tq}\right)^2+\left(\kappa^a_{tq}\right)^2=1$,
$\left(\lambda^{v}_{tq}\right)^2+\left(\lambda^a_{tq}\right)^2=1$.
In the SM, vertices such as those in (\ref{topfcn}) can only be generated at
very strongly GIM suppressed loops.
 
Single top production is more sensitive to top anomalous couplings but
top decays help to disentangle the type of anomalous coupling
involved.
Beam polarization is
very efficient in significantly reducing the background and is therefore particulary important
in limits obtained from single top production.
 The background is essentially
dominated by the $W^+ +{\rm 2 jets}$ final state, with $W^+$ decaying into
$l \nu$ and one jet misidentified as a $b$-jet.

With polarization $(80\%,0)$, the background decreases by a factor of
$1/(1-P_{e^-})\approx5$ while keeping 90\% of the signal.  With $(80\%,-45\%)$
the background is reduced by a factor of
$1/(1-P_{e^-})(1+P_{e^+})\approx9$ and the signal is increased by 20\%
compared to the case of no polarization \cite{Aguilar-Saavedra:2001ab}.
In conclusion, $S/B$ and $S/\sqrt{B}$ are improved by factors of 2.1 and 1.7,
respectively.\footnote{For a discussion of these ratios,
see sect.~\ref{sect:suppr-bck}.}
 
Already with $e^-$ and $e^+$ polarization (80\%, 45\%), as an example, one
improves the 3-$\sigma$ discovery limits on the 
vector ($\gamma^{\mu}$) coupling at
$\sqrt{s}=500$~GeV by a factor of 3 (a factor of 1.7 compared to only electron
polarization) and the limits on the tensor ($\sigma^{\mu\nu}$) coupling at
$\sqrt{s}=800$~GeV by about a factor 2.6 (a factor 1.8 compared to electron
polarization only), see table~\ref{tab_qcd2}.

\begin{table}[htb]
\begin{center}
\hspace*{-.8cm}\parbox{14cm}{}\vspace{.2cm}
\hspace*{-1cm}
\begin{tabular}{|l||c|c|c|}
\hline
& unpolarized beams & $|P_{e^-}|=80\%$ & $(|P_{e^-}|,|P_{e^+}|)=(80\%,45\%)$ \\
\hline\hline
 & \multicolumn{3}{|c|}{$\sqrt{s}=500$~GeV}\\ \hline
 $BR(t\to Zq)(\gamma_{\mu})$ & $6.1\times 10^{-4}$ & $3.9\times 10^{-4}$ &
$2.2\times 10^{-4}$ \\
 $BR(t\to Zq)(\sigma_{\mu\nu})$ & $4.8\times 10^{-5}$ & $3.1\times 10^{-5}$ &
$1.7\times 10^{-5}$ \\
$BR(t\to \gamma q)$ & $3.0\times 10^{-5}$ & $1.7\times 10^{-5}$ &
$9.3\times 10^{-6}$ \\ \hline\hline
 & \multicolumn{3}{|c|}{$\sqrt{s}=800$~GeV} \\ \hline
 $BR(t\to Zq)(\gamma_{\mu})$ & $5.9\times 10^{-4}$ & $4.3\times 10^{-4}$ &
$2.3\times 10^{-4}$ \\
 $BR(t\to Zq)(\sigma_{\mu\nu})$ & $1.7\times 10^{-5}$ & $1.3\times 10^{-5}$ &
$7.0\times 10^{-6}$ \\
$BR(t\to \gamma q)$ & $1.0\times 10^{-5}$ & $6.7\times 10^{-6}$ &
$3.6\times 10^{-6}$\\ \hline
\end{tabular}
\end{center}
\caption[Limits on FCNC in single top production]
{Single top production:
3-$\sigma$ discovery limits on top flavour changing neutral couplings
from top branching fractions at
$\sqrt{s}=500$~GeV with ${\cal L}_{\rm int}=300$~fb$^{-1}$ and 
at $\sqrt{s}=800$~GeV
with ${\cal L}_{\rm int}=500$~fb$^{-1}$ \cite{Aguilar-Saavedra:2001ab}. 
\label{tab_qcd2} }
\end{table}

\begin{table}
\begin{center}
\begin{tabular}{|l||c|c|c|}
\hline
& LHC & ILC, $\sqrt{s}=500$~GeV & ILC, $\sqrt{s}=800$~GeV \\\hline
$BR(t \to Z c)$ $(\gamma_{\mu})$ & $3.6\times 10^{-5}$ 
& $1.9\times 10^{-4}$ & $1.9\times 10^{-4}$ \\
$BR(t \to Z c)$ $(\sigma_{\mu\nu})$ & $3.6\times 10^{-5}$ 
& $1.8\times 10^{-5}$ & $7.2\times 10^{-6}$ \\
$BR(t \to \gamma c)$ & $1.2\times 10^{-5}$ & $1.0\times 10^{-5}$ 
& $3.8\times 10^{-6}$ \\ \hline
\end{tabular}
\end{center}
\caption[LHC and ILC discovery potential for FCN top couplings]{3-$\sigma$ 
discovery limit on top FCN couplings that can be obtained 
from top decay processes at the 
LHC and in single top production at the ILC, 
$\sqrt{s}=500$~GeV and $800$~GeV with 
$(P_{e^-},P_{e^+})=(80\%,60\%)$ 
for one year of operation \cite{Aguilar2}.
One anomalous coupling different from zero
at a time is assumed.
\label{fcn-top2}}
\end{table}

A more recent study was made for $(|P_{e^-}|,|P_{e^+}|)=(80\%,60\%)$
at $\sqrt{s}=500$~GeV with ${\cal L}_{\rm int}=345$~fb$^{-1}$ and at
$\sqrt{s}=800$~GeV with ${\cal L}_{\rm int}=534$~fb$^{-1}$ including
initial state radiation, beamstrahlung and using different kinematical
cuts \cite{Aguilar2}.  Comparison with the limits for FNC couplings
expected at the LHC shows that the LC measurements are complementary
in searches for FCN couplings. Whereas the LHC can be superior in the
discovery potential for $\gamma^{\mu}$ couplings, the ILC at
$\sqrt{s}=800$~GeV with $(80\%,60\%)$ may gain an order of magnitude
for the discovery of $\sigma_{\mu\nu}$ couplings to the $Z$ and the
photon, see table~\ref{fcn-top2}. For comparison with the results of
full simulation studies for the ATLAS and CMS detector with regard to
the LHC discovery potential of FCN couplings, see
also~\cite{Chikovani:2002cg}.

The listed results show that having both beams polarized improves the
results concerning measurements of the top properties and enhances
considerably the discovery potential for deviations from SM
predictions.
\smallskip

{\bf \boldmath
Quantitative results:
The determination of the top vector coupling is improved
by about a factor 3 compared with the case of having only polarized electrons.
The limits for top FCN couplings in single top production
are improved by about a factor 1.8 and the sensitivity to CP-violating
top couplings is improved by about a factor 1.4 when both beams are
longitudinally polarized compared to the case with only polarized electrons,
getting closer to predictions from non-standard models.}

%%%%%%%%%%%%%%%%%%%%%%%%%%%%%%%%%%%%%%%%%%%%%%%%%%%
\section{Standard Model Higgs searches \label{sect:sm-higgs}}
%%%%%%%%%%%%%%%%%%%%%%%%%%%%%%%%%%%%%%%%%%%%%%%%%%%
{\bf \boldmath
One of the major physics goals at the ILC is the precise
analysis of all the properties of the Higgs particle.  For a light Higgs the
two major production processes, Higgs-strahlung $e^+e^- \to H Z$
and $WW$ fusion $e^+e^-\to H \nu \bar{\nu}$, will have similar
rates at $\sqrt{s}=500~\text{GeV}$.  Beam polarization will be
important for background suppression and a better separation of the two
processes. Furthermore, the determination of the general Higgs couplings is
greatly improved when both beams are polarized.}
\smallskip

An accurate study of Higgs production and decay properties can be performed in
the clean environment of $e^+ e^-$ linear colliders in order to experimentally
establish the Higgs mechanism as being responsible for electroweak symmetry
breaking \cite{DESY-Reports}. The study of Higgs particles will therefore
represent a central part of the ILC physics programme.  Beam polarization does
not play a key role in determining the Higgs properties; however, it is very
helpful for separating the production processes, suppressing the dominant
background processes, and improving the accuracy in determining the general
couplings. The use of polarized beams is in this context mainly of statistical
importance. Since the study of the Higgs properties plays one of the major
roles in the ILC programme, we here review the benefits in the Higgs sector of
having both beams polarized simultaneously.

%%%%
\subsection{Separation of production processes}
\begin{figure}
\includegraphics[width=\linewidth]{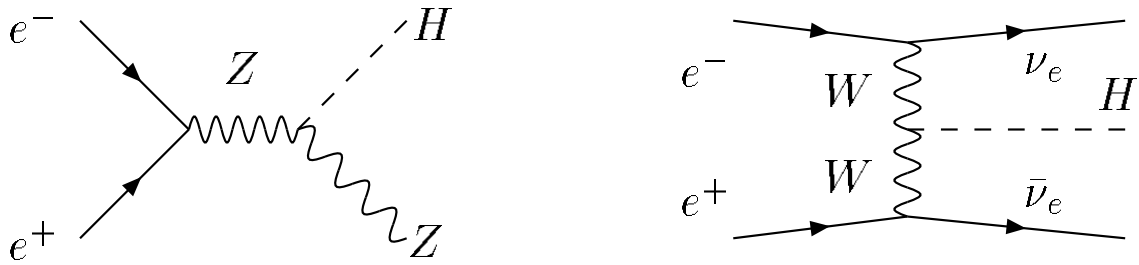}
\caption[Higgs production]{Main production mechanism of the SM Higgs boson at the LC
\label{fig_higgsprod}}
\end{figure}

%%%%
Higgs production at an LC occurs mainly via Higgs-strahlung $e^+ e^-\to HZ$, 
fig.~\ref{fig_higgsprod} left,
and $WW$ fusion, $e^+e^-\to H \nu \bar{\nu}$, fig.~\ref{fig_higgsprod} right.  Polarizing both
beams enhances the signal and suppresses the background.  In
table~\ref{tab_higgs1} the scaling factors, i.e.\ the ratios of
polarized and unpolarized cross sections, are compared for two cases
(1) $(P_{e^-},P_{e^+})=(\pm 80\%,0)$, and (2) $(P_{e^-},P_{e^+})=(\pm 80\%,\mp 60\%)$.

\begin{table}[htb]
\begin{center}
\begin{tabular}{|l||c|c|}
\hline
Configuration 
& \multicolumn{2}{|c|}{Scaling factors} \\
$(P_{e^-}, P_{e^+})$
&$e^+e^-\to H \nu \bar{\nu}$ & $e^+ e^-\to HZ$ \\ \hline
$(+80\%,0)$ & 0.20 & 0.87  \\
$(-80\%,0)$ & 1.80 & 1.13  \\ \hline
$(+80\%,-60\%)$ & 0.08 & 1.26 \\
$(-80\%,+60\%)$ & 2.88 & 1.70  \\ \hline
\end{tabular}
\end{center}
\caption[SM Higgs production scaling factors] {Higgs production scaling
factors, eq.~(\ref{def_scaling}), in the Standard Model at $\sqrt{s}=500$~GeV
for different polarization configurations with regard to the unpolarized 
case~\cite{Desch_obernai,Moortgat-Pick:2001kg}.
\label{tab_higgs1}}
\end{table}
%%%
If a light Higgs with $m_H\le 130$~GeV is assumed, which is the range
preferred by both fits of precision observables in the SM
\cite{Delphi} and predictions of SUSY theories (see
e.g. \cite{Heinemeyer:1998np}), Higgs-strahlung dominates for $\sqrt{s}\lsim
500$~GeV and $WW$ fusion for $\sqrt{s}\gsim 500$~GeV.  At a LC with
$\sqrt{s}=500$~GeV and unpolarized beams, the two processes have
comparable cross sections.
In the $H\nu_e\bar{\nu_e}$ final state, there are
important contributions from both
$HZ$ production and $WW$ fusion. These two contributions exhibit different
shapes of the missing mass distributions, which can be exploited to
obtain an enriched sample of either process (interference must
also be taken into account).
Beam polarization can be used to enhance the $HZ$ contribution with respect to the $WW$
fusion signal and vice versa.
Table~\ref{tab_higgs1} shows that there is a gain of a factor
$(1.26/0.08)/(0.87/0.20)\sim 4$ in the ratio
$\sigma(HZ)/\sigma(H\nu\bar{\nu})$ when left-handed polarized positrons are
used in addition to right-handed polarized electrons.
Thus, the relative contribution of $HZ$ and $WW$ fusion can be
easily extracted from the missing-mass distribution with two different
polarizations without strong model assumptions.
%%%%%%%%%%%%%%%%%%%%%%%%%%%%%%%%%%%%%%%%%%%%%%%%%%%%
\subsection{Suppression of background}\label{sec:212}
\begin{figure}[htb]
\begin{center}
\includegraphics[height=3cm]{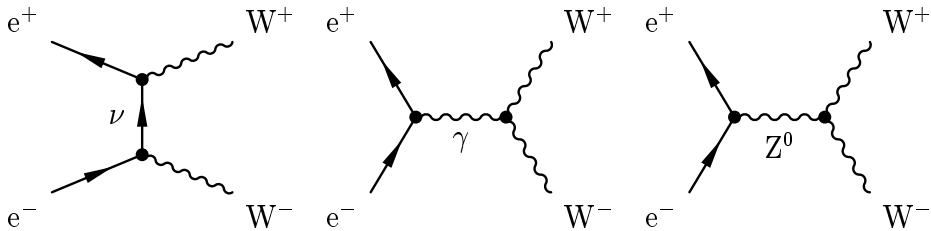}
\end{center}\vspace{-.8cm}
\caption[Feynman graphs for $WW$ production]{ 
Production of $W$ pairs in $\ee$ annihilation.}
\label{fig:wwfeyn}
\end{figure}

\begin{figure}[htb]
\begin{center}
\includegraphics[height=3cm]{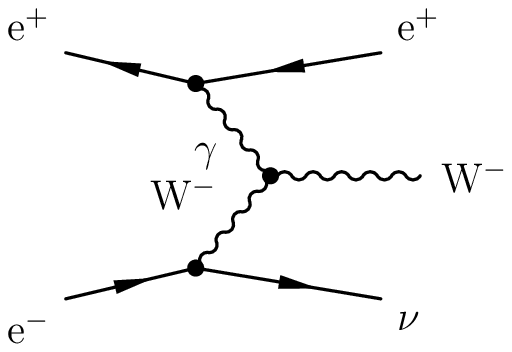}
\hspace{1cm}
\includegraphics[height=3cm]{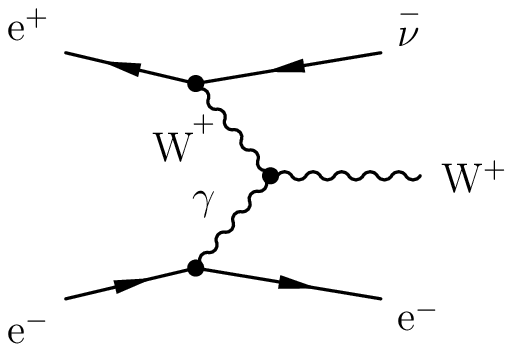}
\end{center}\vspace{-.8cm}
\caption[Feynman graphs for single $W$ production]{ 
Single $W$ production in $\ee$ annihilation.}
\label{fig:swfeyn}
\end{figure}

%%%%%%%%%%%%%%%%%%%%%%%%%%%%%%%%%%%%%%%%%%%%%%%%%%%%
Right-handed electron polarization very efficiently suppresses the
background from $WW$ and single $Z$ production via $WW$ fusion,
$e^+e^- \to Z \nu_e \bar{\nu}_e$, with $Z$ decaying into fermion
pairs. At $\sqrt{s}=500$~GeV the latter is important for a light
Higgs.  For the $WW$ case, fig.~\ref{fig:wwfeyn}, the suppression can
be up to an additional factor 2 if left-handed polarized positrons are
also available compared to the case with only right-handed polarized
electrons.  Positron beam polarization turns out to be also a powerful
tool to suppress the $W$ background from single $W$ production,
$e^+e^- \to W^- e^+ \nu_e$ and $e^+e^- \to W^+ e^- \bar{\nu}_e$, 
see fig.~\ref{fig:swfeyn}.
%%%

%%%

Instead, the advantage of polarization in reducing the $ZZ$ background
is rather limited, see table~\ref{back_WW}.
However, even in that case where the $S/B$ ratio is only slightly
improved, positron polarization in addition to electron polarization
(e.g.\ $|P_{e^-}|=80\%$ and $|P_{e^+}|=60\%$) improves the statistical
significance, see table~\ref{tab_stat}.  In the case of the $HZ$ signal
compared to the $ZZ$ background one gains more than 20\% in
$S/\sqrt{B}$ when using $(P_{e^-},P_{e^+})=(+80\%,-60\%)$ instead of
only $(P_{e^-},P_{e^+})=(+80\%,0)$.

%%%%%%%%%%%%%%%%%%

\begin{table}[htb]
\begin{center}
\begin{tabular}{|c|c|c|}
\hline
Beam polarization & $e^+e^-\to W^+W^-$ & $e^+e^- \to ZZ$ \\
\hline
$(+80\%, 0)$ & 0.20 & 0.76\\ \hline
$(-80\%, 0)$ & 1.80 & 1.25 \\ \hline\hline
$(+80\%, -60\%)$ & 0.10 & 1.05\\\hline
$(-80\%,  +60\%)$ & 2.85 & 1.91 \\
\hline
\end{tabular}
\caption[Scaling factors for $WW$ and $ZZ$ production]{ Scaling 
factors, eq.~(\ref{def_scaling}), 
of $WW$ and $ZZ$ production at $\sqrt{s}=500$~GeV for different 
polarization configurations
with regard to the unpolarized 
case~\cite{Desch_obernai,Moortgat-Pick:2001kg}.
\label{back_WW}
}
\end{center}
\end{table}

%%%%%%%%%%%%%%%%%%%%%%%%%%%%%%%%%%%%%%%%%%%%%%%%%%%%%%%%%%%%%%%%%%%%%%%%%%%%
\subsection[Determination of general  $ZZH$ and $Z\gamma H$ couplings]{
\boldmath Determination of 
general $ZZH$ and $Z\gamma H$ couplings}
%%%%%%%%%%%%%%%%%%%%%%%%%%%%%%%%%%%%%%%%%%%%%%%%%%%%%%%%%%%%%%%%%%%%%%%%%%%%
The accuracy obtainable at the ILC in the determination of the general $ZZH$
and $Z\gamma H$ couplings, was assessed from
$e^+e^-\to HZ\to Hf\bar f$, by using an optimal-observable method which allows
to minimize statistical uncertainties on the 
couplings, \cite{Kniehl_desy00,Hagiwara:2000tk}.
The general effective
$HZV$ interaction Lagrangian considered is parametrized as
\cite{Hagiwara:1993sw}
\begin{equation}
{\cal L}=(1+a_Z)\frac{g_Z m_Z}{2}HZ_\mu Z^\mu
+\frac{g_Z}{m_Z}[b_V HZ_{\mu\nu} V^{\mu\nu}
+c_V(\partial_\mu HZ_\nu-\partial_\nu HZ_\mu)V^{\mu\nu}
+\tilde b_V HZ_{\mu\nu}\,\tilde V^{\mu\nu}],
\end{equation}
where $\tilde V^{\mu\nu}=\epsilon^{\mu\nu\alpha\beta} V_{\alpha\beta}$
is the dual of $V^{\mu\nu}=\partial^{\mu}
V^{\nu}-\partial^{\nu}V^{\mu}$ and $V=\gamma$, $Z$.  It was shown that
beam polarization is essential for determining the sensitivity to the
seven general couplings, the CP-even $a_Z$, $b_Z$, $c_Z$,
$b_{\gamma}$, $c_{\gamma}$ and the CP-odd $\tilde{b}_Z$,
$\tilde{b}_{\gamma}$~\cite{Hagiwara:2000tk}. In particular, to fix the
$Z\gamma H$ couplings beam polarization is essential.  Simultaneous
polarization of the $e^+$ and $e^-$ beams results in an increase in
the sensitivity, so that for $\sqrt{s}=500$~GeV, ${\cal L}_{\rm
int}=300$~fb$^{-1}$ and $(\PE,\PP)=(\pm 80\%,60\%)$ the sensitivity is
improved by 20--30\% compared to the case of $(\pm 80\%,0)$
\cite{Kniehl_desy00,Moortgat-Pick:2001kg}, cf.
table~\ref{tab-generalhiggs}.
\begin{table}
\begin{center}
\begin{tabular}{|l||c|c|c|c|}
\hline
 & \multicolumn{3}{|c|}{$\epsilon_{\tau}=0=\epsilon_b$}
& $\epsilon_{\tau}=50\%$, $\epsilon_b=60\%$ \\ \hline
$(P_{e^-},P_{e^+})$ & $(0,0)$ & $(80\%,0)$ &
$(80\%, 60\%)$ & $(80\%,60\%)$
\\ \hline
Re($b_Z$) & 5.5 & 2.8 & 2.3 & 2.2 \\
Re($c_Z$) & 6.5 & 1.4 & 1.1 & 1.1 \\ \hline
Re($b_{\gamma}$) & 123.2 & 5.2 & 3.6 & 3.4 \\
Re($c_{\gamma}$) & 54.2 & 1.1 & 0.8 & 0.7 \\ \hline
Re($\tilde{b}_Z$) & 10.4 & 9.5 & 7.8 & 5.2 \\
Re($\tilde{b}_{\gamma}$) & 61.8 & 14.5 & 10.1 & 6.3 \\ \hline
Im($b_Z-c_Z$) & 105.5 & 7.0 & 4.9 & 4.6 \\
Im($b_{\gamma}-c_{\gamma}$) & 20.6 & 7.0 & 5.7 & 5.4\\
Im($\tilde{b}_Z$) & 52.1 & 3.2 & 2.2 & 2.2 \\
Im($\tilde{b}_{\gamma}$) & 10.1 & 3.2 & 2.6 & 2.6 \\ \hline
\end{tabular}
{\caption[Determination of general Higgs couplings]{Determination of
general Higgs couplings: Optimal errors in units $[10^{-4}]$ on 
general $ZZ\Phi$ and
$Z\gamma \Phi$ couplings for different identification
efficiencies $\epsilon_{\tau}$,
$\epsilon_b$ and beam polarizations
\cite{Kniehl_desy00,Hagiwara:2000tk,Moortgat-Pick:2001kg}.
 \label{tab-generalhiggs} }}
\end{center}
\end{table}

\smallskip

{\bf \boldmath
Quantitative results: For Higgs masses and c.m.\ energies 
where both Higgs-strahlung and
$WW$ fusion lead to similar production cross sections, beam polarization is
important for the distinction between these processes. The separation is
improved by a factor of about 4 with $(P_{e^-},P_{e^+})=(+80\%,
-60\%)$, with respect to the case with only right-handed polarized
electrons. Furthermore, a factor of 2 can be gained for the $WW$
background suppression. When determining the general $ZVH$
couplings, the sensitivity is increased by about 30\% with respect to the case
where only the electrons are polarized and in general limits of the order of 
$10^{-4}$ can be reached.}

%%%%%%%%%%%%%%%%%%%%%%%%%%%%%%%%%%%%%%%%%%%%%%%%%%%
\section{Triple gauge boson couplings in $WW$ production}
%%%%%%%%%%%%%%%%%%%%%%%%%%%%%%%%%%%%%%%%%%%%%%%%%%%

%%%%%%%%%%%%%%%%%%%%%%%%%%%%%%%%%%%%%%%%%%%%%%%%%%%
\subsection{Impact of longitudinally-polarized beams}
%%%%%%%%%%%%%%%%%%%%%%%%%%%%%%%%%%%%%%%%%%%%%%%%%%%
{\bf \boldmath
In order to test the electroweak gauge group, it is convenient to apply a
general parametrization of the gauge-boson self-interactions, which leads to
14 complex parameters, 6 of them CP-violating. Determining all of them in 
$WW$-pair production represents a strong test of the SM, which predicts only 4
couplings (CP-conserving) to be non-zero.  With both beams longitudinally
polarized the sensitivity to the different triple gauge couplings is strongly
enhanced.}
\smallskip

An important feature of the electroweak Standard Model is the non-Abelian
nature of its gauge group, which gives rise to gauge boson self-interactions,
in particular to the triple gauge couplings (TGCs) $WW\gamma$ and $WWZ$.  The
most general vertex contains altogether 14 complex parameters, six of them 
CP-violating.  The SM predicts only four CP-conserving real couplings to be
non-zero at tree level.

The triple gauge boson vertex $WWV$ ($V=Z$ or $\gamma$ and $g_{WW\gamma}=-e$,
$g_{WWZ}=-e \cot\theta_W$, $e$ denoting the positron charge and $\theta_W$ the 
Weinberg angle) can be described
in the most general form by the effective Lagrangian\cite{Hagiwara:1986vm}:
\begin{eqnarray}
\label{eq:TGClag}
  \frac{{\cal L}^{WWV}}{i g_{WWV}} & = 
& g_1^{V} V^{\mu} \left ( W^{-}_{\mu\nu}W^{+\nu} 
- W^{+}_{\mu\nu}W^{-\nu} \right)
  - \kappa_{V}W^{-}_{\mu}W^{+}_{\nu}V^{\mu\nu}
  - \frac{\lambda_{V}}{m_W^2}V^{\mu\nu} W_{\mu}^{+\rho}W^{-}_{\rho\nu} 
\nonumber \\
& + & i g_4^{V} W^{-}_{\mu}W^{+}_{\nu} (\partial^{\mu}V^{\nu} 
+ \partial^{\nu}V^{\mu}) \nonumber \\
& + & i g_5^{V} \varepsilon^{\mu\nu\rho\sigma} \left[
    (\partial_{\rho}W^{-}_{\mu})W^{+}_{\nu} 
- W^{-}_{\mu}(\partial_{\rho}W^{+}_{\nu}) \right] V_{\sigma} \nonumber\\
& - &\frac{\tilde{\kappa}_V }{2} W^-_\mu W^+_\nu 
\varepsilon^{\mu\nu\rho\sigma} V_{\rho\sigma} 
  - \frac{\tilde{\lambda}_V}{2 m_W^2} W^-_{\rho\mu}{W^{+\mu}}_{\nu}
\varepsilon^{\nu\rho\alpha\beta}V_{\alpha\beta},
\end{eqnarray}
which is parametrized by seven couplings for each $V$.  
Their behaviour under the discrete symmetries C, P and CP
can be used to divide them into four classes.  The three couplings \Cgv{1},
\Ckv{} and \Clv{} conserve C and P, while \Cgv{5} violates C and P but
conserves CP. The couplings \Cgv{4}, \Ckvt{} and \Clvt{} violate CP, but
\Cgv{4} conserves P, while \Ckvt{} and \Clvt{} conserve C.  In the SM at tree
level the couplings are \Cgv{1}= \Ckv = 1, while all others are zero.  For
convenience one introduces the deviations from the Standard-Model values:
$\Cdgg=\Cgg{1}-1$, $\Cdgz=\Cgz{1}-1$, $\Cdkg=\Ckg-1$ and $\Cdkz=\Ckz-1$.

Electromagnetic gauge invariance requires $\Cgg{1}=1$ and reduces the number
of C and P conserving couplings to five.  If one imposed 
\SUU{} on eq.~(\ref{eq:TGClag}), one would obtain
the following relations among the C- and P-conserving couplings, see 
e.g.~\cite{Gounaris:1996uw,Nachtmann:2004ug}
\begin{eqnarray}
\label{eq:su2u1}
\Cdkz & = & -\Cdkg\tan^2\theta_W + \Cdgz \nonumber \\
\Clz  & = & \Clg.
\end{eqnarray}
This is sometimes used in fits with
a reduced number of independent couplings and tests of particular 
models (not used in the results shown here).

A precision measurement of the TGCs at high energies will be a crucial
test of the validity of the SM, given that a variety of new-physics
effects can manifest themselves by deviations from the SM predictions (for
references, see e.g.~\cite{Diehl:2002nj}).  Although no deviation from the
SM has been found for the TGCs from LEP data \cite{Heister:2001qt}, the
bounds obtained are comparatively weak.  The tightest bounds on the
anomalous couplings, i.e., on the differences between couplings and their
SM values are of order $0.05$ for $\Delta g_1^Z$ and $\lambda_{\gamma}$,
of order $0.1$ for $\Delta \kappa_{\gamma}$, and of order $0.1$ to $0.6$
for the real and imaginary parts of C- and/or P-violating couplings.  
These numbers correspond to fits where all anomalous couplings, except 
one, are set to zero.  Moreover, many couplings, e.g.\ the imaginary parts of
CP-conserving couplings, have been excluded from the analyses so
far.

At a future linear \mbox{$e^+e^-$} collider one will be able to study these
couplings with unprecedented accuracy.  A process particularly suitable 
to study for
this purpose is $W$~pair production
$e^+e^-\to W^+W^-\to (f_1\bar f_2)(f_3\bar f_4)$, where the final fermions
are either leptons or quarks, and where both the $WW\gamma$ and the $WWZ$
couplings can be measured at the c.m.s. energy scale.
%%%%%%%%%%%%%%%%%%%%%%%%%%%%%%%%%%%%%%%%%%%%%%%%%%%
\subsection*{Study of TGCs with optimal observables}
%%%%%%%%%%%%%%%%%%%%%%%%%%%%%%%%%%%%%%%%%%%%%%%%%%%%

In \cite{Diehl:2002nj} the prospects to measure the full set of 28 (real) TGCs
was systematically investigated for unpolarized beams as well as for
longitudinal beam polarization, using optimal integrated observables.  These
observables
are constructed to give the smallest possible statistical uncertainties for a
given event distribution \cite{Diehl:1993br,Nachtmann:2004fy}.  In addition,
the above-mentioned discrete symmetries are used to simplify the analysis by
classifying the TGCs and to test the stability of the results.
In $W$~pair production the covariance matrix of these observables
consists of four blocks that correspond to CP-even or CP-odd TGCs and to their
real or imaginary parts.  Within each block all correlations between couplings
are taken into account, and simultaneous diagonalization of the 
covariance matrix
allows to treat all 14 couplings at a time and derive separate limits 
on each one.

\begin{table}[htb]
\begin{center}
\hspace*{-1cm}
\begin{tabular} {|l|rrrrrrrr|}
\hline
$\sqrt{s}=500$~GeV & Re$\,\Delta g_1^{\gamma}$ & Re$\,\Delta g_1^Z$ & 
Re$\,\Delta
\kappa_{\gamma}$ & Re$\,\Delta \kappa_Z$ & Re$\,\lambda_{\gamma}$ &
Re$\,\lambda_Z$ & Re$\,g_5^{\gamma}$ & Re$\,g_5^Z$\\
\hline 
&&&&&&&&\\[-2.5ex]
 No polarization & 6.5 & 5.2 & 1.3 & 1.4 & 2.3 & 1.8 & 4.4 & 3.3\\
$(P_{e^-},P_{e^+})=(\pm 80\%,0)$ & 3.2 & 2.6 & 0.61 & 0.58 & 1.1 & 0.86 & 
2.2 & 1.7\\
$(P_{e^-},P_{e^+})=(\pm 80\%,\mp 60\%)$ & 1.9 & 1.6 & 0.40 & 0.36 & 0.62 & 
0.50 & 1.4 & 1.1\\
$(P_{e^-}^{\rm T},P_{e^+}^{\rm T})=(80\%,60\%)$ & 2.8 & 2.4 & 0.69 & 0.82 & 0.69 & 0.55 & 
2.5 & 1.9\\ \hline \hline
$\sqrt{s}=800$~GeV &&&&&&&&\\
\hline
&&&&&&&&\\[-2.5ex]
 No polarization & 4.0 & 3.2 & 0.47 & 0.58 & 1.1 & 0.90 & 3.1 & 2.5\\
$(P_{e^-},P_{e^+})=(\pm 80\%,0)$ & 1.9 & 1.6 & 0.21 & 0.21 & 0.53 & 0.43 & 
1.6 & 1.3\\
$(P_{e^-},P_{e^+})=(\pm 80\%,\mp 60\%)$ & 1.1 & 0.97 & 0.14 & 0.13 & 0.29 & 
0.24 & 0.97 & 0.82\\
$(P_{e^-}^{\rm T},P_{e^+}^{\rm T})=(80\%,60\%)$ & 1.8 & 1.5 & 0.27 & 0.35 & 0.28 & 0.23 & 
1.7 & 1.3\\ \hline
\end{tabular}\vspace{-.5cm}
\normalsize
\end{center}
\caption[CP-conserving triple gauge couplings]{\label{tab:1} One-$\sigma$
statistical reach in units of $10^{-3}$ on the real parts of CP-conserving
TGCs in the multi-parameter analysis including all anomalous couplings at
\mbox{$\sqrt{s}=$ 500 GeV}, $\Lumint=500~\text{fb}^{-1}$ and
\mbox{$\sqrt{s}=$ 800 GeV}, $\Lumint=1000~\text{fb}^{-1}$, without
and with different beam polarizations~\cite{Diehl:2003qz}.}
\end{table}

The results for the real parts of the CP-conserving TGCs are shown in
table \ref{tab:1}.  Only those events where one $W$ boson decays hadronically
and the other one into $e\nu$ or $\mu\nu$ are considered due to
the favourable reconstruction and branching ratios.  In the case of
longitudinal polarization, the luminosity is assumed to be equally
distributed among both signs and the results are then combined.

At 800~GeV and an  integrated luminosity
of ${\cal L}=1000$~fb$^{-1}$,
all uncertainties (with or without polarization) are smaller
than for $\sqrt{s}=500$~GeV and 500~fb$^{-1}$, reflecting
that the anomalous effects in the cross section would be more pronounced
at the higher energy. This holds in particular 
for \mbox{${\rm Re}\, \Delta \kappa_{\gamma}$}.  For both
c.m.~energies the errors on the couplings decrease by about a factor 2 when
going from unpolarized beams to longitudinal $e^-$ polarization and an
unpolarized $e^+$~beam.  Going from unpolarized beams to polarized $e^-$ and
$e^+$ this gain factor is between 3 and 4 for all couplings, except for ${\rm
Re}\, \Delta \kappa_Z$ at 800~GeV, where it is 4.5.

It has been emphasized \cite{Diehl:1993br} that the following linear
combinations can
be measured with much smaller correlations than for
the parametrization (\ref{eq:TGClag}):
\begin{eqnarray}
g_1^{\rm L} & = & 4\sin^2\theta_W\, g_1^{\gamma} + (2 - 4\sin^2\!\theta_W )\,
\xi\, g_1^Z,    \nonumber \\
g_1^{\rm R} & = & 4\sin^2\theta_W\, g_1^{\gamma} - 4\sin^2\!\theta_W\, \xi\,
g_1^Z,  
\label{eq:lrgz}
\end{eqnarray}
where $\xi = s/(s - m_Z^2)$, and similarly for the other couplings.
The L and R couplings appear in the amplitudes for left-
and right-handed initial $e^-$, respectively.  
Therefore, this parametrization is
more `natural' in the presence of beam polarization than the
conventional one of eq.~(\ref{eq:TGClag}). 
For detailed plots showing the sensitivity to the
TGCs as a function of the degree of longitudinal polarization, we refer
to \cite{Diehl:2002nj}.  There, an extended optimal-observable method
\cite{Diehl:1997ft} has been used, where correlations between TGCs are
eliminated through appropriate energy- and polarization-dependent
reparametrizations.

For the imaginary parts of the CP-conserving couplings, see table
\ref{tab:2}, the linear combinations
\mbox{$\tilde{h}_{\pm} = {\rm Im}(g_1^{\rm R} \pm \kappa^{\rm R})/\sqrt{2}$}
are used instead of ${\rm Im}\,g_1^{\rm R}$ and ${\rm Im}\,\kappa^{\rm R}$~\cite{Diehl:2003qz}.

\begin{table}[htb]
\begin{center}
\leavevmode
\small
\renewcommand{\arraystretch}{1.2}
\begin{tabular} {|l|rrrr|rrrr|}
\hline
%&&&&&&&&\\
$\sqrt{s}=500$~GeV
& Im$\,g_1^{\rm L}$ & Im$\,\kappa^{\rm L}$ & Im$\,\lambda^{\rm L}$ 
& Im$\,g_5^{\rm L}$ &
$\tilde{h}_-$ & $\tilde{h}_+$ & Im$\,\lambda^{\rm R}$ & Im$\,g_5^{\rm R}$\\ \hline 
No polarization & 2.7 & 1.7 & 0.48 & 2.5 & 11 & --- & 3.1 & 17 \\ 
$(P_{e^-},P_{e^+})=(\mp 80\%,0)$ & 2.6 & 1.2 & 0.45 & 2.0 & 4.5 & --- &
1.4 & 
4.3 \\ 
$(P_{e^-},P_{e^+})=(\mp 80\%,\pm 60\%)$ & 2.1 & 0.95 & 0.37 & 1.6 & 2.5 &
--- & 0.75 & 2.3 \\ 
$(P^{\rm T}_{e^-},P^{\rm T}_{e^+})=(80\%,60\%)$ & 2.6 & 1.2 & 0.46 & 2.0 & 3.7 & 3.2 &
0.98 & 4.4\\
\hline\hline
$\sqrt{s}=800$~GeV &&&&&&&&\\ \hline
No polarization & 1.5 & 0.74 & 0.18 & 1.5 & 6.0 & --- & 1.2 & 9.0 \\ 
$(P_{e^-},P_{e^+})=(\mp 80\%,0)$ & 1.5 & 0.60 & 0.17 & 1.3 & 2.4 & --- &
0.54 & 2.7 \\ 
$(P_{e^-},P_{e^+})=(\mp 80\%,\pm 60\%)$ & 1.2 & 0.48 & 0.14 & 1.0 & 1.3 &
--- & 0.29 & 1.4 \\ 
$(P^{\rm T}_{e^-},P^{\rm T}_{e^+})=(80\%,60\%)$ & 1.5 & 0.60 & 0.17 & 1.3 & 2.1 & 2.0 &
0.39 & 2.8\\
\hline
\end{tabular}\vspace{-.5cm}
\end{center}
\normalsize
\caption[Imaginary parts of triple gauge couplings]{\label{tab:2} 
Same as table~\protect\ref{tab:1},
but for the imaginary parts and with the L--R parametrization in
units $10^{-3}$~\cite{Diehl:2003qz}.}
\end{table}

%%%%%%%%%%%%%%%%%%%%%%%%%%%%%%%%%%%%%%%%%%%%%%%%%%%
\subsection*{Simulation study of sensitivity to TGCs} 
%%%%%%%%%%%%%%%%%%%%%%%%%%%%%%%%%%%%%%%%%%%%%%%%%%%
The sensitivity to TGCs for both unpolarized and longitudinally-polarized
beams has been simulated in \cite{menges}, using a spin density matrix method 
and $\cos\theta_W$ distributions, limiting the multi-parameter
fit to a restricted number of couplings (in particular, no imaginary parts).
Concerning the expected statistics, the cases considered are $\sqrt s=500\hskip
3pt {\rm GeV}$,
${\cal L}_{\rm int}=500\hskip 3pt {\rm fb}^{-1}$ and
$\sqrt s=800\hskip 3pt {\rm GeV}$,
${\cal L}_{\rm int}=1\hskip 3pt{\rm ab}^{-1}$. At both 
energies roughly $4\times 10^{6}$
$W$-pair signal events are expected in one or two years runs. As in the
previous analysis, the semileptonic $WW\to q{\bar q} \ell \nu$, 
$\ell=e,\mu,\tau$ decay channel was
used, and in the polarized beams case the luminosity
has been equally shared between the two
considered polarization configurations [($\mp 80$\%,0) for unpolarized 
positrons
and $(- 80\%,+ 60\%)$, $(+80\%,-60\%)$ for both beams polarized]. Also, for maximal 
sensitivity
and to disentangle $WWZ$ from$WW\gamma$ couplings, the data at both
polarizations should be fitted simultaneously.
 
The Monte Carlo generation of 
$W^-W^+$ signals included also effects from initial
state radiation and beamstrahlung, and $W$-pair generators appropriate to
initial polarized beams were used. Other uncertainties that can be relevant to
the measurement, such as those on the $W$-mass and on the beam energy, have
been taken into account. The background is known  from LEP analyses 
 to be very small \cite{Abbiendi:2000ej}, but in any case it 
has been fully accounted
for. Also simulated detector effects (using the detector 
simulation program SIMDET~\cite{Pohl:1999uc}) 
have been included  in the unpolarized 
case in the single-parameter fits
of the coupling constants~\cite{menges}.
                                                                               
\begin{table}[hb]
\begin{center}\hspace{-1.2cm}
\begin{minipage}{\linewidth}
\renewcommand{\thefootnote}{\alph{footnote}}
\renewcommand{\footnoterule}{}
\renewcommand{\arraystretch}{1.2}
\begin{tabular}{|cc|rrrrr|rrrr|}
\hline 
&& \Cdgz & \Cdkg & \Clg  & \Cdkz & \Clz & \Cgz{4} & \Cgz{5} & \Ckzt & \Clzt \\
\hline
500\,\GeV &
Unpolarized   & 3.8 & 0.5 & 1.2 & 0.9 & 1.2 & 8.6 & 2.8 & 6.5 & 1.1 \\
& $(|P_{e^-}|,|P_{e^+}|)=(80\%,0)$ 
& 2.5 & 0.4 &  0.8 & 0.5 &  0.9 & 8.0 & 2.3 & 5.1 & 1.0 \\
& $(|P_{e^-}|,|P_{e^+}|)=(80\%,60\%)$  
& 1.6 & 0.3 &  0.6 & 0.3 &  0.7 & 4.6 & 1.7 & 3.9 &  0.8 \\
\hline
800\,\GeV &
Unpolarized  & 3.9 & 0.3 &  0.5 & 0.5 &  0.5 & 4.2 & 2.9 & 3.0 &  0.5 \\
& $(|P_{e^-}|,|P_{e^+}|)=(80\%,0)$ 
& 2.2 & 0.2 & 0.5 & 0.3 & 0.5 & 3.2 & 2.4 & 2.4 &  0.4 \\
& $(|P_{e^-}|,|P_{e^+}|)=(80\%,60\%)$
& 1.3 & 0.2 &  0.3 & 0.2 &  0.3 & 1.8 & 1.4 & 1.4 &  0.3 \\
\hline
\end{tabular}
\end{minipage}
\end{center}
\caption[Simulated sensitivity for triple gauge couplings] {Expected sensitivity [$10^{-3}$] for
the real parts of different couplings at c.m.\ energies of $500$ and $800\,\GeV$ and
${\cal L}_{\rm int}=500$ and $1000\,\invfb$, respectively. In the case of
polarized beams the luminosity is equally shared between the two
combinations $(P_{e^-},P_{e^+})=(\pm 80\%,\mp 60\%)$~\cite{menges}.}
\label{tab:results:polarization}
\end{table}

The results of this analysis at the generator level for both
unpolarized and polarized beams 
are displayed in table~\ref{tab:results:polarization}; 
they are at the level $10^{-3}-10^{-4}$. The numerical results in this table 
are obtained by allowing
only one coupling to vary freely,
with all others set to their SM values (one-parameter fit). 
They show that both beams
at the highest degree of polarization can substantially improve
the sensitivity expected from electron polarization only. In
particular it turns out that most correlations among the different couplings
can be suppressed using polarized beams. The correlation can be reduced by about a factor two
with $(P_{e^-},P_{e^+})=(\pm 80\%,\mp 60\%)$ compared to $(P_{e^-},P_{e^+})=(\pm 80\%,0)$ 
derived for a two-dimensional fit at $\sqrt{s}=500$~GeV.
For further comparison with multi-parameter fits and correlation matrix 
see~\cite{menges}.
A quantitative discussion of the major
systematic uncertainties has also been performed. 
The systematic uncertainty
on the measurements of all couplings, that results from varying the
experimental beam polarization $\Delta P$ by about $1\%$, is largely
dominant (by a factor up to $5$--$10$ over the statistical
uncertainty, especially at $\sqrt{s}=500$~GeV). This implies that, in
order to fully exploit the high statistical sensitivity, the
polarization must be known very precisely in order to reduce such
systematic uncertainty, say to $\Delta P\sim 0.1\%$--$0.2\%$, so that
$\Delta P_{e^-}/P_{e^-}=0.1-0.2$ \% and $\Delta
P_{e^+}/P_{e^+}=0.2-0.3$ \% would be needed. Such an experimental
accuracy on beam polarizations requires excellent polarimetry or, in
case, the application of the Blondel scheme, where polarizations could
be derived from polarized cross sections, and with presumably very small
correlations with the TGCs.
In case of only polarized electrons the beam polarization can be
also deduced from the left-right asymmetry from the $W$ production in the
forward region (also with rather small  correlations with the TGCs),
cf. section~\ref{sect:colldata}.\smallskip
                                                                               
{\bf Quantitative results: With both beams longitudinally polarized the gain 
in the sensitivity to the triple 
gauge couplings is up to a factor of 1.8 with
respect to the case of only the electron beam polarized. 
}

%%%%%%%%%%%%%%%%%%%%%%%%%%%%%%%%%%%%%%%%%%%%%%%%%%%
\subsection{Use of transversely-polarized beams \label{tgc-trans}}
%%%%%%%%%%%%%%%%%%%%%%%%%%%%%%%%%%%%%%%%%%%%%%%%%%%
\begin{sloppypar}
{\bf Specific azimuthal asymmetries with transversely-polarized beams
may be crucial for sensitive tests of TGCs. Indeed, the imaginary part
of one specific CP-conserving anomalous couplings is accessible with
transversely- but not with lon\-gi\-tudinal\-ly-po\-larized beams.}
\end{sloppypar} 

It was
emphasized in \cite{Fleischer:1993ix} that
transversely-polarized beams are important as a tool for studying
TGCs and longitudinal $W_{\rm L}$, in particular for measuring
relative phases among helicity amplitudes in $WW$ production.

Optimal observables and multi-parameter analysis have been applied to assess
the sensitivity obtainable on TGCs in the case of both electron and positron
transversely polarized, and the results are displayed in the bottom
lines of tables~\ref{tab:1} and \ref{tab:2} \cite{Diehl:2003qz}. It turns out that, for
most couplings (except $\tilde h_+$ to be discussed separately), the expected
uncertainty is approximately of the same size as in the case
of only electron longitudinal polarization, but worse than
obtained in the case of both beams longitudinally polarized. This situation
is common to both c.m.\ energies $\sqrt s=500\hskip 3pt {\rm GeV}$ and
$\sqrt s=800{\rm GeV}$.

Consequently, one can conclude that, at least for the c.m.\ energies and the
degrees of polarization considered here, the best sensitivity (and separate
limits) on both CP-con\-serving and CP-violating couplings are obtained by the
option of both $e^-$ and $e^+$ longitudinally polarized.
 
The notable exception is represented by $\tilde h_+$ which, as shown in
\cite{Diehl:2002nj}, is not measurable from the normalized event
distribution even with longitudinal polarization.
However, with transversely-polarized beams it is possible to measure 
$\tilde h_+$ with a good sensitivity, 
see table~\ref{tab:2} \cite{Diehl:2003qz}.
In the $\gamma-Z$ parametrization, eq.~(\ref{eq:TGClag}), this means that 
the four couplings
${\rm Im}g_1^{\gamma}$, ${\rm Im}g_1^{Z}$, ${\rm Im}\kappa_{\gamma}$ and
${\rm Im}\kappa_{Z}$ are not simultaneously measurable without transverse
polarization.
 
In conclusion, although for most anomalous TGCs the longitudinal polarization of
both beams is most convenient, the measurement of {\it all} possible couplings
requires to spend part of the integrated luminosity of the linear collider also
in the transverse polarization mode.
 
{\bf\boldmath 
Quantitative result: One specific CP-conserving triple gauge coupling,
$\tilde h_+$, is only accessible with both beams transversely polarized.
With polarizations $(P_{e^-}^{\rm T},P_{e^+}^{\rm T})$=(80\%, 60\%), a
one-$\sigma$ statistical uncertainty of about $3.2\times 10^{-3}$ 
could be obtained on $\tilde h_+$ at $\sqrt{s}=500$~GeV.}

%%%%%%%%%%%%%%%%%%%%%%%%%%%%%%%%%%%%%%%%%%%%%%%%%%%
\section{Precision electroweak measurements at GigaZ \label{sec:gigaz}}
%%%%%%%%%%%%%%%%%%%%%%%%%%
\renewcommand{\sweff}{\sin^2\!\theta_W^\text{eff}}
%%%%%%%%%%%%%%%%%%%%%%%%%%
\subsection[Measurement of $\sweff$, application of the Blondel scheme]{Measurement of \boldmath{$\sweff$},
application of the Blondel scheme}
%%%%%%%%%%%%%%%%%%%%%%%%%%%%%%%%%%%%%%%%%%%%%%%

{\bf \boldmath
Extremely sensitive tests of the SM can be performed with the help of
electroweak precision observables. These can be measured with very high
accuracy with the GigaZ option of the ILC, i.e., running with high luminosity
at the $Z$-boson resonance.  
Measuring accurately the
left--right asymmetry allows a determination of the effective weak mixing angle
$\sweff$ with the highest precision.  However, in order to
exploit the gain in statistics at GigaZ, the relative uncertainties on the beam
polarization have to be kept below 0.1\%. This ultimate precision cannot be
reached with Compton polarimetry, but by using a modified Blondel scheme,
which requires both beams polarized.}
\smallskip 

The GigaZ option refers to running the ILC at the $Z$-boson resonance,
yielding about $10^9$ $Z$ events in 50--100 days of running, resulting in the
most sensitive test of the SM ever made, i.e.\ determining the electroweak
precision observables with an unprecedented precision, see
table~\ref{tab_ew-georg} \cite{Heinemeyer:2002jq}.

\begin{table}[htb]
\hspace{1cm}\begin{tabular}{|c|cccc|}
\hline
& SLC/LEP2/Tevatron& Tevatron/LHC & LC & GigaZ/$WW$\\\hline
$m_W$ [Mev] & 34 & 15 & 10  & 7 \\
$\sweff$ $[10^{-5}]$ & 16 & 14--20 & -- & 1.3\\
\hline
\end{tabular}
\caption[Comparison of electroweak precision observables]{
Precision of $m_W$ and the electroweak mixing angle, $\sweff$,
 compared at present and future
experiments~\cite{Heinemeyer:2002jq}. For the measurement of 
$\sweff$ at the GigaZ option the Blondel scheme
with $(|P_{e^-}|,|P_{e^+}|)=(80\%,60\%)$ has been 
applied~\cite{Hawkings:1999ac}.
The precision for the $m_W$ measurement has been derived  
in threshold scans, applying both beams polarized with $(|P_{e^-}|,|P_{e^+}|)=(80\%,60\%)$~\cite{Wilson-mw}.
\label{tab_ew-georg}}
\end{table}

In the SM, the left--right asymmetry $A_{\rm LR}$ can be written in terms of
the effective leptonic electroweak mixing angle:
\begin{equation}
\ALR
=\frac{2(1-4 \sweff)}{1+(1-4 \sweff)^2}.
\label{eq_ew1}
\end{equation}
The statistical power of the data sample can be fully exploited only when
$\Delta A_{\rm LR}({\rm pol})<\Delta A_{\rm LR}({\rm stat})$.
For $10^8$--$10^9$ $Z$s this occurs when
$\Delta P_{\rm eff}<0.1\%$. In this limit
$\Delta\sweff\sim 10^{-5}$, which is an order of magnitude
smaller than the present value. Thus, 
it will be crucial to
minimize the error in the determination of the polarization.

While improvements in Compton polarimetry
achieving a precision $<0.1\%$ may be difficult to attain,
the desired precision should, nevertheless, be possible with
the Blondel scheme \cite{Blondel}, where the electron and
positron polarizations are measured from the polarized
cross sections, see also sect.~\ref{sect:blondel}.
In this scheme, it is not necessary to
know the beam  polarization with such extreme accuracy, since
$A_{\rm LR}$ can be directly expressed via cross sections for
producing $Z$s with longitudinally-polarized beams:
\begin{eqnarray}
\sigma&=&\sigma_{\rm unpol}[1-P_{e^-} P_{e^+}+A_{\rm LR}(P_{e^+}-P_{e^-})],
\label{eq_ew2a}\\
\mbox{\hspace{-1cm}}
A_{\rm LR} &=& \sqrt{\frac{(\sigma_{\rm ++}+\sigma_{\rm +-}-\sigma_{\rm -+}
-\sigma_{\rm --})
(-\sigma_{\rm ++}+\sigma_{\rm +-}-\sigma_{\rm -+}+\sigma_{\rm --})}
{(\sigma_{\rm ++}+\sigma_{\rm +-}+\sigma_{\rm -+}+\sigma_{\rm --})
(-\sigma_{\rm ++}+\sigma_{\rm +-}+\sigma_{\rm -+}-\sigma_{\rm --})}}.
\label{eq_ew2b}
\end{eqnarray}
where the cross sections on the RHS have been introduced
in sect.~\ref{subsec:2}.
In eqs.~(\ref{eq_ew2a}) and (\ref{eq_ew2b}) the absolute polarization values
of the left- and right-handed degrees of beam polarization 
are assumed to be the same.  These assumptions 
have to be checked experimentally by means of polarimetry techniques;
since only relative measurements are needed for these measurements, 
the absolute calibration of the polarimeter cancels and
the uncertainty $\Delta A_{\rm LR}$ is practically
independent of $\Delta P_{e^\pm}/P_{e^\pm}$~\cite{Hawkings:1999ac}.

\begin{figure}[htb]
\begin{picture}(15,16)
\setlength{\unitlength}{1cm}
\put(0,-5.5){\mbox{\includegraphics[width=\linewidth]{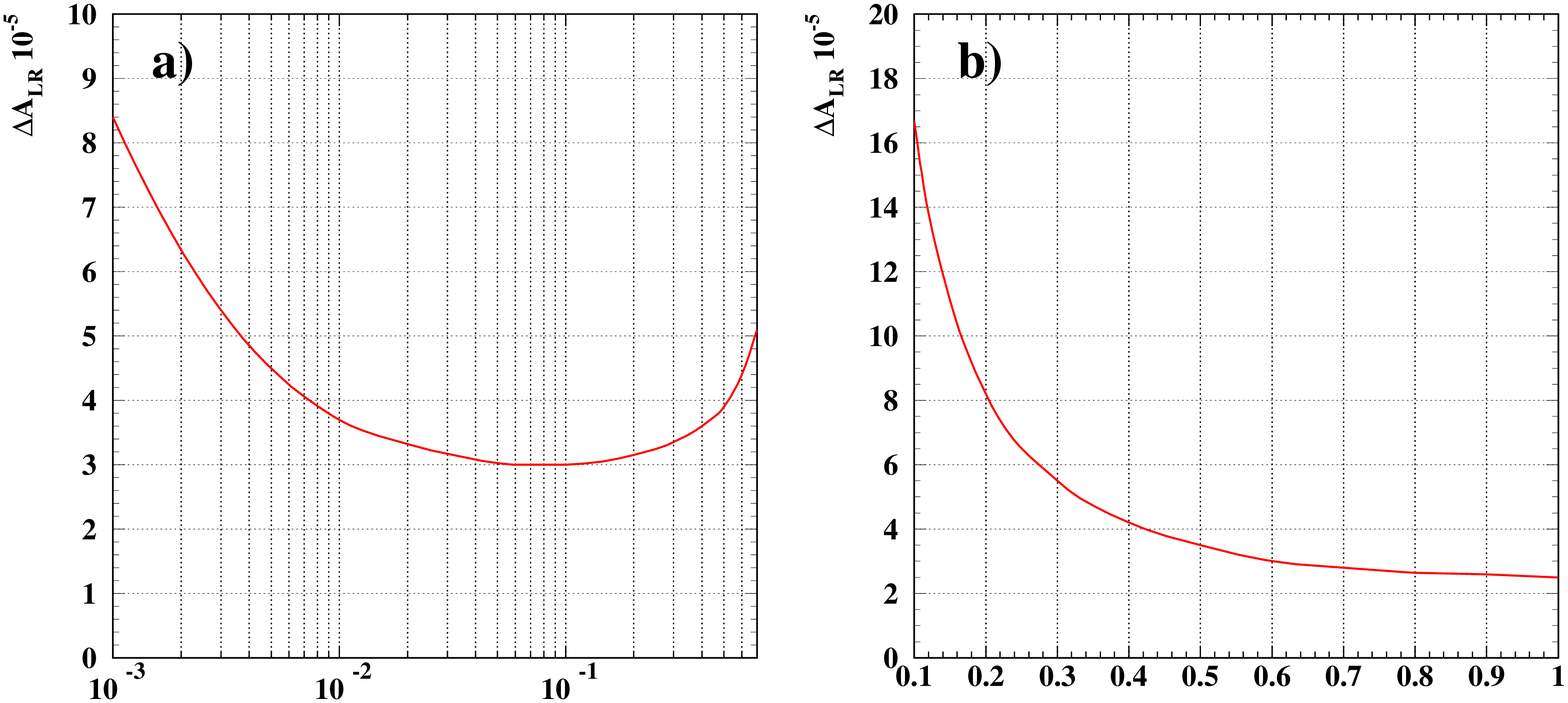}}}
\put(4.7,-5.5){$\scriptstyle  \bf
({\cal L}_{\rm\bf int}^{\bf (++)}+{\cal  L}_{\rm\bf int}^{\bf (--)})/{\cal L}_{\rm\bf int}$}
\put(14.8,-5.4){$\scriptstyle\bf P_{e^+}$}
\end{picture}\vspace{5.2cm}
\caption[GigaZ: Left-right asymmetry]{Test of the electroweak theory: the
statistical error on $A_{\rm LR}$ of $e^+ e^-\to Z\to \ell \bar{\ell}$ at
GigaZ, (a) as a function of the fraction of luminosity spent on the less
favoured polarization combinations $\sigma_{++}$ and $\sigma_{--}$ and (b) its
dependence on $\PP$ for fixed $\PE=\pm 80\%$
\cite{Hawkings:1999ac}.\label{fig_ew1}}
\end{figure}

It can be seen from (\ref{eq_ew2b}) that the Blondel scheme also requires some
luminosity for $\sigma_{++}$ and $\sigma_{--}$. However, 
as shown in figure~\ref{fig_ew1}\,a)
only about 10\% of running time will be needed for these combinations
to reach the accuracy desired for these high-precision measurements.
Fig.~\ref{fig_ew1}\,b)
shows the statistical error on $A_{\rm LR}$ as a function of the positron
polarization for $P_{e^-}=80\%$.
Already with 20\% positron polarization the goal
of $\Delta \sweff < 10^{-4}$ can be reached.
For the comparison of different beam polarization configurations and the gain 
for the $A_{\rm LR}$ measurements, see also \cite{Gideon}.

With the polarization of both beams using the Blondel scheme, i.e.\ 80\% polarization for
electrons and 60\% polarization for positrons, an accuracy of 
$\Delta\sweff = 1.3 \times 10^{-5}$ can be 
achieved in the leptonic final state~\cite{Hawkings:1999ac}. 
If only electron polarization were
available, the accuracy would be about 
$\Delta\sweff = 9.5 \times 10^{-5}$~\cite{moeni} if only
$\Delta P_{e^-}/P_{e^-}=0.5\%$ are achievable. If $P_{e^-}=90\%$ but 
$\Delta P_{e^-}/P_{e^-}=0.25\%$ are assumed, an accuracy of about 
$\Delta\sweff = 5 \times 10^{-5}$~\cite{Rowson:2000qn} may be reachable.

As an example of the potential of the GigaZ $\sweff$
measurement, fig.~\ref{fig_ew2}\cite{Heinemeyer:2002jq}
compares the present experimental accuracy on $\sweff$ and 
$m_W$ from LEP/SLD/Tevatron and the prospective
accuracy from the LHC and from the LC without GigaZ option
with the predictions of the SM and the MSSM. With GigaZ a very sensitive
test of the theory will be possible. 

%%%

\begin{figure}[htb]
\begin{center}
\setlength{\unitlength}{1cm}
\begin{picture}(12,8.5)
\put(2,0.)
{\mbox{\epsfysize=7.5cm\epsffile{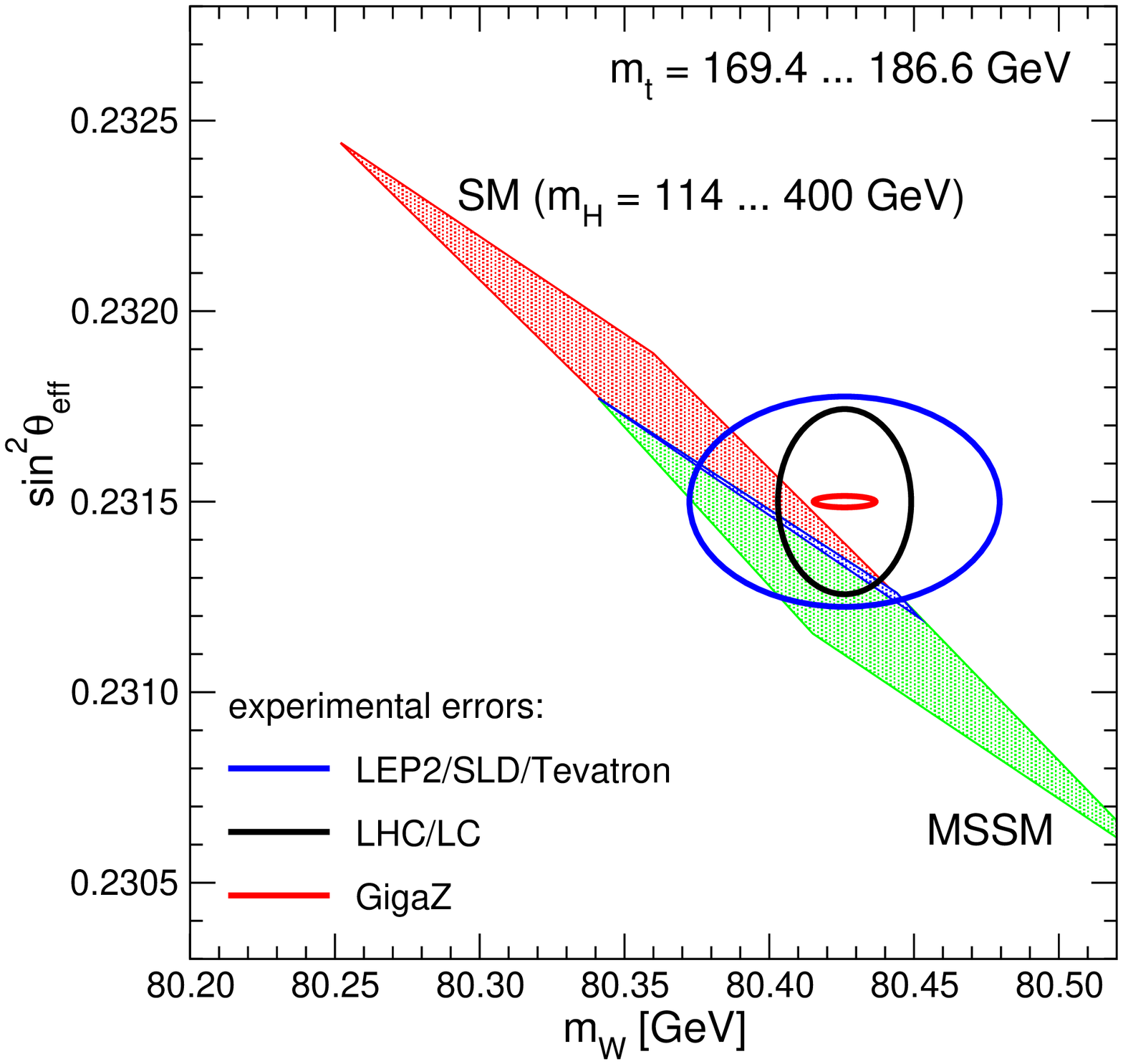}}}
\end{picture}
\caption[GigaZ: $\sin^2\theta_W^{\text{eff}}$ vs. $m_W$]{Prospective
precision on $\sweff$ from $\ALR$:
a high-precision measurement at GigaZ of the left--right asymmetry
$A_{\rm LR}$ and
consequently of $\sin^2\theta_\text{eff}$
allows a test of the electroweak theory at an unprecedented
level. The allowed parameter space of the SM and the MSSM
in the $\sin^2\theta_\text{eff}$--$m_W$ plane is shown together with the
experimental accuracy reachable at GigaZ (applying the Blondel scheme) and polarized threshold scans
for $m_W$. For comparison, the current experimental
accuracy (LEP/SLD/Tevatron) and the prospective accuracy at the LHC
and an LC without GigaZ option (LHC/LC) are also shown
\cite{Weiglein2,Erler:2000jg,Heinemeyer:2002jq}.
\la{fig_ew2}}
\end{center}
\end{figure}

{\bf \boldmath
Quantitative results:
Compared with the case with only the electron beam polarized and 
using Compton polarimetry, 
the effective gain is up to an order of magnitude
in the accuracy for measuring
$\sweff$ with both beams polarized and using the Blondel scheme if
$(|P_{e^-}|,|P_{e^+}|)=(80\%,60\%)$ and $\Delta P_{e^\pm}/ P_{e^\pm}=0.5\%$
are assumed.}

%%%%%%%%%%%%%%%%%%%%%%%%%%%%%%%%%%%%%%%%%%%%%%%%%%%%%%%%%%%%%%%%%%%%%%%%%%%%%%%%%%%%%%%%%%%%%%%%%%%
\subsection[Constraints from $\sweff$ on $m_H$ 
and SUSY parameters]{Constraints 
from \boldmath{$\sweff$} on Higgs-boson masses and SUSY parameters}
%%%%%%%%%%%%%%%%%%%%%%%%%%%%%%%%%%%%%%%%%%%%%%%%%%%%%%%%%%%%%%%%%%%%%%%%%%%%%%%%%%%%%%%%%%%%%%%%%%%
{\bf \boldmath Compared with the case where only the electron beam is
polarized, the polarization of both beams leads to a gain of about one
order of magnitude in the accuracy of the effective weak mixing angle,
$\sweff$.  Within the SM, this has a dramatic effect on the indirect
determination of the Higgs-boson mass, providing a highly sensitive
consistency test of the model that may possibly point towards large
new-physics scales. Within the MSSM, the large increase in the
precision of $\sweff$ will allow to obtain stringent indirect bounds
on SUSY parameters. This will constitute, in analogy to the SM case, a
powerful consistency test of supersymmetry at the quantum level and
may be crucial to constrain SUSY parameters that are not directly
experimentally accessible.}
\smallskip

%%%%%%%%%%%%% F I G U R E %%%%%%%%%%%%%%%%%%%%%%%%%%%%%%%%%%%%
\begin{figure}[htb!]
\begin{center}
\setlength{\unitlength}{1cm}
\begin{picture}(12,8.5)
\put(-2,0)
{\mbox{\epsfig{figure=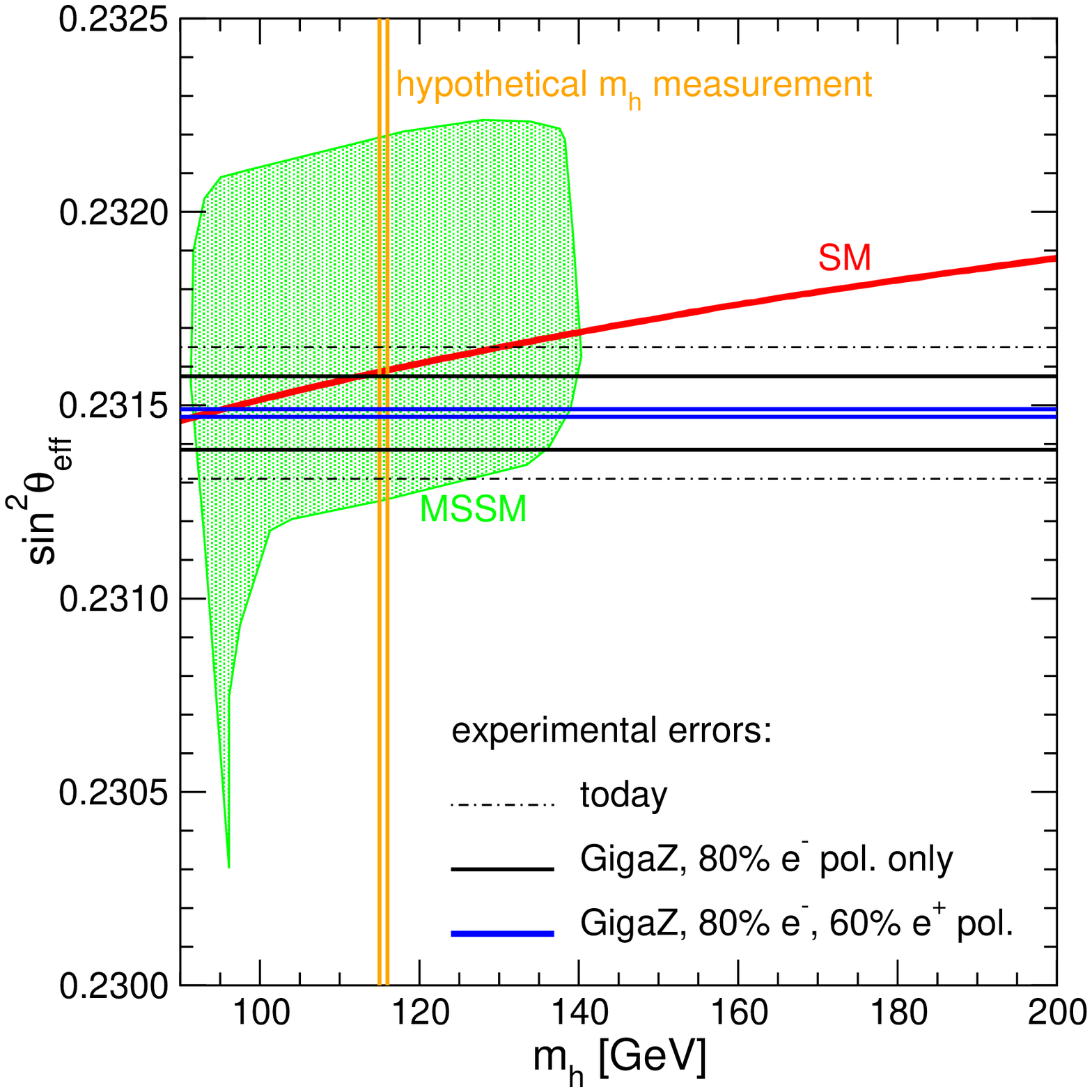, width=7.7cm,height=7.1cm}}}
\put(6.,0.)
{\mbox{\epsfysize=7cm\epsffile{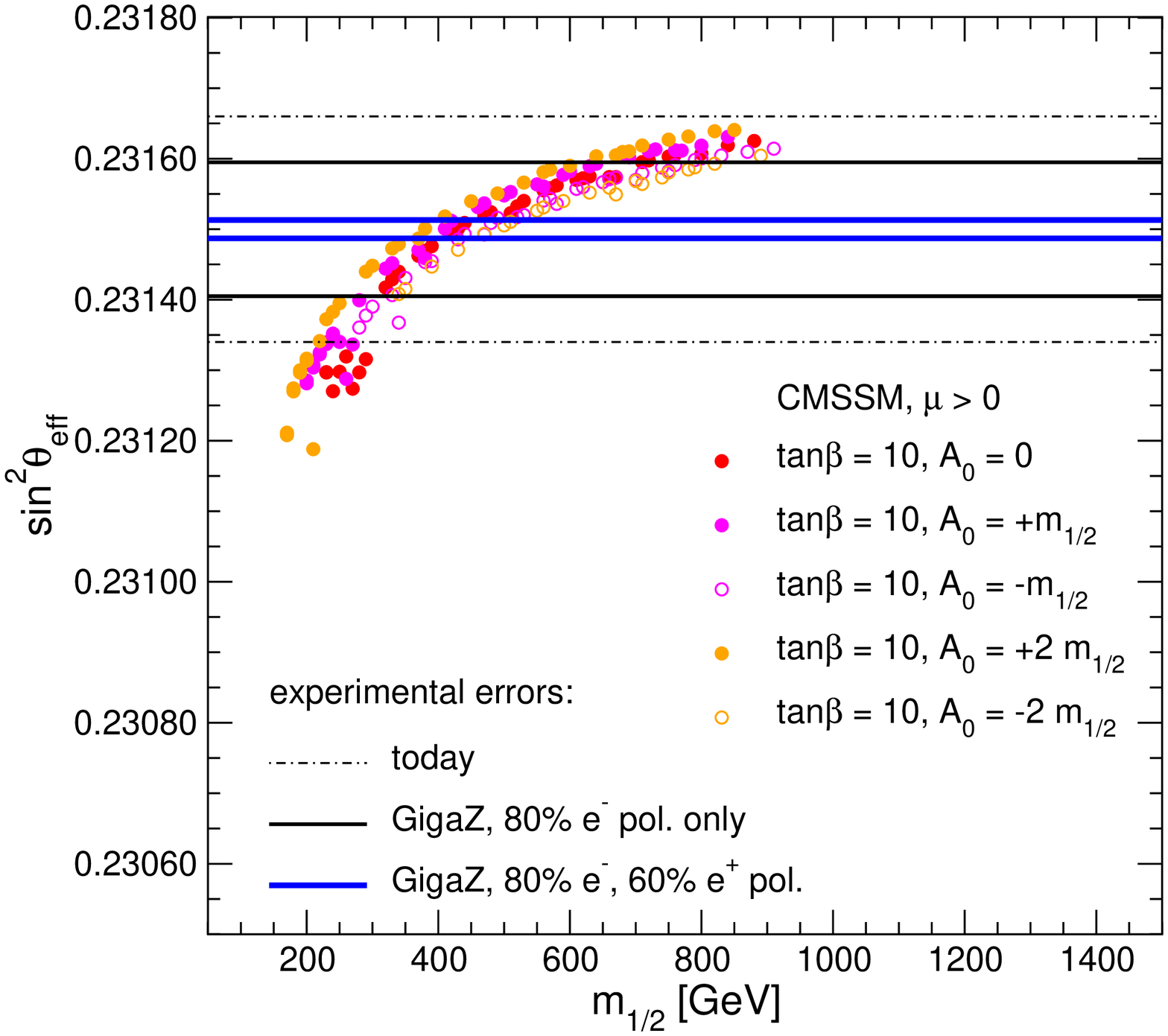}}}
\end{picture}
\caption[GigaZ: leptonic effective mixing angle versus $m_h$ and
$m_{1/2}$]{Left panel: the predictions for $\sweff$ in the SM and the MSSM
are compared with the experimental accuracies
obtainable at GigaZ with an 80\% polarized electron beam only and with
simultaneous polarization of both beams. The present experimental
error on $\sweff$ is also indicated.
The theoretical predictions are given in terms of $m_h$, which denotes
the Higgs-boson mass in the SM and the mass of the lightest CP-even 
Higgs boson in the MSSM, respectively.
The allowed area in the MSSM results from a scan over the relevant SUSY 
parameters entering via loop corrections.
The sensitivity of the $\sweff$ measurement to deviations between
the two models is illustrated for
a hypothetical measurement of 
$m_h = 115.5\pm 0.5 \gev$\cite{Svenni-Georg}.  
Right panel: Allowed
range for the SUSY mass parameter $m_{1/2}$ in a specific model, the
CMSSM~\cite{Svenni-Georg}.  Experimental constraint from LEP searches
as well as bounds from cold dark matter searches have been taken into
account.  
}
\label{fig:mhSW}
\end{center}
\end{figure}
%%%%%%%%%%%%% F I G U R E %%%%%%%%%%%%%%%%%%%%%%%%%%%%%%%%%%%%

At the GigaZ option of the ILC
a precision measurement of the effective leptonic weak mixing angle at
the $Z$-boson resonance, $\sweff$,  will be possible, allowing a very sensitive
test of the electroweak theory~\cite{Erler:2000jg}. 
The impact of the more precise measurement of $\sweff$ for testing the
electroweak theory is illustrated in fig.~\ref{fig:mhSW} (left panel), where the
experimental accuracy (using the current experimental central value of
$\sweff$~\cite{datasummer2003}) is compared with the predictions in the
SM and the Minimal Supersymmetric extension of the Standard Model (MSSM), 
see also sect.~\ref{sect:susy}. 
The theoretical predictions are shown for values of $m_h$ corresponding
to the SM Higgs-boson mass and to the mass of the
lightest CP-even Higgs boson $h$ in the MSSM, respectively. 
In the region where the two models
overlap, $m_h \lsim 140 \gev$~\cite{Heinemeyer:1998np,Allanach:2004rh}, the SM
prediction corresponds to the MSSM result in the limit where all SUSY partners
are heavy. The area in the plot associated with
the MSSM prediction was obtained by
varying all relevant SUSY parameters independently, taking into account the
constraints from the direct search for SUSY particles and the LEP Higgs
search\cite{Barate:2003sz,LEPHiggsMSSM}. 
The MSSM predictions are based on the results described in~\cite{Heinemeyer:2002jq}, 
and the Higgs-mass predictions have been
obtained with {FeynHiggs2.0}~\cite{Heinemeyer:1998yj}. For the 
top-quark mass a value of $m_t=175 \pm 0.1$~GeV has been used, assuming
the prospective linear collider accuracy (using instead a central value of 
$m_t= 178$~GeV would induce a decrease in the $\sweff$ prediction of about
$10^{-4}$).

The impact of the precision on $\sweff$ reachable with polarization of
both beams compared to the case of electron polarization only can be clearly seen
by confronting the SM prediction for $\sweff$ as function of $m_h$ with 
the prospective experimental accuracy.
Requiring the SM prediction to agree with the $\sweff$ measurement at the one-$\sigma$ level
(neglecting the uncertainties from unknown higher-order corrections)
constrains the Higgs-boson mass to an interval of few GeV for the case
of simultaneous polarization of both beams, while a measurement of $\sweff$ based
on electron polarization only leaves an uncertainty of about 
$\pm 25 \gev$ in $m_h$. 

Comparing the indirect constraints on the Higgs-boson mass with a
direct measurement of $m_h$ provides a sensitive test of the
electroweak theory at the quantum level. This is illustrated in
fig.~\ref{fig:mhSW} (left panel), where a hypothetical measurement of
$m_h=115.5 \pm 0.5$~GeV is indicated. The assumed error of $0.5$~GeV
summarizes both the prospective experimental error and the theoretical
uncertainty in the relation between $m_h$ and $\sweff$. The latter is
significantly larger in the MSSM, but for simplicity we use the more
conservative MSSM value for both models.  For the scenario shown in
the plot the measurement of $\sweff$ based on electron polarization
only combined with the $m_h$ measurement would not allow to
distinguish between the predictions of the SM and the MSSM. Both
models would be compatible with the experimental measurements at the
one-$\sigma$ level in this case. A measurement of $\sweff$ with
simultaneous polarization of both beams, on the other hand, would
resolve the different quantum corrections of the two models. The
high-precision measurement of $\sweff$ combined with the $m_h$
measurement would show a large deviation from the SM prediction and
would therefore indicate a clear preference for the MSSM. Within the
MSSM the precision measurement provides a stringent consistency test,
yielding sensitive constraints on the underlying SUSY parameters
entering via quantum corrections.

In fig.~\ref{fig:mhSW} (right panel) the strong constraints 
on the parameter space of new physics models arising from the
precision measurement of $\sweff$ are 
demonstrated~\cite{Ellis:2004tc}. Shown is the prediction of $\sweff$
as a function of $m_{1/2}$, the generic fermionic mass parameter in
the CMSSM (the Constrained MSSM), see
sect.~\ref{sect:susy}. Experimental constraints from LEP searches as
well as bounds from cold dark matter searches have been taken into
account. The unprecedented accuracy in $\sweff$ strongly constrains 
the allowed parameter range for $m_{1/2}$, which is reduced by about a
factor of~5 with $(|P_{e^-}|,|P_{e^+}|)=(80\%,60\%)$ compared with the
case of only polarized electrons $|P_{e^-}|=80\%$.
\smallskip

{\bf \boldmath 
Quantitative results: 
The gain of about one order of magnitude in the accuracy of the 
determination of $\sweff$
applying a Blondel scheme with $(|P_{e^-}|,|P_{e^+}|)=(80\%,60\%)$ and 
$\Delta P_{e^\pm}/P_{e^\pm}=0.5\%$ instead of using only $P_{e^-}=80\%$, 
has strong impacts on bounds on the Higgs boson mass
as well as on bounds on SUSY mass parameters.
The resulting indirect bounds on the mass of the Higgs boson
in the SM also improves by about an order of magnitude.  
Regarding a specific SUSY model the allowed parameter range for 
the specific SUSY mass parameters $m_{1/2}$ is reduced by about a factor five.
}

%%%%%%%%%%%%%%%%%%%%%%%%%%%%%%%%%%%%%%%%%%%%%%%%%%%
\subsection{CP violation in 3-jet and 4-jet decays of the Z-boson 
at GigaZ}
%%%%%%%%%%%%%%%%%%%%%%%%%%%%%%%%%%%%%%%%%%%%%%%%%%%
{\bf \boldmath
An interesting sector where to look for physics beyond the SM is
represented by  CP-violating $Z$ decays to heavy leptons and quarks, in
particular  to final states with $b\bar{b}$ pairs. The high statistics
available at GigaZ should allow a search for signals from non-standard
CP-violating couplings with unprecedented sensitivity. It is also improved by
initial electron and positron longitudinal polarization. 
It is expected that one
can reach an accuracy of about an order of magnitude better than at LEP. }
\smallskip

CP-violating effects in flavour-diagonal $Z$ decays are found to be extremely
small in the SM \cite{zdecay}, therefore this sector seems particularly
suitable for non-standard physics searches. A relevant example is represented
by the processes $Z \rightarrow b \bar{b} g$ and $Z \to b \bar{b} g g$, where
$g$ denotes the gluon, leading to 3-jet and 4-jet final states, respectively
\cite{Bernreuther:1994vg,Nachtmann:2003hg}.  
The CP-violating couplings in the $Z b
\bar{b} g$ vertex and in the $Z b \bar{b} g g$ vertex (the latter related to
the former by QCD gauge invariance), have been searched for at LEP, 
where no significant
deviation from the SM was found \cite{aleph}.

For a model-independent analysis
of CP-violation in the above processes, an 
effective Lagrangian approach is most convenient \cite{zdecay} 
and, accordingly,
one can add to the SM the CP-violating
$Z b \bar{b} g$ four-point vertex:
\begin{eqnarray}
  \label{lcp} 
{ {\cal L}_{\rm CP}(x) = }
  [\; h_{Vb}\: \bar{b}(x)\: T^a\: \gamma^{\nu}\:
  b(x) +  h_{Ab}\: \bar{b}(x)\: T^a\: \gamma^{\nu}\: \gamma_5\: b(x)\; ]\;
  Z^{\mu}(x)\:
  G^a_{\mu\nu}(x)\;\;,
\end{eqnarray}
where $T^a$ are the familiar SU$(3)_{\rm C}$ generators.
Here, $h_{Vb}$ and $h_{Ab}$ are real CP-violating vector and axial-vector 
chirality-conserving coupling constants.  
Due to the quadratic gluon term present in $G^a_{\mu\nu}$, the vertex
eq.~(\ref{lcp}) is involved in both \ZTJ\ and in 
\mbox{$Z \rightarrow$ 4 jets}.
Dimensionless coupling
constants $\widehat{h}_{Vb,Ab}$ can then be defined by using
$m_Z$ as a scale parameter, through
the relations $h_{Vb,Ab} = {e\: g_s}\; \widehat{h}_{Vb,Ab}/({\sin\theta_W\:
\cos \theta_W \:m_Z^2})$, with $g_s$ the QCD coupling constant.

Such chirality-conserving CP-violating interactions (\ref{lcp})
can originate in different scenarios, for instance in multi-Higgs extensions
of the SM \cite{higgs} and in models with excited quarks, typical
of compositeness, where quarks have substructure \cite{excq}.

In the latter scenario, one assumes that 
the $b$ quark has an excited
partner $b'$ with mass $m_{b'}$ and spin $\frac{1}{2}$. 
Both chirality-conserving $Zb'b$ couplings at the GigaZ scale
and chirality-flipping
$b b' g$ couplings, allowed in such composite models, can be expressed via
an effective interaction of the form
\begin{eqnarray}
  \label{lexcq} \nonumber
  {\cal L}'(x) =
  & - & \frac{e}{2 \: \sin \theta_W\: \cos \theta_W}\;
        Z_{\mu}(x)\: \bar{b}'(x)\: \gamma^{\mu} \; (g_V' - g_A' \gamma_5)\;
        b(x)\\
  & - & i \;\frac{g_s}{2 m_{b'}}\; \hat{d}_c \: \bar{b}'(x)\:
  \sigma^{\mu\nu}\:
  \gamma_5\: T^a\: b(x)\: G^a_{\mu\nu}(x) + {\rm h.c.}
\end{eqnarray}
Here $g_V'$, $g_A'$ and $\hat{d}_c$ are complex constants that,
as is conventional for composite models, can be 
expected to be of order unity if the novel dynamics underlying 
compositeness is strong.
It is understood that
$m_{b'} \gg m_Z$, and from virtual $b'$ exchange one derives
for the couplings in eq.~(\ref{lcp}): 
$  \widehat{h}_{Vb} = {\rm Re}(\hat{d}_c \:
  g_A'^*){m_Z^2}/{m_{b'}^2}$ and 
$  \widehat{h}_{Ab} = -{\rm Re}(\hat{d}_c \:
  g_V'^*){m_Z^2}/{m_{b'}^2}$ \cite{higgs}.

Assuming flavour-tagging \cite{TDR},
the $b$-quark momenta  can be reconstructed. 
CP-violating couplings may then be searched for by using
CP-odd observables constructed from the $b$ and $\bar{b}$ quark momenta
\cite{zdecay,Bernreuther:1994vg} (with `3' the cartesian component
in the $Z$ rest frame corresponding to the positron beam):
\begin{equation}
T_{33} = (\widehat{\bf k}_{\bar{b}} - \widehat{\bf k}_b)_3 \; 
(\widehat{\bf k}_{\bar{b}} \times \widehat{\bf k}_b)_3 \;,
\label{ten}
\end{equation}
\vspace*{-0.5cm}
\begin{equation}
V_3 = (\widehat{\bf k}_{\bar{b}} \times \widehat{\bf k}_b)_3 \;.
\label{vec}
\end{equation}
The observable $T_{33}$ transforms as a tensor component, $V_3$ as
a vector component.

To a very good approximation, for $Z\to 3$ jets and $Z\to 4$ jets the tensor
observable $T_{33}$ and the vector one $V_3$ turn out to be sensitive only to
the combinations $ \widehat{h}_b = \widehat{h}_{Ab} g_{Vb} - \widehat{h}_{Vb}
g_{Ab}$ and $ \widetilde{h}_b = \widehat{h}_{Vb} g_{Vb} - \widehat{h}_{Ab}
g_{Ab}$, respectively, where $g_{Vb}=-\frac{1}{2}+\frac{2}{3} \sin^2\theta_W$
and $g_{Ab}=-\frac{1}{2}$\cite{Bernreuther:1994vg}.  Clearly, a non-zero
value of one of the above observables is an unambiguous indicator of 
CP violation. Note that 
for longitudinally-polarized beams, the additional assumption that
the $Z e^+e^-$ vertex is free from chirality-flipping interactions is
necessary.

%%%%%%%%%%%%%%%%%%%%%%%%%%%%%%%%%%%%%%%%%%%%%%%%%%%
\subsection*{Numerical results}
%%%%%%%%%%%%%%%%%%%%%%%%%%%%%%%%%%%%%%%%%%%%%%%%%%%

The sensitivities  
of the observables  (\ref{ten}) and (\ref{vec})
on the couplings $\widehat{h}_b$ and $\widetilde{h}_b$
 depend on the jet resolution
cut \ycut, see also \cite{jade}.
The expectation values of $T_{33}$  do not depend on
initial longitudinal polarization, and therefore 
the sensitivity to  $\widehat{h}_b$ only reflects the statistics.

The one-$\sigma$ uncertainty on the measurement 
of $\widetilde{h}_b$ through the observable (\ref{vec}) is evaluated
on a statistical basis of 
$N=10^9$ $Z$ decays with unpolarized beams. The results for
\ZTJ\ and different choices of beam polarizations are shown in
table~\ref{fig:asym_3jets}  
(the results for \ZJ\ are presented in \cite{Nachtmann:2003hg}).
As indicated by the table, the advantage of initial beam polarization
is significant: for $P_{e^-}=-80\%$ and $P_{e^+}=+60\%$ 
the sensitivity is maximal, more than a
factor of 6 compared to the case of unpolarized beams (the 
improvement due to additional positron polarization leads to a factor of about 
$1.5$ compared to the case where only electrons are polarized).
On the other hand, numerical results show a stability
against changes in the jet resolution parameter \ycut up to 0.1. Notice
also, that the results in table~\ref{fig:asym_3jets} are obtained by assuming,
 for the measurement of the vector observable $V_3$, a $100\%$
$b$--$\bar b$ distinction.
In the case of
no experimental signal at one-$\sigma$, lower bounds on $m_{b'}$ can be
derived. These are also shown in table~\ref{fig:asym_3jets} under the 
assumption ${\rm
Re}(\hat{d}_c \: g_A'^*) = {\rm Re}(\hat{d}_c \: g_V'^*) = 1 $.

As can be seen, the coupling $\widetilde{h}_b$ could be measured 
at GigaZ with an accuracy of order $1.5\times 10^{-3}$ 
(one-$\sigma$ level) from 
\ZTJ 
\phantom{.}(one obtains qualitatively similar results from \ZJ 
\phantom{.} for lower values of \ycut).  
In the case of non-observation of the effect at the one-$\sigma$ level,
the lower bound on the $m_{b'}$ mass, $m_{b'}> 2.2$~TeV, can be derived,
assuming as usual the couplings of the novel compositeness interaction
to be of the strong interaction size. Such bound should be compared with the
best current limits $m_{q'}>775$~GeV \cite{nello-neu} that, however,
applies to $u$- and $d$-quarks but may not exclude lighter $b'$ 
excitations.\footnote{These numbers should be compared to the excited
quark mass limits at the 2-$\sigma$ level. In that case a
measurement of \hb\, \hbn\ has to produce a mean value larger than
$2\, \Delta\widehat{h}_b$, $2\, \Delta\widetilde{h}_b$ 
to be able to claim a non-zero effect. The mass limits at the one-$\sigma$
level given in table~\ref{fig:asym_3jets} have to be divided by a factor
$\sqrt{2}$ to get the limits at the 2-$\sigma$ level.}
Although not sensitive to beam polarization, i.e.\ unpolarized 
beams would be sufficient, one may mention for the case of the observable
$T_{33}$ that the one-$\sigma$ accuracy on $\widehat{h}_b$ 
would be of order 0.004 from \ZTJ. 

In a realistic analysis, also actual reconstruction efficiences and systematic
uncertainties should be taken into account. Nevertheless, assuming them 
to be of the same size as the statistical uncertainties, 
one could conclude that the achievable sensitivities to
\hb\ and \hbn\ are 
one order of magnitude better than those obtained at LEP.  This
can give valuable information on, e.g., the scalar sector 
in multi-Higgs extensions
of the SM, as well as provide stringent tests of models with excited quarks.
\smallskip

\begin{table}[htb]
\begin{center}
\renewcommand{\arraystretch}{1.2}
\begin{tabular}{|c|c|c|c|c|c|c|}
\hline
 \multicolumn{2}{|c}{}  & \multicolumn{5}{|c|}{$(P_{e^-},P_{e^+})$} \\ 
$V_3$ & $y_{cut}$ & $(0,0)$ & $(+80\%,0)$ & $(-80\%,0)$ &
$(+80\%,-60\%)$ & $(-80\%,+60\%)$ \\\hline
$\Delta \widetilde{h}_b$ $[10^{-4}]$& 0.01 & 85 & 27 & 19 & 18 & 14 \\ 
                & 0.1  & 94 & 30 & 21 & 20 & 15 \\ \hline
$m_{b'}$ [TeV]      & 0.01 & 0.91 & 1.6 & 1.9 & 2.0 & 2.3 \\
                  & 0.1  & 0.87 & 1.5 & 1.8 & 1.9 & 2.1 \\ \hline 
\end{tabular}
\caption[Sensitivity of CP-violating observables and excited quark mass
limits]{Accuracy on the vector observable $V_3$ on 
$\Delta\widetilde{h}_b$ and corresponding
lower limits on the excited quark mass $m_{b'}$ at
the one-$\sigma$ level.~\cite{Schwani}.
\label{fig:asym_3jets}}
\end{center}
\end{table}
%%%%%%%%%%%%%%%%%%%%%%%%%%%%%%end%%figure%%%%%%%%%%%%%%

{\bf \boldmath
Quantitative example: Having both beams polarized with $|P_{e^-}|=80\%$,
$|P_{e^+}|=60\%$ improves the sensitivity by up to a factor 6 compared with
using unpolarized beams and by about 20--35\% compared with the case when only
electrons are polarized.}

%%%%%%%%%%%%%%%%%%%%%%%%%%%%%%%%%%%%%%%%%%%%%%%%%%%
\chapter{Searches for new physics
with polarized \boldmath{$e^-$} and \boldmath{$e^+$} beams 
\label{npphysics}}
%%%%%%%%%%%%%%%%%%%%%%%%%%%%%%%%%%%%%%%%%%%%%%%%%%%

%%%%%%%%%%%%%%%%%%%%%%%%%%%%%%%%%%%%%%%%%%%%%%%%%%%
\section{Supersymmetry \label{sect:susy}}
%%%%%%%%%%%%%%%%%%%%%%%%%%%%%%%%%%%%%%%%%%%%%%%%%%%
\subsection{Introduction and choice of SUSY scenarios
\label{sect:introsusy}}
%%%%%%%%%%%%%%%%%%%%%%%%%%%%%%%%%%%%%%%%%%%%%%%%%%%

The importance of the polarization of both beams at the ILC is here
demonstrated for supersymmetry, which is one of the best motivated
possibilities for New Physics (NP).  With supersymmetry the
unification of the U(1), SU(2) and SU(3) gauge couplings is possible
and this new symmetry also stabilizes the Higgs mass with respect to
radiative corrections.  If nature is supersymmetric at the electroweak
scale, there is a priori a large number of parameters specifying
different scenarios.

The Minimal Supersymmetric Standard Model (MSSM) is the minimal
extension of the SM particle sector to incorporate supersymmetry. In
addition to the particles of the SM, the MSSM contains their
supersymmetric partners: sleptons $\tilde{\ell}^{\pm}$,
$\tilde{\nu}_{\ell}$ ($\ell=e$, $\mu$, $\tau$), squarks $\tilde{q}$,
charginos $\tilde{\chi}^{\pm}_{1,2}$ and neutralinos
$\tilde{\chi}^0_i$, $i=1,\ldots,4$. Two complex Higgs doublets
$(H^0_1,H^+)$, $(H^-,H^0_2)$ are needed to generate the masses of up-
and down-quarks. They lead to five physical Higgs bosons, the neutral
$h$, $H$ (CP-even), $A$ (CP-odd) and the charged $H^{\pm}$ particles.

Both the SM and the SUSY partners are described in common multiplets
and carry the same quantum numbers---with the exception of the spin
quantum number, which differs by half a unit. Since the SUSY partners
are not degenerate in mass with their SM partners, SUSY has to be a
broken symmetry. In a most general parametrization, soft 
supersymmetry-breaking terms
with many new parameters are introduced.  Due to the electroweak
symmetry breaking and the SUSY breaking, the interaction eigenstates of 
the fermions, gauginos and higgsinos $\tilde{W}^{\pm}$, $\tilde{H}^{\pm}_{1,2}$
($\tilde{\gamma}$, $\tilde{Z}$, $\tilde{H}^0_{1,2}$), i.e.\, the
SUSY partners of the charged (neutral) gauge and Higgs bosons, mix to
form mass eigenstates $\tilde{\chi}^{\pm}_{1,2}$
($\tilde{\chi}^0_{1,\ldots,4}$).  The masses and couplings of the
charginos and neutralinos are determined by the corresponding mass
matrices, which depend on the U(1) (SU(2)) gaugino mass parameters $M_1$
($M_2$), the higgsino mass parameter $\mu$ and the ratio of the Higgs
expectation values $\tan\beta=v_2/v_1$, see e.g. \cite{haberkane}.

Corresponding to the two chirality states of the leptons and quarks
one has the left and right scalar partners $\tilde{\ell}_L$,
$\tilde{q}_L$ and $\tilde{\ell}_R$, $\tilde{q}_R$. The mass matrices
of the sfermions depend on scalar mass parameters $M_Q$, $M_U$, $M_D$,
$M_L$, $M_E$
trilinear couplings $A_{\ell}$, $A_{q}$ and $\mu$ and
$\tan\beta$. Mixing effects between the R- and L-scalar states are
expected to be most important for the third generation; the
corresponding mass eigenstates are called $\tilde{t}_{1}$,
$\tilde{b}_1$, $\tilde{\tau}_1$ and $\tilde{t}_2$, $\tilde{b}_2$,
$\tilde{\tau}_2$.

In order to distinguish between SUSY and SM particles a new quantum
number R-parity is introduced, R=$-1$ for SUSY particles, and $+1$ for
SM particles. If R-parity is conserved, the SUSY particles can only be
produced in pairs and the lightest SUSY particle has to be stable and
represents the final particle of all decay chains of SUSY particles. In
all studies shown here, the lightest SUSY particle is assumed to be
the lightest neutralino $\tilde{\chi}^0_1$, which is also a good
candidate for the cold dark matter particle.

Since in SUSY also many new sources of CP violation could occur, one
ends up, even in the MSSM with conserved R-parity, with 105 new 
parameters.  With specific model assumptions about the SUSY breaking
mechanism and mass unifications, the number of free parameters is
strongly reduced. In the so-called minimal supergravity (mSUGRA) model,
for example, one has only three parameters and one sign.  However, one
should keep in mind, that at future experiments at the LHC and the LC,
one has---after detecting signals expected by SUSY---to determine the
 parameters as model-independently as possible and to confirm the
underlying assumptions.

%%%%%%%%%%%%%%%%%%%%%%%%%%%%%%%%%%%%%%%%%%%%%%%%%%%%%%%%%%%%%%%%%
\subsection*{Choice of SUSY scenarios \label{sect:susyscenarios}}
%%%%%%%%%%%%%%%%%%%%%%%%%%%%%%%%%%%%%%%%%%%%%%%%%%%%%%%%%%%%%%%%%

A priori it can not be stated whether one specific point in the
multi-parameter space of SUSY is less probable than others.  There
exist mass limits from the direct searches for Higgs and SUSY particles 
at LEP and the Tevatron and indirect bounds from tests of 
electroweak precision observables and
of searches for $g_{\mu}-2$~\cite{Ellis:2001yu} and 
$b\to s \gamma$~\cite{Buras:1998ra}
from low-energy experiments. Indirect limits for the CP-violating
phases can be derived from the EDM of the electron, neutron and atomic  
systems\cite{Altarev:1992cf}.  
Further indirect bounds for the MSSM parameter space 
can be derived from dark 
matter searches \cite{Belanger:2004hk}.  
The chosen point has to be
checked case-by-case whether it violates any experimental bounds.  
If a specific SUSY breaking scheme has been assumed, as e.g.
mSUGRA,  exclusion bounds in the corresponding parameter space can be
derived due to the restricted small number of parameters. 
However, such choices reduce the variety of possible
signatures considerably.

In order to give an overview of the many effects of beam
polarization in searches for supersymmetry the chosen SUSY scenarios
have not been restricted to a specific SUSY breaking scheme and
unification assumptions.  This approach has led to the consideration,
in this report, of a large variety of different SUSY scenarios.  In
some examples polarizing both beams is absolutely necessary for
analyzing the properties of particles and couplings.  In other studies
the polarization of both beams is needed for quantitative reasons and
leads to better statistics, i.e., higher cross sections and better
background suppression, which can be decisive for a discovery.
Therefore, having both beams polarized is crucial to face the many
challenges in supersymmetry and to resolve and determine precisely
this candidate for a new theory.

Table~\ref{tab-susyscenarios} provides an overview of
different parameters sets, together with references to the topic that is
addressed in the study.

{\small
\begin{table}\hspace{-1.5cm}
\renewcommand{\arraystretch}{1.2}
\begin{tabular}{|l|cccccc|l|l|}
\hline
 & 
$|M_1|$ & $\varphi_{M_1}$ & $ M_2$ & $|\mu|$ & $\varphi_{\mu}$ & $\tan\beta$ &
 scalar sector parameters & topic of study \\ \hline
\multicolumn{8}{|l|}{MSSM scenarios without CP-violating phases:} &\\
S1 & 150 & 0 & 210 & 400 & 0 & 20 & $m_{\tilde{e}_{R,L}}=195$, 200~GeV & 
selectron properties\\
S2 & 103 & 0 & 232 & 403 & 0 & 10 & $m_{\tilde{e}_{R,L}}=187$, 223~GeV & 
\hspace{.5cm} selectron proporties\\
S3\footnote{S3$\equiv$SPS3\cite{Allanach:2002nj}} & 163 & 0 & 311 & 509 & 0 & 10 &
$m_{\tilde{\mu}_{R,L}}=178$, 287~GeV & 
smuon masses\\
S4 & 99 & 0 & 193 & 140 & 0 & 20 & $m_{\tilde{\tau}_1}=155$~GeV & stau properties\\
   & &&&&&& $|\cos\theta_{\tilde{\tau}}|=0.08$ & \\ 
S5 & -- & -- & -- & -- &-- & -- & $m_{\tilde{t}_1}=200$~GeV & stop properties\\
& &&&&&& $|\cos\theta_{\tilde{t}}|=0.4$ and $0.66$& \\
S6\footnote{S6$\equiv$SPS1a~\cite{Allanach:2002nj}}& 
GUT & 0 & 193 & 352 & 0 & 10 & $m_{\nu_e,\tilde{e}_{L}}=186$, 202~GeV & 
charginos\\
S7 & 90 & 0 & 350 & 140 & 0 & 20 & $m_{\tilde{e}_{L,R}}$ variable &
neutralinos\\ 
S15 &
 100 & 0 & 200 & 1000 & 0 & [2,30] & 
$M_{\tilde{Q}}=M_{\tilde{U}}=350$,& 
heavy Higgs\\
&&&&&&& $A_t=A_b=700+\mu/\tan\beta$& \\
&&&&&&& $m_A=[400,800]$~GeV& \\ \hline
\multicolumn{8}{|l|}{MSSM scenarios with CP-violating phases:} & 
CP asymmetries:\\ 
S8 & 200 & $\pi/5$ & 400 & 240 & 0 & 10 & $m_{\tilde{e}_{R,L}}=220$, 372~GeV & 
$\tilde{\chi}^0_2$: 2-body decay\\
S9 & 100 & 0 & 200 & 250 & 0 & 5 & $m_{\tilde{\tau}_{1,2}}=143$, 210~GeV & 
$\tilde{\chi}^0_2$: 2-body decay\\
&&&&&&& $|A_{\tau}|=1500$, $\varphi_{A_{\tau}}=\pi/2$ & 
\phantom{$\tilde{\chi}^0_2$:} into $\tau$; 
$\varphi_{A_{\tau}}\neq 0$\\
S10 & 100 & $\pi/5$ & 200 & 250 & 0 & 5 & $m_{\tilde{\tau}_{1,2}}=144$, 
209~GeV &  
$\tilde{\chi}^0_2$: 2-body decay\\
&&&&&&& $A_{\tau}=250$, $\varphi_{A_{\tau}}=0$ & 
\phantom{$\tilde{\chi}^0_2$:} into $\tau$; $\varphi_{M_1}\neq 0$ \\
S11 & 150 & $\pi/5$  & 300 & 200 & 0 & 10 & $m_{\tilde{e}_{R,L}}=224$, 268~GeV & 
$\tilde{\chi}^0_{2}$: 3-body decay\\
S12 & GUT & $\pi/2$ & 
\multicolumn{2}{c}{$M_2$--$|\mu|$ variable}& 0 & 3 & $m_{\tilde{e}_{R,L}}=150$, 
400~GeV & 
$\tilde{\chi}^0_1\tilde{\chi}^0_2$  
with $P_{e^-}^{\rm T}$, $P_{e^+}^{\rm T}$\\ \hline
\multicolumn{8}{|l|}{Scenarios in extended SUSY models:} &\\ 
S13~a) & 195 & 0 & 300 & 350 & 0 & 20 & 
$m_{\tilde{e}_{R,L}}=143$, 202~GeV & 
distinction:\\
S13~b) & 270 & 0 & 381 & -- & 0 & 20 & 
$m_{\tilde{e}_{R,L}}=143$, 202~GeV, & 
MSSM$\leftrightarrow$NMSSM\\
&&&&&&& $\mu_{\rm eff}=350$, $\kappa=0.152$ & \\ 
S14 & -- &-- & -- & -- &-- &-- & $m_{\tilde{\nu}_e}=650$, $\Gamma_{\tilde{\nu}_e}=1$ & 
R-parity violation:\\
&&&&&&& $\lambda_{131}=0.05$ & spin-0 in s-channel\\ \hline
\end{tabular}
\caption[Parameters of the chosen SUSY scenarios]{SUSY parameters of
the scenarios studied in this report: 
the U(1) (SU(2)) gaugino mass parameters $M_1$ ($M_2$),
the higgsino mass parameter $\mu$, the ratio of the Higgs vacuum
expectation values $\tan\beta=v_2/v_1$, the mass of the CP-odd Higgs
$m_{A}$, the slepton masses $m_{\tilde{\ell}_{L,R}}$, the trilinear
couplings of the 3rd generations $A_t$, $A_b$, $A_{\tau}$ and the
possible non-vanishing phases $\varphi_{M_1}$, $\varphi_{\mu}$,
$\varphi_{A_{t,b,\tau}}$. All mass parameters and trilinear couplings are
given in GeV.  In case that unification between the U(1) and SU(2)
parameters is assumed, the absolute value of $M_1$ is given by
$|M_1|=\frac{5}{3}\tan^2\theta_W M_2$ and is denoted by `GUT'.  Listed
are only those parameters which are relevant for the presented study.
The large number of scenarios reflects the variety of
possible different signatures in the MSSM; the scenario number reflects
sequence of corresponding plots in this chapter
($^{*}$$\equiv$SPS3 and
$^{\dagger}$$\equiv$SPS1a~\cite{Allanach:2002nj}).
\label{tab-susyscenarios}}
\end{table}
}

%%%%%%%%%%%%%%%%%%%%%%%%%%%%%%%%%%%%%%%%%%%%%%%%%%%
\subsection[Determination of selectron properties]
{Determination of selectron properties 
\label{sect:susysel}}
%%%%%%%%%%%%%%%%%%%%%%%%%%%%%%%%%%%%%%%%%%%%%%%%%%%

{\bf \boldmath 
In order to test whether the SUSY partners of the electrons/positrons
carry the same chiral quantum numbers as their SM partners one has to separate
the scattering process from the annihilation process. With both beams
polarized the production vertices in the $t$- and $u$-channel can be 
analysed independently.  Another
important test of the theory is to show that the SUSY Yukawa couplings 
are equal to the gauge couplings.  Polarized positrons are
needed for such model tests, in particular 
in scenarios where even a fully polarized
electron beam is insufficient.}
\smallskip

\begin{figure}[htb]
\setlength{\unitlength}{1cm}
\begin{picture}(12,4.5)
\put(3,1){\mbox{\epsfysize=2.cm\epsffile{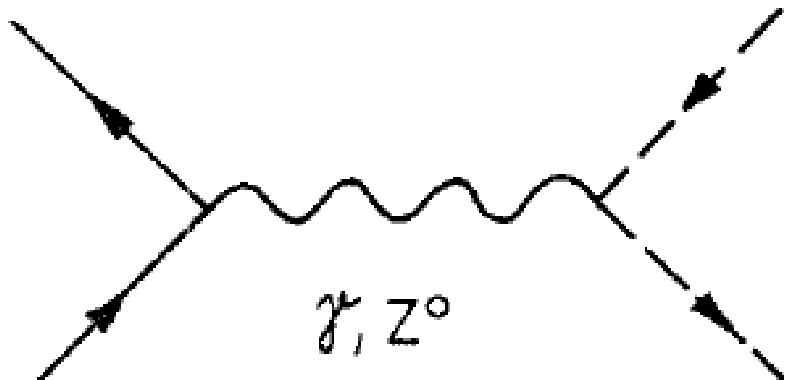}}}
\put(2.5,1){\small $e^+$}
\put(2.5,3){\small $e^-$}
\put(7.5,1){\small $\tilde{e}^+_{L,R}$}
\put(7.5,3){\small $\tilde{e}^-_{L,R}$}
\put(11.5,.5){\mbox{\epsfysize=4.cm\epsffile{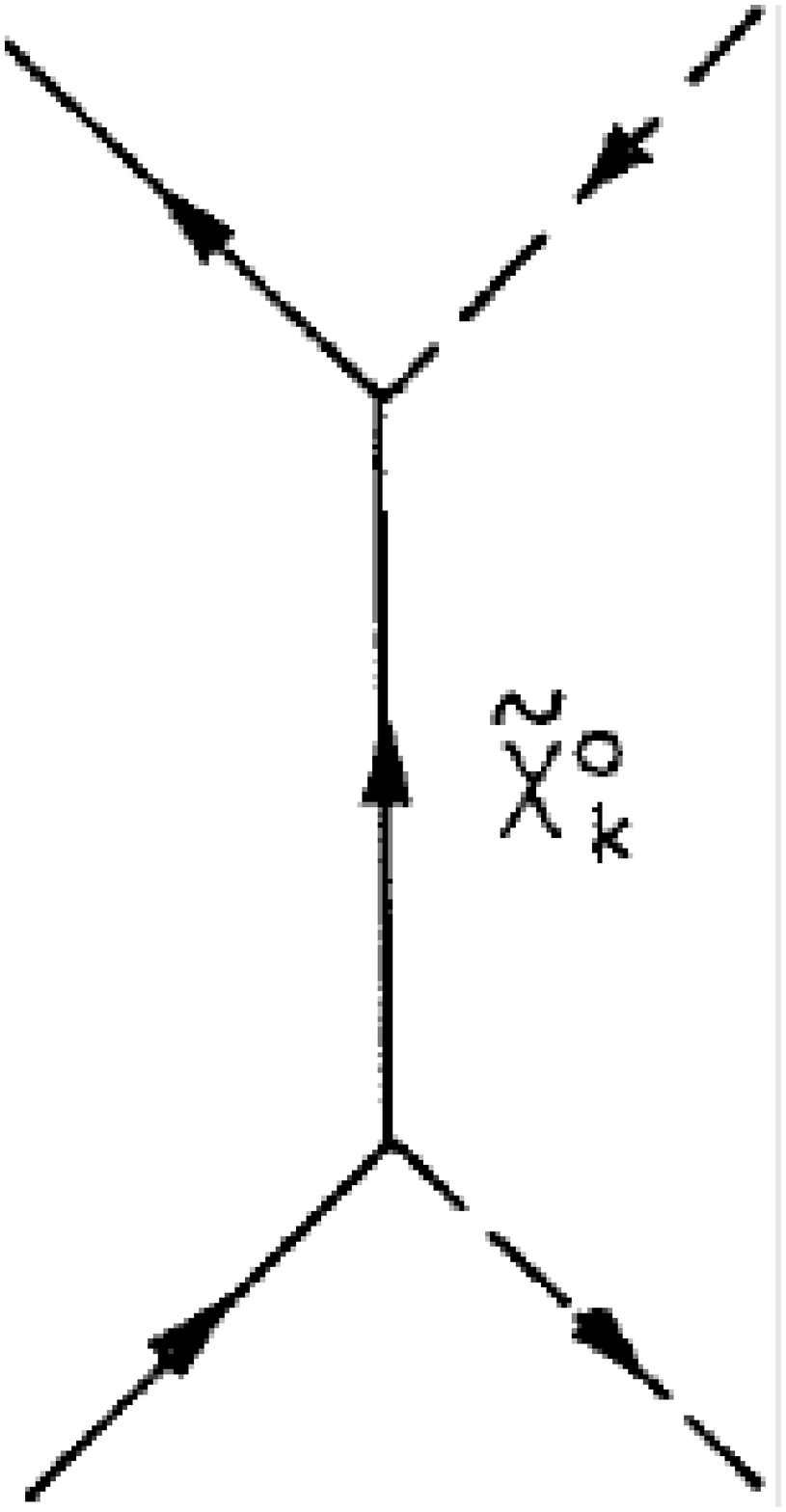}}}
\put(11,.5){\small $e^+$}
\put(11,4.5){\small $e^-$}
\put(14,.5){\small $\tilde{e}^+_{L,R}$ }
\put(14,4.5){\small $\tilde{e}^-_{R,L}$ }
\end{picture}
\vspace{-.5cm}
\caption[Selectron  production]{Selectron production: $\gamma$, 
$Z$-exchange in the $s$-channel and 
$\tilde{\chi}^0_1$,$\ldots,\tilde{\chi}^0_4$-exchange in the $t$-channel.
\label{feyn-selectron}}
\end{figure}

In this section selectron production, $e^+ e^- \to \tilde{e}_{L,R}^+
\tilde{e}^-_{L,R}$, with polarized beams is studied. The process
occurs via $\gamma$, $Z$ exchange in the $s$-channel and via neutralino
 exchanges, $\tilde{\chi}^0_{1,2,3,4}$, in the $t$-channel, see
fig.~\ref{feyn-selectron}. In the $t$-channel both pair
production, $\tilde{e}^+_L \tilde{e}^-_L$, $\tilde{e}^+_R
\tilde{e}^-_R$, as well as associated production, $\tilde{e}^+_L
\tilde{e}^-_R$, $\tilde{e}^+_R \tilde{e}^-_L$, is possible,
whereas in the $s$-channel only pairs, 
$\tilde{e}^+_L \tilde{e}^-_L$, $\tilde{e}^+_R \tilde{e}^-_R$, can be 
produced. In the
MSSM at tree-level this sector depends  on the scalar masses and, due
to the exchange of all neutralinos in the t-channel, on the
gaugino/higgsino mixing parameters $M_{1,2}$, $\varphi_{M_1}$, $\mu$,
$\varphi_{\mu}$ and $\tan\beta$.

In the following the impact of beam polarization for determining
a) the quantum numbers $L$, $R$ and b) the Yukawa couplings is studied.
%%%
%%%                             

%%%%%%%%%%%%
\subsection*{Chiral quantum numbers \label{sect:selqu}}
%%%%%%%%%%%%
Supersymmetry associates scalars to chiral (anti)fermions  
\begin{equation}
e^-_{\rm L,R}\leftrightarrow \tilde{e}^-_{\rm L,R} \quad\text{and}\quad
e^+_{\rm L,R}\leftrightarrow \tilde{e}^+_{\rm R,L}.
\label{eq_selnumber}
\end{equation} 
In order to prove this association it is necessary to have both beams
polarized \cite{Blochinger:2002zw}.  The association can be directly
tested only in the $t$-channel, as can be inferred from 
fig.~\ref{feyn-selectron}.  Polarized beams serve to separate this
channel from the $s$-channel and enhance the cross section of just
those SUSY partners of the initial chiral $e^-_{\rm L,R}$ and
$e^+_{\rm L,R}$ given by the beam polarization, see
eq.~(\ref{eq_selnumber}).  This is demonstrated by isolation of
$\tilde{e}^+_{\rm L} \tilde{e}^-_{\rm R}$ by the RR configuration of
the initial beams in an example where the selectron masses are close
together, namely $m_{\tilde{e}_{\rm L}}=200$~GeV, $m_{\tilde{e}_{\rm
R}}=195$~GeV so that both $\tilde{e}_{\rm L}$, $\tilde{e}_{\rm R}$
decay via the same channels, $\tilde{e}_{\rm L,R}\to
\tilde{\chi}^0_1 e$. The decay products can be separated e.g.\ via
their different energy spectra and charge separation.  At the LC it is
then possible to measure the selectron masses with an expected
accuracy of typically a few hundred MeV~\cite{TDR}.
In addition, all SM background events, e.g.,
those from $W^+W^-$ production, are
strongly suppressed with the RR configuration.
 The other SUSY parameters correspond to the scenario S1 in 
table~\ref{tab-susyscenarios}.

\begin{figure}[htb]
\begin{center}
     \vspace*{5.2cm}
\begin{picture}(8,10)
\setlength{\unitlength}{1cm}
\put(-8.5,-1){\mbox{\includegraphics[height=.25\textheight]{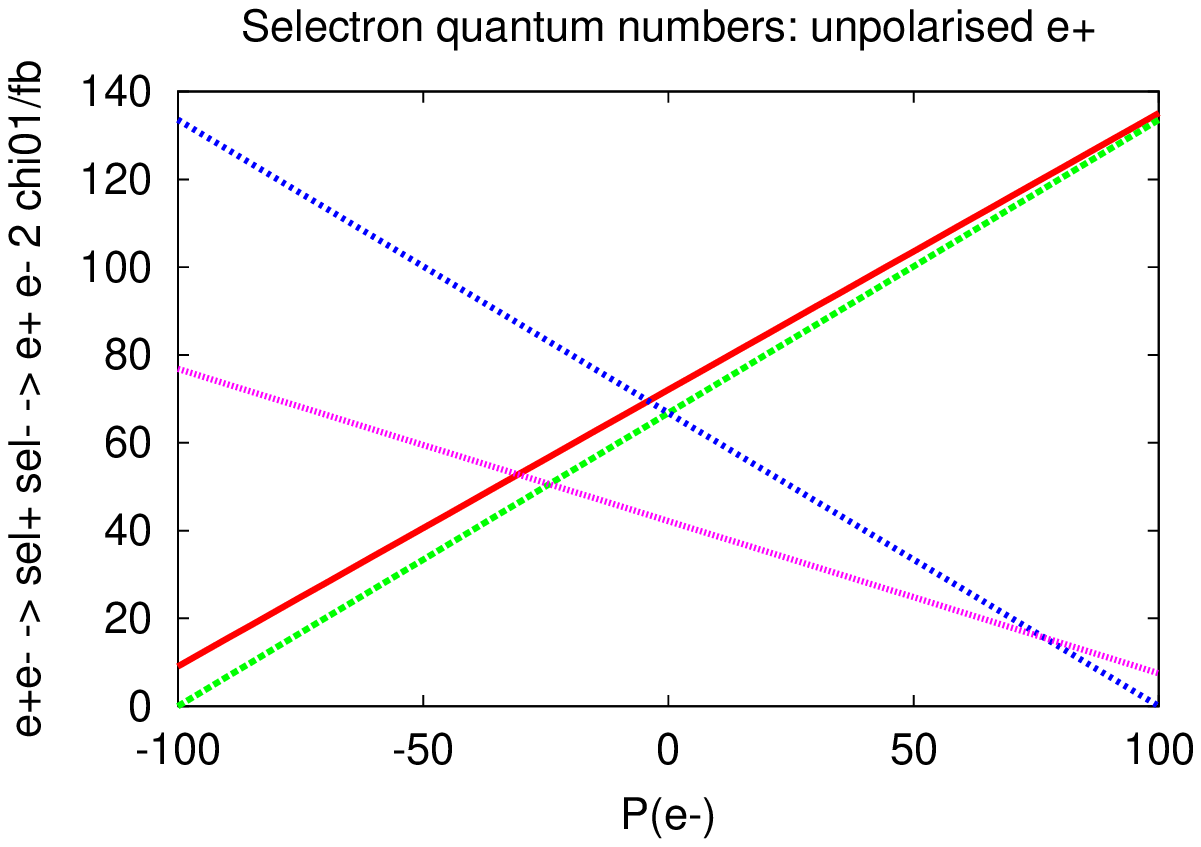}}}
\put(-8.5,4.2){\mbox{\includegraphics[height=.05\textheight,width=.8\textheight]{box}}}
\put(-8.5,-1){\mbox{\includegraphics[height=7cm,width=.5cm]{box}}}
\put(-8.5,-1.3){\mbox{\includegraphics[height=.8cm,width=10cm]{box}}}
\put(-4,-1){\small $P_{e^-} [\%]$}
\put(-7.8,4.4){\small $\sigma\times$BR [fb]}
\put(-4.5,4.4){\small $P_{e^+}=0$}
\put(-1.1,2.3){\small \makebox(0,0)[br]{{\color{Green}
$\tilde{e}^+_{\rm L} \tilde{e}^-_{\rm R}$}}}
\put(-4.6,1.8){\small \makebox(0,0)[br]{{\color{Lila}
$\tilde{e}^+_{\rm L} \tilde{e}^-_{\rm L}$}}}
\put(-4.1,2.6){\small \makebox(0,0)[br]{{\color{Blue}
$\tilde{e}^+_{\rm R} \tilde{e}^-_{\rm L}$}}}
\put(-2.5,2.9){\small \makebox(0,0)[br]{{\color{Red}$
\tilde{e}^+_{\rm R} \tilde{e}^-_{\rm R}$}}}
\put(-6.2,3.7){\small $\sqrt{s}=500$~GeV}
\put(.5,-1){\mbox{\includegraphics[height=.25\textheight]{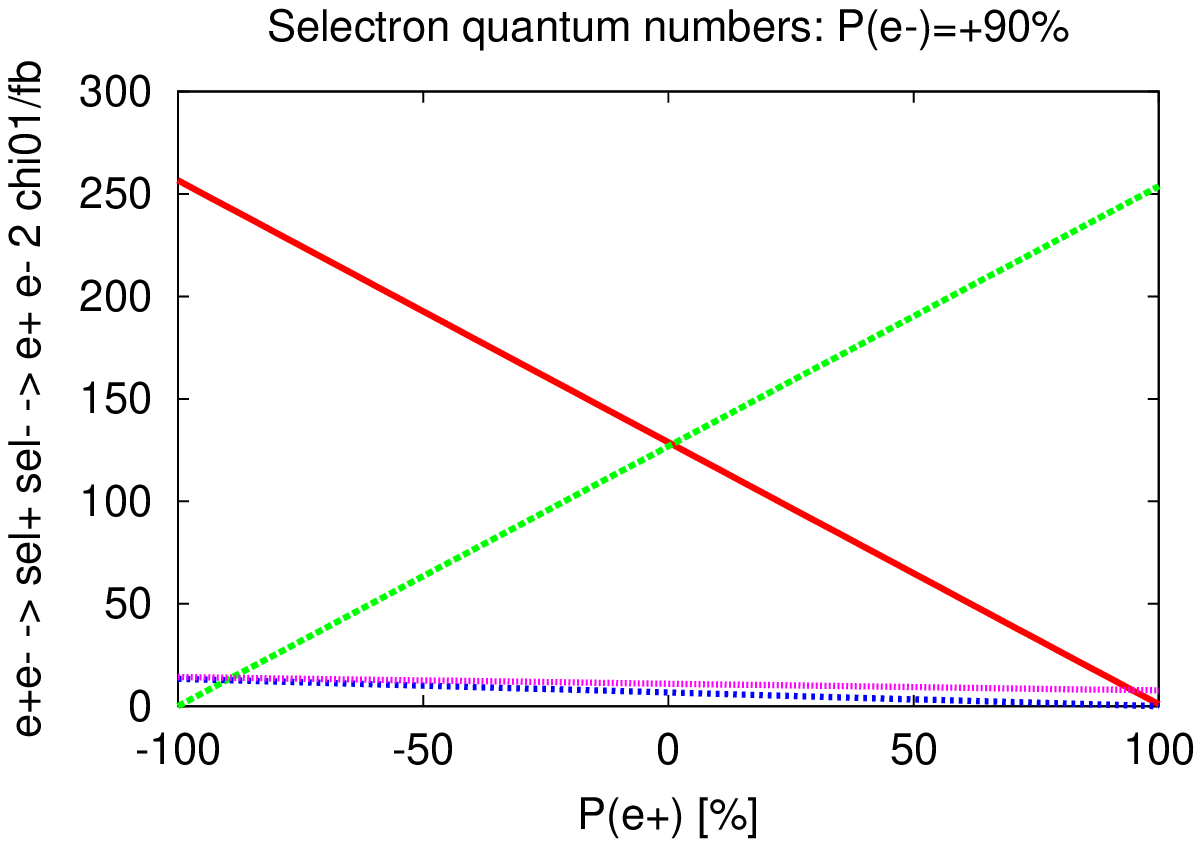}}}
\put(-.5,4.2){\mbox{\includegraphics[height=.05\textheight,width=.8\textheight]{box}}}
\put(.5,-1){\mbox{\includegraphics[height=7cm,width=.5cm]{box}}}
\put(-.5,-1.3){\mbox{\includegraphics[height=.8cm,width=10cm]{box}}}
\put(5,-1){\small $P_{e^+} [\%]$}
\put(1.2,4.4){\small $\sigma\times$BR [fb]}
\put(4.2,4.4){\small $P_{e^-}=+90\%$}
\put(6.3,2.4){\small \makebox(0,0)[br]{{\color{Green}
$\tilde{e}^+_{\rm L} \tilde{e}^-_{\rm R}$}}}
\put(3,2.3){\small \makebox(0,0)[br]{{\color{Red}$
\tilde{e}^+_{\rm R} \tilde{e}^-_{\rm R}$}}}
\put(2.5,3.7){\small $\sqrt{s}=500$~GeV}
\end{picture}\vspace{1cm}
\end{center}\vspace{-.5cm}
\caption[Test of $\tilde{e}^{\pm}$ quantum numbers] 
{\label{fig-sel-var-el} Separation of the selectron pair
$\tilde{e}_{\rm L}^+\tilde{e}^-_{\rm R}$ in $e^+ e^-\to \tilde{e}^+_{\rm L,R}
\tilde{e}^{-}_{\rm L,R}\to e^+e^- 2 \tilde{\chi}^0_1$ is not possible with 
electron polarization only (left panel). If, 
however, both beams are polarized,  the cross sections 
(right panel) differ and the RR configuration separates the pair
$\tilde{e}_{\rm L}^+\tilde{e}^-_{\rm R}$ \cite{Gudi-sel}. 
The SUSY parameters are chosen 
as in scenario S1, table~\ref{tab-susyscenarios}. }
\end{figure}

The importance of having both beams polarized is demonstrated 
in fig.~\ref{fig-sel-var-el}, which exhibits the isolation of the
$\tilde{e}^+_{\rm L} \tilde{e}^-_{\rm R}$ pair. 
Even
extremely high right-handed electron polarization, $P_{e^-}\ge +90\%$, is not
sufficient by itself to disentangle the pairs
$\tilde{e}_{\rm L}^+\tilde{e}_{\rm R}^-$ and $\tilde{e}_{\rm R}^+ \tilde{e}_{\rm R}^-$ and to
test their association to the chiral quantum numbers, since both cross
sections are numerically very close, as seen in
fig.~\ref{fig-sel-var-el} (left panel).  Only with right-handed
polarizations of both beams, the pair $\tilde{e}^+_{\rm L} \tilde{e}^-_{\rm R}$
can be separated, as seen in fig.~\ref{fig-sel-var-el} (right panel).

Note that the $t$- ($s$-wave) and $s$-channel ($p$-wave) 
production could also be separated via threshold scans \cite{Peskin},
where sufficient running time at different energies close to the
threshold is required.  It is, however, also necessary to have both beams
polarized in that case to test whether indeed the couplings of the produced
selectrons uniquely correspond to the chirality of the
electrons/positrons, respectively, as in
eq.~(\ref{eq_selnumber}).

%%%%%%%%%%%%%%%
\subsection*{Yukawa couplings }
%%%%%%%%%%%%%%%

As a consequence of supersymmetry, the SU(2) and U(1) SUSY Yukawa couplings
have to be identical to the corresponding SM gauge couplings.  Assuming that
the masses and mixing parameters of the neutralinos are known, 
the production cross sections of $\tilde{e}_{\rm
R}^+\tilde{e}_{\rm R}^-$ and $\tilde{e}_{\rm L}^+ \tilde{e}^-_{\rm R}$ can be
exploited to derive the Yukawa couplings. In \cite{Freitas:2003yp} a 
one-$\sigma$
uncertainty of $0.2\%$ ($1.2\%$) in the determination of the U(1) (SU(2)) 
Yukawa 
couplings has
been derived for the SUSY reference scenario SPS1a~\cite{Allanach:2002nj}.  The
study was done at $\sqrt{s}=500$~GeV, ${\cal L}_{\rm int}=500$~fb$^{-1}$, including
specific cuts to reduce the SM background and taking also into account effects
from beamstrahlung and initial-state radiation (ISR). With
$(|P_{e^-}|,|P_{e^+}|)=(80\%,50\%)$ the result is improved by a factor of 1.4
compared with the case of $(80\%,0)$.

In this analysis performed in the SPS1a scenario, the chirality of the produced
selectrons can be distinguished by their decay modes, since L-selectrons can
decay into the second-lightest neutralino $\tilde{\chi}^0_2$, while for the
R-selectrons only the decay channel $\tilde{e}_{\rm R}^\pm \to e^\pm \tilde{\chi}^0_1$
is open. For a slightly heavier gaugino mass $M_{1/2}$
and smaller scalar mass $m_0$, however, both selectron states have identical
decay modes, $\tilde{e}_{\rm R,L}^\pm \to e^\pm \tilde{\chi}^0_1$. In this case
the different combinations of $\tilde{e}_{\rm R}$ and $\tilde{e}_{\rm L}$ can
only be distinguished by the initial beam polarization. If one provides the
relative contributions of the different produced selectron pairs from theory,
the use of electron polarization alone would be sufficient to measure
both the SU(2)  and U(1) SUSY Yukawa couplings.
 
\begin{figure}[t]
\centering{
\epsfig{file=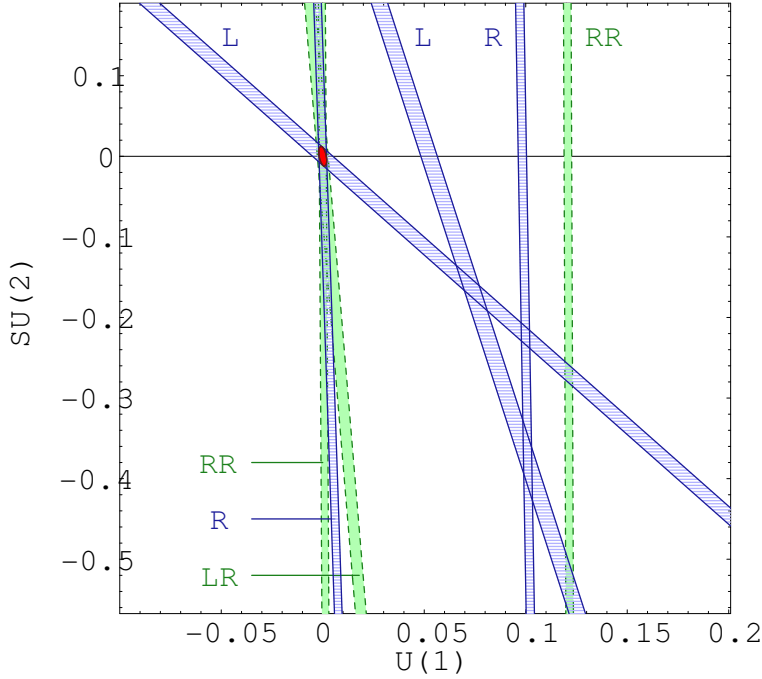,width=10cm} 
}
\caption[Measurement of $\tilde{e}^{\pm}$ 
Yukawa couplings]{1$\sigma$ bounds on the 
determination of the supersymmetric U(1) and
SU(2) Yukawa
couplings between $e^+$, $\tilde{e}_{\rm R,L}^+$ and $\tilde{\chi}^0_i$
from selectron cross-section measurements.
The blue (shaded) bands indicate results from measurements using
only electron beam polarization for two values
$P_{e^-} = +90\%$ (R) and $P_{e^-} = -90\%$ (L). 
The light green bands add
information from measurements with both beams
polarized for the values $(P_{e^+},P_{e^-}) = (-60\%,+90\%)$ (LR) and
$(+60\%,+90\%)$ (RR).
Combining all constraints leads to the dark red region.
The errors correspond to an integrated luminosity of 100 fb$^{-1}$ for each
polarization combination~\cite{Ayres}. 
The SUSY parameters are chosen as in scenario S2,
table~\ref{tab-susyscenarios}.
}
\label{fig:yuk}
\end{figure}

Without using this theoretical information, it is necessary to have both beams
polarized for a measurement of the Yukawa couplings. This
is illustrated in fig.~\ref{fig:yuk} for scenario S2, cf. 
table~\ref{tab-susyscenarios}. The use of only $e^-$ beam polarization
leaves a four-fold ambiguity in the determination of the Yukawa couplings,
which can be resolved by including cross-section measurements with simultaneous
polarization of the $e^+$ and $e^-$ beams. Combining this information, the U(1)
and SU(2) Yukawa couplings can be determined with a precision of 0.2\% and 
1.2\%, respectively, see fig.~\ref{fig:yuk}. The results shown take into
account the selectron decay distributions, including 
SM and SUSY backgrounds that have
been reduced by appropriate cuts, beamstrahlung, ISR and the most important
systematic uncertainties (see  \cite{Freitas:2003yp} for details).

{\bf \boldmath
Quantitative example: The above analysis shows that even an extremely high
degree of electron polarization, say $P_{e^-}\ge 90\%$, would be
insufficient to test the chiral quantum numbers associated to the
scalar $\tilde{e}^{\pm}$. Also, a measurement of the
Yukawa couplings, which is important to prove their equality to the gauge
couplings in SUSY, might not succeed if only polarized electrons were available.}

%%%%%%%%%%%%%%%%%%%%%%%%%%%%%%%%%%%%%%%%%%%%%
\bigskip

%%%%%%%%%%%%%%%%%%%%%%%%%%%%%%%%%%%%%%%%%%%%%%%%%%%
\subsection{Smuon mass measurement above
the production threshold \label{smuon-back}}
%%%%%%%%%%%%%%%%%%%%%%%%%%%%%%%%%%%%%%%%%%%%%%%%%%%

{\bf Smuon production occurs only via \boldmath{$\gamma$} and 
\boldmath{$Z$} exchange. In order to 
measure the masses,
the suppression of the SM \boldmath{$WW$} process is  necessary. 
The polarization of both beams may be decisive to observe all kinematical  
edges and measure 
\boldmath{$m_{\tilde{\mu}_{\rm L,R}}$} in the continuum. 
This is also important to 
optimize threshold scans.}
\smallskip

Smuons are only pair-produced via $\gamma$, $Z$ exchange in the
 $s$-channel, $e^+e^-\to \tilde{\mu}^+_{L,R}\tilde{\mu}^-_{L,R}$,
 cf. fig.~\ref{feyn-selectron}.  The production process depends at
 tree-level only on the mass parameters of the scalar sector. In
 threshold scans at the linear collider the smuon masses can be
 determined very precisely to the order of a few 100~MeV
 \cite{TDR}. Since, however, threshold scans need running time, i.e.\
 it requires luminosity specific around this energy, it is important
 to measure the masses already rather precisely far beyond the
 threshold, in the continuum. To achieve a precision measurement it is
 important to suppress efficiently background processes. For smuon
 production, $W^+W^-$ final states represent one of the worst
 backgrounds; however, they can easily be suppressed with right-handed
 electron/left-handed positron beams. The scaling factors are listed
 in table~\ref{back_WW}, see section~\ref{sect:sm-higgs}.
 
The background suppression may be decisive for the detection of the new
particles and 
the accurate measurement of their masses
$m_{\tilde{\mu}}$ in the continuum.  Since
$\tilde{\mu}^+\tilde{\mu}^-$ production only proceeds via 
$s$-channel $\gamma$, $Z$ exchange, 
the initial beam configurations
LR ($P_{e^-}<0$, $P_{e^+}>0$) and RL ($P_{e^-}>0$, $P_{e^+}<0$) are
favoured.  A case study has been made \cite{uriel} for the scenario
S3, cf.  table~\ref{tab-susyscenarios}, where the masses are \bequ
m_{\tilde{\mu}_R}=178.3 \mbox{ GeV},\quad m_{\tilde{\mu}_L}=287.1
\mbox{ GeV}.  \eequ
\begin{table}[htb]
\begin{center}
\begin{tabular}{|c|c|c|c|}
\hline
Polarization & $\sigma(\tilde{\mu}^+_R\tilde{\mu}^-_R)$ [fb]
& $\sigma(\tilde{\mu}^+_L\tilde{\mu}^-_L)$ [fb]
& $\sigma(W^+ W^-)$ [fb]\\ \hline
$(-80\%, -80\%)$ & 11.44 & 5.06 & 1448 \\ \hline
$(-80\%, +80\%)$ & 21.23 & 37.74 & 12995 \\ \hline
$(+80\%, -80\%)$ & 82.99 & 8.37 & 198 \\ \hline
$(+80\%, +80\%)$ & 11.44 & 5.06 & 1448 \\ \hline\hline
$(-80\%, 0)$  & 16.34 & 21.40 & 7241 \\ \hline
$(+80\%, 0)$  & 47.21 & 6.72 & 824 \\ 
\hline
\end{tabular}
\caption[WW background in smuon production]{Cross sections for
$e^+e^-\to\tilde{\mu}^+\tilde{\mu}^-$  at $\sqrt{s}=750$~GeV for the scenario
S3~\cite{uriel}.  One observes a
large reduction of the $W^+W^-$ cross section when the electron is
right-handed and the positron is left-handed. This helps significantly in
observing $\tilde{\mu}_L$.
\label{tab_smuon}
}
\end{center}
\end{table}

\begin{figure}[htb]
\begin{picture}(8,8)
\setlength{\unitlength}{1cm}
\put(0,0){\mbox{\includegraphics[height=.24\textheight,width=.35\textheight]
{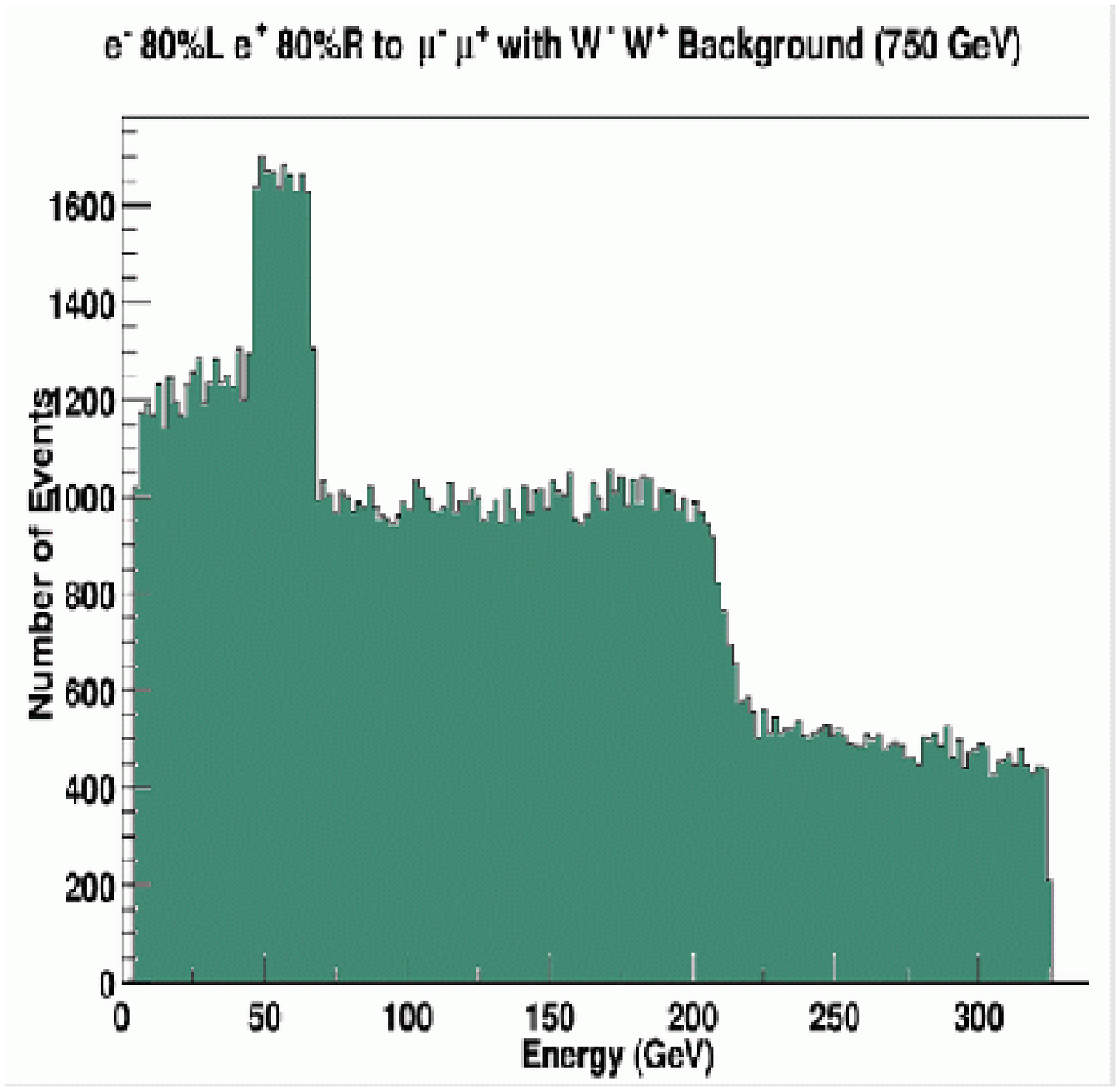}}}
\put(0,5.3){\mbox{\includegraphics[height=.05\textheight,width=.8\textheight]{box}}}
\put(-.1,-1){\mbox{\includegraphics[height=7cm,width=.5cm]{box}}}
\put(0,-.1){\mbox{\includegraphics[height=.5cm,width=10cm]{box}}}
\put(1.9,-.1){\small Energy spectrum of $\mu^+\mu^-$ [GeV]}
\put(-.3,5.3){\small  $\mu^+\mu^-$ events (incl.\ $W^+W^-$)}
\put(3.1,4.5){\small $(P_{e^-},P_{e^+})=(-80\%,+80\%)$}
\put(5.3,5.3){\small $\sqrt{s}=750$~GeV}
\put(9,0){\mbox{\includegraphics[height=.235\textheight,width=.34\textheight]
{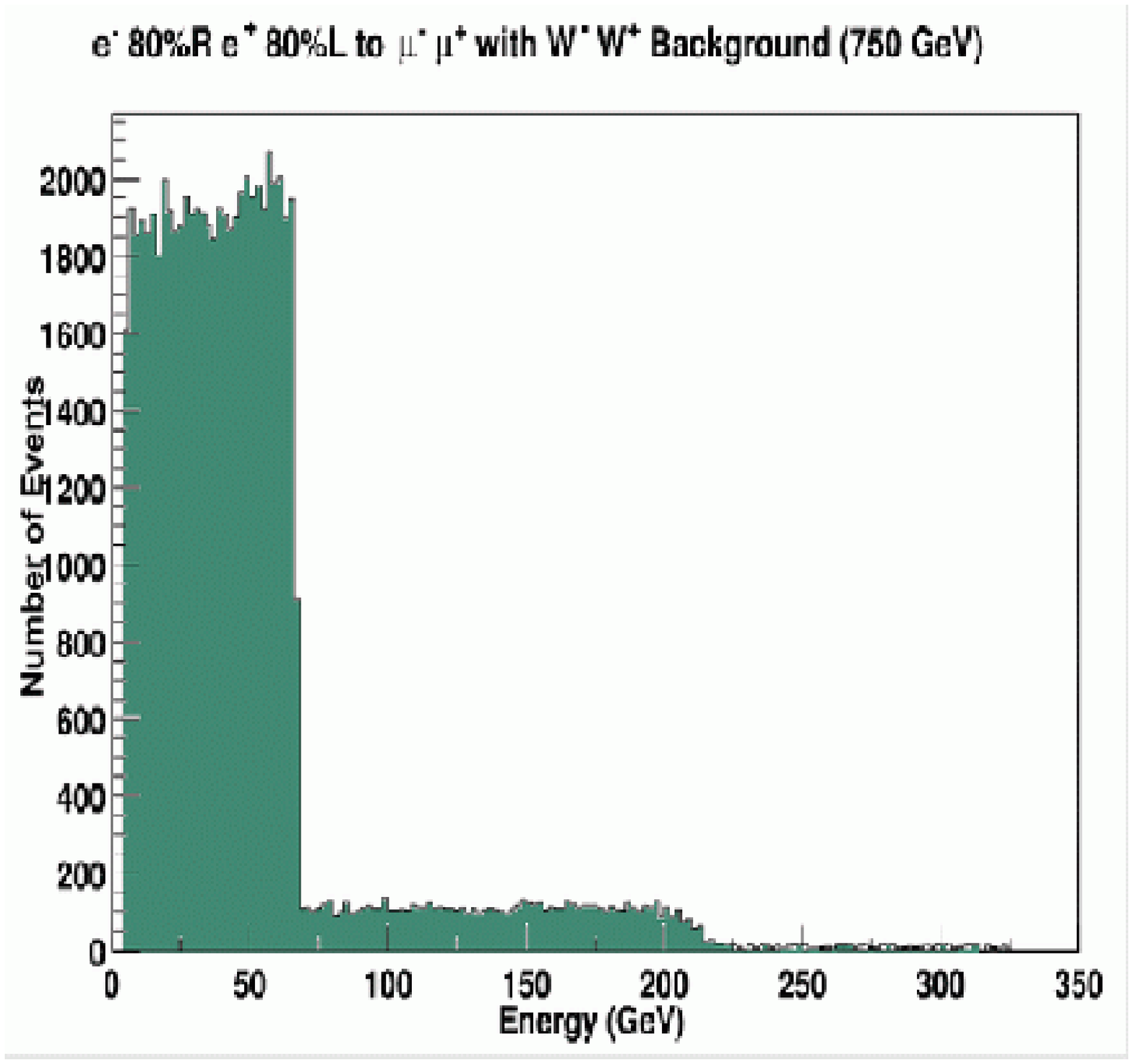} }}
\put(9,5.3){\mbox{\includegraphics[height=.05\textheight,width=.8\textheight]{box}}}
\put(8.85,-1){\mbox{\includegraphics[height=7cm,width=.5cm]{box}}}
\put(9,-.1){\mbox{\includegraphics[height=.5cm,width=10cm]{box}}}
\put(10.8,-.1){\small Energy spectrum of $\mu^+\mu^-$ [GeV]}
\put(8.9,5.3){\small  $\mu^+\mu^-$ events (incl.\ $W^+W^-$)}
\put(14.3,5.3){\small $\sqrt{s}=750$~GeV}
\put(11.7,4.5){\small $(P_{e^-},P_{e^+})=(+80\%,-80\%)$}
\end{picture}
\vspace{.1cm}
\caption[Smuon mass measurement: muon energy spectrum]{Energy spectrum 
of muons from 
$\tilde{\mu}_{L,R}$ 
decays into $\mu\tilde{\chi}^0_1$ final states, including the
$W^+W^-$ background decaying into $\mu\nu$ final states
in the scenario S3, cf. table~\ref{tab-susyscenarios},
for two combinations of beam polarizations for
$\sqrt{s}=750$~GeV and ${\cal L}_{\rm int}=500$~fb$^{-1}$~\cite{uriel}.
\label{sigmaWW} }
\end{figure}

The cross sections are
shown in table~\ref{tab_smuon}, where the considerable reduction in the $WW$
production cross section for right-handed electrons and left-handed positrons
can be noted.  One gains a factor of about 2.6 in the ratio $S/\sqrt{B}$
with $(P_{e^-},P_{e^+})=(+80\%,-80\%)$ compared to $(+80\%,0)$.
In fig.~\ref{sigmaWW} the expected muon energy distribution is
shown for an integrated luminosity of 500~fb$^{-1}$ at $\sqrt{s}=750$~GeV and
for the polarization configurations $(P_{e^-},P_{e^+})=(-80\%,+80\%)$ (left
panel) and $(+80\%,-80\%)$ (right panel).  The background
from $W^+W^-$ decaying into the $\mu \nu$ final state is included. Only with
polarized $e^-$ and $e^+$ beams can both muon-energy edges, at around 65
and 220~GeV, be reconstructed. The slepton masses can be determined in
the continuum up to a few GeV uncertainty.  This shows the real importance of
positron polarization for a clear observation of the 
low-energy edge associated to the 
$\tilde{\mu}_R$, which cannot 
be clearly seen unless the positron is polarized
\cite{uriel}.

\vspace*{5.5cm}

\smallskip

{\bf \boldmath 
Quantitative examples:
The most important background to $\tilde{\mu}$ pair production is 
$WW$ pair production.
Compared with the case of only the electron beam polarized, the signal 
gains about a factor 
1.8 and the background is suppressed by 
about a factor of 4 with $(P_{e^-},P_{e^+})=(+80\%,-80\%)$ compared to
$(+80\%,0)$. 
With both beams polarized, a rather accurate measurement of the smuon masses 
is possible already in the continuum, which can then be
used to devise possible 
threshold scans.
}

%%%%%%%%%%%%%%%%%%%%%%%%%%%%%%%%%%%%%%%%%%%%%%%%%%%
\subsection{Determination of third-generation sfermion parameters
\label{sect:thirdgen}}
%%%%%%%%%%%%%%%%%%%%%%%%%%%%%%%%%%%%%%%%%%%%%%%%%%%
{\bf \boldmath
The advantages of having both beams polarized in third-generation 
sfermion production
are the larger cross sections and a more precise determination of
masses and mixing angles.}
\smallskip

In the third generation of sfermions, Yukawa terms give rise 
to a mixing between the 
`left' and `right' states $\tilde f_{\rm L}$ and $\tilde 
f_{\rm R}$ ($\tilde f = \tilde t, \tilde b, \tilde\tau$).
The mass eigenstates are
$\tilde f_1^{} = \tilde f_{\rm L}^{} \cos\theta_{\tilde f} + 
\tilde f_{\rm R}^{}\sin\theta_{\tilde f}$, and
$\tilde f_2^{} = \tilde f_{\rm R}^{} \cos\theta_{\tilde f} - 
\tilde f_{\rm L}^{}\sin\theta_{\tilde f}$, 
with $\theta_{\tilde f}$ the sfermion mixing angle.

\begin{figure}[htb]
\setlength{\unitlength}{1cm}
\begin{picture}(12,5)
\put(1.5,-.3){\mbox{\epsfysize=6.cm\epsffile{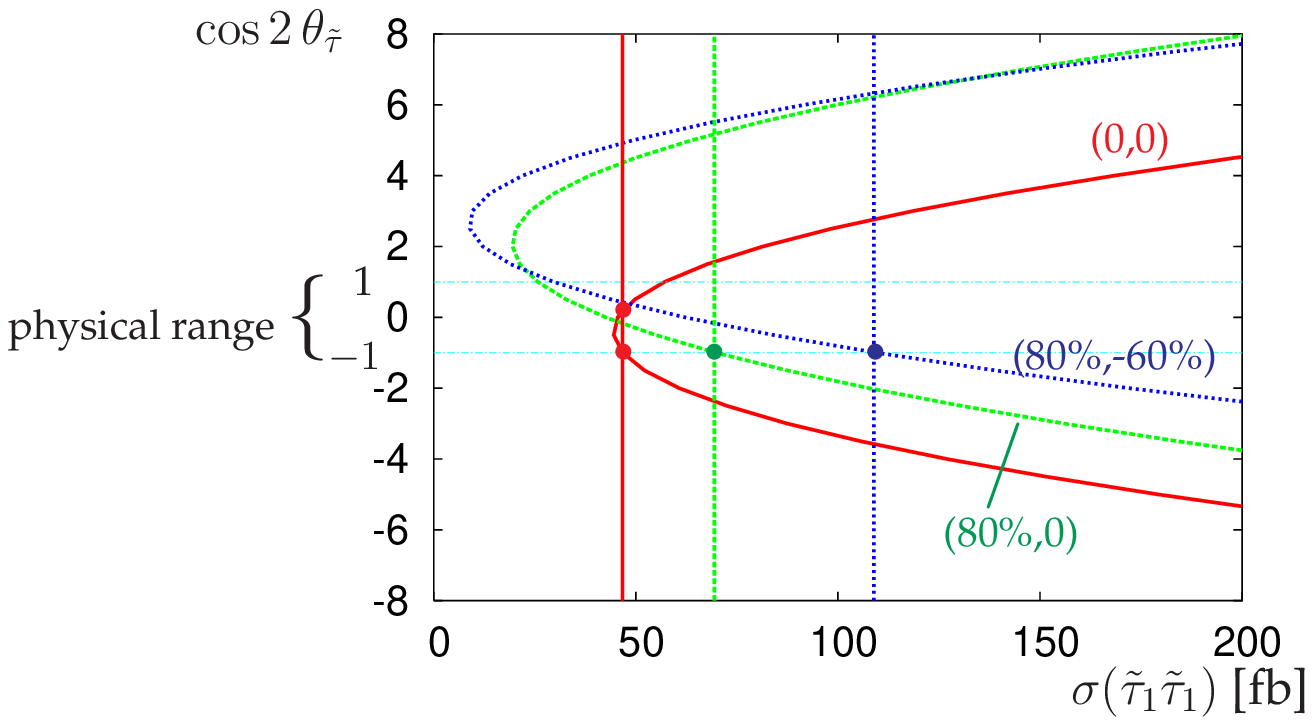}}}
\end{picture}
\vspace{-.5cm}
\caption[Determination of the stau mixing angle]{\label{susy-stau-angle} { Mixing parameter $\cos
2\theta_{\tilde{\tau}}$ vs.\ cross section $\sigma(e^+ e^- \to
\tilde{\tau}_1\tilde{\tau}_1)$ at $\sqrt{s}=500~$GeV for beam polarizations
as indicated.  The vertical lines indicate the
predicted cross sections~\cite{boos-stau}. 
The SUSY parameters are chosen as in 
scenario S4, cf. table~\ref{tab-susyscenarios}.}}
\end{figure}

In the following phenomenological studies of 
third-generation sfermions in $e^+e^-$ annihilation at $\sqrt{s} = 500$~GeV 
are summarized.
Information on the mixing angle can be obtained by
measuring production cross sections with different combinations of
beam polarizations.  It has been shown in
\cite{Bartl:1997yi,Bartl:2000kw,boos-stau} that beam polarization is important
to resolve ambiguities, see 
fig.~\ref{susy-stau-angle}. For the unpolarized case, two
values of $\cos 2\theta_{\tilde{\tau}}$ 
($\theta_{\tilde{\tau}}$ being the mixing angle) are consistent with the
cross sections (red lines). However, the use of polarized beams allows
a single solution (green and blue lines) to be identified.  
Moreover, 
the simultaneous polarization of both beams is useful for enhancing
$\Peff$, eq.~(\ref{eq_peff}), together with the signal and thus for reducing 
uncertainties~\cite{boos-stau}.  
This might be particularly important for $\tilde{\tau}/\tau$ analyses, 
which will be difficult to perform, since $\tau$s decay also
into 2$\pi$ and 3$\pi$ final states.  In~\cite{boos-stau}
a complete detector simulation of signal and background 
(using the program SIMDET~\cite{Pohl:1999uc}) for the scenario S4, 
table\ref{tab-susyscenarios}, including 
angle cuts, QED radiation and beamstrahlung and using 
$(P_{e^-},P_{e^+})=(+80\%,-60\%)$, has been performed. 
The result was that $m_{\tilde{\tau}_1}$ could even be measured
to a precision of 500~MeV, the mixing angle $\cos2 \theta_{\tilde{\tau}}$ up to $1\%$ 
and the polarization of the $\tau$, $P_{\tau}$, up to $6\%$ at the ILC.
Compared to the case with only polarized electrons of $|P_{e^-}|=80\%$
the polarization of both beams leads in the determination of
$m_{\tilde{\tau}_1}$ and $P_{\tau}$ approximately
to a reduction of the uncertainty by about a factor 1.6.
Since the determination of the mixing angle $\cos 2 \theta_{\tilde{\tau}}$ is mainly
determined by systematics positron polarization leads in this context only to an 
improvement of O(10\%).

In the following, the production of light stops is discussed in detail.
The reaction $e^+ e^- \to \tilde{t}_i \bar{\tilde t}_j$, $i,j=1,2$ proceeds
via $\gamma$ and $Z$ exchange in the $s$-channel. The $\tilde t_i$
couplings depend on the stop-mixing angle $\theta_{\tilde t}$.
In Figs.~\ref{fig:stop11pol}\,a,\,b, contour lines of
the cross section $\sigma (e^+e^- \to \tilde t_1 \bar{\tilde t}_1)$ are shown
as functions of the beam polarizations $P_{e^-}$ and
$P_{e^+}$ at $\sqrt{s}=500$~GeV for
$m_{\tilde t_1} = 200$~GeV, with (a)  $\cos\theta_{\tilde t} = 0.4$  and
(b) $\cos\theta_{\tilde t}= 0.66$, see scenario S5, 
table~\ref{tab-susyscenarios}.
Initial-state radiation (ISR) and one-loop SUSY-QCD corrections
are included (for details, see \cite{Bartl:2000kw,Bartl:1997yi}).
The white windows show the range of polarizations
$|{P_{e^-}}| < 0.9$ and $|{P_{e^+}}| < 0.6$.
As can be seen, one significantly
increases the cross section by 
up to a factor of about $1.6$ with $(P_{e^-},P_{e^+})=(+90\%,-60\%)$
compared to $(+90\%,0)$.
Moreover, beam polarization strengthens the $\cos\theta_{\tilde t}$ dependence
and can thus be essential for the determination of the mixing angle.
Corresponding cross sections for the production
of sbottoms, staus and $\tau$-sneutrinos are presented in
\cite{Bartl:2000kw}.

{\setlength \unitlength{1cm}
\begin{figure}[htb]
\vspace{5.5cm}
\begin{center}
\begin{picture}(8,5)
\setlength{\unitlength}{1cm}
\put(-4.5,-1){\mbox{\includegraphics[height=.35\textheight]{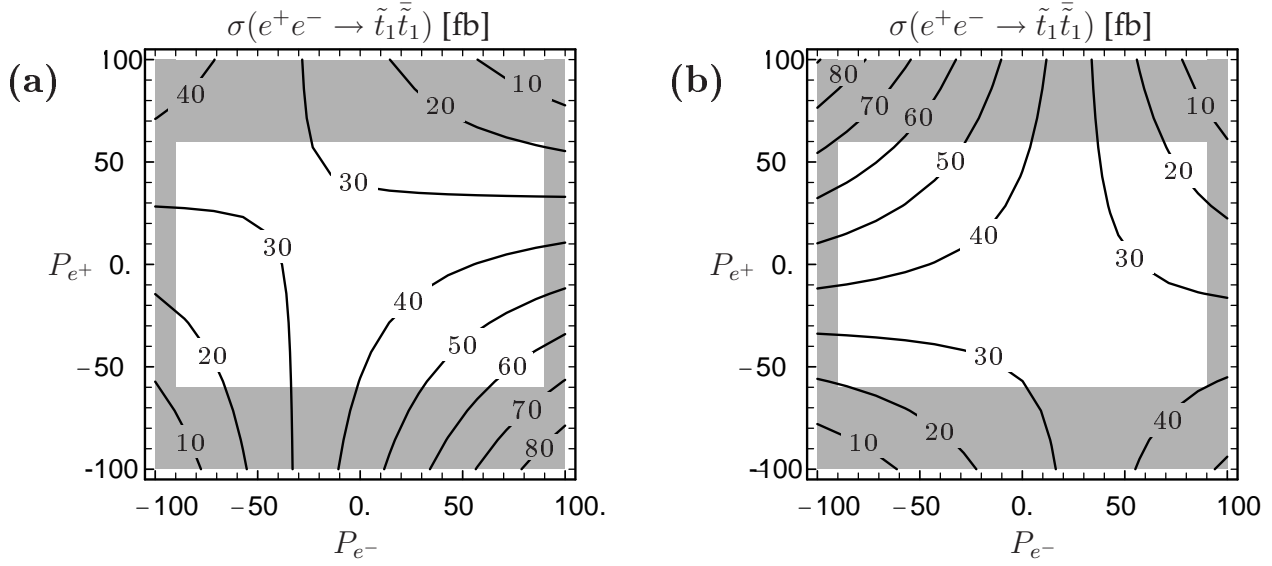}}}
\put(-4.5,-.7){\mbox{\includegraphics[height=.05\textheight,width=.8\textheight]{box}}}
\put(-4.2,0){\mbox{\includegraphics[height=7cm,width=.8cm]{box}}}
\put(4.6,0){\mbox{\includegraphics[height=7cm,width=.8cm]{box}}}
\put(-4.5,-5){\mbox{\includegraphics[height=.8cm,width=10cm]{box}}}
\put(9,-.2){ $P_{e^-}$}
\put(5,3.5){ $P_{e^+}$}
\put(-0,-.2){ $P_{e^-}$}
\put(-3.8,3.5){ $P_{e^+}$}
\put(-1.3,6.7){$\sigma(e^+e^-\to\tilde{t}_1\bar{\tilde{t}}_1)$ [fb]}
\put(7.5,6.7){$\sigma(e^+e^-\to\tilde{t}_1\bar{\tilde{t}}_1)$ [fb]}
\end{picture}
\vspace{-.5cm}
\end{center}
\caption[Production of $\tilde{t}_1\tilde{\bar{t}}_1$ with polarized beams]
{Dependence of
$\sigma(e^+e^-\to\tilde t_1\tilde{\bar{t}}_1)$ on
degree of electron and positron polarization at $\sqrt{s}=500$ GeV,
for $m_{\tilde t_1}=200$ GeV,
$|\cos\theta_{\tilde t}|=0.4$ in (a) and $|\cos\theta_{\tilde t}|=0.66$ in (b), cf. scenario S5, 
table~\ref{tab-susyscenarios} \cite{Bartl:2000kw}.
}
\label{fig:stop11pol}
\end{figure}
}

\begin{figure}[t]
\setlength \unitlength{1mm}
\begin{center}
\epsfig{figure=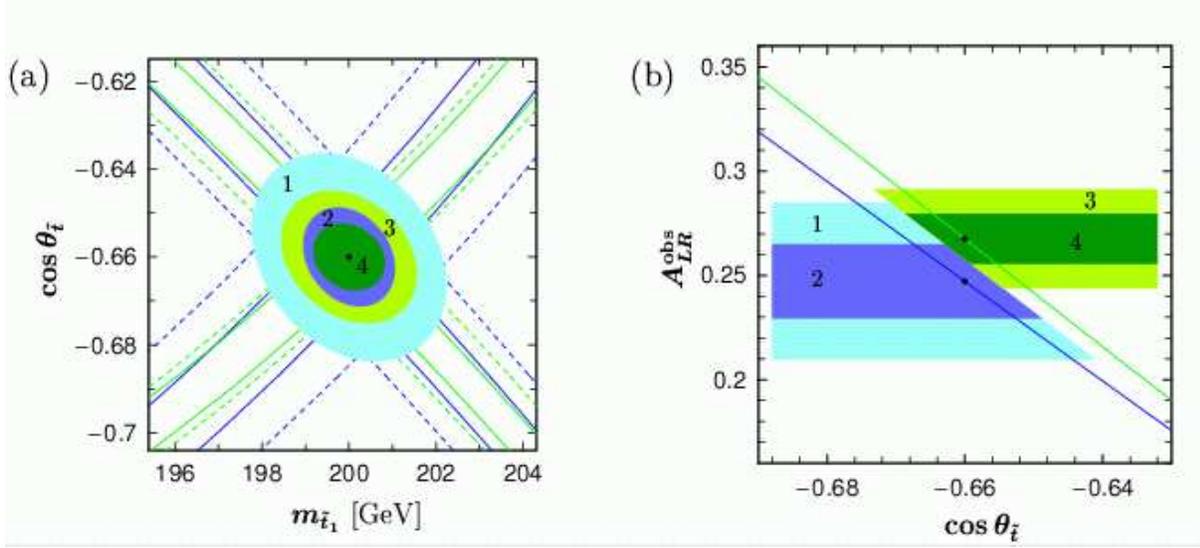,width=16cm}
\end{center}
\vspace{-8mm}
\caption[Determination of stop parameters via cross sections and $A_{\rm LR}$]{
(a) Cross section error bands
and 68\% CL error ellipses for determining
$m_{\tilde t_1}$
and $\cos\theta_{\tilde t}$ 
from cross section measurements for the four cases
given in Table~\ref{table:pol-err};
the dashed lines are for ${\cal L}_{\rm int}=100\;{\rm fb}^{-1}$ and
the full lines for ${\cal L}_{\rm int}=500\;{\rm fb}^{-1}$, the blue (green)
lines are for cases 1 and 2 (cases 3 and 4).
(b) Error bands for the 
determination of $\cos\theta_{\tilde t}$ from $A_{\rm LR}^{\rm obs}$.
In both plots $m_{\tilde t_1}=200$ GeV, $\cos\theta_{\tilde t}=-0.66$,
$\sqrt{s}=500$ GeV.~\cite{Sabine-Helmut}. The SUSY parameters are chosen as in scenario
S5, cf.\ table~\ref{tab-susyscenarios}.
\label{fig:pol-err}}
\end{figure}

The precision that can be obtained from cross section measurements on
the parameters of the stop
sector is estimated
using the parameter point S5, table~\ref{tab-susyscenarios},
with $m_{\tilde t_1}=200$ GeV and $\cos\theta_{\tilde t}=-0.66$
as an illustrative example.
The four cases
$(P_{e^-}, P_{e^+}) = (\mp 0.9, 0)$ and
$(P_{e^-}, P_{e^+}) = (\mp 0.9, \pm 0.6)$
are studied
with integrated
luminosities of ${\cal L}_{\rm int}$ = 50~fb$^{-1}$  and ${\cal L}_{\rm int}$ = 250~fb$^{-1}$
for each polarization. 
Again, one-loop  SUSY-QCD and ISR corrections are included.
A $1\%$ 
uncertainty in the polarizations of the $e^+$ and $e^-$ beams has been assumed 
as well as a theoretical uncertainty in the cross sections of 1\%, taking full
error propagation into account. The anticipated experimental precisions on
$\sigma_{-+}$ and $\sigma_{+-}$ are based on the Monte Carlo study
of \cite{Berggren:1999ss}.
Fig.~\ref{fig:pol-err}a shows the error bands for the cross sections
and the corresponding 68\% CL error ellipses in the
$m_{\tilde t_1}$--$\cos\theta_{\tilde t}$ plane for the four cases
mentioned above. 

The mixing angle can be determined even more precisely
from the left-right asymmetry
$A_{\rm LR}^{\rm obs} = (\sigma_{-+} - \sigma_{+-})/(\sigma_{-+} + \sigma_{+-})$,
cf. eq.~(\ref{eq-def-alr}),
because here the kinematical dependence on $m_{\tilde t_1}$ drops out.
The error bands for $A_{LR}$ for the four cases as a function of
$\cos\theta_{\tilde t}$ are shown in Fig.~\ref{fig:pol-err}b.
Table~\ref{table:pol-err} summarizes the cross sections and statistical errors
as well as the resulting errors
on $m_{\tilde t_1}$ and $\cos\theta_{\tilde t}$~\cite{Sabine-Helmut}.

% --- Table: stop mass and mixing angle determination ---
\begin{table}[h!]
$\begin{array}{c|ccccccc|ccc}
{\rm case} & {\cal L}_{\rm int} & P_{e^-} & P_{e^+}  & \sigma_{-+} & \sigma_{+-}& 
\Delta \sigma_{-+}^{\rm stat.} &
\Delta \sigma_{+-}^{\rm stat.} & \Delta m_{\tilde t_1} &
\begin{array}{c} {\rm Fig.~\ref{fig:pol-err}a} \\
\Delta \cos\theta_{\tilde t} \end{array} &
\begin{array}{c} {\rm Fig.~\ref{fig:pol-err}b} \\
\Delta \cos\theta_{\tilde t} \end{array} \\
\hline
1 & 100~{\rm fb}^{-1} & \mp 0.9 & 0       & 44~{\rm fb} & 27~{\rm fb} & 4.7\% &
6.3\% & 1.1\% &  3.6\% & 2.3\% \\
2 & 500~{\rm fb}^{-1} & \mp 0.9 & 0       & 44~{\rm fb} & 27~{\rm fb} & 2.1\% &
2.8\% & 0.5\% &  1.8\% & 1.1\% \\
3 & 100~{\rm fb}^{-1} & \mp 0.9 & \pm 0.6 & 69~{\rm fb} & 40~{\rm fb} & 3.1\% &
4.4\% & 0.8\% &  2.3\% & 1.4\% \\
4 & 500~{\rm fb}^{-1} & \mp 0.9 & \pm 0.6 & 69~{\rm fb} & 40~{\rm fb} & 1.4\% &
2.0\% & 0.4\% &  1.1\% & 0.7\%
\end{array}$
\caption[4 cases of error propagation to production
of $\tilde{t}_1\tilde{t}_1$ with polarized beams]{
Parameters for the cases studied, cross sections, assumed statistical errors
and resulting precisions on $m_{\tilde t_1}$ and $\cos\theta_{\tilde t}$
corresponding to Fig.~\ref{fig:pol-err}~\cite{Sabine-Helmut} and SUSY scenario 
S5, cf.\ table~\ref{tab-susyscenarios}.
\label{table:pol-err}}
\end{table}

{\bf\boldmath 
Quantitative example: In the cases studied in this section,
the cross sections are enhanced by a factor of about 1.6
with both beams polarized, $(|P_{e^-}|,|P_{e^+}|)=(90\%,60\%)$,
as compared to the case with only $(|P_{e^-}|,P_{e^+})=(90\%,0)$.
This gives up to 40\% improvement in the determination of the stop
mass and mixing angle. }

%%%%%%%%%%%%%%%%%%%%%%%%%%%%%%%%%%%%%%%%%%%%%%%%%%%
\subsection{Chargino and neutralino production \label{sect:charprod}}
%%%%%%%%%%%%%%%%%%%%%%%%%%%%%%%%%%%%%%%%%%%%%%%%%%%
{\bf \boldmath
In chargino and neutralino production, a complicated interplay takes 
place between the $s$-channel amplitudes and the amplitudes originating 
from scalar exchange in 
the $t$- and $u$-channels.
Consequently, only polarized electrons may not be sufficient for
a model-independent determination of the relevant MSSM parameters, 
and positron polarization would also be needed.}
\smallskip

\begin{figure}[htb]
\setlength{\unitlength}{1cm}
\begin{picture}(12,3.6)
\put(1.0,0){\mbox{\epsfysize=4.cm\epsffile{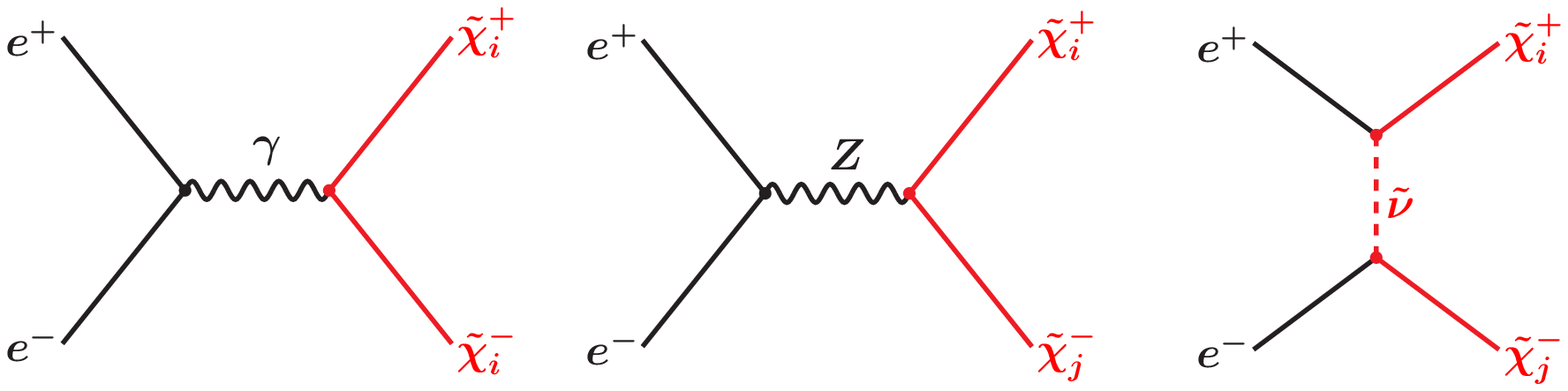}}}
\end{picture}
\vspace{-.5cm}
\caption[Chargino  production]{Chargino production: $\gamma$, 
$Z$-exchange in the $s$-channel and $\tilde{\nu}_e$-exchange 
in the $t$-channel.
\label{fig:Feynman-chargino}}
\end{figure}
                                                                                
\begin{figure}[htb]
\setlength{\unitlength}{1cm}
\begin{picture}(12,3.6)
\put(3.0,0){\mbox{\epsfysize=4.cm\epsffile{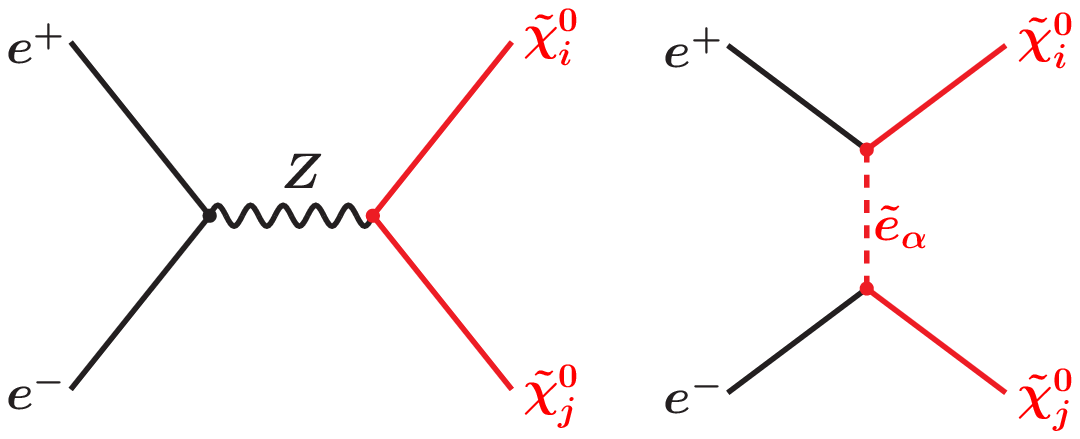}}}
\end{picture}
\vspace{-.5cm}
\caption[Neutralino  production]{Neutralino production:
$Z$-exchange in the $s$-channel and $\tilde{e}_L$, $\tilde{e}_R$-exchange
in the $t$- and $u$-channel. 
\label{fig:Feynman-neutralino}}
\end{figure}

In the following the fermionic sector of the
charginos, $\tilde{\chi}^{\pm}_{1,2}$, and neutralinos,
 $\tilde{\chi}^{0}_{1,\ldots,4}$ is studied. 
It is expected that at least some of the
charginos and neutralinos are sufficiently light to be directly observed
at $\sqrt{s}=500$~GeV. Since these particles are mixtures of the 
gaugino and higgsino interaction states their analysis will give direct access
to the SUSY breaking parameters $M_1$, $M_2$, $\mu$ and 
$\tan\beta$~\cite{gaugino-parameters}.
Beam polarization may be useful for the determination of the parameters
and couplings of the gaugino/higgsino sector, 
in particular if only part of the spectrum turns out to be
kinematically accessible \cite{Choi:2001ww}.  

Charginos (neutralinos) are produced in pairs 
via exchange of the $\gamma$, $Z$  ($Z$) gauge bosons in the $s$-channel 
and via scalar SUSY particle exchange, $\tilde{\nu}_e$ ($\tilde{e}_{L,R}$), 
in the $t$-channel ($t$- and $u$-channel), see 
figs.~\ref{fig:Feynman-chargino} and \ref{fig:Feynman-neutralino}.
The production processes depends therefore at tree-level
on the parameters $M_2$, $\mu$, $\tan\beta$
(in the case of $\tilde{\chi}^0_j$: also on $M_1$) and on the mass
parameters of the exchanged scalar particles.
If the beams were fully polarized, only
the LR and RL beam-polarization configurations would contribute to
the production processes. The cross sections in the configurations LL
and RR are non-vanishing for partially 
polarized beams. Since charginos and neutralinos are mixtures of gauginos
(which couple only 
to scalar SUSY partners $\tilde{\nu}_e$ and $\tilde{e}_{L,R}$, respectively) 
and higgsinos (which couple only to the $Z$ boson in the $s$-channel)
interesting features can be exposed when applying all possible
polarization configurations for the different chargino and neutralino 
pair productions~\cite{gudi2}. 

\begin{figure}[htb]
\setlength{\unitlength}{1cm}
\begin{picture}(10,6)(0,0)
	\put(-.3,0){\includegraphics{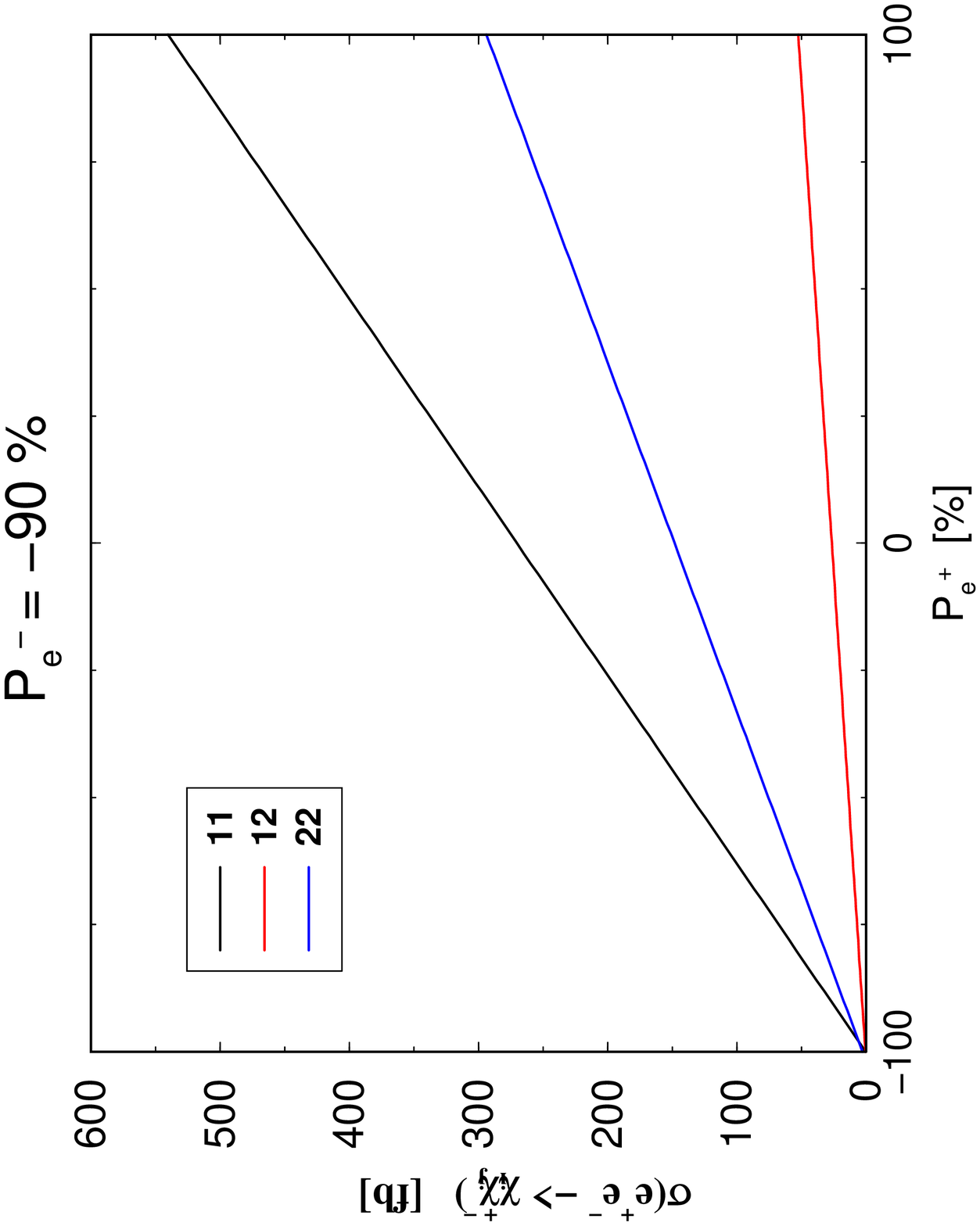}}
	\put(7.7,0){\includegraphics{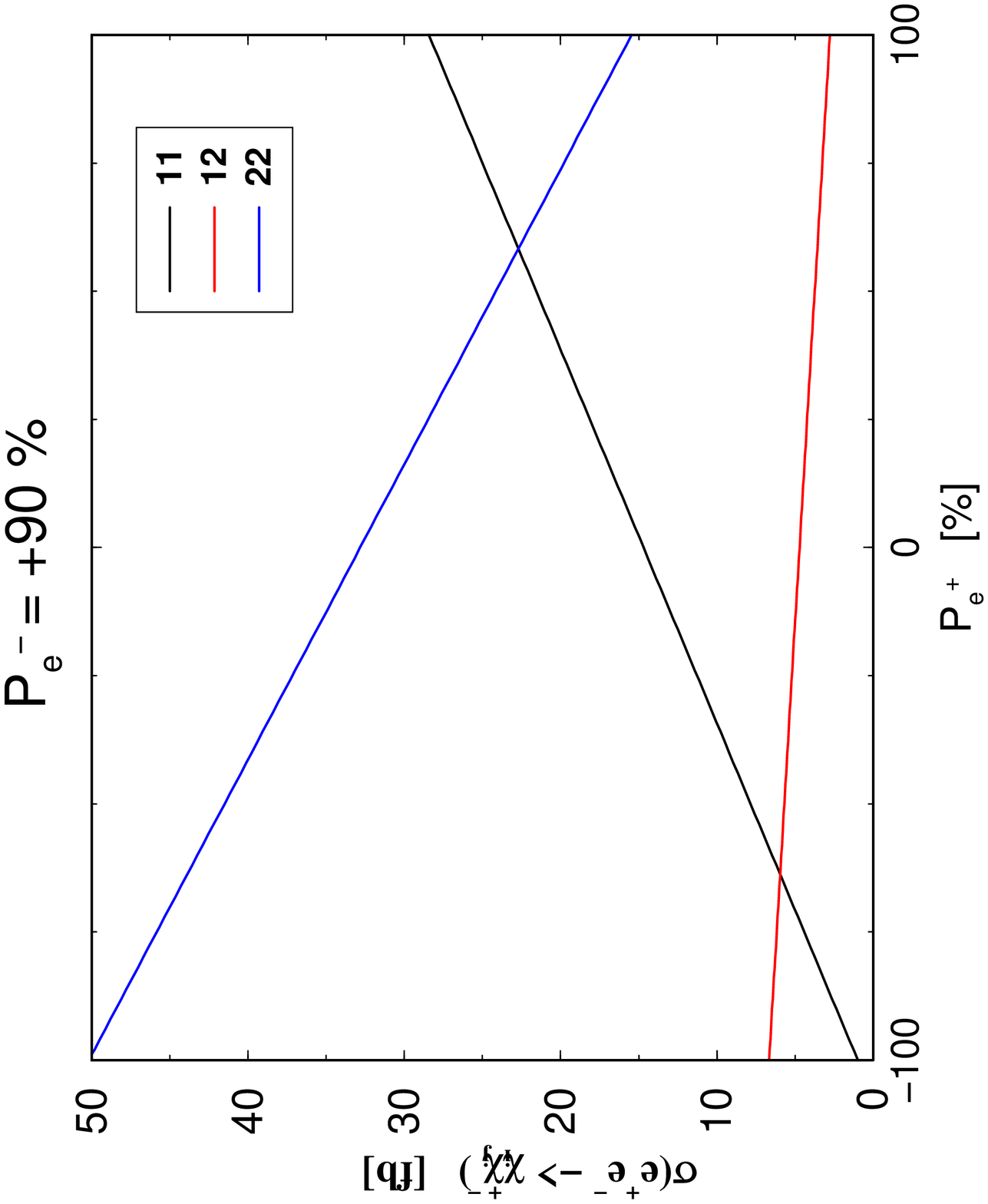}}
\end{picture}
\vspace{-.2cm}
\caption[Chargino production cross sections with polarized beams]
{Chargino production: cross sections for $e^+e^-\to \tilde{\chi}^-_i
\tilde{\chi}^+_j$ at $\sqrt{s}=1$ TeV for fixed electron polarization
$P_{e^-}= -90$\% (left) and $P_{e^-}= +90$\% (right) as a function of positron 
polarization~\cite{Jan}. 
The SUSY parameters
are chosen as in scenario S6, see table~\ref{tab-susyscenarios}.}
\label{fig:charg}
\end{figure}

\begin{sloppypar}
Beam polarization can significantly enhance the signal and therefore improve
the $S/\sqrt{B}$ ratio. In the case of chargino pair production, as can be
seen in the left panel of fig.~\ref{fig:charg} (for the scenario S6, 
see table~\ref{tab-susyscenarios}) 
the cross section for ($-90\%$, $+60\%$) is enhanced
by a factor of $1.6$ compared to the ($-90\%, 0$) case. For
right-polarized electrons the enhancement is much weaker in this scenario.  
Nevertheless,
polarized beams can be used to disentangle the chargino and sneutrino
parameters, exploiting the fact that the $\tilde{\nu}_e$ exchange affects only
the amplitude with left-chiral electrons (and right-chiral positrons).  If the
incoming beams were 100\% polarized, one could switch the $\tilde{\nu}_e$
exchange on and off, and analyse the chargino system alone or with
$\tilde{\nu}_e$ included, and thus perform an independent determination of
chargino and $\tilde{\nu}_e$ parameters. With realistic beam polarization, the
$\tilde{\nu}_e$ exchange affects all amplitudes depending on the degree of
polarization, as seen in fig.~\ref{fig:charg}. Consequently, the polarization
of both beams provides a unique analysing tool.
\end{sloppypar}

The case of neutralinos is even more interesting since the structure 
in the $t$- and $u$-channels is richer.  
As can be seen in fig.~\re{fig_susy2}a for S6, 
see table~\ref{tab-susyscenarios}, 
the cross sections for pair production are also enhanced by
a factor 1.6 for ($-90\%$, $+60\%$) with respect 
to the case ($-90\%$, $0$). For
right-polarized electrons, similar results are obtained, see e.g.\
$\sigma(\tilde{\chi}^0_3\tilde{\chi}^0_4)$, in fig.~\re{fig_susy2}b.  In this
configuration an even greater advantage of polarizing both beams with
different signs is the suppression of the dominant $WW$ background.
One should note, however, that the obtained
enhancement can strongly depend 
on the considered scenario, see table~\ref{tab_polneut}.

\vspace*{.5cm}                                                                   
\begin{figure}[htb]
\setlength{\unitlength}{1cm}
\begin{picture}(12,5)
\put(0.,-1){\mbox{\epsfysize=6.5cm\epsffile{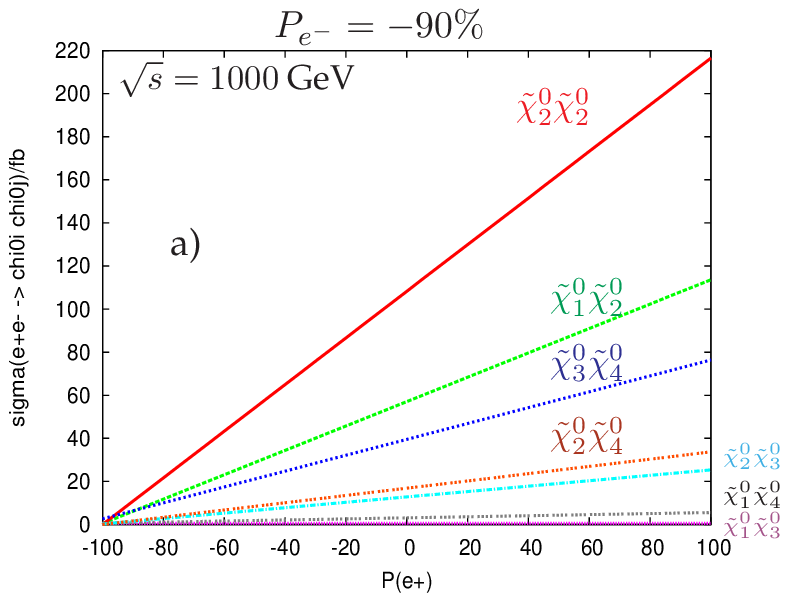}}
 \mbox{\epsfysize=6.5cm\epsffile{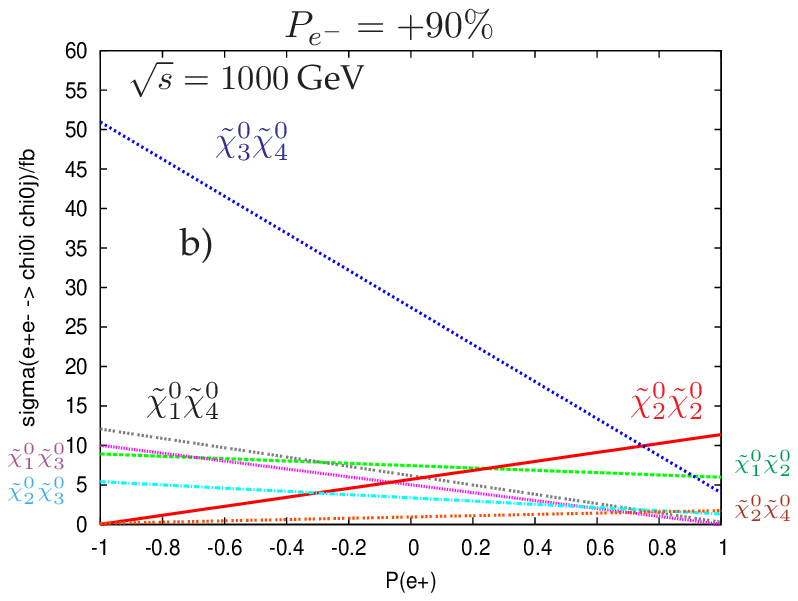}}}
\put(-8.5,-1.2){\mbox{\includegraphics[height=7cm,width=.5cm]{box}}}
\put(-.2,-.2){\mbox{\includegraphics[height=7cm,width=.8cm]{box}}}
\put(8.,.8){\mbox{\includegraphics[height=5.5cm,width=.8cm]{box}}}
\put(-.5,-.7){\mbox{\includegraphics[height=.4cm,width=10cm]{box}}}
\put(8.5,-.7){\mbox{\includegraphics[height=.4cm,width=10cm]{box}}}
\put(3.5,-.7){\small $P_{e^+} [\%]$}
\put(12,-.7){\small  $P_{e^+} [\%]$}
\put(-.5,4.5){\small $\sigma$ [fb]}
\put(7.9,4.5){\small $\sigma$ [fb]}
\end{picture}
\vspace{.5cm}
\caption[Neutralino production cross section with polarized
beams]{Neutralino production: Selected cross sections for $e^+ e^-\to
\tilde{\chi}^0_{i}\tilde{\chi}^0_{j}$ at $\sqrt{s}=1$~TeV in
the scenario S6, see table~\ref{tab-susyscenarios}, 
where the two lightest neutralinos are
gaugino-like and the two heavier ones are higgsino-like, shown for
fixed electron polarization and variable positron polarization,
a) $P_{e^-}=-90\%$, and b) $P_{e^-}=+90\%$~\cite{Gudi-neut}. \la{fig_susy2}}
\end{figure}

Particularly interesting is the behaviour of the
$\tilde{\chi}^0_2\tilde{\chi}^0_2$ cross section, which is enhanced by a
factor of 1.6 even for the exceptional configuration $(+90\%,+60\%)$ 
with respect 
to the case ($+90\%, 0)$. This effect is due to the lack of mixing,
the $\tilde{\chi}^0_2$ being nearly pure wino, and thus 
 not completely suppressed for partial polarization
$(+90\%,+60\%)$.  In the case of completely polarized beams, the RR cross
section for $\tilde{\chi}^0_2\tilde{\chi}^0_2$ would vanish, since---as
mentioned before---only a V-A interaction occurs in this process.

Neutralino production is very sensitive also to the mass of the exchanged
particles, $\tilde{e}_{\rm L}$ and 
$\tilde{e}_{\rm R}$.  The ordering of the cross
sections for different polarization configurations (with
$(|P_{e^-}|,|P_{e^+}|)=(80\%,60\%)$) depends on the character of
the neutralinos as well as on the masses of the exchanged particles:
\smallskip

\noindent
$\bullet$ Pure higgsino (only $s$-channel):\\
\bequ
\sigma_{-+}>\sigma_{+-}>\sigma_{-0}>\sigma_{00}>\sigma_{ +0}>\sigma_{--}>\sigma_{++}
\la{eq_susy1a}
\eequ
$\bullet$ Pure gaugino and $m_{\tilde{e}_{\rm L}}\gg m_{\tilde{e}_{\rm R}}$ 
(only $t$- and $u$-channels):\\
\bequ
\sigma_{+-}>\sigma_{+0}>\sigma_{00}>\sigma_{++}>\sigma_{--}>\sigma_{-0}>\sigma_{-+}
\la{eq_susy1b}
\eequ
$\bullet$ Pure gaugino and 
$m_{\tilde{e}_{\rm L}}\ll m_{\tilde{e}_{\rm R}}$ 
(only $t$- and $u$-channels):\\
\bequ
\sigma_{-+}>\sigma_{-0}>\sigma_{00}>\sigma_{--}>\sigma_{++}>
\sigma_{+0}>\sigma_{+-},
\la{eq_susy1c}
\eequ
with denotation as defined in eq.~(\ref{eq_intro1}), sect.\ref{subsec:long}.
If only the electron beam is polarized, then the orderings of 
the cross sections
for the cases (\re{eq_susy1a}) and (\re{eq_susy1c}) remain the same.
If, however, both beams are simultaneously polarized,
the three cases could be distinguished from one another~\cite{gudi2}.

If the slepton masses are known, the fundamental SUSY parameters,
$M_1$, $M_2$, $\mu$ and $\tan\beta$ can easily be determined from
measurements of polarized cross sections of the two lightest neutralinos 
and the lightest chargino~\cite{Choi:2001ww}. 
To give an example of the involved interplay between gaugino/higgsino 
mixing character,
the influence of the virtual slepton masses and the dependence of beam 
polarization,
polarized cross sections $\sigma(e^+e^-\to\tilde{\chi}^0_1\tilde{\chi}^0_2)$ 
for different $m_{\tilde{e}_{L,R}}$ are listed in table~\ref{tab_polneut}
for the SUSY parameters chosen as in scenario S7, 
see table~\ref{tab-susyscenarios}. 

Considering the large number of possible independent SUSY parameters, the
availability of different observables using the polarization of both beams is
a decisive tool for the kind of analysis described here.

\begin{table}[htb]
{\small
\begin{center}
\begin{tabular}{|c|c|cc|cc|}
\hline
($m_{\tilde{e}_{\rm L}}$, $m_{\tilde{e}_{\rm R}}$)&
$\sigma(e^+e^- \to \tilde{\chi}^0_1 \tilde{\chi}^0_2)$ &
\multicolumn{4}{c|}{Scaling factors: $\sigma^{\rm pol}/\sigma^{\rm unpol}$} \\
$[$GeV$]$ & [fb] unpolarized 
& $(-90\%,0\%)$ & $(+90\%,0\%)$ & $(-90\%,+60\%)$ & $(+90\%,-60\%)$ \\\hline
 (200, 200)   & 102 & 0.4 & 1.6 & 0.6 & 2.5 \\
 (500, 500)   &  29 & 0.5 & 1.5 & 0.7 & 2.4 \\
 (1000, 1000) & 7.4 & 0.7 & 1.3 & 1.1 & 2.0 \\
 (1500, 1500) & 4.0 & 0.9 & 1.1 & 1.3 & 1.8 \\
 (2000, 2000) & 2.9 & 1.0 & 1.0 & 1.5 & 1.6 \\
 (500, 1500)  & 8.7 & 1.4 & 0.6 & 2.3 & 0.8 \\
 (1500, 500)  &  24 & 0.2 & 1.8 & 0.3 & 2.8 \\
 \hline
\end{tabular}
\end{center} 
}
\caption[Polarized neutralino cross sections]{Polarized cross sections
$\sigma(e^+e^-\to \tilde{\chi}^0_1\tilde{\chi}^0_2)$ at $\sqrt{s}=500$~GeV
with all possible beam-polarization configurations for ($P_{e^-}$,
$P_{e^+}$)=($\mp 90\%$, $\pm 60\%$) and for different selectron masses
$m_{\tilde{e}_{L,R}}$, cf.\cite{Gudi-neut}. 
The SUSY parameters are given by scenario S7, 
table~\ref{tab-susyscenarios}. Cross sections
in the configuration RR and LL are $<1$~fb and not listed. \label{tab_polneut}}
\end{table}

{\bf \boldmath
Quantitative examples:
One can gain up to a factor of about 1.8 for the
production cross section with both beams polarized with respect 
to the case where only electrons are polarized.  To determine the 
fundamental parameters model-independently 
it is necessary to observe cross sections with differently
polarized beams. Parameters might be such that having only one beam polarized
would not unambiguously determine the parameters.}

%%%%%%%%%%%%%%%%%%%%%%%%%%%%%%%%%%%%%%%%%%%%%%%%%%%
\subsection{CP violation in neutralino production and decay
\label{sect:cpasy}}
%%%%%%%%%%%%%%%%%%%%%%%%%%%%%%%%%%%%%%%%%%%%%%%%%%%
{\bf \boldmath 
With both beams polarized, the CP-violating effects can be enhanced by a 
considerable factor, with respect to the case  of only electrons polarized.}
\smallskip

Particularly interesting in supersymmetry is the study of new CP-violating 
sources. In the following, suitable observables for
exploring CP violation in the gaugino/higgsino sector are shown.

The 
interaction Lagrangian for the neutralino and chargino sectors of the MSSM, 
see e.g.\ \cite{Bartl:2003tr},
 allows for some  SUSY parameters to be complex: 
the gaugino mass parameter $M_1$ ($M_2$ is assumed to be real), the higgsino
mass parameter $\mu$, and ---involving third generation sfermions--- the 
trilinear coupling parameter $A_{\tau}$ in the
stau sector,  
\begin{equation}
\mu = |\mu| \, e^{ i\,\varphi_{\mu}},\quad 
M_1 =|M_1| \, e^{ i\,\varphi_{M_1}},\quad
A_{\tau} = |A_{\tau}| \, e^{i\,\varphi_{A_{\tau}}}.
\end{equation}
The physical phases 
can cause large CP-violating effects already at tree level
\cite{oshimo,Choi:1999cc,Bartl:2003tr,Bartl:2004ut,Bartl:2004vi,Kittel:2004kd,Kittel:2005rp,Bartl:2003gr,
Choi:2003pq,Bartl:2004jj,charginopaper}.  In the following, the
dependence on the beam polarization of these CP-violating effects in
neutralino production and decay is reviewed \cite{Bartl:2003kn}.

In the following neutralino production
\begin{eqnarray} \label{production}
   e^+ + e^-&\to&
	\tilde{\chi}^0_2+\tilde{\chi}^0_j, \quad j=1,3, 
\end{eqnarray}
with either leptonic two-body decays 
\begin{eqnarray} \label{decay_1}
   \tilde{\chi}^0_2&\to& \tilde{\ell} + \ell_1 \quad\mbox{and}\quad 
\tilde{\ell}\to\tilde{\chi}^0_1+ \ell_2   
   \end{eqnarray}
or the corresponding leptonic three-body decays
\begin{equation}
\tilde{\chi}^0_2\longrightarrow\tilde{\chi}^0_1 + \ell_1 + \ell_2,
\label{decay_2}
\end{equation}
with $\ell_{1,2}= e,\mu,\tau$ is studied.
The spin correlations between
production 
and decays
of the neutralinos
lead to CP-violating effects 
already at tree level and allow the definition of several
CP-odd asymmetries. In the following only the decays of the $\tilde{\chi}^0_2$ are studied.

It is convenient to introduce the triple product $ {\mathcal T} = ({\bf
p}_{e^-} \times {\bf p}_{\ell_2})\cdot {\bf p}_{\ell_1}$, where ${\bf
p}_{e^-}$, ${\bf p}_{\ell_1}$ and ${\bf p}_{\ell_2}$ are the 
momenta of the initial 
$e^-$ beam and the two final leptons $\ell_1$ and $\ell_2$, respectively. A
T-odd asymmetry of the cross section $\sigma$ for the processes
(\ref{production})--(\ref{decay_2}) can then be defined as:
\begin{equation} \label{Tasymmetry}
{\mathcal A}_{\rm T} = 
 \frac{\sigma({\mathcal T}>0) - \sigma({\mathcal T}<0)}
    {\sigma({\mathcal T}>0) + \sigma({\mathcal T}<0)} =
 \frac{{\int} \mathrm{sign} \{ \mathcal{T} \}
  |{\cal M}|^2 d\mathrm{Lips}} {{\int} |{\cal M}|^2 d\mathrm{Lips}}. 
\end{equation}
If final-state phases and finite width effects of exchanged particles are neglected,
${\mathcal A}_{\rm T}$ is CP-violating (CP-odd)
due to CPT invariance.
$\mathcal{A}_\mathrm{T}$ is proportional to the difference between the number of
events with the final lepton $\ell_1$ above 
the plane spanned
by ${\bf p}_{e^-}$ and ${\bf p}_{\ell_2}$
and those below it.
The dependence of  ${\mathcal A}_{\rm T}$ on 
$\varphi_{M_1}$ and  $\varphi_{\mu}$ has been analysed 
in \cite{oshimo,Choi:1999cc,Bartl:2003tr,Kittel:2005rp,Bartl:2004jj}, see sections a) and 
b) below.

When the neutralino decays into a $\tau$ lepton,
\begin{equation}
\tilde{\chi}^0_i\to\tilde{\tau}_k^{\pm} \tau^{\mp},\quad k=1,2,
\label{eq_taudecay}
\end{equation}
the transverse $\tau^-$ and $\tau^+$
polarizations $P_{\tau^-}$ and $P_{\tau^+}$, perpendicular to the
plane formed  by the $\tau$ and $e^-$ momenta, are T-odd and
give rise to the CP-odd observable 
\begin{eqnarray} \label{ACP}
{\mathcal A}_{\rm CP}=\frac{1}{2}(P_{\tau^-}-P_{\tau^+}),
\end{eqnarray}
which is also sensitive to $\varphi_{A_{\tau}}$.  ${\mathcal A}_{\rm
CP}$ has been discussed in \cite{Kittel:2005rp,Bartl:2003gr} for
several MSSM scenarios.  In order to measure the asymmetries
${\mathcal A}_{\rm T}$ and ${\mathcal A}_{\rm CP}$ in the leptonic
two-body decays of the neutralino, eq.~(\ref{decay_1}), the lepton
$\ell_1$ from the neutralino decay and the lepton $\ell_2$ from the
$\tilde{\ell}$ decay, have to be distinguished.  have to be
distinguished.  This can be accomplished by measuring the different
energy spectra of the $\tau$s~\cite{Bartl:2003tr}.

To measure the asymmetries
${\mathcal A}_{\rm T}$ and ${\mathcal A}_{\rm CP}$,
it is important that
both these asymmetries and the cross
sections $\sigma$ are large.
In the following,
the impact of  longitudinally-polarized $e^+$ and $e^-$ beams of a linear 
collider in the 500 GeV range 
on the measurement of ${\mathcal A}_{\rm T}$, ${\mathcal A}_{\rm CP}$
and $\sigma$ is discussed.
\medskip

%%%%%%%%%%%%%%%%%%%%%%%%%%%%%%%%
\subsubsection*{a) Two-body decays}

\begin{figure}[htb]
\setlength{\unitlength}{1cm}
\begin{picture}(10,8)(0,0)
	\put(0,0){\includegraphics{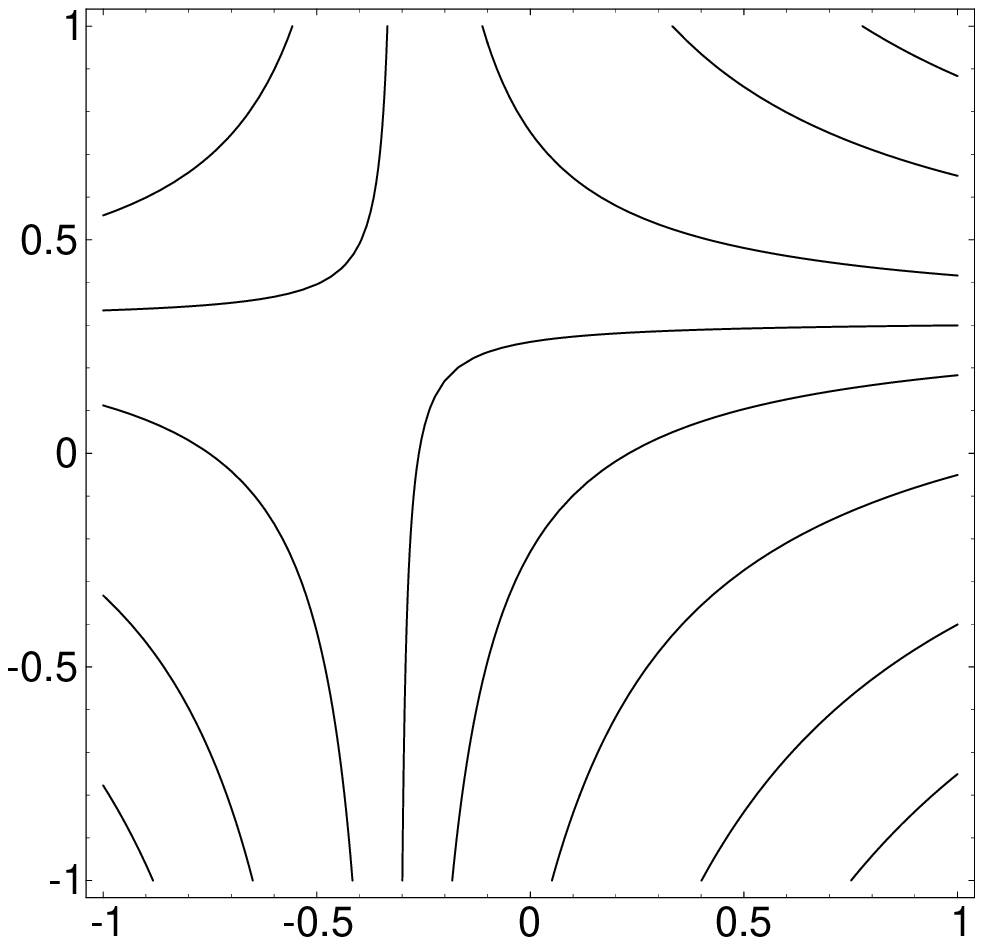}}
	\put(1.5,7.5){\fbox{$\sigma(e^+\,e^- \to\tilde{\chi}^0_1 
				\tilde{\chi}^0_1 \ell_1 \ell_2 )$ in fb}}
	\put(6.5,-.3){$P_{e^-}$}
	\put(0.2,7.3){$P_{e^+}$}
	\put(1.2,1.2){4}
	\put(1.5,2.1){12}
	\put(2.0,3.6){20}
	\put(2.2,5.2){24}
	\put(3.5,3.9){24}
	\put(1.3,6.1){28}
	\put(4.3,3.1){28}
	\put(5.2,2.3){36}
	\put(6.,1.5){48}
	\put(6.6,0.8){60}
	\put(3.8,5.3){20}
	\put(5.2,5.8){12}
	\put(6.3,6.3){4}
   \put(8,0){\includegraphics{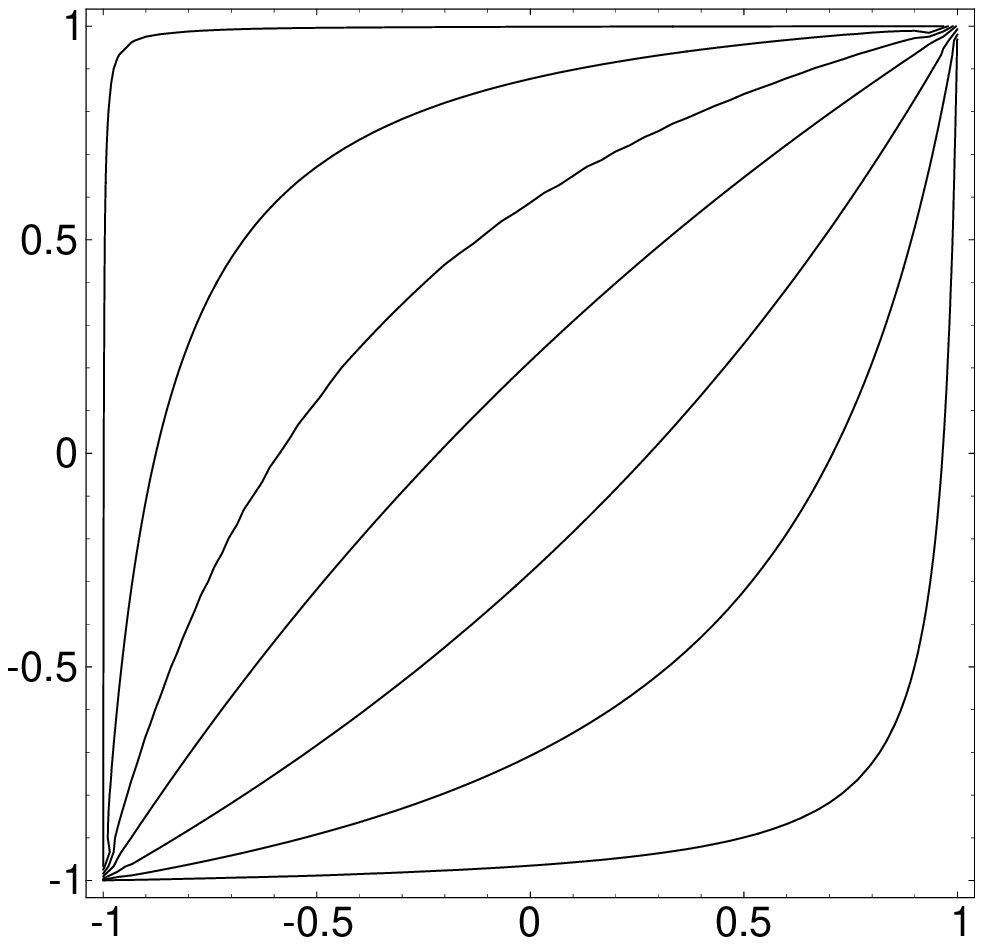}}
	\put(11.,7.5){\fbox{${\mathcal A}_{\rm T}$ in \% }}
	\put(14.5,-.3){$P_{e^-}$}
	\put(8.2,7.3){$P_{e^+}$}
	\put(9.2,6.3){-4.5}
	\put(10.2,5.2){-3}
	\put(11.0,4.4){0}
	\put(11.6,3.8){3}
	\put(12.5,3){6}
	\put(13.2,2.1){8}
	\put(14.5,1.){9}
\end{picture}
\vspace*{.5cm}
\caption[T-odd asymmetry for neutralino two-body decays]{ Contour
lines of $\sigma$ and ${\mathcal A}_{\rm T}$ in neutralino production,
$e^+e^-\to\tilde{\chi}^0_1\tilde{\chi}^0_2$ and the subsequent decay,
eq.~(\ref{decay_1}) for scenario S8, table~\ref{tab-susyscenarios},
with the CP-violating phase $\varphi_{M_1}=0.2
\pi$~\cite{Bartl:2003kn}.
\label{plot_1}}
\end{figure}

\begin{figure}[htb]
\setlength{\unitlength}{1cm}
\begin{picture}(10,8)(0,0)
	\put(0,0){\includegraphics{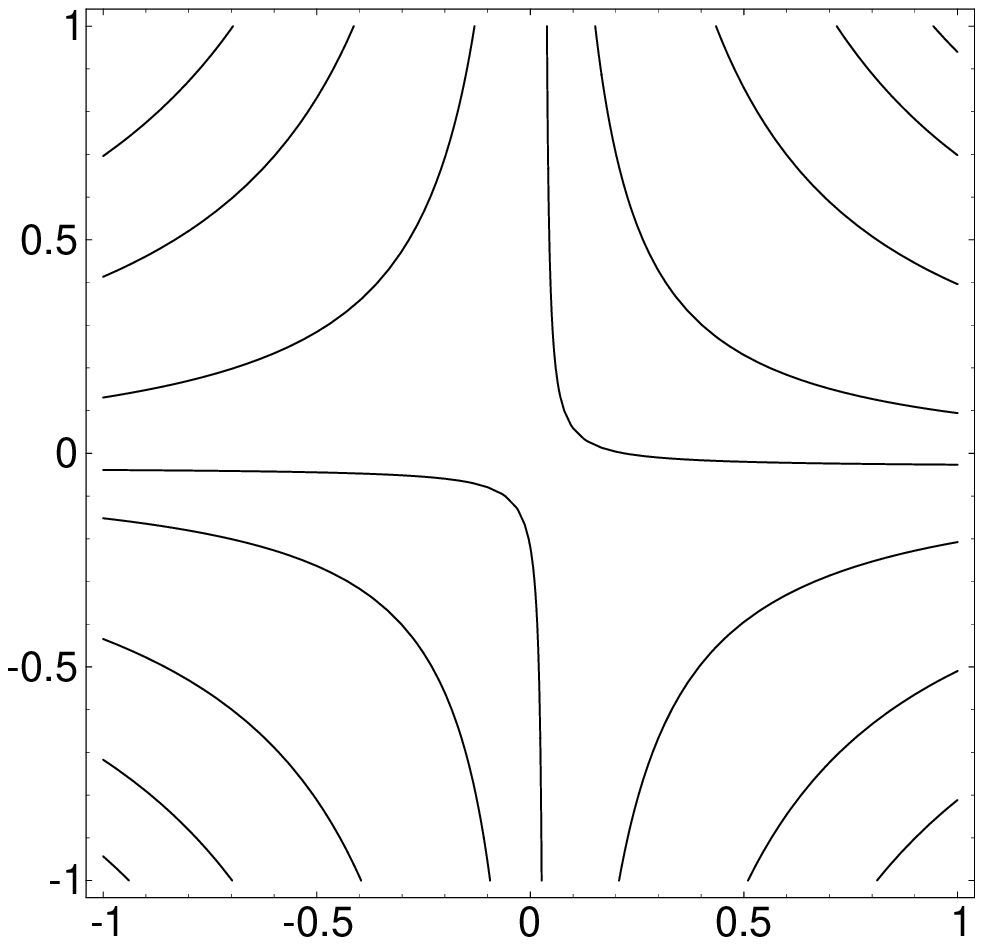}}
	\put(1.5,7.5){\fbox{$\sigma(e^+\,e^- \to\tilde{\chi}^0_1 
				\tilde\tau^+_1 \tau^-)$ in fb}}
	\put(6.5,-.3){$P_{e^-}$}
	\put(0.2,7.3){$P_{e^+}$}
	\put(1.1,0.8){1}
	\put(1.6,1.2){5}
	\put(2.2,1.8){10}
	\put(3.0,2.6){15}
	\put(3.1,3.7){17}
	\put(3.,4.7){20}
	\put(2.2,5.5){25}
	\put(1.5,6.0){30}
		\put(5.,3.3){17}
		\put(4.8,4.3){15}
		\put(5.6,5.2){10}
		\put(6.3,5.9){5}
		\put(6.7,6.4){1}
		\put(5.3,2.1){20}
	\put(6.1,1.3){25}
	\put(6.6,0.7){30}
   \put(8,0){\includegraphics{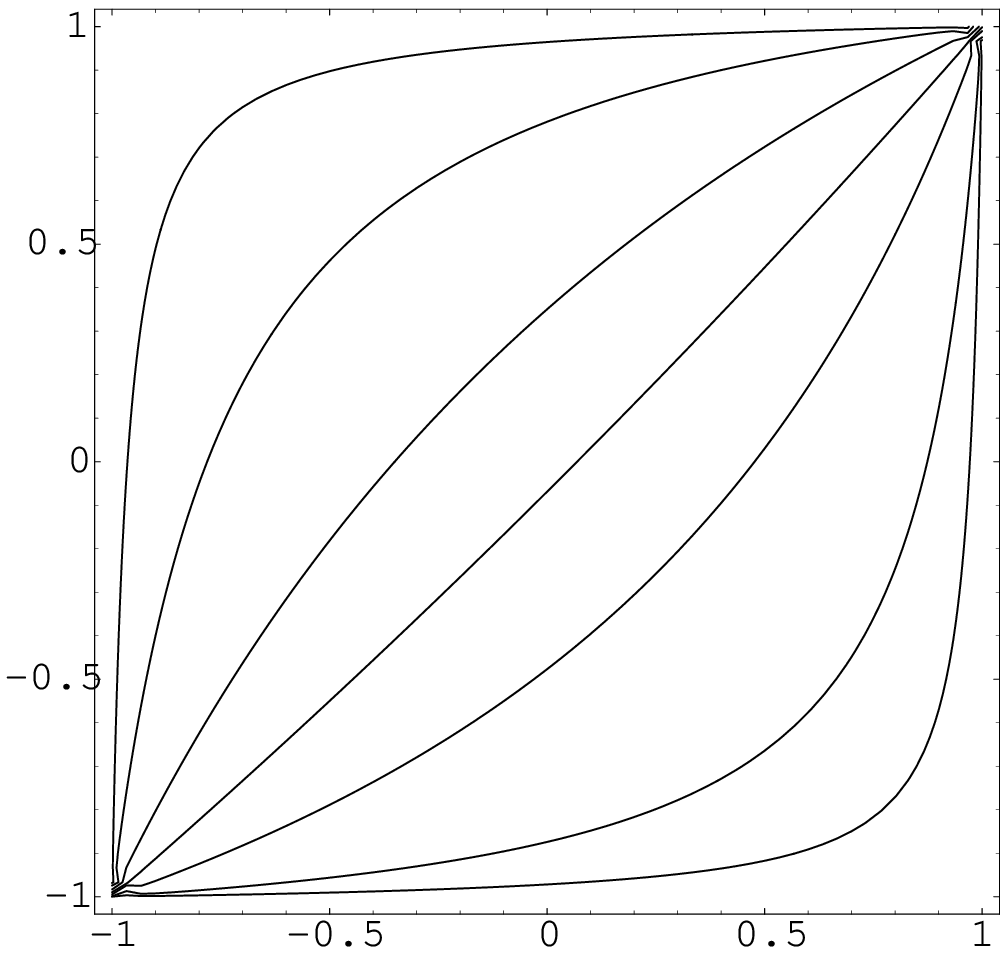}}
	\put(11.,7.5){\fbox{${\mathcal A}_{\rm CP}$ in \% }}
	\put(14.5,-.3){$P_{e^-}$}
	\put(8.2,7.3){$P_{e^+}$}
	\put(9.,6.4){-14.5}
	\put(10.,5.4){-12}
	\put(11.0,4.4){-6}
	\put(11.8,3.7){0}
	\put(12.5,3){6}
	\put(13.3,2.1){12}
	\put(14.3,.8){13.5}
\end{picture}
%}
\vspace*{.5cm}
\caption[CP-odd asymmetry for neutralino decays into tau leptons (I)]{
Contour lines of $\sigma$ and ${\mathcal A}_{\rm CP}$ of the process,
$e^+e^-\to\tilde{\chi}^0_1\tilde{\chi}^0_2$, with subsequent decay,
eq.~(\ref{eq_taudecay}) for the scenario S9,
table~\ref{tab-susyscenarios}, with the CP-violating phase
$\varphi_{A_{\tau}}=\pi/2$~\cite{Bartl:2003kn}.
\label{plot_2}}
\end{figure}

\begin{figure}[htb]
\setlength{\unitlength}{1cm}
\begin{picture}(10,8)(0,0)
	\put(0,0){\includegraphics{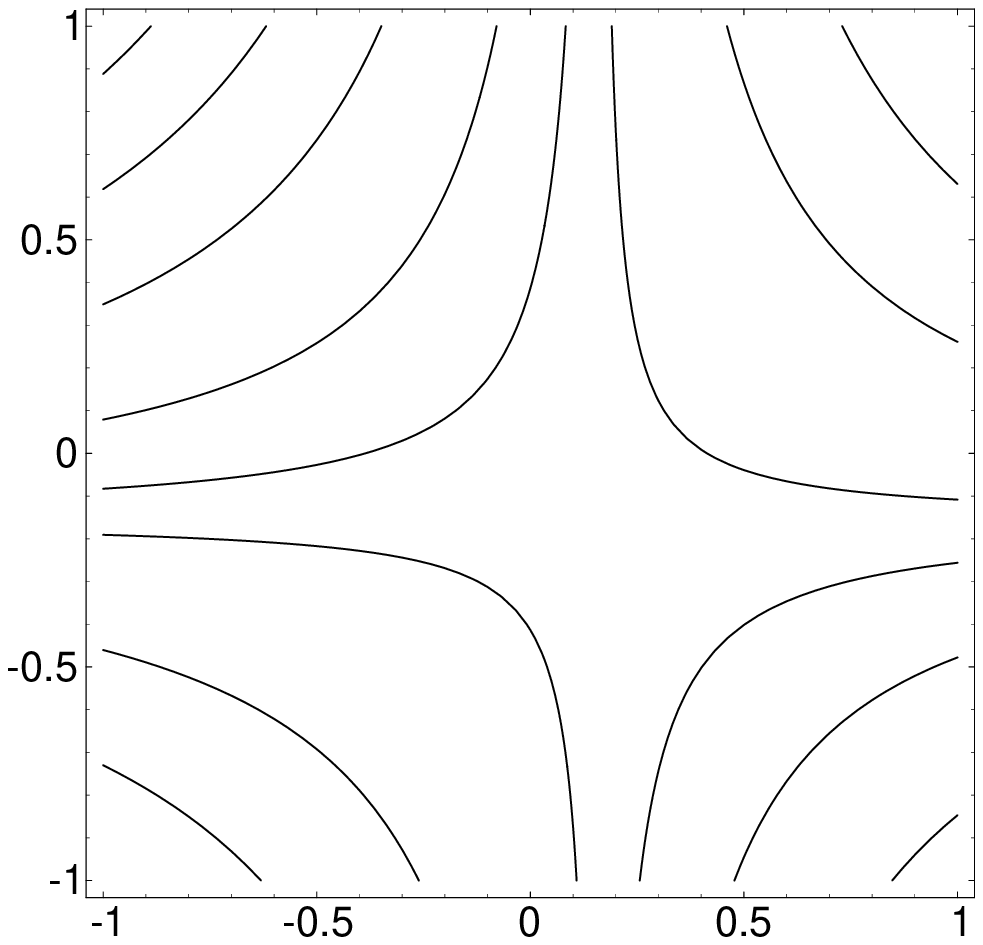}}
	\put(1.5,7.5){\fbox{$\sigma(e^+\,e^- \to\tilde{\chi}^0_1 
				\tilde\tau^+_1 \tau^-)$ in fb}}
	\put(6.5,-.3){$P_{e^-}$}
	\put(0.2,7.3){$P_{e^+}$}
	\put(1.6,1.2){5}
	\put(2.4,1.7){10}
	\put(3.4,2.1){15}
	\put(3.6,3.8){17}
	\put(2.9,4.6){20}
	\put(2.1,5.2){25}
	\put(1.5,5.7){30}
	\put(1.1,6.3){35}
		\put(4.9,4.3){15}
		\put(5.6,5.){10}
		\put(6.3,5.9){5}
		\put(5.4,2.1){17}
	\put(6.1,1.4){20}
	\put(6.7,0.7){25}
   \put(8,0){\includegraphics{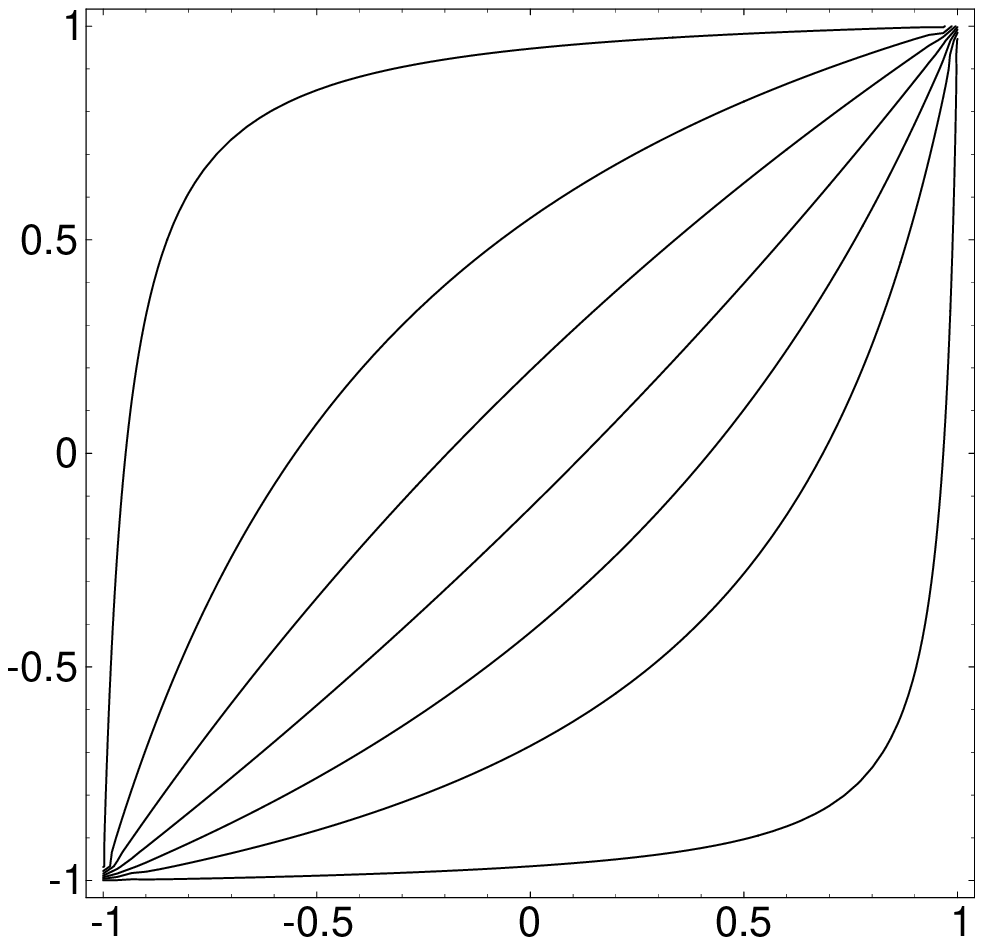}}
	\put(11.,7.5){\fbox{${\mathcal A}_{\rm CP}$ in \% }}
	\put(14.5,-.3){$P_{e^-}$}
	\put(8.2,7.3){$P_{e^+}$}
		\put(9.4,6.1){-12}
	\put(10.9,5.){-9}
	\put(11.4,4.3){-6}
	\put(11.9,3.7){-3}
	\put(12.5,3.2){0}
	\put(13.3,2.2){3}
	\put(14.4,.9){6.5}
\end{picture}
%}
\vspace*{.5cm}
\caption[CP-odd asymmetry for neutralino decays into tau leptons (II)]{
Contour lines of $\sigma$ and ${\mathcal A}_{\rm CP}$ 
of the process, $e^+e^-\to\tilde{\chi}^0_1\tilde{\chi}^0_2$, with subsequent
decay, eq.~(\ref{eq_taudecay})
for the scenario S10, table~\ref{tab-susyscenarios}, with the 
CP-violating phase 
$\varphi_{M_1}=0.2 \pi$~\cite{Bartl:2003kn}.
\label{plot_3}}
\end{figure}

The above asymmetries ${\mathcal A}_{\rm T}$ and ${\mathcal A}_{\rm CP}$
are shown in figs.~\ref{plot_1}--\ref{plot_3} 
for $e^+e^-\to \tilde{\chi}^0_1\tilde{\chi}^0_2$ at
$\sqrt{s}=500$~GeV, with the decays (\ref{decay_1}) and (\ref{eq_taudecay}). 

In fig.~\ref{plot_1} the dependence
of the cross sections 
$\sigma(e^+e^-\to\tilde{\chi}^0_1\tilde{\chi}^0_2) \times
{\rm BR}(\tilde \chi^0_2\to\tilde\ell_R\ell_1)\times
{\rm BR}(\tilde\ell_R\to\tilde\chi^0_1\ell_2)$
and of ${\mathcal A}_{\rm T}$ on the beam polarizations  
is shown \cite{Bartl:2003tr,Bartl:2003kn}.  The contributing
CP-violating phases in that process with $\ell_{1,2}\neq \tau$ are $\varphi_{M_1}$ and
$\varphi_{\mu}$. Since the experimental results from the
EDM~\cite{Altarev:1992cf} put stringent bounds particularly on
$\varphi_{\mu}$, we choose only $\varphi_{M_1}\neq 0$. The SUSY
parameters are chosen according to scenario S8,
table~\ref{tab-susyscenarios}, and lead to $\mathrm{BR}(\tilde
\chi^0_2\to\tilde\ell_R\ell_1) = 0.63$ (summed over both signs of
electric charge) and $\mathrm{BR}(\tilde\ell_R \to
\tilde\chi^0_1\ell_2) = 1$.  Note that both ${\mathcal A}_{\rm T}$ and
$\sigma$ are considerably enhanced for positive electron and negative
positron polarizations.

This choice of polarization enhances the contributions
of the right-slepton exchange in the neutralino production,
eq.~(\ref{production}), and reduces that of the left 
slepton exchange \cite{gudi1,gudi2}. 
While the contributions of right and left slepton exchanges 
enter $\sigma$ with the 
same signs, they enter
${\mathcal A}_{\rm T}$ with  opposite signs, which accounts for
the sign change of ${\mathcal A}_{\rm T}$ in fig.~\ref{plot_1}.

In figs.~\ref{plot_2} and \ref{plot_3}, contour lines of  the $\tau$
polarization asymmetry ${\mathcal A}_{\rm CP}$, eq.~(\ref{ACP}), are shown. 
Since in those cases the decay into $\tau$s is studied one additional 
CP-violating phase contributes, $\varphi_{A_{\tau}}$.
Since the asymmetry ${\mathcal A}_{\rm CP}$ is very sensitive to
both phases $\varphi_{A_{\tau}}$ and $\varphi_{M_1}$  the dependence
of  
$\sigma= 
\sigma(e^+e^-\to\tilde\chi^0_1\tilde\chi^0_2 ) \times
{\rm BR}(\tilde \chi^0_2\to\tilde\tau_1^+\tau^-)$ and ${\mathcal A}_{\rm CP}$
on the beam polarization is shown  
in fig.~\ref{plot_2} for the case $\varphi_{A_{\tau}}=\pi/2$ in scenario S9,
and  in fig.~\ref{plot_3}
for $\varphi_{M_1}=0.2\pi$ in scenario S10, see table~\ref{tab-susyscenarios}.

In scenario S9
a large value of $|A_{\tau}|=1500$ GeV has been chosen because ${\mathcal
A}_{\rm CP}$ increases with increasing $|A_{\tau}| \gg |\mu|\tan\beta$
\cite{Bartl:2003gr,Bartl:2003kn}.  For unpolarized beams the asymmetry is 1\%. 
If only the electron beam is polarized, $(P_{e^-},P_{e^+})=(+80\%,0)$, 
the asymmetry reaches values of about $\pm 10\%$. 
It achieves
values larger than $\pm 13\%$ if both beams are
polarized with opposite signs. 
Furthermore, if both beams are polarized,
one also gains in statistics.
The cross section $\sigma=
\sigma(e^+e^-\to\tilde\chi^0_1\tilde\chi^0_2 ) \times
{\rm BR}(\tilde \chi^0_2\to\tilde\tau_1^+\tau^-)$, shown
in fig.~\ref{plot_2} with 
${\rm BR}(\tilde \chi^0_2\to\tilde\tau_1^+\tau^-)=0.22$,
is very sensitive to the 
beam polarization and varies between 1 fb and 30 fb.

In scenario S10 the phase $\varphi_{M_1}\neq 0$ is considered, which
influences the neutralino $\tilde{\chi}^0_{1,2}$ masses and couplings.
Therefore CP-violating effects occur both in the production as well as
in the decay process.  The neutralino branching ratio is ${\rm
BR}(\tilde\chi^0_2\to\tilde\tau_1^+\tau^-)=0.19$ for the considered
scenario.  Despite the small phases, ${\mathcal A}_{\rm CP}$ reaches
values up to $-12\%$ for negative $e^-$ and positive $e^+$ beam
polarizations.

\medskip

%%%%%%%%%%%%%%%%%%%%%%%%%%%%%%%%%%%%%%%%
\subsubsection*{b) Three-body decays}

In this subsection 
the influence of longitudinal 
$e^-$ and $e^+$ beam polarizations
on the T-odd asymmetry ${\mathcal A}_{\rm T}$, eq.~(\ref{Tasymmetry}),
is analysed~\cite{Bartl:2004jj}.
The centre-of-mass energy is $\sqrt{s}=500$ GeV and
the phases are chosen as 
$\varphi_{M_1}=0.2\pi$ and $\varphi_{\mu}=0$.
\setlength{\unitlength}{1cm}
\begin{figure}[ht!]
\begin{picture}(17,10)
\put(0.5,0){\mbox{\epsfig{file=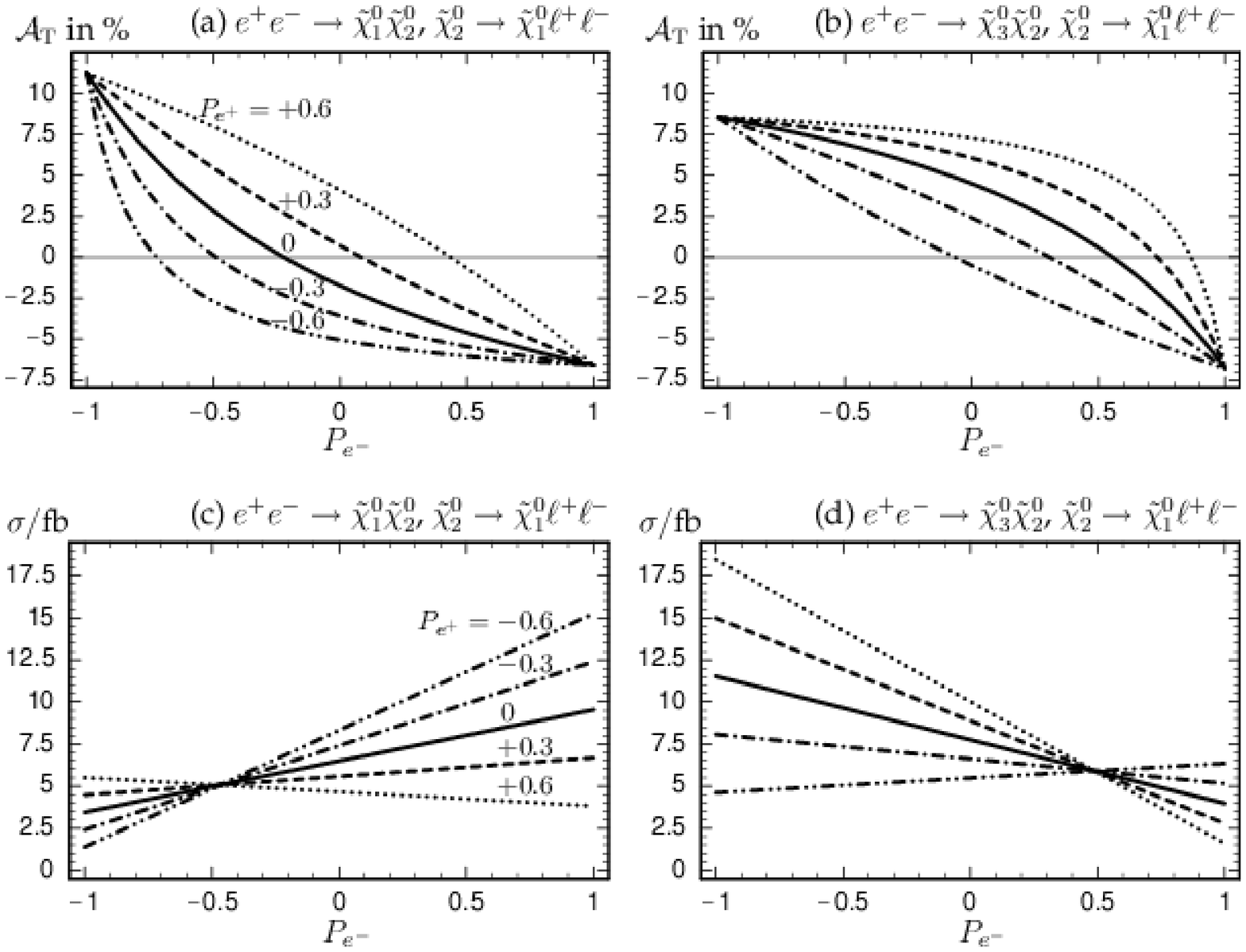,width=15cm,height=11cm}}}
\end{picture}
\caption[T-odd asymmetry for neutralino three-body decays]
{\label{fig:ToddNeutralino3body}
(a), (b): The T-odd asymmetry ${\mathcal A}_{\rm T}$, eq.~(\ref{Tasymmetry}),
for $e^+ e^- \to \tilde{\chi}^0_i \tilde{\chi}^0_2$, (a) $i=1$ and (b) $i=3$,
with subsequent leptonic three-body decay $\tilde{\chi}^0_2 \to
\tilde{\chi}^0_1 \ell^+ \ell^-$, and (c), (d) the corresponding cross sections
$\sigma=\sigma(e^+ e^- \to \tilde{\chi}^0_i \tilde{\chi}^0_2) \times
\mathrm{BR}(\tilde{\chi}^0_2 \to \tilde{\chi}^0_1 \ell^+ \ell^-)$, (c) $i=1$
and (d) $i=3$, summed over $\ell = e,\mu$, as a function of the $e^-$ beam
polarization $P_{e^-}$ for different $e^+$ beam polarizations $P_{e^+}$ and
$\sqrt{s}=500$ GeV in the scenario S11, table~\ref{tab-susyscenarios}~\cite{Bartl:2004jj}.}
\end{figure}

In figs.~\ref{fig:ToddNeutralino3body}(a) and (b) ${\mathcal A}_{\rm
T}$ is shown as a function of $P_{e^-}$ for different
$e^+$ beam polarizations, $-60\% \le P_{e^+} \le +60\%$.
The chosen SUSY parameters, cf.\ S11 in table~\ref{tab-susyscenarios}, 
lead to a large
mixing between the gaugino and higgsino components of $\tilde{\chi}^0_1$,
$\tilde{\chi}^0_2$ and $\tilde{\chi}^0_3$\cite{Bartl:2004jj}.  
For
both production processes in fig.~\ref{fig:ToddNeutralino3body}
${\mathcal A}_{\rm T}$ is positive (negative) for
$P_{e^-} \to -1$ ($P_{e^-} \to +1$).  For these polarizations the
$\tilde{e}_L$ ($\tilde{e}_R$) contributions to the neutralino
spin density matrix 
are dominant.  For $(P_{e^-},P_{e^+}) = (-90\%,0)$ and $(+90\%,0)$ 
the result is ${\mathcal A}_{\rm T} = +9.0\,\%$ and $-6.2\,\%$ 
for $\tilde{\chi}^0_1 \tilde{\chi}^0_2$ production and ${\mathcal
A}_{\rm T} = +8.3\,\%$ and $-4.8\,\%$ for
$\tilde{\chi}^0_3 \tilde{\chi}^0_2$ production, respectively.  Additional
positron beam polarization of opposite sign only slightly enhances these
asymmetries.  However, it considerably enhances the corresponding cross
sections, hence the
expected rates to measure ${\mathcal A}_{\rm T}$ by a factor of about $1.5$,
see figs.~\ref{fig:ToddNeutralino3body}(c) and (d).

The relative statistical error on ${\mathcal A}_{\rm T}$ is given by
$\delta {\mathcal A}_{\rm T} = 
 \Delta {\mathcal A}_{\rm T}/{\mathcal A}_{\rm T} =
 {\cal N}_{\sigma}/({\mathcal A}_{\rm T}\sqrt{N})$ with 
 ${\cal N}_{\sigma}$, the number of standard deviations and $N = \sigma\mathcal{L}_{\rm int}$, the 
number of events for a total integrated
luminosity $\mathcal{L}_{\rm int}$.
Assuming $\delta {\mathcal A}_{\rm T} \approx 1$ for 
${\mathcal A}_{\rm T}$ to be measurable
it follows that
${\cal N}_{\sigma} \approx |{\mathcal A}_{\rm T}| \sqrt{\sigma \mathcal{L}}$
\cite{Bartl:2003tr}.
In table~\ref{tab:ToddNeutralino3body},  ${\mathcal A}_{\rm T}$,
$\sigma$ and the corresponding standard deviations ${\cal N}_{\sigma}$ of
${\mathcal A}_{\rm T}$ are given for $\mathcal{L}_{\rm int} = 500~\textrm{fb}^{-1}$
and several sets of beam polarizations. 
With $(|P_{e^-}|,|P_{e^+}|) = (90\%,60\%)$
the number of standard deviations 
${\cal N}_{\sigma}$ increases by up to a factor 1.4
for $\tilde{\chi}^0_1 \tilde{\chi}^0_2$ production 
and by up to a factor 1.6 for
$\tilde{\chi}^0_3 \tilde{\chi}^0_2$ production in scenario S11, table~\ref{tab-susyscenarios},
in comparison to only polarized electrons.

\begin{table}[htb]
{\small
\renewcommand{\arraystretch}{1.1}
\begin{tabular}{|l||c|c|c|c|c||c|c|c|c|c|}
\hline
 & \multicolumn{5}{c||}{$e^+ e^- \to \tilde{\chi}^0_1 \tilde{\chi}^0_2$,
     $\tilde{\chi}^0_2 \to \tilde{\chi}^0_1 \ell^+ \ell^-$}
 & \multicolumn{5}{c|}{$e^+ e^- \to \tilde{\chi}^0_3 \tilde{\chi}^0_2$,
     $\tilde{\chi}^0_2 \to \tilde{\chi}^0_1 \ell^+ \ell^-$}
 \\ \hline
$(P_{e^-},P_{e^+})$ & 
  $(0,0)$ & $(-,0)$ & $(-,+)$ & $(+,0)$ & $(+,-)$ &
  $(0,0)$ & $(-,0)$ & $(-,+)$ & $(+,0)$ & $(+,-)$ \\ \hline
${\mathcal A}_{\rm T}$ [\%] & 
   $-1.8$ & $9.0$   & $10.6$  & $-6.2$  & $-6.5$
 & $4.5$  & $8.3$   & $8.5$   & $-4.8$  & $-6.3$ \\
$\sigma$ [fb]
  & $6.5$ & $3.7$   & $5.4$   & $9.2$  & $14.6$
  & $7.8$ & $11.2$  & $17.6$  & $4.3$  & $6.3$   \\
${\cal N}_{\sigma}$         
  & $1.0$ & $3.9$   & $5.5$   & $4.2$  & $5.6$
  & $2.8$ & $6.2$   & $8.0$   & $2.2$  & $3.5$   \\
\hline
\end{tabular}
}
\caption[T-odd asymmetry for neutralino three-body decays]
{\label{tab:ToddNeutralino3body}
The T-odd asymmetries ${\mathcal A}_{\rm T}$, eq.~(\ref{Tasymmetry}),
are listed for $e^+ e^- \to \tilde{\chi}^0_i \tilde{\chi}^0_2$, $i=1,3$,
with subsequent leptonic three-body decay
$\tilde{\chi}^0_2 \to \tilde{\chi}^0_1 \ell^+ \ell^-$.
 The corresponding cross section
  $\sigma = \sigma(e^+ e^- \to \tilde{\chi}^0_i \tilde{\chi}^0_2) \times 
   \mathrm{BR}(\tilde{\chi}^0_2 \to \tilde{\chi}^0_1 \ell^+ \ell^-)$,
summed over $\ell = e,\mu$,
and the standard deviations
${\cal N}_{\sigma}= |{\mathcal A}_{\rm T}| \sqrt{\sigma \mathcal{L}_{\rm int}}$
are given for $\Lumint = 500~\textrm{fb}^{-1}$, $\sqrt{s}=500$~GeV and
several beam polarizations $(P_{e^-},P_{e^+})=(\pm 90\%, \pm 60\%)$
in the scenario S11, table~\ref{tab-susyscenarios}~\cite{Bartl:2004jj}.
}
\end{table}

{\bf \boldmath Quantitative examples: In neutralino production with
subsequent two-body and three-body decays the cross sections are
enhanced by a factor of about 1.5 with both beams polarized compared
with only polarized electrons.  The CP-odd and T-odd asymmetries are
enhanced by up to a factor of about 1.3 and their measurability 
by up to a factor of about 1.6 with both
beams polarized compared with the case of only polarized electrons.}

%%%%%%%%%%%%%%%%%%%%%%%%%%%%%%%%%%%%%%%%%%%%%%%%%%%
\subsection{SUSY CP asymmetries with transversely-polarized beams}
%%%%%%%%%%%%%%%%%%%%%%%%%%%%%%%%%%%%%%%%%%%%%%%%%%%
{\bf \boldmath
Only if both beams are polarized, it is possible to exploit 
transversely-polarized $e^-$ and $e^+$ beams 
in chargino/neutralino production and decay. 
Both CP-even and CP-odd azimuthal asymmetries can be 
constructed which give access to determine
CP-violating phases in a broad range of the MSSM parameter space.}
\smallskip

Transversely-polarized beams offer the possibility of detailed studies 
of the effects of CP violation. They
give access to azimuthal asymmetries that can be
defined directly in terms of products of final particle momenta, without the
need to measure final-state polarizations. For $(V,A)$ interactions, due
to the negligible electron mass, new observables involving
transversely-polarized beams are available only if both beams are
polarized. In this section the effects  of transversely-polarized beams
in the chargino~\cite{Bartl:2004xy} and neutralino~\cite{neutralinopaper} processes
\begin{alignat}{2}
& e^+e^- \to \tilde{\chi}^+_1 \tilde{\chi}^-_j,&\quad &\mbox{with}\quad
\tilde{\chi}^-_j\to \tilde{\nu}_{\ell} \ell^- \quad\mbox{and}\quad
\tilde{\chi}^-_j\to W^- \tilde{\chi}^0_1,\quad j=1,2,
\label{tr-char1}\\
& e^+e^- \to \tilde{\chi}^0_1 \tilde{\chi}^0_j,&\quad &\mbox{with}\quad
 \tilde{\chi}^0_j\to\tilde{\ell}^{\pm}_{L,R}\ell^{\mp}_1\to \ell^{\mp}_1 
\ell^{\pm}_2\tilde{\chi}^0_1,\quad j=2,3,4,
\label{tr-neut1}
\end{alignat}
are summarised.

\begin{sloppypar}
With transversely-polarized beams new CP-sensitive observables such
as CP-odd triple product correlations as well as CP-even azimuthal asymmetries
can be exploited. CP-odd observables are necessary  
to determine the underlying interactions unambiguously.
Only 
two complex parameters enter the neutralino/chargino sector:    
the higgsino mass parameter $\mu$ and the gaugino mass parameter
$M_1$. 
\end{sloppypar}

%%%%%%%%%%%%%%%%%%%%%%%%%%%%%%%%%%%%%%%%%%%%%%%%%%%
\subsection*{Transverse beam polarization in chargino production}
%%%%%%%%%%%%%%%%%%%%%%%%%%%%%%%%%%%%%%%%%%%%%%%%%%%
In the case of chargino production and decay, eq.~(\ref{tr-char1}),
T-odd asymmetries based on triple products of the electron/positron
momentum, transverse polarization vector and the momentum of the
outgoing lepton or $W$ vanish \cite{Bartl:2004xy}. 
Therefore, T-odd observables could only be constructed in that case
by analyzing both decays of $\tilde{\chi}^{\pm}_1$ and of
$\tilde{\chi}^{\mp}_2$, taking into account the spin-spin
correlations between both decaying charginos~\cite{Bartl:2004xy}.

On the other hand, transverse polarizations allow to construct CP-even
azimuthal asymmetries\cite{soffer,Choi:2000ta}, which are sensitive to
the phases $\varphi_{M_1}$ and $\varphi_{\mu}$:
\begin{equation}
A_{\phi}=\frac{1}{\sigma}\,
\left[\int_+ \frac{{\rm d}\sigma}{{\rm d}\phi}\,{\rm d}\phi
- \int_- \frac{{\rm d}\sigma}{{\rm d}\phi}\,{\rm d}\phi\right],
\end{equation} 
where $\phi$ is the azimuthal angle of the observed final $\ell$ or
$W^-$, eq.~(\ref{tr-char1}), $\int_{\pm}$ refers to the integrated
phase space hemisphere above (below) the scattering plane and ${\rm
d}\sigma/{\rm d}\phi$ denotes the differential cross section for
production and decay including the complete spin correlations of
the charginos. It has been shown for different scenarios 
in~\cite{Bartl:2004xy} and for the process (\ref{tr-char1})
that this CP-even asymmetry is a promising observable for
the determination of the phase $\varphi_{M_1}$ 
(up to a two-fold ambiguity) and can reach up to about 20\%.

%%%%%%%%%%%%%%%%%%%%%%%%%%%%%%%%%%%%%%%%%%%%%%%%%%%
\subsection*{Transverse beam polarization in neutralino production}
%%%%%%%%%%%%%%%%%%%%%%%%%%%%%%%%%%%%%%%%%%%%%%%%%%%

Contrary to the case of chargino production (Dirac fermions),
it is possible to construct T-odd triple product correlations in the
production of neutralinos (Majorana fermions)\cite{neutralinopaper}.

Consider first CP-odd observables in neutralino production
with transverse $e^{\pm}$ beam polarization, eq.~(\ref{tr-neut1}). 
The $e^-$ direction is chosen as the $z$ axis and
the direction of its transverse polarization vector
as the $x$ axis. If the angle between the transverse
polarization vectors of the $e^+$ and the $e^-$ is $\pi/2$,
with $\phi_+=\pi/2$ and $\phi_-=0$), the
CP-violating contributions can be isolated with the
following integration (this $A_{\rm CP}(\theta)$ should not be confused
with the ${\mathcal A}_{\rm CP}$ of eq.~(\ref{ACP})):
\begin{equation}\label{trans2}
A_{\rm CP}(\theta)=\frac{1}{\sigma}
\left[\int^{\pi/4}_{0} -\int^{3\pi/4}_{\pi/4}
+\int^{5\pi/4}_{3\pi/4}- \int^{7\pi/4}_{5\pi/4}
+\int^{2\pi}_{7\pi/4}
\right]
\frac{{\rm d}^2\sigma}{{\rm d}\phi \, {\rm d}\cos\theta} {\rm d}\phi~.
\end{equation}
Because of the Majorana nature of the neutralinos the CP-violating
terms of the $t$- and $u$-channel contributions cancel each other, if integrated
over the whole range of $\theta$. Therefore, the CP-odd asymmetry is defined as
\begin{equation}\label{trans3}
A_{\rm CP}=
\left[\int^{\pi/2}_{0} -\int^{\pi}_{\pi/2}\right]
A_{\rm CP}(\theta)\,{\rm d}\cos\theta~.
\end{equation}
In order to measure $A_{\rm CP}$ it is necessary to reconstruct the
directions of the neutralinos, this can be done by analysing the subsequent two-body decays in
eq.~(\ref{tr-neut1}).

\begin{figure}[htb]
\setlength{\unitlength}{1cm}
\begin{center}
\begin{picture}(15,12.5)
\put(-5.3,-11.5){\mbox{\epsfig{figure=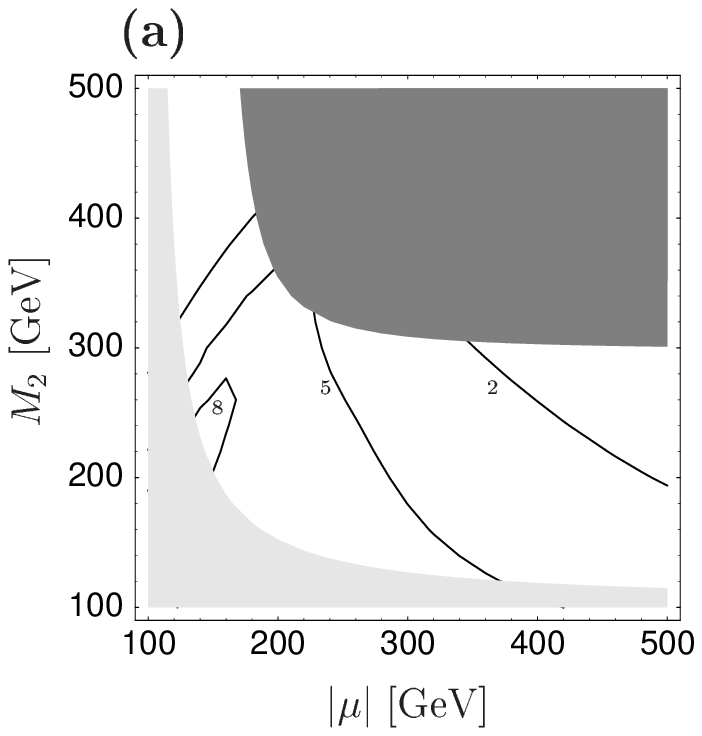,height=27cm,width=19.4cm}}}
\put(2.7,-11.5){\mbox{\epsfig{figure=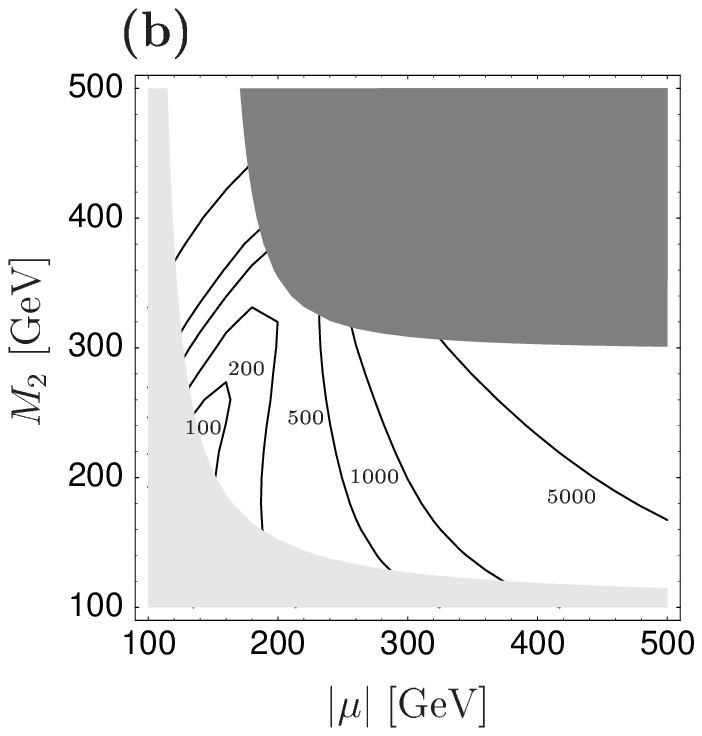,height=27cm,width=19.4cm}}}
\put(3.,12.3){$A_{\rm CP}[\%]$}
\put(11,12.3){${\cal L}_{\rm int}$[fb$^{-1}$]}
\end{picture}
\end{center}
\vspace{-6.7cm}
\caption[CP-odd asymmetry for neutralino production
with transverse beam polarization]
{Contours of (a) the CP-odd asymmetry $A_{\rm CP}$, Eq.~(\ref{trans3}), for 
neutralino production, $e^+e^- \to \tilde{\chi}^0_1 \tilde{\chi}^0_2$,
at $\sqrt{s}=500$~GeV 
with  transverse beam polarization 
$(P^{\rm T}_{e^-}, P^{\rm T}_{e^+})=(100\%,100\%)$ 
in the $|\mu|$--$M_2$ plane for 
the scenario S12, table~\ref{tab-susyscenarios},
and (b) the luminosity $\mathcal{L}_{\rm int}$
needed to measure the asymmetry at the
5-$\sigma$ level with
a degree of transverse polarization 
$(P^{\rm T}_{e^-},P^{\rm T}_{e^+})=(80\%,60\%)$.
The light gray region is experimentally excluded because 
$m_{\tilde{\chi}^\pm_1}$ has to be larger then the exclusion bound
104~GeV (LEP limit). The dark gray region
is excluded  because 
$m_{\tilde{\chi}^0_1}$ has to be smaller than $m_{\tilde{e}_R}$ 
(since $m_{\tilde{\chi}^0_1}$ is assumed to be the stable LSP)~\cite{neutralinopaper}.
}
\label{fig:trans1}
\end{figure}

Fig.~\ref{fig:trans1}a shows contours of
$A_{CP}$ in $e^+e^- \to \tilde{\chi}^0_1\tilde{\chi}^0_2$ 
at $\sqrt{s}=500$~GeV in the $|\mu|$--$M_2$ SUSY parameter  plane with
the other MSSM parameters as given in S12, table~\ref{tab-susyscenarios}.
The CP-odd asymmetry $A_{\rm CP}$, eq.~(\ref{trans3}), is sizable
for large mixing between the gaugino and the higgsino components
of $\tilde{\chi}^0_1$ and $\tilde{\chi}^0_2$.
If the beams were fully transversely polarized,
$(P_{e^-}^{\rm T}, P^{\rm T}_{e^+})=(100\%,100\%)$, then
$A_{CP}$ could reach up to about $8.8\%$.

Since the asymmetries are here only of the order of a few percent, it is important to know
how much luminosity $\Lumint$ is required for a measurement with sufficient significance.
It can be estimated from the relation \cite{Bartl:2003tr,neutralinopaper}
\begin{equation}
\Lumint=({\cal N}_{\sigma})^2/[(P_{e^-}^{\rm T} P_{e^+}^{\rm T} 
A_{\rm CP})^2  \, \sigma],
\label{sig-estimate}
\end{equation}
where ${\cal N}_{\sigma}$ denotes the number of standard deviations and $\sigma$ 
the corresponding
neutralino cross section.
In fig.~\ref{fig:trans1}b is shown
how much luminosity is required with 
$(P_{e^-}^{\rm T},P_{e^+}^{\rm T})=(80\%,60\%)$ for a discovery with 5-$\sigma$. 
For the maximum value of the asymmetry $A_{\rm CP}$
the required luminosity $\mathcal{L}_{\rm int}$ is about $80~\mbox{fb}^{-1}$.

The reconstruction of the direction of the neutralinos is not
necessary if a CP-odd asymmetry is defined
with respect to the azimuth of 
the final leptons $\ell_{1,2}$ \cite{neutralinopaper}. The 
derived asymmetries 
are of the same order of magnitude as in the case studied before.
At a LC an arbitrary and independent
relative orientation of both transversely-polarized beams
could be studied which has not received much attention in the 
literature so far 
and provides high 
flexibility and a larger amount of observables~\cite{neutralinopaper}.
\medskip
 
{\bf \boldmath
Quantitative examples:
With transversely-polarized $e^-$ and $e^+$ beams 
new CP-sensitive observables could be observed.
In chargino production and decay CP-even azimuthal asymmetries
of the order of 10\% are obtained. In neutralino production
CP-odd as well as CP-even asymmetries can be constructed.
At $\sqrt{s}=500$~GeV with $(P^{\rm T}_{e^-},P^{\rm T}_{e^+})=(80\%,60\%)$
and an integrated luminosity ${\cal L}_{\rm int}=500~\mbox{fb}^{-1}$, the
discovery of a CP-odd asymmetry $A_{CP}$ at the 5-$\sigma$ level 
is possible for a broad range of parameters.}

%%%%%%%%%%%%%%%%%%%%%%%%%%%%%%%%%%%%%%%%%%%%%%%%%%%
\subsection{Non-minimal SUSY models}
%%%%%%%%%%%%%%%%%%%%%%%%%%%%%%%%%%%%%%%%%%%%%%%%%%%
\subsubsection{Extended neutralino sector
\label{sect:singlino}}
%%%%%%%%%%%%%%%%%%%%%%%%%%%%%%%%%%%%%%%%%%%%%%%%%%%
{\bf \boldmath
Extensions of the MSSM particle sector, e.g.\ by adding 
Higgs singlets or new gauge bosons, enlarge the neutralino
sector. Cross sections for the 
production of singlino-dominated neutralinos are typically suppressed.
The polarization of both beams may be crucial
for a) observing signals  with production of singlino-dominated neutralinos 
and b) distinguishing between
the MSSM and the extended model.    Furthermore
the light neutralino cross sections may show  opposite dependences on polarization, 
so that a model distinction becomes possible.}
\smallskip

Non-minimal extensions of the particle sector of the MSSM are
characterized by an additional singlet superfield $S$ with vacuum
expectation value $x$ and a trilinear coupling $\lambda$ relating the
singlet superfield and the two doublet Higgs superfields in the
superpotential.  In the Next-to-Minimal Supersymmetric Standard Model
(NMSSM) \cite{Franke:1995tc,Franke:1994hj, Franke:1995tf,
Ellwanger:1997jj,Gunion:2004si,Ellwanger:2005uu}, the superpotential contains also a
trilinear term of the singlet superfield with coupling $\kappa$,
\begin{equation}
W \subset \lambda H_1 H_2 S+\frac{1}{3} \kappa S^3.
\label{eq_defnmssm}
\end{equation}
In \es-inspired models~\cite{Hewett:1987jn} one has in addition
to the singlet superfield one extra
neutral gauge boson $Z'$. There is an
additional gaugino mass parameter $M'$ related to this extra U(1) gauge
factor.  

Dominant singlet higgsino (singlino) 
component in the lightest neutralino 
exist for large vacuum expectation values $x \gtrsim 1$~TeV \cite{Hesselbach:2001ri}.

%%%%%%%%%%%%%%%%%%%%%%%%%%%%%%%%%%%%%%%%%%%%%%%%%%%
\subsubsection*{Production of singlino-dominated neutralinos}
%%%%%%%%%%%%%%%%%%%%%%%%%%%%%%%%%%%%%%%%%%%%%%%%%%%
Since the singlino component does not couple to gauge bosons,
gauginos, (scalar) leptons and (scalar) quarks, cross sections for
the production of the exotic neutralinos are generally small
\cite{Franke:2001nx, Hesselbach:2002nv, Moortgat-Pick:1999bg,
 Hesselbach:1999qd}.
However, they may be produced at a high luminosity
$e^+e^-$ linear collider with cross sections
sufficient for detection,
which can still be enhanced by using the polarization of both beams. 
For a wide range of $x$ values 
the associated production of the
singlino-dominated neutralino yields detectable cross sections.
In the NMSSM \cite{Franke:1995tc} a singlino-dominated neutralino
$\tilde{\chi}^0_S$ with mass
$\approx 2 \kappa x$ decouples from the other neutralinos  
for large $x \gg |M_2|$ in
the neutralino mixing matrix,
while the other neutralinos $\tilde{\chi}^0_{1,\ldots,4}$
have MSSM character.
In  an \es-inspired model with one extra neutral 
gauge boson $Z'$ and
one additional singlet superfield, that contains six neutralinos, 
a nearly pure light 
singlino-like $\tilde{\chi}^0_S$ exists for very large values $|M'| \gg x$  
with mass $\approx 0.18 \,x^2/|M'|$ in zeroth approximation
\cite{deCarlos:1997yv,Franke:2001nx,Hesselbach:2001ri}.

\begin{figure}[htb]
\setlength{\unitlength}{1cm}
\begin{picture}(16,12)
\put(0,6.4){\epsfig{file=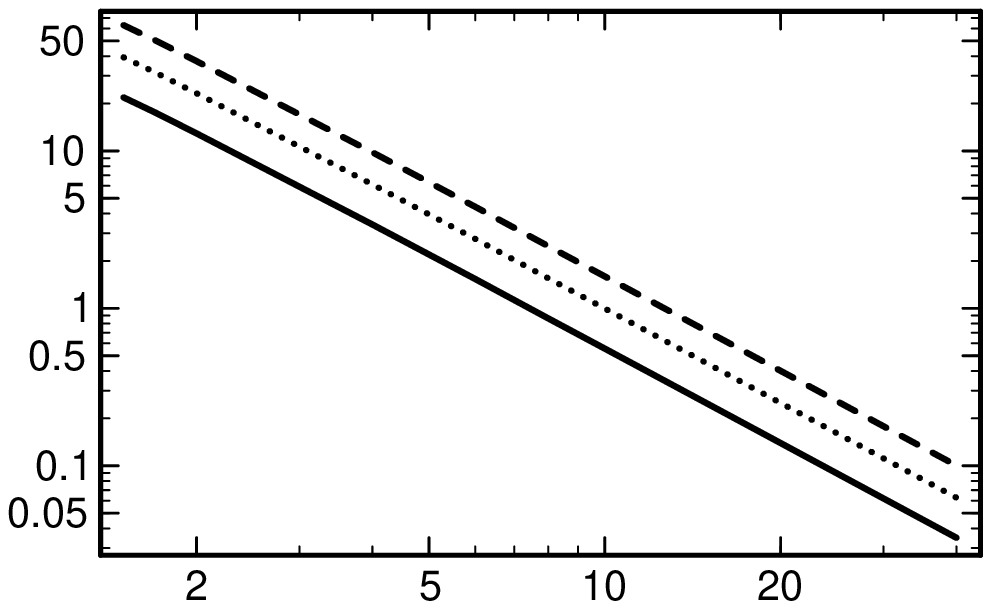,scale=.75}}
\put(6.4,6.2){$x$ [TeV]}
\put(0.1,11.3){$\sigma$ [fb]}
\put(2.1,11.3){NMSSM}
\put(4.9,11.4){$m_{\tilde{\chi}^0_S}=70$ GeV}
\put(8.35,6.4){\epsfig{file=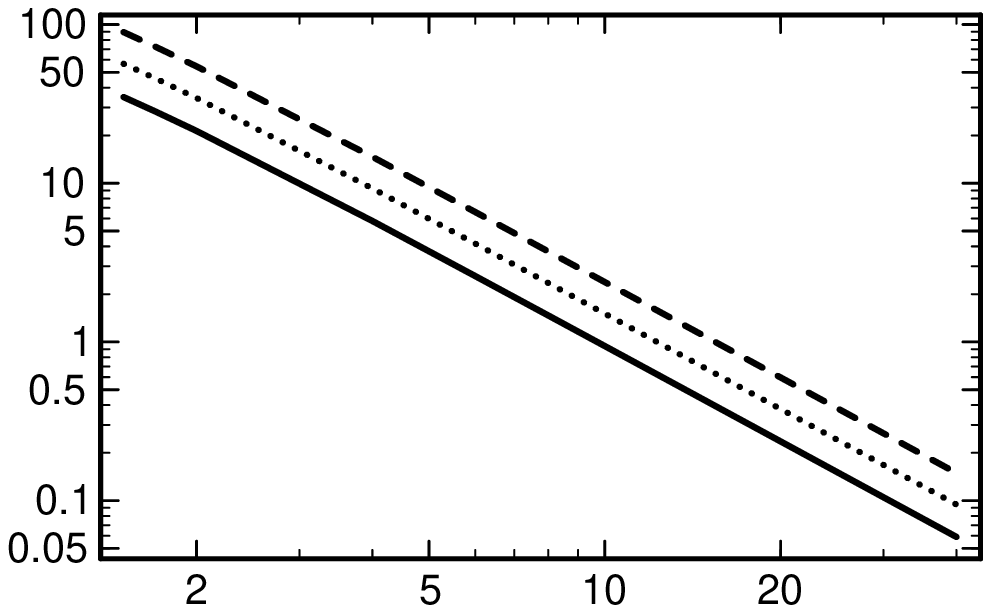,scale=.75}}
\put(14.8,6.2){$x$ [TeV]}
\put(8.5,11.35){$\sigma$ [fb]}
\put(10.4,11.3){NMSSM}
\put(13.1,11.4){$m_{\tilde{\chi}^0_S}=120$ GeV}
\put(0,.4){\epsfig{file=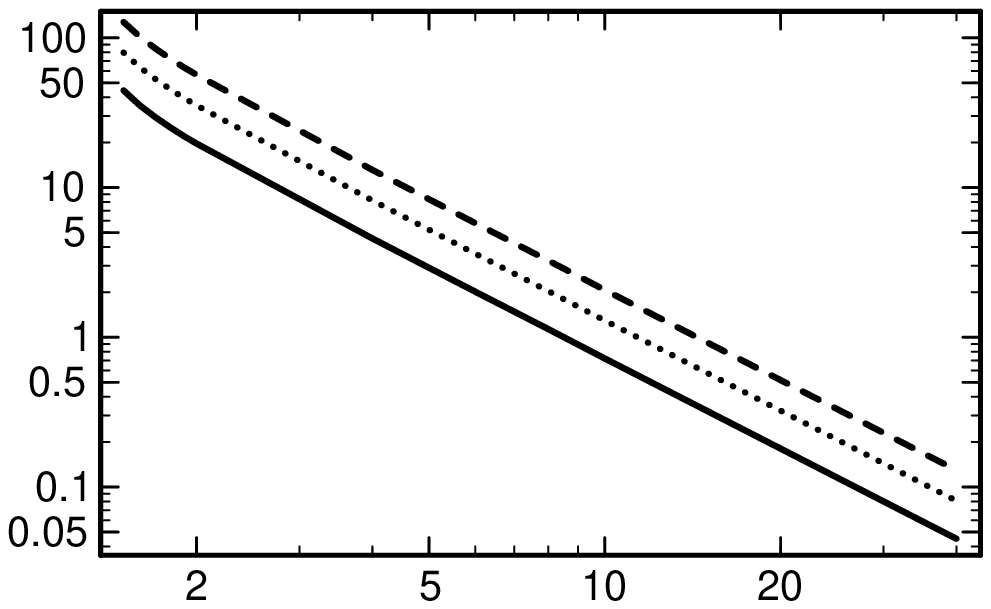,scale=.75}}
\put(6.4,.2){$x$ [TeV]}
\put(0.1,5.3){$\sigma$ [fb]}
\put(2.6,5.3){\es}
\put(4.9,5.4){$m_{\tilde{\chi}^0_S}=70$ GeV}
\put(8.35,.4){\epsfig{file=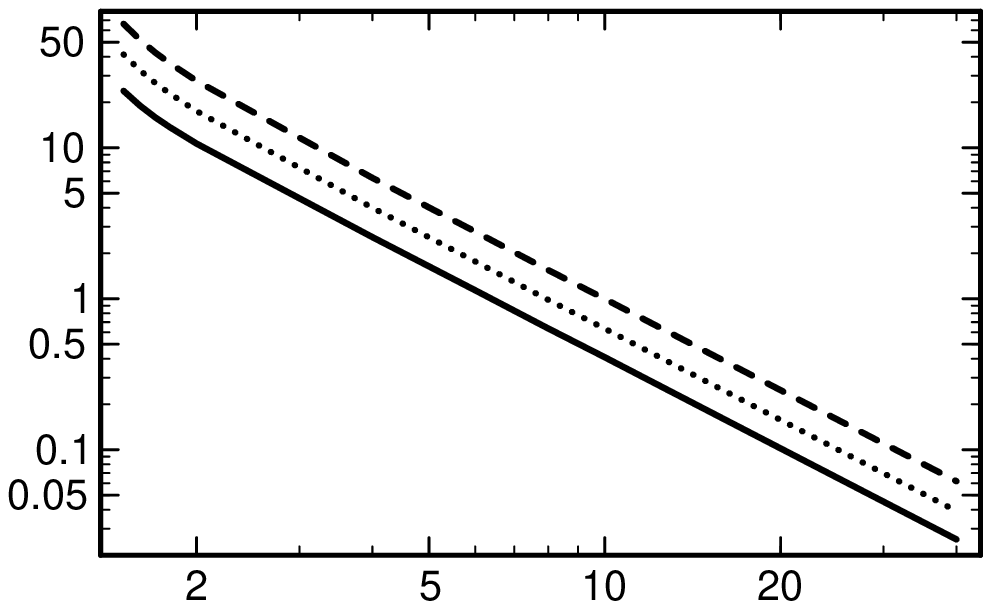,scale=.75}}
\put(14.8,.2){$x$ [TeV]}
\put(8.5,5.3){$\sigma$ [fb]}
\put(10.8,5.3){\es}
\put(13.1,5.4){$m_{\tilde{\chi}^0_S}=120$ GeV}
\end{picture}
\caption[Singlino-dominated neutralino 
production]{\label{singlinoproduction}Cross sections for the production of 
a singlino-dominated neutralino $\tilde{\chi}^0_S$ via $e^+ e^- \to
\tilde{\chi}^0_S \tilde{\chi}^0_1$ at $\sqrt{s} = 500$~GeV,
for unpolarized beams (solid) and beam polarizations of
$(P_{e^-},P_{e^+})=(+80\%,0)$ (dotted) and
$(+80\%,-60\%)$ (dashed),
in the NMSSM and an \es{}-inspired model. The SUSY parameters
are chosen corresponding to scenario S6 with
$\mu_\mathrm{eff} = \lambda x = 352$~GeV, cf\ table~\ref{tab-susyscenarios}. 
The mass of the $\tilde{\chi}^0_S$ is fixed at $70$~GeV and $120$~GeV by
the parameters $\kappa$ (NMSSM) and $M'$ (\es\ model)~\cite{Franke:2001nx}.}
\end{figure}

In fig.~\ref{singlinoproduction} the cross section for associated production 
of the
singlino-dominated $\tilde{\chi}^0_S$ together with the lightest MSSM-like
neutralino $\tilde{\chi}^0_1$ is shown. 
The scenarios considered are such that the MSSM-like
neutralinos have similar masses and mixing character as in the
`typical mSUGRA' scenario S6, cf.\ table~\ref{tab-susyscenarios}, 
with a chosen $\mu_{\rm eff}=\lambda x=352$~GeV.
The resulting neutralino masses are
$96$, $177$, $359$ and $378$~GeV.
The cross sections are shown for unpolarized beams and beam
polarizations $(P_{e^-},P_{e^+})=(+80\%,0)$ and $(+80\%,-60\%)$.  
Electron
beam polarization $P_{e^-}=+80\%$ enhances the cross section by a factor $1.5$
to $1.8$, while additional positron beam polarization $P_{e^+}=-60\%$ gives a
further enhancement factor of about $1.6$.  Assuming a cross
section of $1$~fb to be sufficient for discovery, the singlino-dominated
neutralino can be detected with unpolarized beams for $x< 7.4$~TeV ($9.7$ TeV)
in the NMSSM with $m_{\tilde{\chi}^0_S} = 70$~GeV ($120$~GeV) and for $x<
8.5$~TeV ($6.4$ TeV) in the \es{} model.  For a polarized electron beam, the
reach in $x$ is enhanced to $x< 10.0$~TeV ($12.3$~TeV) in the NMSSM and $x<
11.4$~TeV ($7.9$~TeV) in the \es{} model, and for both beams polarized to
$x< 12.6$~TeV ($15.5$~TeV) in the NMSSM and $x< 14.4$~TeV ($10.0$~TeV) in the
\es{} model; see table~\ref{tab_nmreach}. 

With both beams polarized the cross sections
are enhanced by a factor $2.4$--$2.9$ with respect to unpolarized beams,
depending on the scenario. This enhances the reach for the singlino-dominated
neutralinos to singlet vacuum expectation values as large as $15$~TeV.

\begin{table}[htb]
{\small
\begin{center}
$x$ reach [TeV]  \\
\begin{tabular}{|c||c|c||c|c|}
\hline
& \multicolumn{2}{|c||}{NMSSM}& \multicolumn{2}{|c|}{\es}\\
$(P_{e^-},P_{e^+})$ & $m_{\tilde{\chi}^0_S}=70$~GeV &  $m_{\tilde{\chi}^0_S}=120$~GeV &
 $m_{\tilde{\chi}^0_S}=70$~GeV &  $m_{\tilde{\chi}^0_S}=120$~GeV \\
\hline
$(0, 0)$ & $7.4$  & $9.7$  & $8.5$   & $6.4$  \\
$(+80\%, 0)$ & $10.0$   & $12.3$  & $11.4$   
& $7.9$  \\  
$(+80\%,-60\%)$ & $12.6$  & $15.5$  & $14.4$  & 
$10.0$  \\
\hline
\end{tabular}
\end{center}
} % end small
\caption[Accessible parameter ranges in
NMSSM and  $\mathrm{E_6}$-inspired models]{
Accessible range of the singlet vacuum expectation value $x$ under
the discovery assumption of 
$\sigma(e^+e^-\to \tilde{\chi}^0_S \tilde{\chi}^0_i)\ge 1$~fb~\cite{Franke:2001nx}.
The SUSY parameters
are chosen corresponding to scenario S6 with
$\mu_\mathrm{eff} = \lambda x = 352$~GeV, cf\ table~\ref{tab-susyscenarios}. 
The mass of the $\tilde{\chi}^0_S$ is fixed at $70$~GeV and $120$~GeV by
the parameters $\kappa$ (NMSSM) and $M'$ (\es\ model)~\cite{Franke:2001nx}.
\label{tab_nmreach} }
\end{table}

%%%%%%%%%%%%%%%%%%%%%%%%%%%%%%%%%%%%%%%%%%%%%%%%%%%
\subsubsection*{\bf Distinction between MSSM and NMSSM \label{sect:distinct}}
%%%%%%%%%%%%%%%%%%%%%%%%%%%%%%%%%%%%%%%%%%%%%%%%%%%

There are regions of the SUSY parameter space where the distinction
between the MSSM and the NMSSM is very difficult (see for instance
\cite{Moortgat-Pick:2005vs} and references therein).  If the whole
neutralino sector is kinematically accessible, sum rules for the production cross
sections show a different energy dependence in the MSSM and the NMSSM
\cite{Choi:2001ww}. If, however, only a part of the spectrum is
accessible, polarization effects play an important role. As an
example, consider the case where only the two lightest
neutralinos are accessible at $\sqrt{s}=500$~GeV with, however, very
low cross sections in the MSSM as well as in the NMSSM.  Only with
polarized beams the cross sections are sufficiently enhanced to become 
measurable. Furthermore, the different dependence on polarization
allows a distinction between the models, as exemplified in
table~\ref{nmssm-pol}.  The cross sections for both beams polarized
are enhanced by a factor of 1.6 with respect to the case of only
electrons polarized.

\begin{table}[htb]
\begin{center}
$\sigma(\tilde{\chi}^0_1\tilde{\chi}^0_2)$ [fb]  \\
\begin{tabular}{|c|c|cc|cc|}
\hline
& Unpolarized 
& $(-90\%,0)$ & $(+90\%,0)$ &
$(-90\%,+60\%)$ & $(+90\%,-60\%)$ \\ \hline
MSSM & 0.6 & 0.8 & 0.5 & 1.3 & 0.7 \\
NMSSM & 0.5 & 0.1 & 0.8 & 0.1 & 1.4 \\
\hline
\end{tabular}
\end{center}
\caption[Neutralino cross sections in different models]{Cross 
section for the process $e^+e^-\to\tilde{\chi}^0_1\tilde{\chi}^0_2$ at
$\sqrt{s}=500$~GeV for beam polarization configuration 
$(P_{e^-},P_{e^+})$.  The light-neutralino masses are given by
$m_{\tilde{\chi}^0_1}=189$~GeV and $m_{\tilde{\chi}^0_2}=267$~GeV
for the scenario S13 a) (MSSM parameters) and S13 b) (NMSSM parameters), 
cf.\ table~\ref{tab-susyscenarios}.
The heavy neutralinos, the sleptons
and the light chargino are assumed to be kinematically inaccessible~\cite{Moortgat-Pick:1999bg}.
\label{nmssm-pol}}
\end{table}

{\bf \boldmath
Quantitative examples: In extensions of the MSSM
including an additional Higgs singlet or a new gauge boson, the polarization of both beams
enhances the rates of the neutralino cross sections by about
a factor 1.6, as well as the reach on the singlet vacuum expectation value 
by about
a factor 1.3, up to about 15 TeV, with respect to the case where only
electrons are polarized.}

%%%%%%%%%%%%%%%%%%%%%%%%%%%%%%%%%%%%%%%%%%%%%%%%%%%
\subsubsection{R-parity-violating SUSY \label{sect:rparity}}
%%%%%%%%%%%%%%%%%%%%%%%%%%
{\bf \boldmath
If R-parity is not conserved, scalar neutral
particles, such as sneutrinos, can be produced in the $s$-channel giving rise to spectacular
signals. With left-polarized
electrons and left-polarized positrons, the signal uniquely indicates the
production of a spin-zero particle.}
\smallskip

\begin{figure}[htb]
\setlength{\unitlength}{1cm}
\begin{picture}(12,4.5)
\put(.7,.5){\mbox{\epsfysize=4.cm\epsfxsize=15cm\epsffile{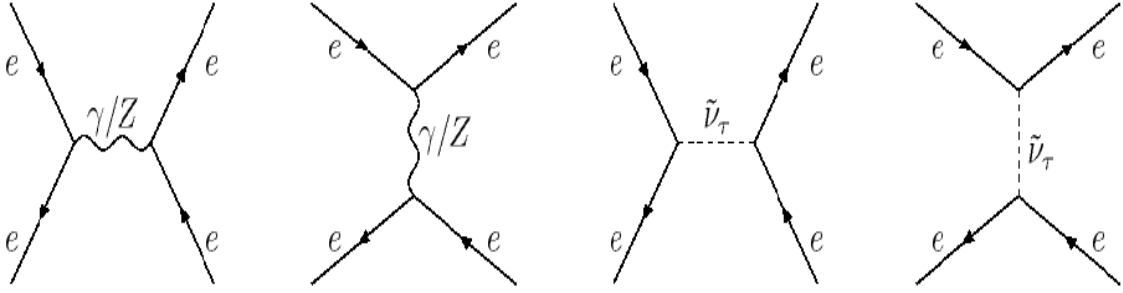}}}
\end{picture}
\caption[R-parity violation in Bhabha scattering]{\label{fig-bhabha} 
Diagrams for Bhabha scattering
  $e^+e^- \rightarrow e^+ e^-$ including $s$- and $t$-channel exchange
  of $\tilde{\nu}_{\tau}$ ($\lambda_{131} \ne 0$).}
\end{figure}

\begin{figure}[htb]
\setlength{\unitlength}{1cm}
\begin{picture}(12,5.5)
\put(3,.5){\mbox{\epsfysize=4.cm\epsffile{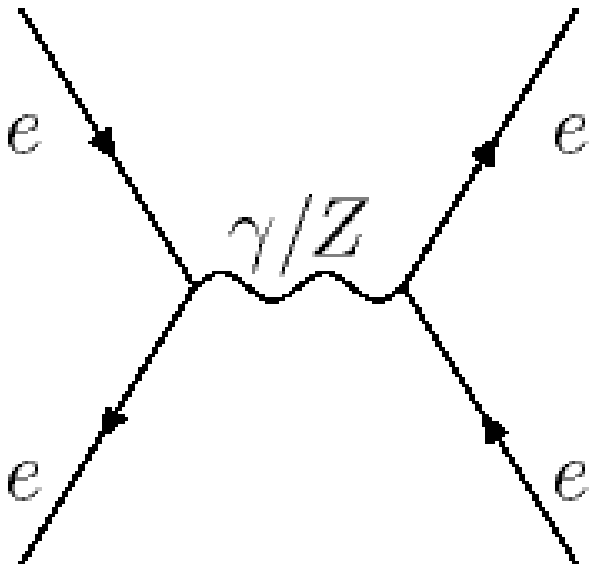}}}
\put(6.9,1){\mbox{\includegraphics[height=.25\textheight,width=.5\textheight]{box}}}
\put(7.1,1.3){ $\mu^+$}
\put(7.,3.5){ $\mu^-$}
\put(9.5,.5){\mbox{\epsfysize=4.cm\epsffile{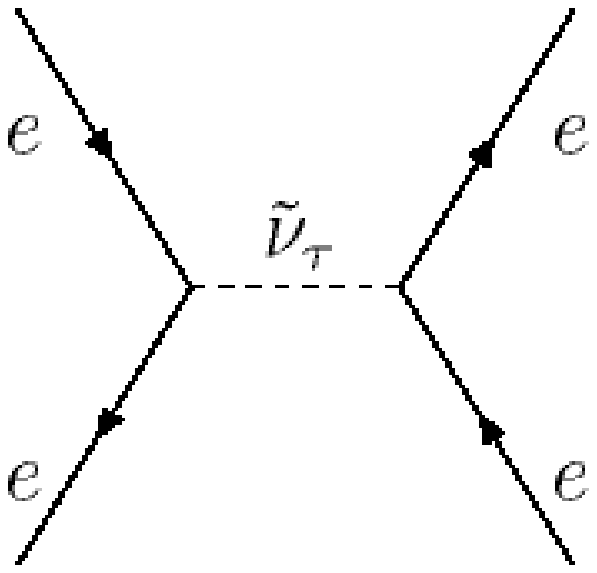}}}
\put(13.3,0.3){\mbox{\includegraphics[height=.25\textheight,width=.5\textheight]{box}}}
\put(13.5,1.3){ $\mu^+$ }
\put(13.5,3.5){ $\mu^-$ }
\end{picture}
\vspace{-.5cm}
\caption[R-parity-violation in muon production]{
Diagrams for muon production
$e^+e^- \rightarrow \mu^+ \mu^-$ with $s$-channel exchange
of $\tilde{\nu}_{\tau}$ ($\lambda_{131} \ne 0$, $\lambda_{232} \ne 0$).
\label{fig-mumu}}
\end{figure}

\begin{figure}[htb]
\vspace{-3cm}
\setlength{\unitlength}{1cm}
\begin{minipage}{10cm}
\psfig{file=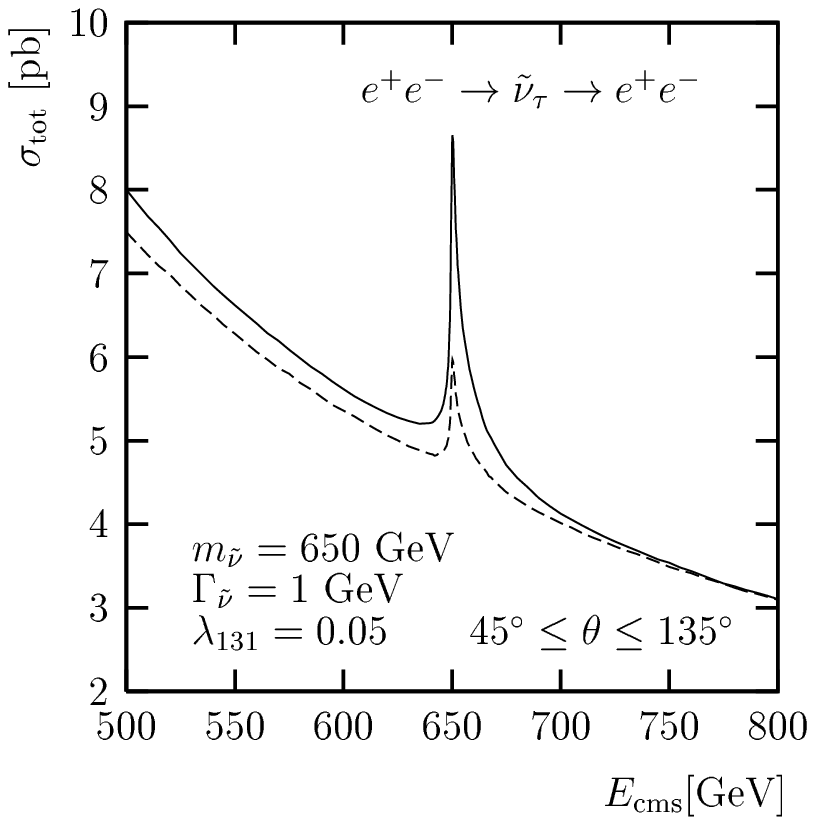,width=1.5\textwidth}
\end{minipage}\hspace{2cm}
\begin{minipage}{10cm}
\begin{picture}(5,5)
\put(-4.5,-8.1){\mbox{\psfig{file=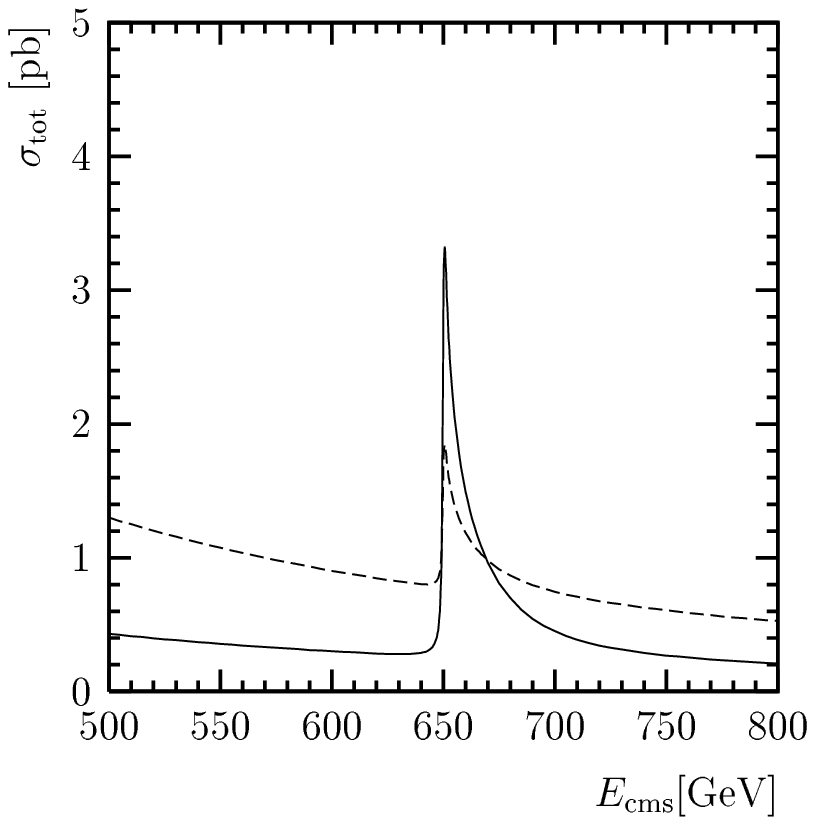,width=1.5\textwidth}}}
\put(-.8,8.9){\scriptsize $e^+e^-\to\tilde{\nu}_{\tau}\to \mu^+\mu^-$}
\put(-.4,8.4){\scriptsize $\lambda_{131}=0.05=\lambda_{232}$}
\end{picture}
\end{minipage}
\vspace{-11.5cm}
\caption[Sneutrino production in R-parity violating SUSY]{Sneutrino
production in the R-parity-violating model. Resonance production
 $e^+ e^-\to \tilde{\nu}_{\tau}$ 
interfering with Bhabha scattering (left panel) 
and resonance production for 
$e^+e^-\to \tilde{\nu}_{\tau} \to \mu^+ \mu^-$ (right panel)
for
different configurations of beam polarization:
$(P_{e^-},P_{e^+})=(-80\%,+60\%)$ (dashed), $(-80\%,-60\%)$ (solid).
An angular cut of $45^{\circ}\le \theta\le 135^{\circ}$ has been
applied in Bhabha scattering \cite{Spiesi}.
\label{fig_susy5}}
\end{figure}

The assumption that R-parity is conserved has no strong theoretical 
justification. Allowing R-parity-violation, in general, 
one has to introduce in the superpotential the interaction terms that violate
lepton and/or baryon number. For illustration purposes we restrict our discussion
to the trilinear lepton number violating terms~\cite{Hall:1983id}:
\begin{equation}
W_{\rp}=\lambda_{ijk} 
L^i_L L^j_L \bar{E}^k_R+\lambda'_{ijk} L^i_L Q^j_L \bar{D}^k_R,
\label{eq_sub-rp}
\end{equation}
where 
$L_L$ ($Q_L$) denotes the left-handed doublets of leptons (quarks) and
$E_R$ ($D_R$) the right-handed singlets of the charged leptons (down-type quarks),
and $i,j,k$ stand for generation numbers.
Respecting the current 
experimental bounds for the couplings \cite{lfv-limits},
the following SUSY processes, that receive the $s$-channel $\tilde{\nu}_{\tau}$ exchange,  
\begin{eqnarray}
e^+ e^- \to \tilde{\nu}_{\tau} \to e^+e^- \label{eq_bhabharp}\\
e^+ e^- \to \tilde{\nu}_{\tau}\to \mu^+\mu^- \label{eq_mumurp},
\end{eqnarray}
are possible,  could be observed at the ILC.
The contributing diagrams are shown in figs.~\ref{fig-bhabha} and
\ref{fig-mumu}, cf. also~\cite{magda}.
The striking effects of the spin-0 s-channel exchange can be enhanced 
using the LL configuration of beam 
polarization. The sneutrino production channel
in eq.~(\ref{eq_bhabharp}) receives a strong background from Bhabha scattering
($s$- and $t$-channel $\gamma/Z$ exchange), as well as also from the $t$-channel
sneutrino exchange. Therefore
the spin-0 verification with beam polarization may be weakened for this process
compared to the process in eq.~(\ref{eq_mumurp}).
The situation is illustrated in fig.~\ref{fig_susy5} for the
R-parity-violating couplings $\lambda_{131}=0.05$, $\lambda_{232}=0.05$,
scenario S14 in table~\ref{tab-susyscenarios}. 
The role of the same-sign initial beam polarization in enhancing the
signal over the background is evident, 
(cf.\ also table~\ref{tab_susy1}~\cite{Spiesberger}).

%%%                  

For the same final states, the $s$-channel 
$Z'$ exchange with $m_{Z'}\sim m_{\tilde{\nu}}$ would manifest itself
by a similar peak in the cross section if the beams were RL or LR polarized since
this configuration would be preferred in that case~\cite{Rizzo:1998vf}.
Therefore, beam polarization of both beams provides an alternative approach
for verifying the spin of the exchanged objects and distinguishing the models
without final-state analysis.

\begin{table}[htb]
\begin{center}
Cross section $\sigma$ [pb] \\
\begin{tabular}{|c||c|c||c|c|}
\hline
$(P_{e^-},P_{e^+})$ & $e^+e^-\to e^+ e^-$ with &  Bhabha  & 
$e^+e^-\to \mu^+ \mu^-$ with & $e^+e^-\to \mu^+ \mu^-$\\
& $e^+ e^-\to \tilde{\nu}_{\tau}\to e^+ e^-$& & 
$e^+ e^-\to \tilde{\nu}_{\tau}\to \mu^+ 
\mu^-$  & \\ 
\hline
unpolarized & 7.17   & 4.50  & 2.50 & 0.44\\
$(-80\%, 0)$ & 7.32   & 4.63 & 2.58 & 0.52\\
$(-80\%,-60\%)$ & 8.66   & 4.69 & 3.33& 0.28 \\
$(-80\%,+60\%)$ & 5.97    & 4.58  & 1.85 & 0.78 \\
\hline
\end{tabular}
\caption[Sneutrino production rates vs.\ 
 Bhabha- and $\mu^+\mu^-$-scattering]
{Cross sections for sneutrino production  in
$e^+ e^-\to \tilde{\nu}_{\tau}\to e^+ e^-$
and $e^+ e^-\to \tilde{\nu}_{\tau}\to \mu^+ \mu^-$
for different degrees of polarization. The study was made at 
$\sqrt{s}=650$~GeV for $m_{\tilde{\nu}}=650$~GeV,
$\Gamma_{\tilde{\nu}}=1$~GeV, an angular cut of 
$45^{\circ}\le \theta \le 135^{\circ}$ and
the R-parity-violating couplings $\lambda_{131}=0.05$ and
$\lambda_{232}=0.05$, respectively\cite{Spiesi}.
\label{tab_susy1}}
\end{center}
\end{table}

In the above example both the production and decay of $\tilde{\nu}_{\tau}$ violates
R-parity. However, sneutrino can decay via R-parity conserving coupling, which may be the 
dominant decay mode. In this case we get another interesting class of 
R-parity violating processes in which 
single SUSY particle production can occur~\cite{Chemtob:1998jn}. 
For example,
single chargino (neutralino) production,
\begin{equation} 
e^+e^-\to \tilde{\chi}^{\pm}_1\mu^{\mp},
\end{equation}
occurs via sneutrino (charged slepton) exchange in the $s$- and $t$-channels.
The chargino $\tilde{\chi}^{\pm}_1$ decays subsequently into
$\ell^{\pm} \tilde{\chi}^0_1\nu$ and $\tilde{\chi}^0_1$ into 3 leptons.
The process requires the same lepton flavour violating couplings as 
in the example shown above.
The characteristic feature of this process is that it 
requires same sign helicities, the LL configuration.\smallskip

{\bf \boldmath
Quantitative example: Electron
polarization with $P_{e^-}=-80\%$
enhances the signal, i.e.\ Bhabha scattering or muon production 
including scalar neutrino exchange in the $s$-channel, 
only slightly by about $2$-$3\%$, whereas the simultaneous
polarization of both beams with $(P_{e^-},P_{e^+})=(-80\%,-60\%)$
produces an increase of about $20$-$30\%$. 
}

%%%%%%%%%%%%%%%%%%%%%%%%%%%%%%%%%%%%%%%%%%%%%%%%%%$
\subsection{Production of heavy Higgs bosons 
in the MSSM \label{sect:heavyhiggs}}
%%%%%%%%%%%%%%%%%%%%%%%%%%%%%%%%%%%%%%%%%%%%%%%%%%%
{\bf \boldmath
Searches for heavy SUSY  Higgs particles can be
extremely challenging for both the LHC and the ILC.  Exploiting
single Higgs-boson production in $e^+e^- \to \nu \bar{\nu} H$ extends the kinematical
reach considerably.  However, in the decoupling region, $m_A\gg m_Z$, the 
suppressed couplings of the heavy Higgs boson to SM gauge bosons
lead to very small rates. This difficulty could be attenuated by accumulating a very high integrated
luminosity, together with a further enhancement of the signal cross section by polarizing both
beams.} 
\smallskip

The possibility to enhance cross sections with beam polarization can
be very important for detecting processes with a low rate. 
An example has been worked out in 
\cite{Hahn:2002gm}, where the production of the heavy neutral CP-even
Higgs boson $H$ of the MSSM was studied. 
The MSSM Higgs sector is characterized by two parameters at tree-level, the mass of the pseudoscalar
Higgs boson, $m_A$, and $\tan\beta$. The CP-even Higgs bosons $h$, $H$ share their couplings to 
the gauge bosons, while the couplings of the CP-odd Higgs boson $A$ to two gauge bosons
vanishes.

Over large parts of the whole parameter 
space, i.e.\ $m_A\gg m_Z$, the lightest CP-even Higgs boson, $h$, of the MSSM is 
SM-like, and the coupling of the heavy CP-even Higgs 
boson to two gauge bosons is suppressed~\cite{haberkane}. 
In this case only 
the pair production channel $e^+e^- \to HA$ contributes at full strength. 
Since for large values of $m_A$ the heavy Higgs bosons $A$ and $H$ are
approximately mass degenerate, $m_A \approx m_H$, the pair production
channel $e^+e^- \to HA$ is limited by kinematics to the region 
$m_H < \sqrt{s}/2$. If the rare process 
$e^+e^- \to \nu \bar\nu H$, governed by the suppressed
coupling of $H$ to two gauge bosons, can be exploited, 
the kinematic reach of the linear collider can be 
extended. 

In \cite{Hahn:2002gm} it was shown that higher-order contributions
to the couplings of the heavy Higgs boson to the gauge bosons
can remedy the suppression. This leads to considerably higher 
cross sections in certain domains of the MSSM parameter space and
makes the process potentially accessible at the linear collider. This requires a
high integrated luminosity and polarized beams. The cross section
is enhanced for left-handed electrons and
right-handed positrons. While an 80\% polarization of the
electron beam results in a cross section that is enhanced by a
factor 1.8, the polarization of both beams, i.e.\ 80\% polarization
for electrons and 60\% for positrons, would roughly yield
an enhancement by a factor of 2.9. 
Therefore, with polarization of both beams, this process may be measurable.
With 
the ILC running at the high energy of 1~TeV, 
the enhancement of
the cross section by the beam polarization can extend the
kinematic reach by roughly 100~GeV, see
fig.~\ref{heavyhiggs} (right), with respect to the case of unpolarized
beams, fig.~\ref{heavyhiggs} (left). Here the SUSY parameters are chosen
as in scenario S15, cf.\ table~\ref{tab-susyscenarios}. 
It has been assumed
that at least about
20 heavy Higgs boson events 
have to be observed in order to establish this channel. This corresponds to an
integrated luminosity of the order of $2~{\rm ab}^{-1}$ for a production  
cross section larger than $0.1$~fb 
(left dashed and dark regions in fig.~\ref{heavyhiggs}).

\begin{figure}[htb]
\begin{picture}(15,6.2)
    \put(1.1,0.1){\epsfig{file=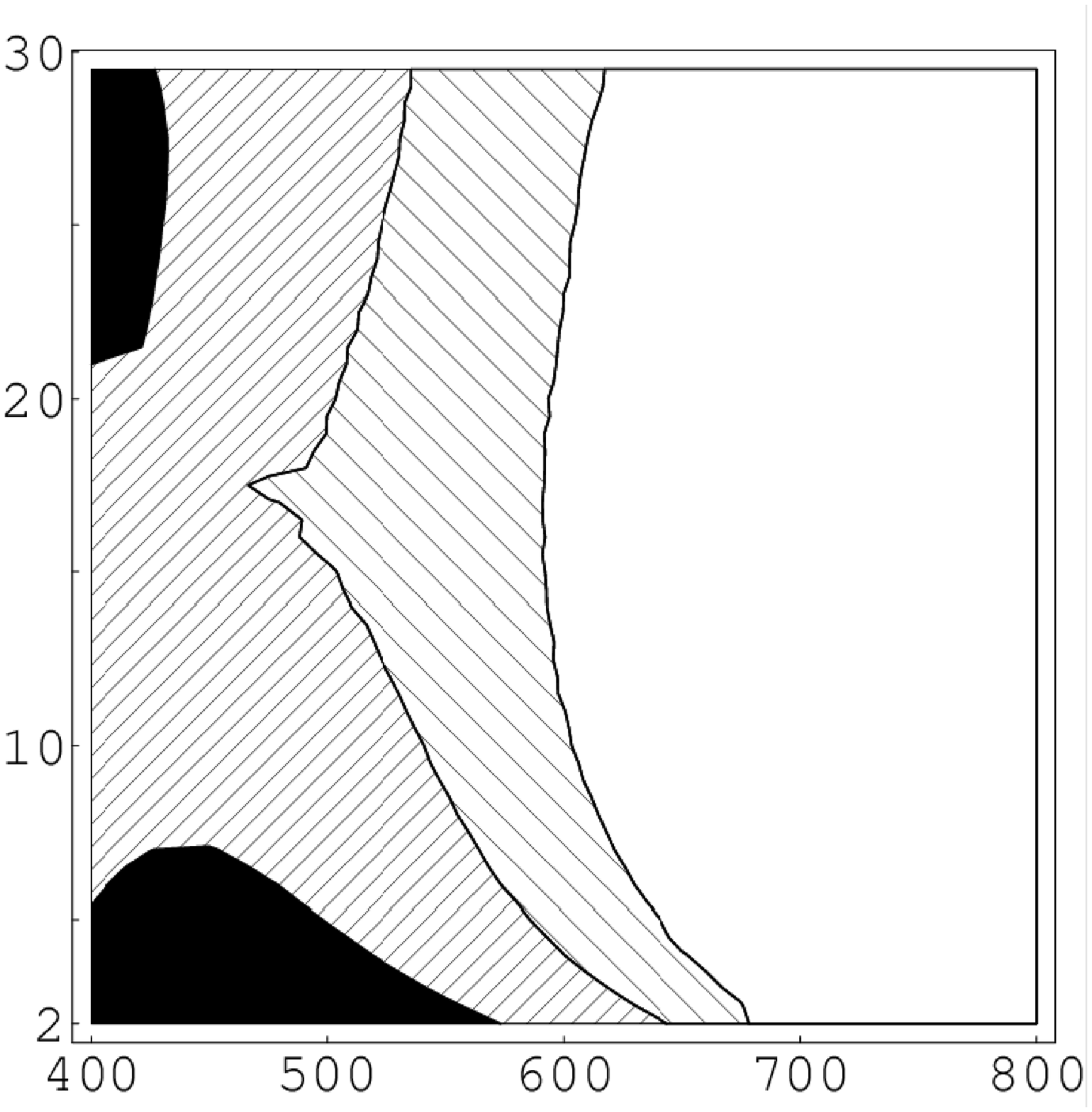,width=6.3cm}}
    \put(6,-.3){$m_A$ [GeV]}
    \put(0.1,5.8){$\tan\beta$}
    \put(9.8,0){\epsfig{file=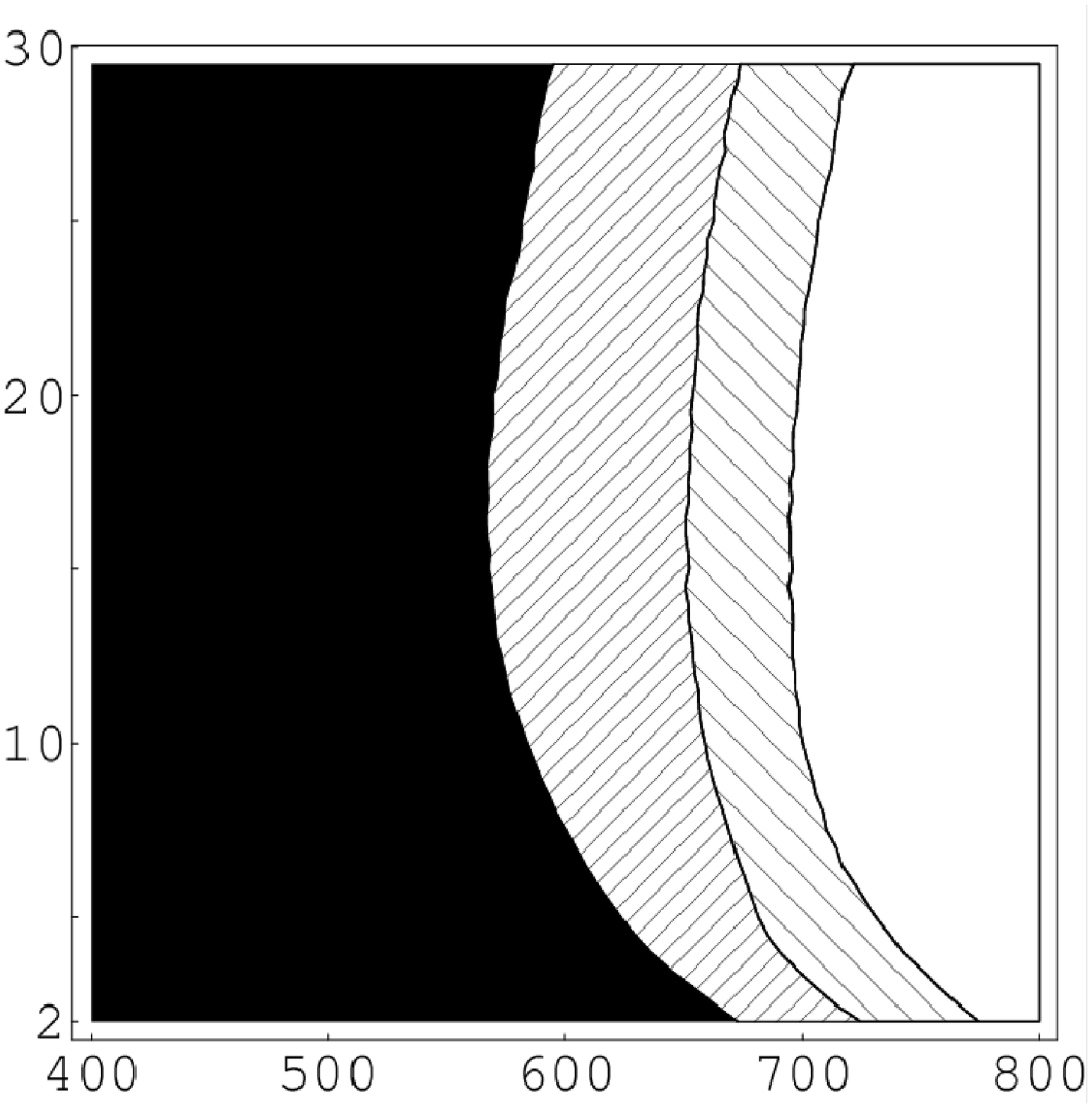,width=6.2cm}}
    \put(8.6,5.8){$\tan\beta$}
    \put(14.5,-.4){$m_A$ [GeV]}
\end{picture}\vspace{.2cm}
\caption[Heavy SUSY Higgs production]{Cross section of heavy Higgs production  
$\sigma(e^+e^-\to \nu\bar{\nu} H)$ for the scenario S15, table~\ref{tab-susyscenarios},
in the $m_A$--$\tan\beta$ plane
with $\sigma>0.05$~fb (black), $>0.02$~fb (right dashed) and
$>0.01$~fb (left dashed) for unpolarized beams (left) and with both beams
polarized, $(P_{e^-},P_{e^+})=(-80\%,+60\%)$ (right) at $\sqrt{s}=1$~TeV~\cite{Hahn:2002gm}. 
\label{heavyhiggs} }
\end{figure}  

{\bf \boldmath
Quantitative example:
Using $(P_{e^-},P_{e^+})=(-80\%,+60\%)$ instead of $(-80\%,0)$,
one gains about a factor $1.6$ for the signal cross sections.
Therefore 
the polarization
of both beams could be decisive to accumulate enough statistics to
observe signals of the heavy SUSY Higgs bosons.}

%%%%%%%%%%%%%%%%%%%%%%%%%%%%%%%%%%%%%%%%%%%%%%%%%%%%%%%%%%%%%%%%%%%%%
\section{Effective contact-interactions and heavy gauge bosons
\label{sect:indirectsearches}}
%%%%%%%%%%%%%%%%%%%%%%%%%%%%%%%%%%%%%%%%%%%%%%%%%%%%%%%%%%%%%%%%%%%%%

Effective contact-interactions (CI) represent a general tool for 
parametrizing at `low-energy' the effects of non-standard dynamics
characterized by exchanges, among the SM particles, of very high-mass states,
much higher than the available accelerator energy. Manifestations
of such new interactions can therefore be searched for only through deviations
of cross sections from the SM predictions, and `indirect' bounds
on the new energy-scales and coupling constants can be derived. Denoting
by $\Lambda$ the above-mentioned new large scales, for dimensional
reasons the deviations of the observables from the SM predictions can be 
suppressed by powers of the ratio $\sqrt{s} /\Lambda$ which should be smaller
that unity for the effective theory to be a reliable description. Such
effects are therefore expected to be small, and favoured at high-energy
and high-luminosity machines.

%%%%%%%%%%%%%%%%%%%%%%%%%%%%%%%%%%%%%%%%%%%%%%%%%%%
\subsection{Analysis of four-fermion contact-interactions
\label{sect:intro}}
%%%%%%%%%%%%%%%%%%%%%%%%%%%%%%%%%%%%%%%%%%%%%%%%%%%
{\bf \boldmath
Longitudinal beam polarization of both beams is
decisive to derive model-independent bounds on the different possible
couplings. With both beams polarized, the error in $\Peff$ is reduced and
the accuracy of the $A_{\rm LR}$ measurement is considerably enhanced
and more observables can be defined. 
The systematic errors can be significantly
reduced with both beams polarized, which is of crucial value.}
\smallskip

In the following it is focused on the production of Standard Model 
fermion-pairs. Although such processes are not primarily devoted at the LC
to the search of new phenomena, they guarantee a good sensitivity to 
exchanges of heavy mass scales 
$\Lambda\gg m_{W,Z}$ thanks to the clear signature of the final states in 
the detector and to the available high statistics.

For the fermion-pair production process
\begin{equation}
e^+ + e^-\to f+\bar{f}, \label{proc}
\end{equation}
the general, $SU(3)\times SU(2)\times U(1)$ 
symmetric $eeff$ CI Lagrangian (dimension $D=6$) with 
helicity-conserving and flavour-diagonal fermion 
currents, was proposed in \cite{Eichten:1983hw}:       
\begin{equation}
{\cal L}_{\rm CI}
=\frac{1}{1+\delta_{ef}}\sum_{i,j}g^2_{\rm eff}\hskip
2pt\epsilon_{ij}
\left(\bar e_{i}\gamma_\mu e_{i}\right)
\left(\bar f_{j}\gamma^\mu f_{j}\right).
\label{lagra}
\end{equation}
In eq.~(\ref{lagra}), $i,j={\rm L,R}$ denote left- or right-handed helicities
(see below), generation and colour indices have been suppressed, and the CI
coupling constants are parametrized in terms of corresponding mass scales as
$\epsilon_{ij}=\eta_{ij}/\Lambda^2_{ij}$ with $\eta_{ij}=\pm 1,0$, depending on
the chiral structure of the individual interactions. Also, conventionally
$g^2_{\rm eff}=4\pi$ is assumed, as a reminder that, in the 
case of compositeness, the
new interaction would become strong at $\sqrt s$ of the order of
$\Lambda_{ij}$.  Obviously, deviations from the SM and upper bounds or
exclusion ranges for the CI couplings can be equivalently expressed as lower
bounds and exclusion ranges for the corresponding mass scales $\Lambda_{ij}$.

A general, model-independent analysis of the process (\ref{proc}) 
in terms of the interaction (\ref{lagra}) must simultaneously 
account for all CI couplings as free, non-vanishing parameters. 
On the other hand, one would like to set, from data,
separate constraints on the $\Lambda$s.

A simplifying procedure is to assume non-zero values for only one of the
couplings (or one specific combination of them) at a time when all others are
set to zero, to test {\it specific} CI models only.

With longitudinally-polarized beams, the analysis of contact interactions
can make use of observables integrated over the polar scattering angle: the
(unpolarized) total cross section $\sigma_{0}$ and forward--backward
asymmetry $A_{\rm FB}$, the left--right asymmetry $A_{\rm LR}$ and left--right
forward--backward asymmetry $A_{\rm LR,FB}$.  These are defined 
in this subsection in the notation
of~\cite{Schrempp:1987zy}, which differs from the notation in
eqs.~(\ref{eq_intro1})--(\ref{eq_intro2}), (\ref{eq_intro3}), (\ref{eq_intro4})
in section~\ref{subsec:1}: the first and second subscripts refer to the
helicities of the incoming and outgoing fermion respectively.  Therefore the
components are denoted with $\hat{\sigma}$:
\begin{equation}
\sigma_{0}
=\frac{1}{4}\left[\hat{\sigma}_{\rm LL}+\hat{\sigma}_{\rm LR}
+\hat{\sigma}_{\rm RR}+\hat{\sigma}_{\rm RL}\right],
\label{sigth}
\end{equation}
\begin{equation}
A_{\rm FB}=
\frac{3}{4}\,
\frac{\hat{\sigma}_{\rm LL}-\hat{\sigma}_{\rm LR}+\hat{\sigma}_{\rm RR}-\hat{\sigma}_{\rm RL}}
     {\hat{\sigma}_{\rm LL}+\hat{\sigma}_{\rm LR}+\hat{\sigma}_{\rm RR}+\hat{\sigma}_{\rm RL}},
\label{afbth}
\end{equation}    
\begin{equation}
A_{\rm LR}=
\frac{\hat{\sigma}_{\rm LL}+\hat{\sigma}_{\rm LR}-\hat{\sigma}_{\rm RR}-\hat{\sigma}_{\rm RL}}
     {\hat{\sigma}_{\rm LL}+\hat{\sigma}_{\rm LR}+\hat{\sigma}_{\rm RR}+\hat{\sigma}_{\rm RL}},
\label{alrth}
\end{equation}
\begin{equation}
A_{\rm LR,FB}=
\frac{3}{4}\,
\frac{\hat{\sigma}_{\rm LL}-\hat{\sigma}_{\rm RR}+\hat{\sigma}_{\rm RL}-\hat{\sigma}_{\rm LR}}
{\hat{\sigma}_{\rm LL}+\hat{\sigma}_{\rm RR}+\hat{\sigma}_{\rm RL}+\hat{\sigma}_{\rm LR}}.
\label{alrfbth}
\end{equation}
The deviations of measurements of these observables from the SM predictions 
are expressed in terms of the CI, $\epsilon_{ij}$, 
of eq.~(\ref{lagra}).

It can be  seen from eqs.~(\ref{sigth}) and (\ref{afbth}) 
that, with unpolarized beams, the CI couplings  could not be individually
constrained within finite ranges, but only mutual correlations could be
derived. With longitudinal beam polarization, the two additional
physical observables, (\ref{alrth}) and (\ref{alrfbth}), are available 
to obtain finite, model-independent bounds on all CI 
couplings~\cite{Babich:2001nc,Babich:2001ik,Babich:2000kx}.

In principle, the polarization of the electron beam
alone  would be sufficient to achieve model-independent results.
The polarization of both beams increases the
cross sections (see section 1.3)
and improves the 
sensitivity to the new parameters which  in general scales for  
dimension D=6 operators such as (\ref{lagra}), with
 \begin{equation}
\frac{m_X}{g_X}\sim \sqrt{\Delta_{\rm stat} \sigma}\sim ({\cal L}_\text{int} 
\cdot s)^{1/4},
\label{eq_scale1}
\end{equation} 
taking into account statistical errors only.  Further, with both beams
polarized, the error of the effective polarization, $P_\text{eff}$ (see also
section~\ref{subsec:2}), is substantially reduced and involves a higher accuracy of
the $A_{\rm LR}$ measurement: 
 \begin{equation} \Delta A_{\rm LR}= \sqrt{(\Delta_\text{stat}
A_{\rm LR})^2+(\Delta_\text{sys} A_{\rm LR})^2} =\sqrt{ \frac{1-P^2_\text{eff} 
A^2_{\rm LR}}{N
P^2_\text{eff}} + A^2_{\rm LR} \left(\frac{\Delta P_\text{eff}}{P_\text{eff}}\right)^2}.
\label{eq-alr2}
\end{equation}

%%%%%%%%%%%%%%%%%%%%%%%%%%%%%%%%%%%%%%%%%%%%%%%%%%%%%%%%%%%%%%%%%%%%%%%%%%%%%
\subsubsection*{Expected sensitivities to 
contact interactions  \label{sect:ff}}
%%%%%%%%%%%%%%%%%%%%%%%%%%%%%%%%%%%%%%%%%%%%%%%%%%%%%%%%%%%%%%%%%%%%%%%%%%%%%
The analysis of CI with $f\ne e,t$ in ref.~\cite{Babich:2001nc} demonstrates
the sensitivity to $\Lambda$ depending on the linear collider parameters. 
To assess the relative roles of statistical and
systematic uncertainties, the time-integrated luminosity $\Lumint$ is varied
from $50$ to $500\ \mbox{fb}^{-1}$, with uncertainty
$\Delta\Lumint/\Lumint=0.5\%$, and a cut
$|\cos\theta|\le 0.99$ is assumed. The polarization of electron and positron
beams is considered as $(|P_{e^-}|,|P_{e^+}|)=(80\%,0)$ and 
$(|P_{e^-}|,|P_{e^+}|)=(80\%,60\%)$, 
with the uncertainties $\Delta P_{e^-}/P_{e^-}=\Delta P_{e^+}/P_{e^+}=0.5\%$. 
 
The model-independent bounds on the mass scales $\Lambda_{ij}$
in the final $b\bar b$ and $c \bar c$ cases,
allowed by these experimental uncertainties,
are shown in fig.~\ref{Fig:bc}. 
Particle identification efficiencies of 60\% and 30\% are assumed
in the $b\bar b$ and $c\bar c$ channels, respectively \cite{Damerell:1996sv}.
Thick lines in fig.~\ref{Fig:bc} correspond to
$(|P_{e^-}|,|P_{e^+}|)=(80\%,60\%)$ while thin lines correspond
to $(|P_{e^-}|,|P_{e^+}|)=(80\%,0)$.

%%%%%%%%%%%%%%%%%%%%%%%%%%%%%%%%%%%%%%%%%%%%%%%%%%%
\begin{figure}[htb]
\refstepcounter{figure}
\label{Fig:bc}
\addtocounter{figure}{-1}
\begin{center}
\setlength{\unitlength}{1cm}
\begin{picture}(10.0,6.5)
\put(-3.5,0){
\mbox{\epsfysize=7cm\epsffile{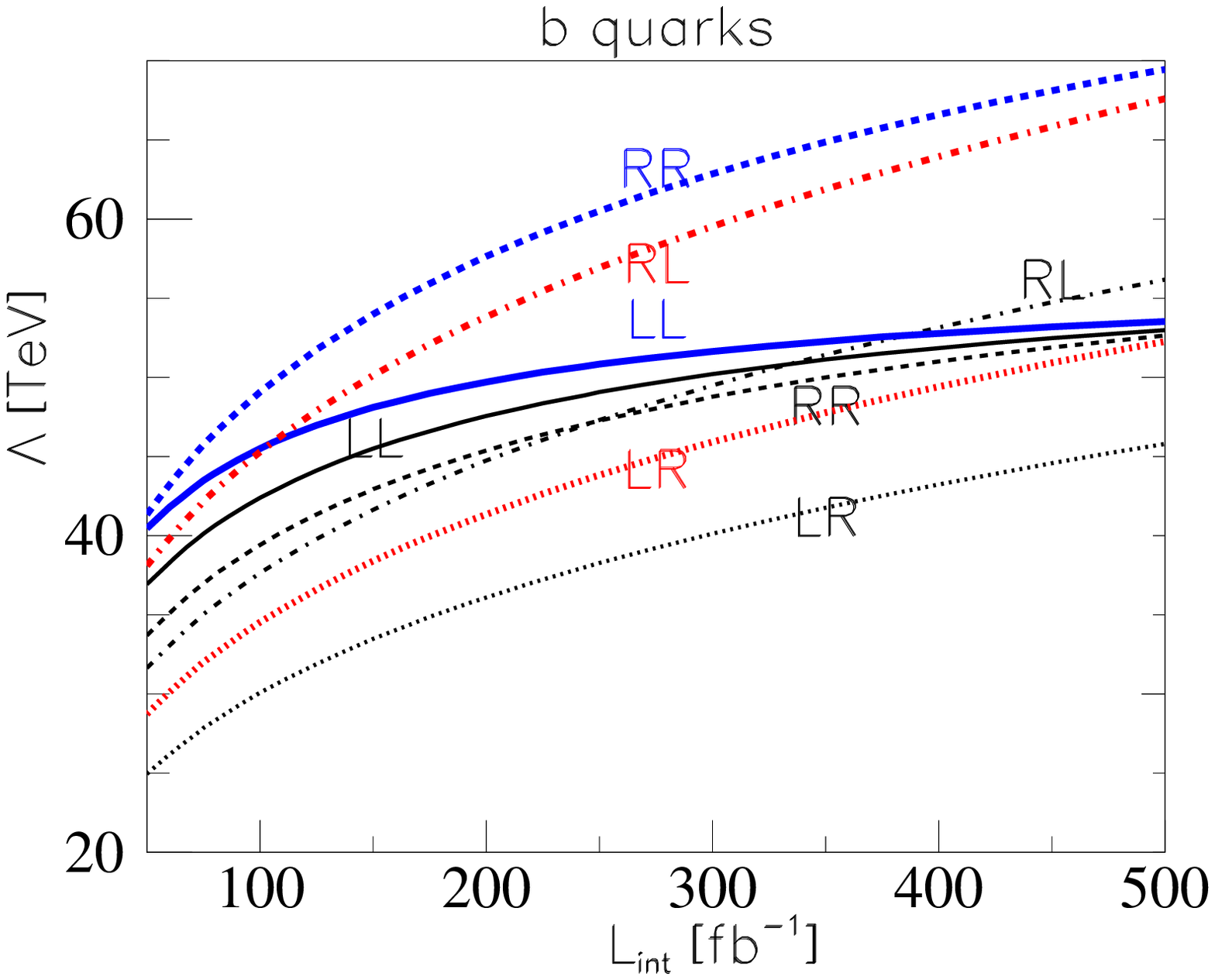}}
\mbox{\epsfysize=7cm\epsffile{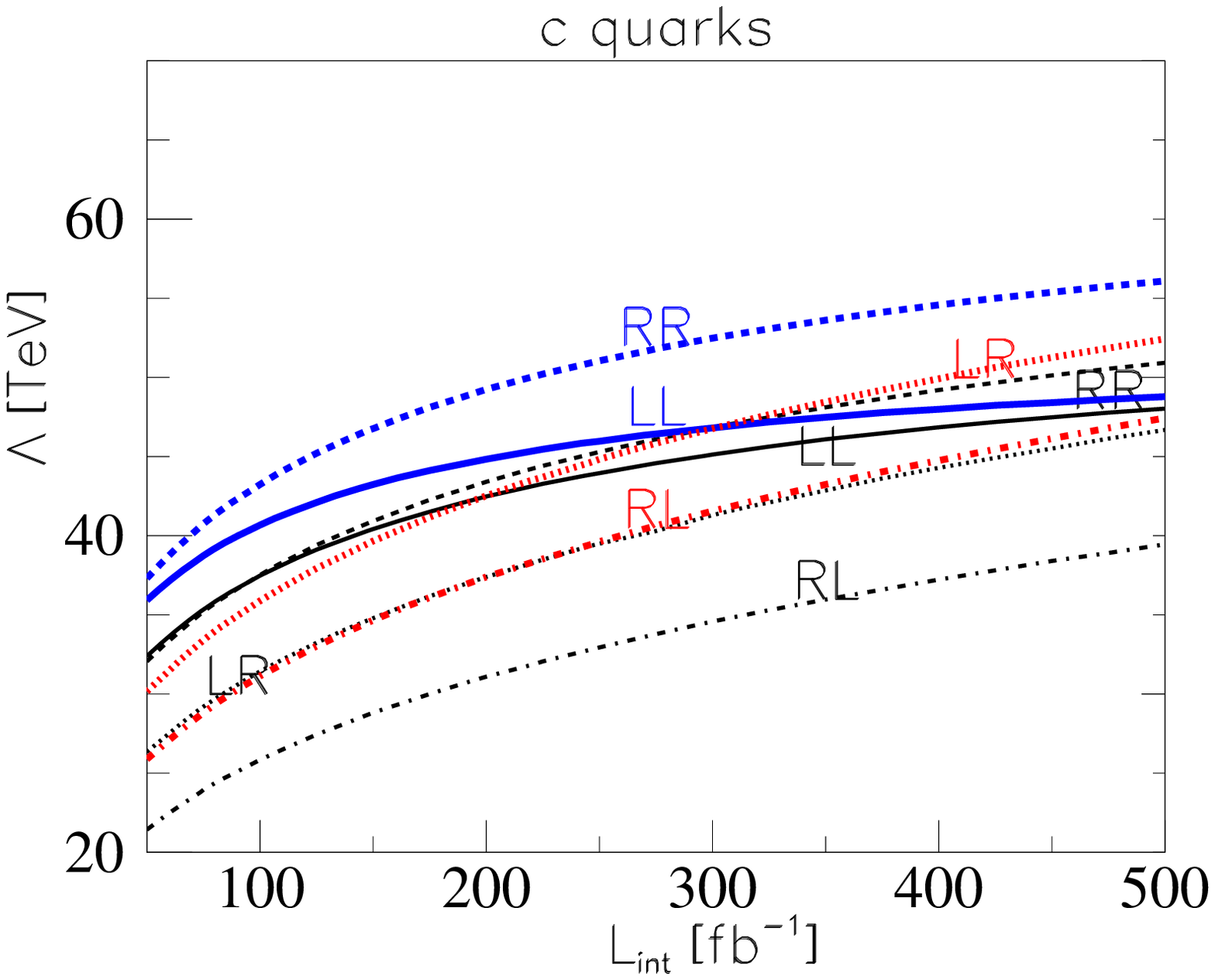}}}
\end{picture}
\vspace*{-4mm}
\caption[Scale of contact interactions vs. integrated luminosity] {Lower bounds (95\% C.L.) on the
scale of CI, $\Lambda$, at $\sqrt{s}=500$~GeV vs.\ the integrated luminosity,
$\Lumint$, for $b\bar{b}$ and $c\bar{c}$ final states taking into account the
four helicity combinations.  Thin lines: $(|P_{e^-}|,|P_{e^+}|)=(80\%,0)$,
thick lines: $(|P_{e^-}|,|P_{e^+}|)=(80\%,60\%)$~\cite{Babich:2001nc}.}
\end{center}
\end{figure}
%%%%%%%%%%%%%%%%%%%%%%%%%%%%%%%%%%%%%%%%%%%%%%%%%%%

In fig.~\ref{fig-sabine} the expected sensitivities for different models
of contact interactions in $e^+e^-\to b \bar{b}$, $c \bar{c}$ are shown,
including systematic ($\Delta_{\rm syst}=0\%$, 0.5\%, 1.0\%), luminosity
($\Delta\Lumint=0.2\%$, $0.5\%$) and polarization uncertainties ($\Delta
P_{e^-}/P_{e^-}=\Delta P_{e^+}/P_{e^+} =0\%$, $0.5\%$).  It can clearly be
seen that the reduction of systematic errors will be decisive to extend the
reach on the $\Lambda$s.  This study was done for $\sqrt{s}=800$~GeV \cite{LC-TH-2001-007}.

\begin{figure}[htb]
  \begin{center}
    \epsfig{file= 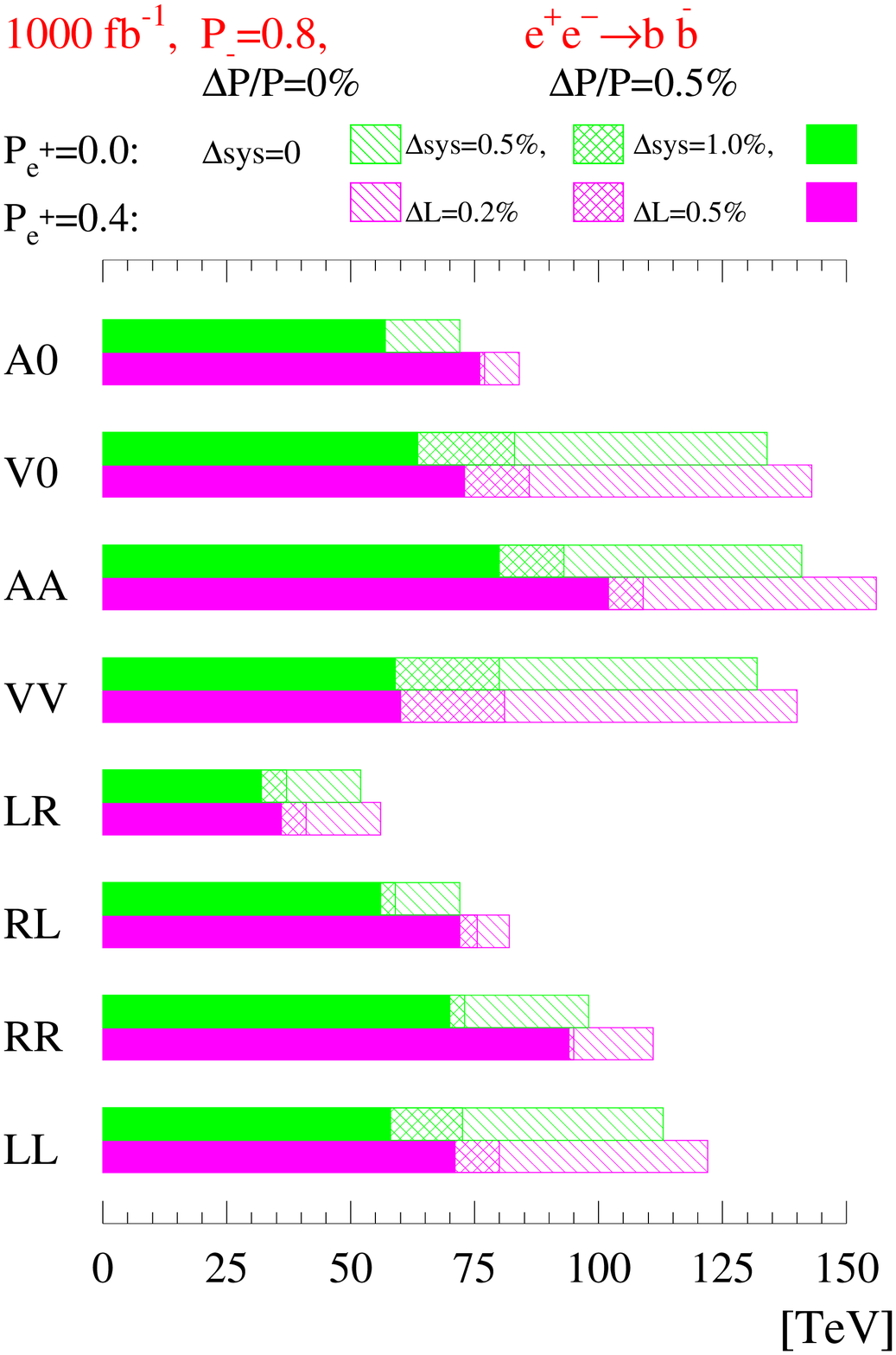,width=6cm,height=8cm}
\hspace{2cm}
    \epsfig{file= 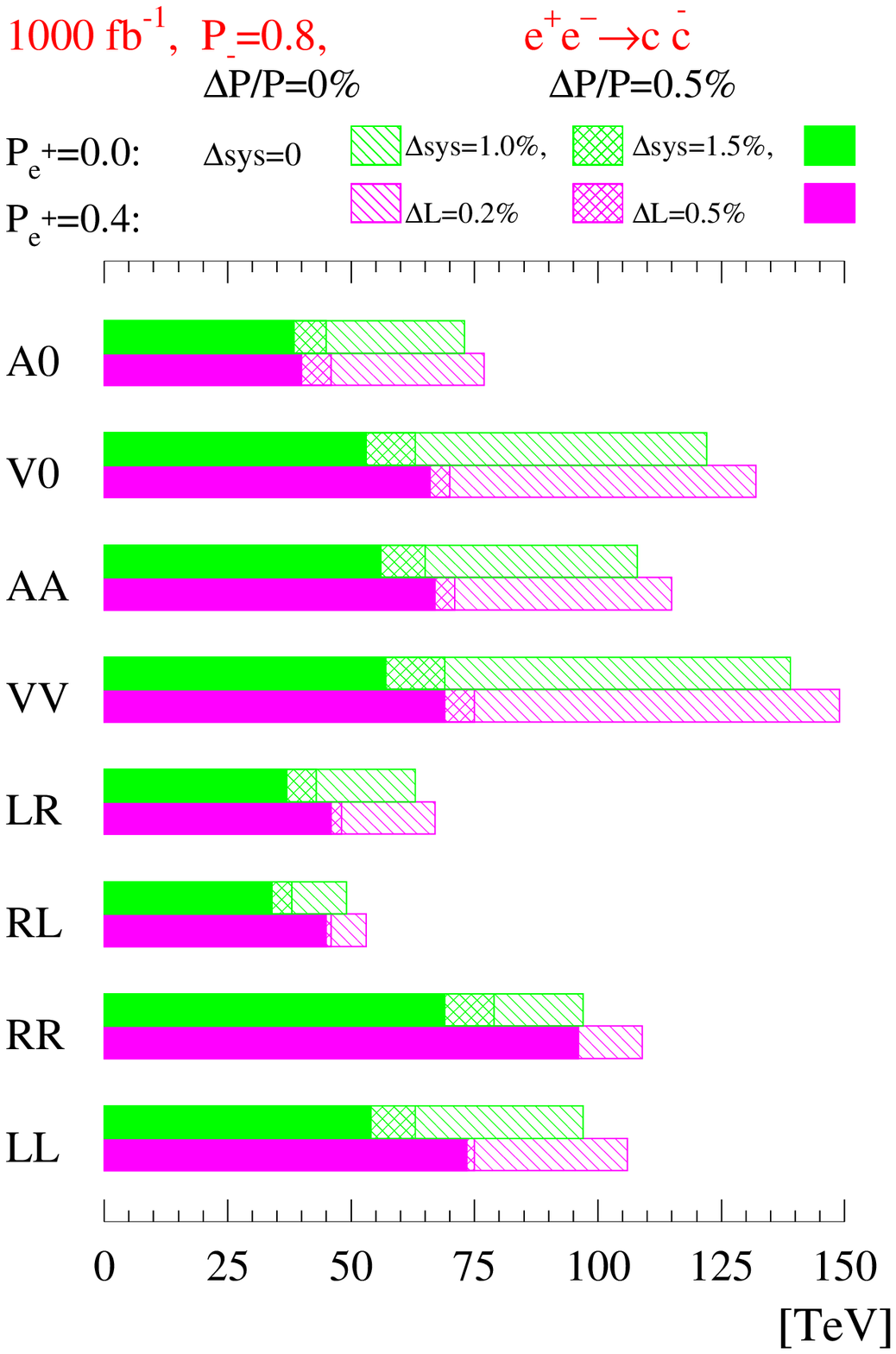,width=8cm,height=8cm}
  \end{center}
\vspace{-1cm}
\caption[Limits ob contact interaction]{ Limits on
contact interactions from $e^+e^- \to b \bar{b}$ and $e^+e^-\to c
\bar{c}$ without positron polarization (green) and with {$40 \%$}
polarization (margenta) for the different uncertainty scenarios
$\Delta P_{e^{\pm}}/P_{e^{\pm}}=0$, $\Delta {\rm sys}=0$, $\Delta L_{\rm int}=0$
(single hatched), $\Delta P_{e^{\pm}}/P_{e^{\pm}}=0.5\%$, $\Delta
{\rm sys}=0.5\%$, $\Delta L=0.2\%$ (double hatched) and $\Delta
P_{e^{\pm}}/P_{e^{\pm}}=0.5\%$, $\Delta {\rm sys}=1.0\%$, $\Delta L_{\rm int}=0.5\%$
(filled) at $\sqrt{s}=500$~GeV and ${\cal L}_{\rm int}=1$~ab$^{-1}$
\cite{LC-TH-2001-007}.}
\label{fig-sabine}
\end{figure}

%%%%%%%%%%%%%%%%%%%%%%%%%%%%%%%%%%%%%%%%%%%%%%%%%%%
\subsubsection*{Contact-interaction analysis in 
Bhabha scattering \label{sect:bhamol}}
%%%%%%%%%%%%%%%%%%%%%%%%%%%%%%%%%%%%%%%%%%%%%%%%%%%

With $\delta_{ef}=1$ the four-fermion contact-interaction Lagrangian of 
eq.~(\ref{lagra}) is relevant to the Bhabha scattering process 
$e^+ e^-\to e^+  e^-$, 
where $\gamma$ and $Z$ bosons are exchanged in both $s$- and $t$-channels.
It turns out that modifications of the pure $t$-channel contribution to the 
cross section, 
$\hat\sigma_{{\rm LR},t}$, depend on the single CI parameter 
($\epsilon_{\rm LR}=\epsilon_{\rm RL}$), while the combinations of
helicity cross sections, $\dd\hat\sigma_{\rm R}$ and  $\dd\hat\sigma_{\rm L}$, 
contribute to the $s$-channel exchange and  
depend on {\it pairs} 
of parameters, ($\epsilon_{\rm RR}$,$\epsilon_{\rm LR}$) and 
($\epsilon_{\rm LL}$,$\epsilon_{\rm LR}$), respectively.

The change of the polarization of each beam allows
the separate measurements of the polarized differential cross
sections $\dd\sigma_{++}$, $\dd\sigma_{+-}$ and $\dd\sigma_{-+}$,
cf. discussion following eq.~(\ref{eq_intro1})\cite{Pankov:2002qk}. These
differential cross sections represent a system of linear equations of
the helicity cross sections $\dd\hat{\sigma}_{\rm R}$,
$\dd\hat{\sigma}_{\rm L}$ and $\hat{\sigma}_{{\rm LR},t}$ and allow
the CI couplings $\epsilon_{\rm LL}$, $\epsilon_{\rm RR}$ and
$\epsilon_{\rm LR}$ to be disentangled.
Equations~(\ref{sigth})--(\ref{alrfbth}) show that this kind of
model-independent analysis requires both $e^+$ and $e^-$ polarized.
Fig.~\ref{Fig:fig-bhabha} shows as an example the result of a $\chi^2$
analysis assuming that no deviation from the SM within the
experimental uncertainty (statistical and systematic) is measured in
$\dd\hat{\sigma}_{\rm L}$, $\dd\hat{\sigma}_{\rm R}$ and
$\dd\hat{\sigma}_{{\rm LR},t}$ for $\Lumint{(e^+e^-)}=50\
\mbox{fb}^{-1}$, $(|P_{e^-}|,|P_{e^+}|)=(80\%,60\%)$,
$\Delta\Lumint/\Lumint=\Delta P_{e^-}/P_{e^-}=\Delta
P_{e^+}/P_{e^+}=0.5\%$.

%%%%%%%%%%%%%%%%%%%%%%%%%%%%%%%%%%%%%%%%%%%%%%%%%%%
\begin{figure}[htb]
\refstepcounter{figure}
\label{Fig:fig-bhabha}
\addtocounter{figure}{-1}
\begin{center}
\setlength{\unitlength}{1cm}
\begin{picture}(10.0,7.3)
\put(-2.5,0.)
{\mbox{\epsfysize=7.5cm\epsffile{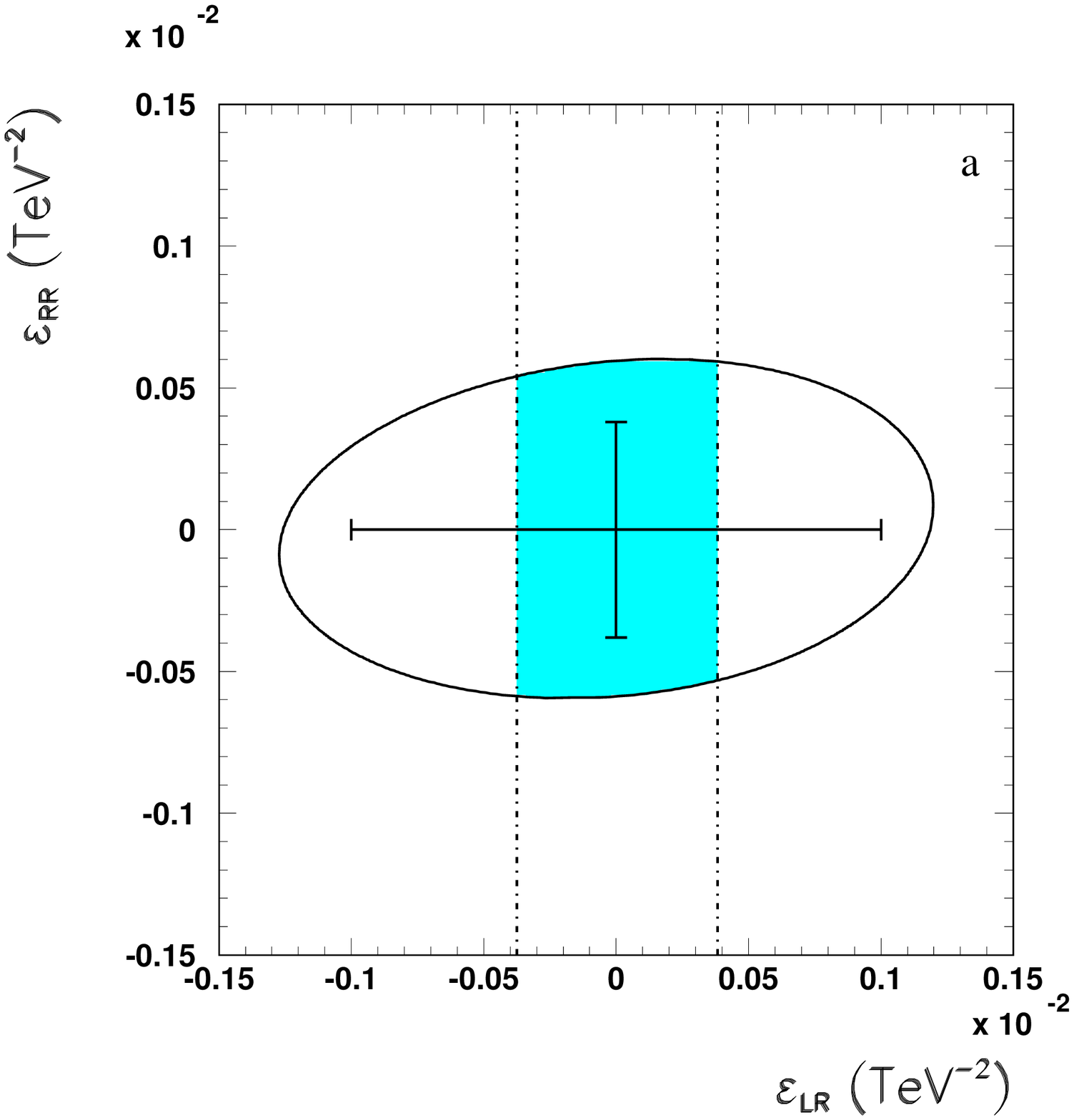}}
 \mbox{\epsfysize=7.5cm\epsffile{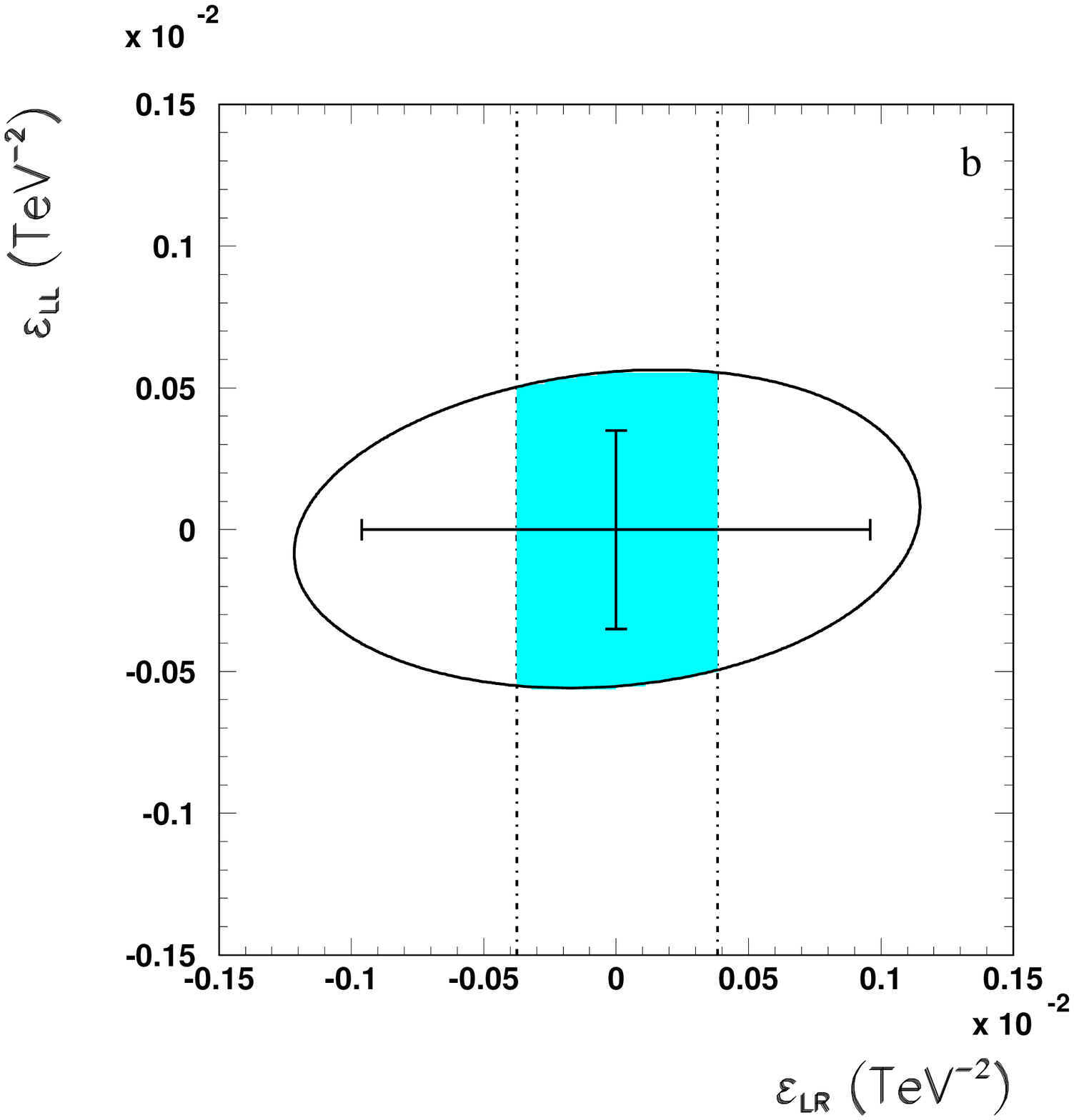}}}
\end{picture}
\caption[95\% C.L. range for contact-interaction couplings] {Allowed areas at $95\%$ C.L. in the
planes $(\epsilon_{\rm LR},\epsilon_{\rm RR})$ and $(\epsilon_{\rm
LR},\epsilon_{\rm LL})$ obtained from $\hat{\sigma}_{\rm R}$ and
$\hat{\sigma}_{\rm L}$ in $e^+e^-\to e^+e^-$ at $\sqrt s=0.5$ TeV,
$\Lumint(e^+e^-)=50\ {\rm fb^{-1}}$, $(\vert \PE\vert,\vert \PP\vert) =(80\%, 60\%)$.  
Vertical dashed lines indicate the range allowed for
$\epsilon_{\rm LR}$ by $\hat{\sigma}_{{\rm LR},t}$.  The crosses indicate the
constraints obtained by taking only one non-zero parameter at a time instead
of two simultaneously non-zero and independent parameters~\cite{Pankov:2002qk}. }
\end{center}
\end{figure}
%%%%%%%%%%%%%%%%%%%%%%%%%%%%%%%%%%%%%%%%%%%%%%%%%%%%%%%%%%%%%%%%%%%%%%
A comparison with M{\o}ller scattering \cite{Pankov:2002qk} 
shows that only
in the case where ${\Lumint}(e^-e^-)$ is not too low are Bhabha and M{\o}ller
scattering complementary as regards the sensitivity to individual
couplings in a model-independent data analysis.

%%%%%%%%%%%%%%%%%%%%%%%%%%%%%%%%%%%%%%%%%%%%%%%%%%%
\subsubsection*{Sensitivity to neutral extra gauge bosons \label{sect:zprime}}
\begin{figure}[htb]
\begin{center}
\begin{picture}(10,6.5)
\setlength{\unitlength}{1cm}
\put(-1.2,5.8){\footnotesize $\sqrt{s}=500$~GeV}
\put(-3,0){\epsfig{file=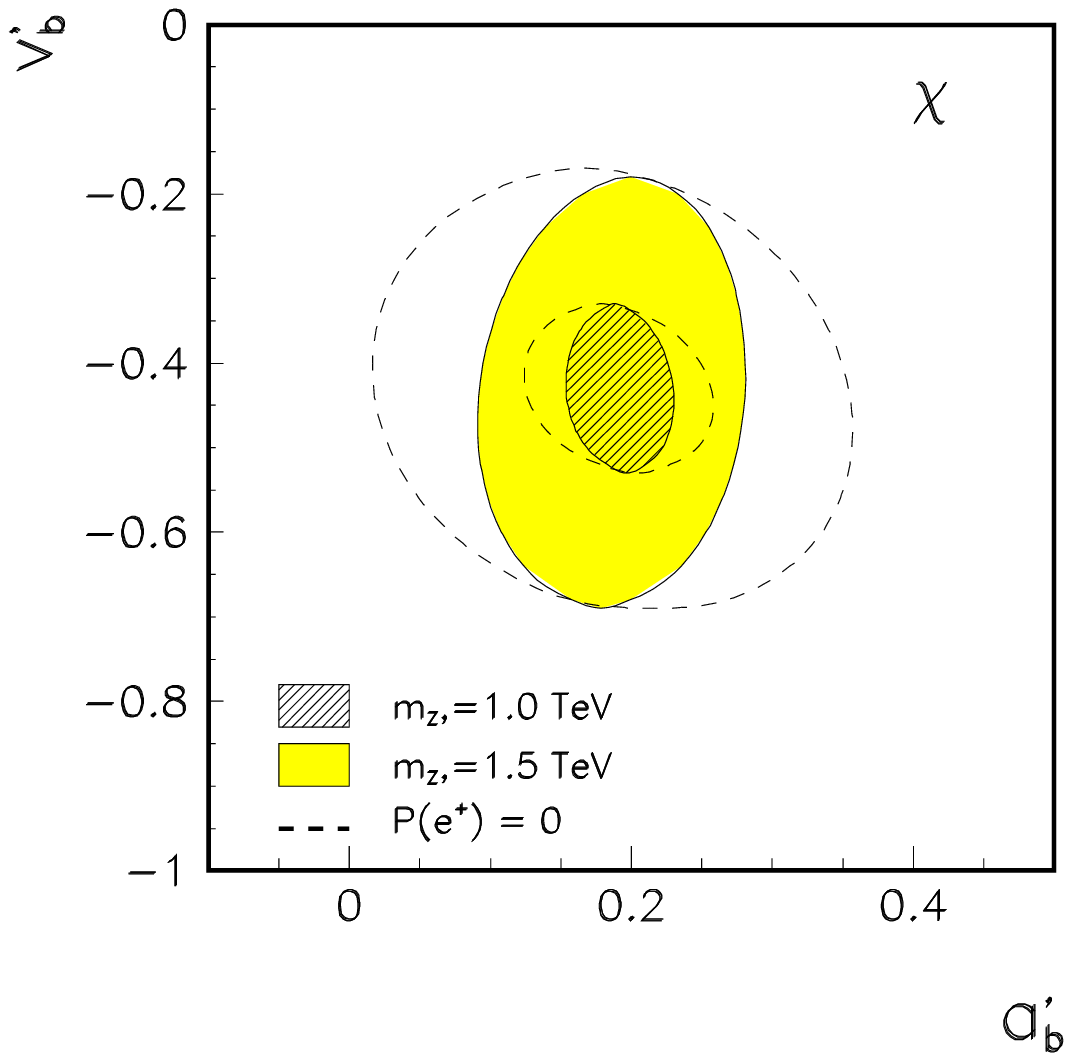,width=7cm,height=6.5cm}}
\put(5,.7){\epsfig{file=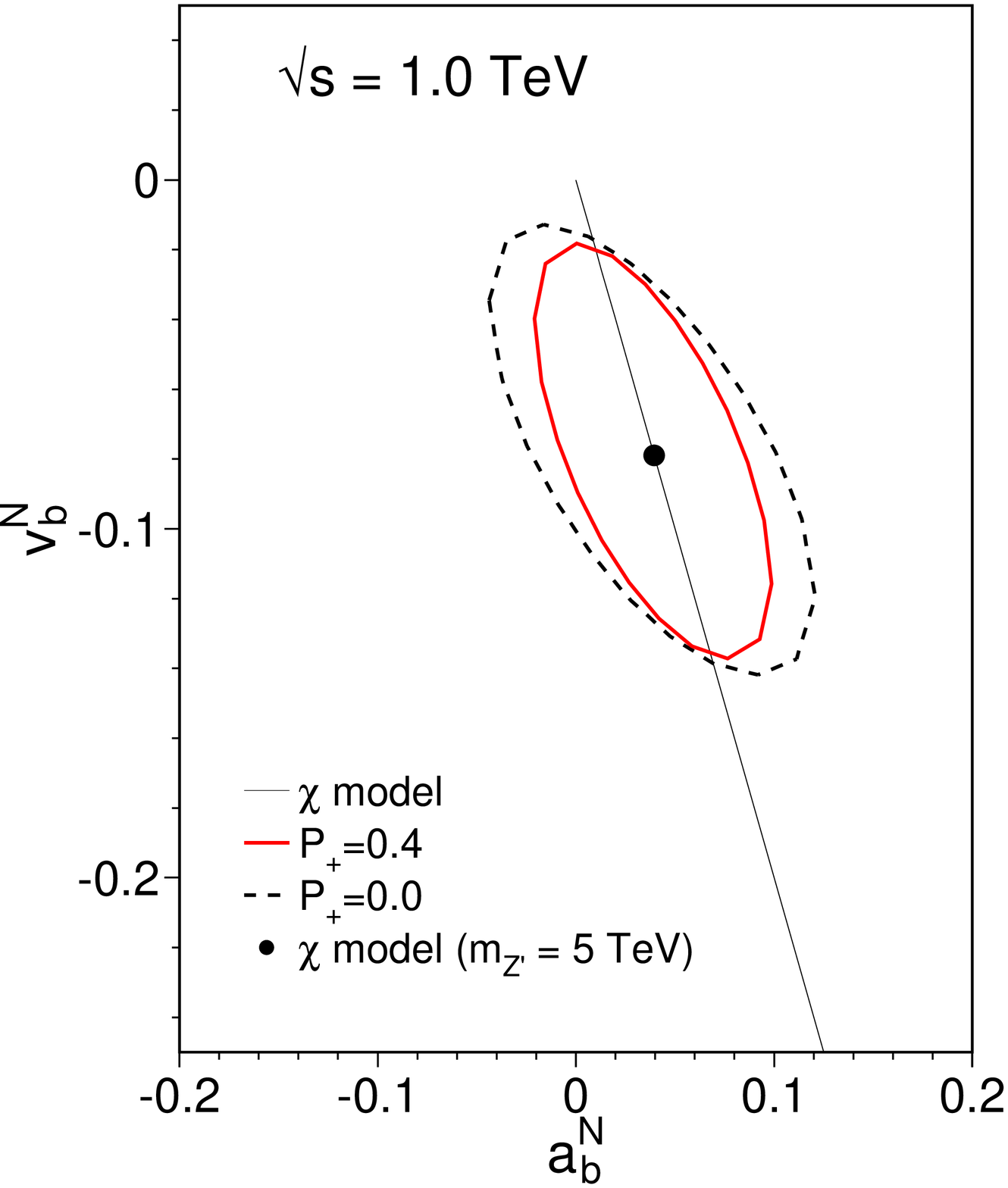,width=7cm,height=5.5cm}}
\end{picture}
\end{center}
\vspace{-1.3cm}
\caption[Determination of $Z'$ couplings and masses 
with polarized beams]{Left: 95\% C.L. contours
for the axial ($a^{'}_b$) and vector ($v^{'}_b$) couplings of the 
$Z'$ for $M_{Z'}=1.0$, 1.5~TeV in the $\chi$ model
with $\sqrt{s}=500$~GeV and ${\cal L}_{\rm int}=500$~fb$^{-1}$. The dashed lines
correspond to $P_{e^+}=0$ \cite{LC-TH-2000-006}; 
$a'_b$, $v'_b$ denote the couplings of $Z'$ to the $b$-quarks.
Right: Expected
resolution power (95\% C.L.) to reconstruct $b\bar{b}$ couplings of 
$Z'$ ($m_{Z'}=5$~TeV)
realized in the $\chi$ model based on the measurement of $b\bar{b}$ final
states. Here, the configuration $(P_{e^-},P_{e^+})=(80\%,40\%)$ is compared with
$(P_{e^-},P_{e^+})=(80\%,0)$. 
The $Z'$ mass is assumed to be unknown in this case~\cite{LC-TH-2001-007}.
The normalized couplings are defined as 
$a^{\rm N}_b=a'_b \sqrt{s/(m^2_{Z'}-s)}$ and 
$v^{\rm N}_b=v'_b\sqrt{s/(m^2_{Z'}-s)}$.
  \label{fig-sabine2}}
\end{figure}

%%%%%%%%%%%%%%%%%%%%%%%%%%%%%%%%%%%%%%%%%%%%%%%%%%%
Extra neutral gauge bosons $Z'$ can be probed by their virtual effects on
cross sections and asymmetries by replacing $\epsilon_{ij} \rightarrow g'_i
g'_j/(s-M_{Z'}^2)$.  For energies below a $Z'$ resonance, measurements of
fermion-pair production are sensitive only to this ratio of $Z'$ couplings 
and
$Z'$ mass.  Therefore, limits on the $Z'$ mass can be obtained only in
dependence on a model with given $Z'$ couplings 
(for reviews see, e.g., \cite{zp,Cuypers:1996jk}).  For example, for the
well-known E$_6$ and
 LR models, mass sensitivities between $4 \times \sqrt{s}$ and
$14 \times \sqrt{s}$ are reached.  Thus, the LC operating at $\sqrt{s}=800$~GeV
may exceed the sensitivity of the LHC (which is about 4--5 TeV) to a potential
$Z'$ in some models. Conversely, if a $Z'$ will be detected at the LHC its
origin can be found by determining the $Z'$ couplings, see
fig.~\ref{fig-sabine2} (left) \cite{LC-TH-2000-006}. While positron-beam
polarization improves only slightly the resolution power for $Z'$ models in
the case of leptonic final states, it will be important for the measurement of
the $Z'$ couplings to quarks.  The polarization of both beams is also
important to enhance the resolution power considering hadronic final
states. In fig.~\ref{fig-sabine2} (right) the reconstruction of the $Z'$ model
is demonstrated without knowledge of the $Z'$ mass based on $b\bar{b}$ final
states. In these analyses the crucial point is the fact that the systematic
errors can be significantly reduced when both beams are polarized
\cite{LC-TH-2000-006,LC-TH-2001-007}.
\smallskip

{\bf \boldmath
Quantitative examples: With the polarization of both beams 
the sensitivity to the
new physics scale can be increased by a factor of up to about 1.3 with respect
to the case with only polarized electrons. A similar gain is reached for the
resolution power for $Z'$ studies in different models. 
The determination of Z' couplings can be substantially improved 
by up to about a factor 1.5. The crucial systematic errors
can be significantly reduced when both beams are polarized.}
\smallskip

%%%%%%%%%%%%%%%%%%%%%%%%%%%%%%%%%%%%%%%%%%%%%%%%%%%
\subsection{Transversely-polarized beams and leptoquark searches }
%%%%%%%%%%%%%%%%%%%%%%%%%%%%%%%%%%%%%%%%%%%%%%%%%%%

{\bf \boldmath
With transversely-polarized beams, CP-conserving and CP-violating
azimuthal asymmetries of the final-state top quark in $e^+e^-\to t \bar{t}$
can be used to probe a scalar leptoquark model.}
\smallskip

It was pointed out in refs.~\cite{Hikasa,Ananthanarayan:2003wi} that transverse
polarization can play a unique role in isolating chirality-violating
couplings, such as scalar or tensorial ones, 
to which longitudinally-polarized beams are
not sensitive.  The interference of new chirality-violating
contributions with the chirality-conserving standard model (SM)
couplings give rise to terms in the angular distribution proportional
to $\sin\theta\,\cos\phi$ and $\sin\theta\,\sin\phi$, where $\theta$
and $\phi$ are the polar and azimuthal angles of a final-state
particle. Chirality-conserving new couplings, on the other hand,
produce interference contributions proportional to
$\sin^2\theta\,\cos2\phi$ and
$\sin^2\theta\,\sin2\phi$. Chirality-violating contributions do not
interfere with the chirality-conserving SM contribution with
unpolarized or longitudinally-polarized beams when the electron mass
is neglected. Hence transverse polarization would enable the
measurement of chirality-violating couplings through the azimuthal
distributions and the unambiguous distinction from chirality-conserving
new interactions.

An example~\cite{Rindani:2004ue} is represented by a 
specific model where chirality violation appears already at the tree level,
by extending the SM gauge group by a
$SU(2)_L$ doublet of scalar leptoquarks (LQ), which couples only to
first-generation leptons and third-generation quarks.  Since the couplings
of leptoquarks to the third generation quarks are relatively weakly
constrained \cite{cdf}, their effect from $t$-channel exchange
in $e^+ e^-\to t \bar{t}$ can be
non-negligible.  The leptoquark couplings of both left
and right chiralities are included in the model, $g_L$ and $g_R$,
and they are also
allowed to be complex. Thus, in principle,
also the possibility of CP violation can be kept
open.  The reader is referred to \cite{buch} for a general discussion
of leptoquark models, and to
\cite{tana} for a brief review of quantum numbers. 

Assume transverse polarizations $P_{e^-}^{\rm T}$ and $P_{e^+}^{\rm T}$ that are
(anti)parallel to each other.  The differential cross section for the
process $e^+e^-\to t \bar{t}$ is here given by the sum of the SM
contribution $\sigma_{\rm SM}$, the pure leptoquark contribution
$\sigma_{\rm LQ}$, and the contribution $\sigma_{\rm int}$ from the
interference of the leptoquark contribution with the SM
contribution. The interference term between the SM $Z$ and the
leptoquark contribution contains terms linear in $P_{e^-}^{\rm T}$ and
$P_{e^+}^{\rm T}$ that are proportional to $\sin\theta\,\cos\phi$ and
$\sin\theta\,\sin\phi$. These terms are proportional to the real and
imaginary  parts of the
chirality-violating couplings $g_Rg_L^*$, respectively, cf. the
analytical expressions in \cite{Rindani:2004ue}, thus, in principle,
also their relative phases can be measured.

The interference terms that are bilinear in $P_{e^-}^{\rm T}$ and $P_{e^+}^{\rm T}$ 
(thus requiring both beams to be polarized) and
proportional to $\sin^2\theta\sin 2\phi$ and $\sin^2\theta\cos 2\phi$
contain $\vert g_L\vert^2$ or $\vert g_R\vert^2$. These terms, however, 
can be studied also using longitudinal polarization.

The interference terms proportional to $\sin\theta\cos\phi$
and $\sin\theta\sin\phi$ are linear in $P_{e^-}^{\rm T}$ and $P_{e^+}^{\rm T}$
(thus not strictly requiring both beams polarized), and are
proportional to $\Re(g_R\,g_L^*)$ and the CP-violating
$\Im(g_R\,g_L^*)$, respectively.

The chirality-violating terms can be isolated by studying the following
azimuthal asymmetries, where $\theta$ is integrated over with a
cut-off $\theta_0$ in the forward and backward directions:
\begin{equation}
A_1(\theta_0) = \frac{1}{r\sigma(\theta_0)} \int_{-\cos\theta_0}^{\cos\theta_0}
	d\cos\theta\left[ \int_0^\pi d\phi - \int_\pi^{2\pi} d\phi \right]
\frac{d\sigma}{d\Omega},
\end{equation}
\begin{equation}
A_2(\theta_0) = \frac{1}{\sigma(\theta_0)} \int_{-\cos\theta_0}^{\cos\theta_0}
	d\cos\theta\left[ \int_{0}^{\pi /2} d\phi - \int_{\pi /2}^{3\pi /2}d\phi 
	+ \int_{3\pi/2}^{2\pi} d\phi\right]
\frac{d\sigma}{d\Omega},
\end{equation}
where 
\begin{equation}
\sigma(\theta_0) = \int_{-\cos\theta_0}^{\cos\theta_0}
d\cos\theta
\int_0^{2\pi} d\phi
\;\frac{d\sigma}{d\Omega}.
\end{equation}
It turns out that $A_1(\theta_0)$ and
$A_2(\theta_0)$ differ only in the factors 
$(P_{e^-}^{\rm T}-P_{e^+}^{\rm T})\,\Im(g_Rg_L^*)$ and $(P_{e^-}^{\rm T}+P_{e^+}^{\rm T})\,\Re(g_Rg_L^*)$.

A numerical study has been done for 
$\sqrt{s} = 500$ GeV and $\Lumint=500$~fb$^{-1}$.
Furthermore, 
$(P_{e^-}^{\rm T},P_{e^+}^{\rm T})=(80\%,-60\%)$, $g_L=1/\sqrt{2}$, 
$g_R=i/\sqrt{2}$ 
which corresponds to maximal CP violation in the leptoquark couplings
and a leptoquark mass $M_{\rm LQ}=1$~TeV.
The asymmetry $A_1(\theta_0)$ is in this case of the order of $4\times
10^{-3}$, and is not very sensitive to the cut-off $\theta_0$.
In the case without CP violation, $g_L=1/\sqrt{2}$ and
$g_R=1/\sqrt{2}$, the sign of $P_{e^+}^{\rm T}$ is
chosen positive to maximize the asymmetry and one obtains the same 
size for $A_2(\theta_0)$.

\begin{figure}[htb]
\centering
\begin{tabular}{ll}
\hskip -2cm 
\epsfig{file=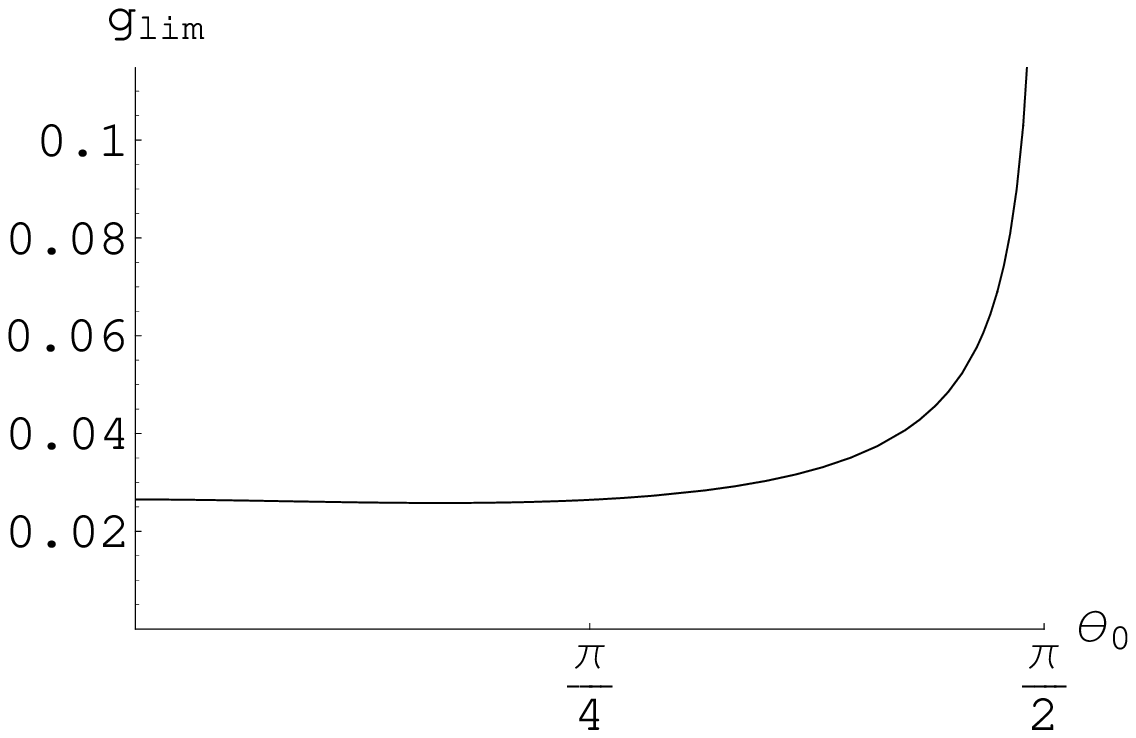,height=5cm}&
\hskip -2cm 
\epsfig{file=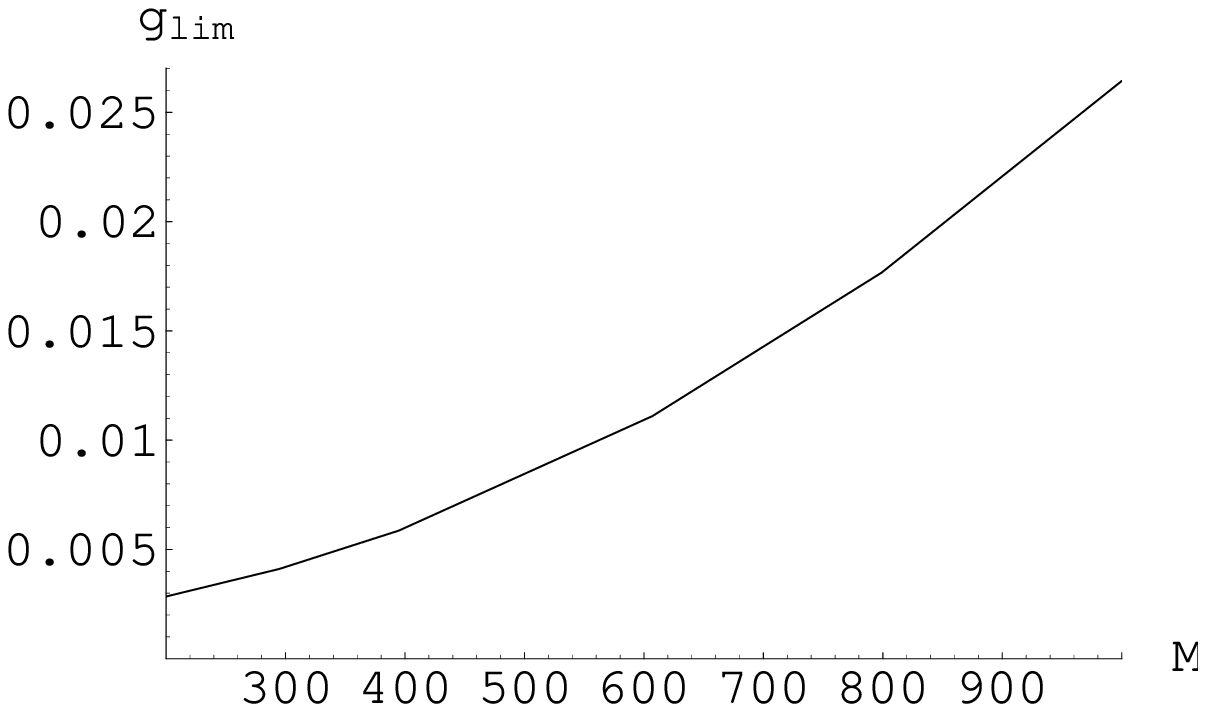,height=5cm}
\end{tabular}
\caption[Limit on chirality-violating couplings with azimuthal 
asymmetries]{
The 90\% CL limit $g_{\rm lim}$ that can be obtained on ${\rm Re} (g_Rg_L^*)$
or ${\rm Im} (g_Rg_L^*)$ respectively from $A_1$ or $A_2$ for 
$\sqrt{s}=500~\text{GeV}$ and $\Lumint=500~\text{fb}^{-1}$ vs.\ $\theta_0$  (left panel)
and $M[GeV]$ (right panel)~\cite{Rindani:2004ue}. \label{fig_lq}}
\end{figure}

Fig.~\ref{fig_lq} shows the 90\% C.L.\
limit $g_{\rm lim}$ that can
be put on the combinations ${\rm Im}(g_Rg_L^*)$ (in the maximal CP violation
case) and ${\rm Re}(g_Rg_L^*)$ (in the CP conservation case).
This limit is obtained by equating the asymmetry to 1.64/$\sqrt{N_{\rm SM}}$,
where $N_{\rm SM}$ is the number of SM events, $N_{\rm SM}= \sigma_{\rm SM}
(\theta_0)\,{\cal L}_{\rm int}$. The possible limit $g_{\rm lim}$ on ${\rm Re}(g_Rg_L^*)$ 
or ${\rm Im}(g_Rg_L^*)$ is
about $2.5\times10^{-2}$ for most values of $\theta_0\leq\pi/4$ 
and $M_{\rm LQ}=1000$ GeV.  

These are `direct' limits.
In the case where the leptoquark has both left- and right-handed
couplings, strong indirect limits exist \cite{limits,Herczeg:2003ag},
the most stringent ones coming from dipole moments of the electron.
Requiring the contribution to the electron 
$g_e-2$ coming from one-loop diagrams
with top and leptoquark internal lines 
to be less than the experimental uncertainty of $8\times 10^{-12}$ gives
${\rm Re}(g_Rg_L^*)/  \left(M_{\rm LQ}/{\rm TeV}\right)^2  < 0.1$,
and therefore the 
limits obtainable from the asymmetry $A_2$ are clearly 
more stringent.
Conversely, the direct limit obtainable from $A_1$, viz., ${\rm Im} 
(g_Rg_L^*) < 2.5\times10^{-2}$ (for $M_{\rm LQ}=1$ TeV)
is not competitive with the indirect constraint
derived from the experimental limit on the
electric dipole moment $d_e$ of about $10^{-27}$ e cm,
which leads to 
$ {\rm Im}(g_Rg_L^*)/ \left(M_{\rm LQ}/{\rm TeV}\right)^2  < 10^{-6}.$

In conclusion, azimuthal asymmetries single out
chirality-violating couplings in scalar-leptoquark models. 
Longitudinal beam
polarization can only put limits on the absolute
values of the left and right chiral couplings. 
In the considered example the limit that can be put
on the real part of the product of the 
couplings $g_Rg_L^*$ is about four times better
than the indirect limit from the $g_e-2$ of the electron.
\smallskip

{\bf \boldmath
Quantitative examples: The CP-conserving asymmetry provides a 4
times better limit for the chirality-violating couplings than the indirect
most stringent limit coming from magnetic dipole moments.  The 
CP-violating asymmetry could also directly probe
CP-violating phases in the leptoquark sector, however, indirect limits from
the electron electric dipole moment are superior.}

%%%%%%%%%%%%%%%%%%%%%%%%%%%%%%%%%%%%%%%%%%%%%%%%%%%
\section{Models of gravity in extra dimensions \label{new-physics}}
%%%%%%%%%%%%%%%%%%%%%%%%%%%%%%%%%%%%%%%%%%%%%%%%%%%

%%%%%%%%%%%%%%%%%%%%%%%%%%%%%%%%%%%%%%%%%%%%%%%%%%%
\subsection{Direct graviton production \label{sect-edback}}
%%%%%%%%%%%%%%%%%%%%%%%%%%%%%%%%%%%%%%%%%%%%%%%%%%%
{\bf \boldmath
A signature for direct graviton production, envisaged in 
formulations of gravity with extra spatial dimensions,
is a relatively
soft photon and missing energy.  The major background process is $\gamma \nu
\bar{\nu}$ production. It is possible to determine the fundamental 
mass scale of gravity as well as
the number of extra dimensions independently.  Background suppression with
right-handed electrons and left-handed positrons is extremely important,
and the discovery reach is significantly enhanced.}
\smallskip

Scenarios of gravity in extra spatial dimensions are currently being 
considered
with great attention in the context of the hierarchy problem between
the Planck and the Fermi mass scales~\cite{add}. The basic idea of 
the scenario of Arkani-Hamed, Dimopoulos and Dvali (ADD), 
is that only gravity can propagate in a bulk
with $3$+$d$ ($d\ge 2$) spatial dimensions compactified
to a radius $R$, while SM particles live in the usual four-dimensional
space. The corresponding fundamental mass scale $M_D$, related to the
four-dimensional Planck mass scale by 
\begin{equation}
G_N^{-1}=8 \pi R^d M_D^{2+d},
\label{def_gravscale}
\end{equation}
with $G_N$ the Newton constant, can for $R$ of the sub-millimeter
size be of the TeV order hence in the
sensitivity reach of highest energy colliders.

In the usual four-dimensional space, the characteristic manifestation
of the  massless graviton propagating in $4+d$ extra dimensions is represented
by the propagation (and emission) of a tower of massive Kaluza-Klein (KK)
graviton excitation states.
Accordingly,
KK gravitons behave as massive, neutral particles very weakly 
(gravitationally) interacting with SM particles, their emission at colliders 
would typically be signalled by events with large missing transverse energy. 
The emission process is calculable in terms of an effective theory, where the 
number of extra dimensions $d$ and the fundamental gravitational mass scale 
$M_D$ appear as parameters~\cite{Giudice:1998ck,Han:1998sg}. 
Such a 'low-energy' 
effective interaction can give reliable results only for c.m. energy 
smaller than $M_D$, depending on the actual value 
of $d$\footnote{Numerous
alternative model realizations of gravity in extra dimensions exist, 
with different features from the ADD regarding, e.g., the KK graviton
spectrum, the $R$ and $d$-admitted values, and even the possibility
that also  SM matter could live in higher-dimensional spaces. For
a phenomenological review of models and the current experimental situation,
see e.g.~\cite{Cheung:2004ab}}., 
and is not supposed to be able to predict the model-dependent, and 
unknown, physical effects at scales above $M_D$. These effects should be 
related to uncalculated, higher dimensional, operators correcting 
the effective theory, and might mimic graviton emission signals.   

At the linear collider, a candidate reaction for directly producing 
gravitons is $e^+e^-\to \gamma G$. The differential cross section is 
($x_\gamma=2E_\gamma/\sqrt s$, $z=\cos\theta$): 
\begin{equation}
\frac{d\sigma}{dx_\gamma dz}=\frac{\alpha}{64 s}\frac{2\pi^{d/2}}{\Gamma(d/2)}
\left(\frac{\sqrt s}{M_D}\right)^{d+2} 
x_\gamma\left(1-x_\gamma\right)^{d/2-1} F_1(x_\gamma,z),
\label{graviemission}
\end{equation}
where the function $F_1$ is explicitly given 
in~\cite{Giudice:1998ck}. 
By its power-dependence on the ratio $\sqrt s/M_D$, eq.~(\ref{graviemission}) 
clearly shows its effective-interaction character: at fixed $d$ the cross 
section increases with the c.m. energy and decreases for larger $M_D$ while, 
recalling that $\sqrt s/M_D$ must be smaller than unity for the 
effective-theory to be reliable, larger $d$-values result in smaller cross 
sections, hence in decreasing statistics. Also, eq.~(\ref{graviemission}) shows 
that for $d>2$ the photon spectrum tends to concentrate towards small energy, 
so that relatively 'soft' photons should be expected.

The major SM background is determined by $e^+e^-\to \gamma\nu\bar\nu$. The 
contribution from $e^+e^-\to\gamma Z\to\gamma\nu\bar\nu$ can easily be  
eliminated by cutting out the $E_\gamma$ region around the corresponding 
$Z$-peak, but there remains a significant, continuous, distribution in 
$E_\gamma$ from $e^+e^-\to\gamma\nu\bar\nu$ that has similar behaviour as the 
signal. This part of the background must therefore be calculated by 
appropriate simulation codes and subtracted in order to use the eventual 
excess single-photon cross sections to derive information on $d$ and $M_D$. 
In this regard, since the neutrino coupling is only left-handed, the 
background has nearly maximal polarization asymmetry and, consequently, 
polarized electron and positron beams should be extremely effective in 
suppressing the $\gamma\nu\bar\nu$ channel. 

\begin{figure}[htb]
\setlength{\unitlength}{1cm}
\begin{picture}(12,8)
\put(4,0.0)
{\mbox{\epsfysize=8.2cm\epsffile{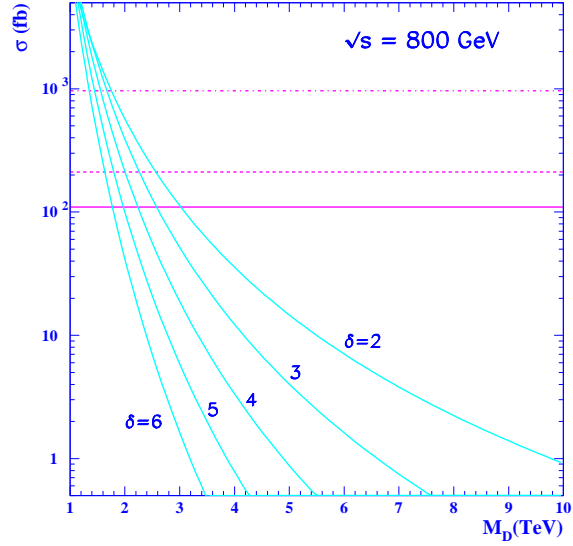}}}
\end{picture}
\caption[Graviton production cross sections vs. $M_D$]{Total 
cross sections for
$e^+e^-\to \gamma G$ at $\sqrt{s}=800$~GeV as a function of the scale $M_D$
and of the number $d$ of extra dimensions. The cross section is for 
$(P_{e^-},P_{e^+})=(80\%,-60\%)$. The three
horizontal lines indicate the background cross sections from $e^+e^-\to \nu
\bar{\nu} \gamma$ for both beams polarized (solid), only electron beam
polarized (dashed) and no polarization (dot-dashed). The signal cross
sections are reduced by a factor of 1.48 for the latter two scenarios
\cite{graham}.  \label{fig1_graham}}
\end{figure}

An example is presented in fig.~\ref{fig1_graham} which shows, 
for different $d$ and $M_D$,  
the signal $e^+e^-\to\gamma G$ cross section at 
$\sqrt{s}=800$~GeV, taking into account also various acceptance 
cuts, ISR and beamstrahlung effects \cite{graham}. The 
background $\gamma\nu\bar\nu$ for various choices of $e^+$ and $e^-$ 
polarizations is also reported, and the comparison with the signal clearly 
shows the strong reduction power obtainable from beam polarization, such 
that the LC potential for exploring the graviton emission is greatly enhanced.

The corresponding 5-$\sigma$ discovery reach on the fundamental gravitational 
scale $M_D$ for various numbers of extra dimensions, at 
$\sqrt s=800$~GeV and 
${\cal L}_{\rm int}=1$~ab$^{-1}$, is shown for both unpolarized 
and polarized beams in table~\ref{tab_ed} \cite{graham}. 
Here the benefits of high 
degree initial polarizations are quite evident.

\begin{table}[htb]
\begin{center}
\begin{tabular}{|r|c|c|c|}
\hline
$M_D$ [TeV] & $(|P_{e^-}|,|P_{e^+}|)=(0,0)$ & $(|P_{e^-}|,|P_{e^+}|)=(80\%,0)$ 
& $(|P_{e^-}|,|P_{e^+}|)=(80\%,-60\%)$ \\ \hline
$d=2$ &   4.48 &   6.27 &   7.86 \\
3 &   3.54 &   4.63 &   5.55 \\
4 &   2.91 &   3.64 &   4.23 \\ 
5 &   2.47 &   3.00 &   3.41 \\ \hline
\end{tabular}
\caption[Discovery reach for large extra dimensions]{Discovery (5-$\sigma$) 
reach $M_D$ for direct graviton production $e^+e^-\to\gamma G$ 
with ${\cal L}_{\rm int}=1$~ab$^{-1}$
at $\sqrt{s}=800$~GeV, for various numbers of extra dimensions 
$d$~\cite{graham}. The major background,
$e^+e^-\to \nu\bar{\nu}\gamma$, can be efficiently suppressed with beam
polarization.  
\label{tab_ed}}
\end{center}
\end{table}

If extra dimensions are the origin of the anomalous single-photon rates, a  
determination of the number of extra dimensions could in principle be 
attempted by measuring the excess (polarized) cross sections at different 
c.m. energies and making a fit to the expected behaviour 
$\sigma\sim (\sqrt s)^d$; some numerical estimates of the sensitivity are 
presented in ref.~\cite{graham}. Hopefully, the result for $d$ should be an 
integer number in order to interpret excess events as graviton emission. In 
conclusion, studying single-photon plus missing energy events at the LC with 
different c.m. energies and beam polarizations can be essential not only for 
model-independent tests of gravity in extra dimensions, but also to help 
distinguishing graviton emission signals from effects of higher dimensional 
operators. 

Studies of graviton emission can also be performed at hadron colliders, e.g.,  
at the LHC the leading experimental signal is expected from inclusive  
$pp\to{\rm jet}+\mbox{$\not\!E$}_{\rm T}$ 
from $qg\to qG$, $q{\bar q}\to gG$ and 
$gg\to gG$. The situation in this case is complicated by the fact that the 
parton c.m. sub-energy $\sqrt{\hat s}$ can take different values, and the 
elementary production sub-processes may well-occur in regions where 
$\hat s\gg M_D^2$ and the 'low-energy' effective-theory approach is not 
applicable to predict the excess 'monojet' cross sections.\footnote{At the LC 
with foreseen c.m. energies in the TeV region, the condition $\sqrt s/M_D<1$ 
assuring the applicability of the effective-theory is, conversely, naturally 
fulfilled.} Thus, on the one hand $M_D$ must be not too small, and on the 
other hand $d$ must be not too large. In practice a range in $d$ and 
$M_D$ must be pre-determined phenomenologically, by truncation procedures in 
$\hat s$ of the parton-model cross sections and cuts on other kinematical 
variables. Also, initial polarization is not available for background 
suppression. Typically, the $M_D$ ranges in which a $5\sigma$ discovery 
potential may be achieved at the LHC with 
${\cal L}_{\rm int}=100\hskip 3pt{\rm fb}^{-1}$, are (in TeV): 
$4.0-7.5$, $4.5-5.9$ and $5-5.3$ for $d=2$, 3 and 4, respectively 
\cite{Vacavant:2000wz}. For higher values of $d$ the effective theory 
may break down, whereas they may still be in the reach of the ILC.

One can conclude, from all the considerations exposed above, that the LC with 
polarized beams has a really important role, also as a complement to LHC, 
in searches and tests of gravity in extra dimensions.   
\smallskip 

{\bf \boldmath
Quantitative examples: The detection of direct graviton signals critically 
depends
on the suppression of the dominant background process
$\gamma\nu\bar{\nu}$.  Compared with the case of only electrons
polarized, the background process is suppressed by a factor of about 2 with
60\% positron polarization whereas the signal is enhanced by a factor of about
1.5.  Furthermore, the discovery reach is enhanced by 
an amount depending on the number of extra dimensions, for example by 
about 25\% (14\%) for $d=2$ ($d=5$).}

%%%%%%%%%%%%%%%%%%%%%%%%%%%%%%%%%%%%%%%%%%%%%%%%%%%
\subsection{Signatures of extra dimensions in fermion pair 
production \label{sect:graviton}}
%%%%%%%%%%%%%%%%%%%%%%%%%%%%%%%%%%%%%%%%%%%%%%%%%%%
{\bf \boldmath
Indirect signals of TeV-scale gravity propagating in large, compactified,
extra spatial dimensions can be probed in the framework of effective 
contact interactions. With both beams longitudinally polarized, the reach on the
relevant mass scales and their identification over indirect effects from
different kinds of non-standard interactions, can be improved by 10--15\%
with respect to the cases of no polarization or only electron polarization.
Furthermore, transverse beam polarization allows an unambiguous distinction
between different realizations of extra-dimensional space-time.}
\smallskip

Two typical model examples of the ADD~\cite{add} and
Randall-Sundrum (RS)~\cite{Randall:1999ee} scenarios are discussed in the following.

In the ADD model, the exchanged tower of KK graviton excitations has
an evenly spaced, and almost continuous, mass spectrum with steps
$\Delta m\propto 1/R$. The summation over the KK states results
in an effective $D=8$ graviton-exchange interaction, which can be written in
the notation of~\cite{Hewett:1998sn}:

\begin{equation}
{\cal L}^{\rm ADD}=i\frac{4\lambda}{M_H^4}T^{\mu\nu}T_{\mu\nu},
\label{dim-8}
\end{equation}
with $T_{\mu\nu}$ the stress-energy tensor of SM particles, $M_H$ a cut-off
on the summation over the KK spectrum expected of the TeV order, and
$\lambda=\pm1$. The exchange of spin-2 graviton fields introduces, in the
helicity amplitudes for $e^+e^-\to f\bar f$, additional terms with new and
characteristic angular dependences \cite{Cullen:2000ef,Hewett:1998sn}. 
Such deviations can be
parametrized phenomenologically by the coupling:
\begin{equation}
f_G=\frac{\lambda s^2}{4\pi\alpha_{\rm em}M_H^4}.
\label{f_g}
\end{equation}
The (high) dimensionality $D=8$ of (\ref{dim-8}) implies the suppression
of deviations from the SM cross section originating from graviton-exchange
by the large power $(\sqrt s/M_H)^4$.  This should be compared to the case
of the four-fermion contact-interaction scales $\Lambda$, where the effects
are suppressed by only $(\sqrt s/\Lambda)^2$, so that numerically
lower bounds on $M_H$ can be expected on statistical grounds. 
Indeed, the search reach on $M_H$ scales, 
according to previous arguments, as
\begin{equation}
M_H\sim [s^{(d -5)} 
{\cal L}_\text{int}]^{1/(2d-8)}= (s^3\cdot{\cal L}_\text{int})^{1/8}.
\label{eq_scale-add}
\end{equation}

In the simplest version of the RS scenario 
\cite{Randall:1999ee,Davoudiasl:1999tf} the setting
is a five-dimensional space-time, the KK excitations are not equally 
spaced and
the characteristic feature is the existence of a spectrum of narrow spin-2
resonance states with masses expected in the TeV range. 
Formally, this can be obtained from (\ref{dim-8}) by the replacement
\begin{equation}
\frac{\lambda}{M_H^4}\to\frac{-1}{8\Lambda_\pi^2}
\sum_n\frac{1}{s-m_n^2+im_n\Gamma_n},
\label{rs}
\end{equation}
where $\Lambda_\pi$ is of the TeV order and $m_n$ ($\Gamma_n$) are the masses
(widths) of the TeV scale KK excitations. (In the applications presented here,
widths are neglected in the evaluation of the relevant cross sections
\cite{Rizzo:2002pc}.)  

The spin-2 nature of these graviton exchanges can be explicitly verified
through the direct analysis of the resonances themselves, were they actually
produced. Conversely, with the linear collider energy below the production
threshold, the indirect signals of $s$-channel exchange of such massive
graviton fields through deviations of the 
$e^+e^-\to f\bar f$ cross sections from the SM
predictions can be tested, and distinguished by appropriately defined
observables, from different possible new physics sources of deviations, such as
the `conventional' 4-fermion contact interactions
\cite{Eichten:1983hw, Babich:2000kx, Pankov:2002qk}.

%%%%%%%%%%%%%%%%%%%%%%%%%%%%%%%%%%%%%%%%%%%%%%%%%%%
\subsection*{Identification of graviton-exchange effects}
%%%%%%%%%%%%%%%%%%%%%%%%%%%%%%%%%%%%%%%%%%%%%%%%%%%
The discovery and the identification reaches on, respectively, $\Lambda$ and
$M_H$ can be assessed by analysing either the dependence of cross sections on
the polar angle, or some suitably defined asymmetries
among integrated differential distributions. 
The sensitivity to new-physics effects can be described by
\begin{itemize}
\item[a)] 
the discovery reach, which gives   
values for $\Lambda$ or $M_H$ 
up to which a deviation from the Standard Model predictions can be observed,  
and 
\item[b)] the identification reach corresponding to values for  
$\Lambda$ or  $M_H$ up to which the particular models producing
the deviation can be differentiated from each other.
\end{itemize}
The 
identification reach at the ILC on $M_H$ at 5-$\sigma$ 
from $e^+e^-\to f\bar f$ is shown in the left panel of
table~\ref{tab-ed1}, for different luminosities and longitudinal polarization
configurations. Systematic uncertainties on luminosity
and beam polarization, in addition to the statistical ones,
have been taken into account. 

A clear signature of graviton exchange arises in
the differential angular distributions, in the left--right
asymmetry and in the center--edge asymmetries (see, e.g.,
\cite{Hewett:1998sn,Osland:2003fn,Rizzo:2002pc,LC-TH-2001-007}).  In
the left panel of table~\ref{tab-ed1}, the values of $M_H$,
indicated as a 5$\sigma$ identification reach, represent the
resolution power for distinguishing signals of graviton 
exchange (\ref{f_g}) from effects
of contact interactions (\ref{lagra}). The results include the
combination of the channels $f=\mu,\tau, b,c$ and are a good
example of the benefits of positron longitudinal polarization.

%%%%%%%%%%%%%%%%%%%%%%%%%%%%%%%%%%%%%%%%%%%%%%%%%%%
\subsection{Use of transversely-polarized beams for graviton searches}
%%%%%%%%%%%%%%%%%%%%%%%%%%%%%%%%%%%%%%%%%%%%%%%%%%%
{\bf \boldmath If both beams are transversely polarized, additional
azimuthal asymmetries sensitive to spin-2 graviton exchange can be
defined.  Although the identification reach on the scale $M_H$ is
similar to that obtained from just longitudinal electron and positron
polarization, the  specific, azimuthal asymmetries allow an
unambiguous separation of the indirect manifestations of ADD- and
RS-like extra-dimensional models.}
\smallskip

With reference to the Introduction, eq.~(\ref{eq_general}), the
$\phi$-dependent part of the matrix element squared for $e^+e^-\to
f\bar f$ will now be considered.  In the study reviewed here
only terms bilinear  in the transverse polarization
contribute, and the two transverse polarization vectors are assumed to
be antiparallel.  In the observables considered in the previous
subsection, this contribution to the differential cross section has
been eliminated by the integration over the full range of $\phi$.

\begin{figure}[htb]
\hspace{-.5cm}
\begin{picture}(18,5.5)
\setlength{\unitlength}{1cm}
\put(.3,0){\mbox{\includegraphics[width=5.5cm,angle=90]{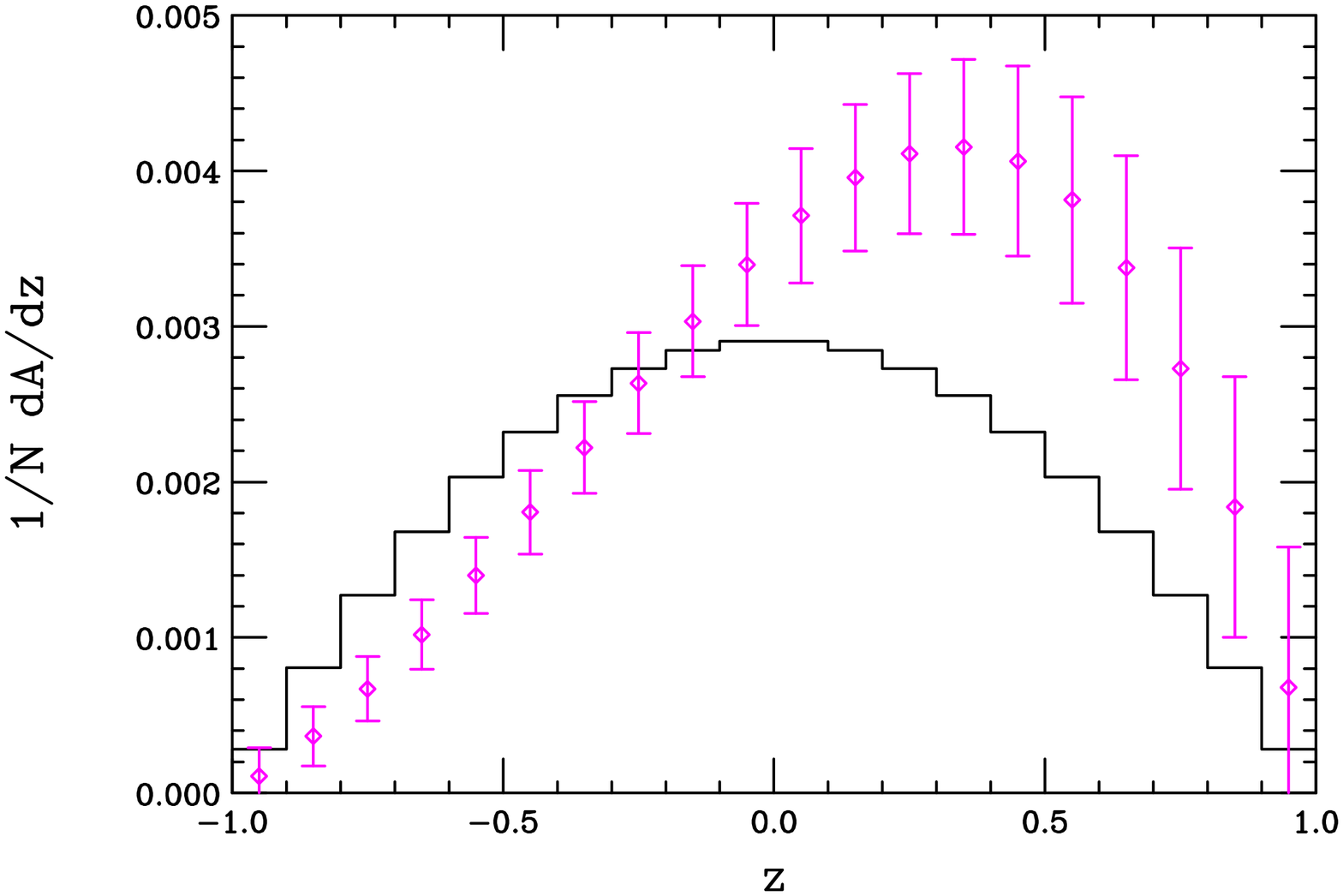}}}
\put(8.8,0){\mbox{\includegraphics[width=5.5cm,angle=90]{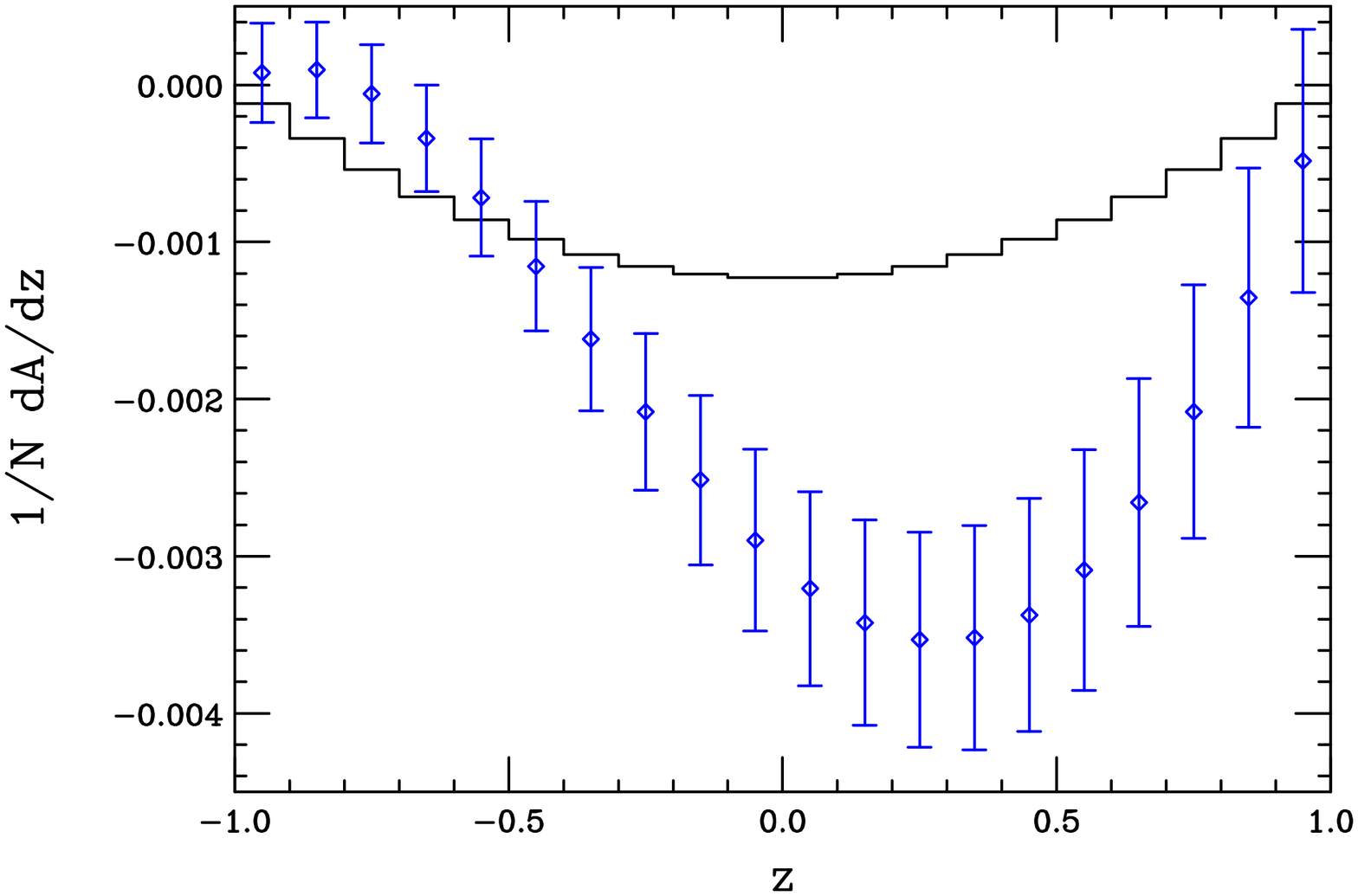}}}
\put(2,4.8){\small $e^+e^-\to c \bar{c}$}
\put(7.2,4.7){\small \color{Lila} ADD}
\put(5,2.8){\small SM}
\put(12.5,4.8){\small $e^+e^-\to b \bar{b}$}
\put(11.5,2){\small \color{Blue} ADD}
\put(13.2,4){\small SM}
\end{picture}
%}
\vspace*{-0.7cm}
\caption[Differential azimuthal asymmetries in the ADD model]{Differential
azimuthal asymmetry distribution for $e^+e^-\to f\bar f$, i.e.  $c \bar{c}$
(left) and $b \bar{b}$ (right), at a 500 GeV LC assuming a luminosity of
$500$~fb$^{-1}$, $z=\cos\theta$.  The histograms are the SM predictions while
the data points assume the ADD model with $M_H=1.5$ TeV;
$(P_{e^-}^{\rm T},P^{\rm T}_{e^+})=(80\%,60\%)$~\cite{Rizzo:2002ww}.\label{fig-ed1}}
\end{figure}

The following azimuthal asymmetry can be defined \cite{Rizzo:2002ww}:
\begin{equation}
\frac{1}{N}\frac{dA^T}{d \cos\theta}
=\frac{1}{\sigma}\,
\left[\int_+\frac{d\sigma}{d\cos\theta d\phi}-
\int_-\frac{d\sigma}{d\cos\theta d\phi}\right],
\label{at}
\end{equation}
where $\theta$ denotes the polar angle and $\int_{\pm}$ indicates
integrations over regions where $\cos{2\phi}$ takes on positive or negative
values. Examples of the $z=\cos\theta$ dependence of the expression
(\ref{at}) in the SM and in the ADD scenario are shown in
fig.~\ref{fig-ed1}, for final states and transverse polarization
configurations as described in the caption.

From eq.~(\ref{at}), the azimuthal forward--backward asymmetry, sensitive to
$\cos\theta$-odd terms can be defined:
\begin{equation}
A_{\rm FB}^T=
\frac{1}{N}\left[\int_{\cos\theta\geq 0}-\int_{\cos\theta\leq 0}\right]d \cos\theta
\frac{d A^T}{d \cos\theta}.
\label{atfb}
\end{equation}
For both for the SM and the 4-fermion contact-interactions
 one finds $A_{\rm FB}^T=0$, since
$N^{-1}dA^T/d\cos\theta\propto 1-\cos^2\theta$ is even in $\cos\theta$
in these cases. Conversely, the spin-2 graviton exchange introduces
$\cos\theta$-odd terms, so that $A_{\rm FB}^T$ is non-zero for this
kind of new physics that, accordingly, can be identified by this
observable. The corresponding expected identification reach on $M_H$
is reported in the right panel of table~\ref{tab-ed1}
\cite{Rizzo:2002ww}.  Concerning the values of transverse
polarizations, 100\% efficiency of the spin rotators have been
assumed.

Actually, as an alternative to $A_{\rm FB}^T$, the distinction of graviton
exchange from other new-physics effects might be attempted by a direct fit to
the $\cos\theta$ dependence of $N^{-1}dA^T/d\cos\theta$, emphasizing the
deviations from the $1-\cos^2\theta$ behaviour. 
Examples for the reach and for the identification reach
on $M_H$, the latter being potentially as high as $10\sqrt s$, are shown in
the left panel of table~\ref{tab-ed2}, up to very high energies. 
It gives the values of $M_H$
corresponding to deviations from the SM-predicted angular behaviour
that could be observed on statistical grounds.
However, notice that for high luminosities, those values are expected to
be reduced by the systematic effects, which become dominant.

%%%%%%%%%%%%%%%%%%%%%%%%%%%%%%%%%%%%%%%%%%%%%%%%%%%

\subsection*{Distinguishing among models with graviton exchange}
%%%%%%%%%%%%%%%%%%%%%%%%%%%%%%%%%%%%%%%%%%%%%%%%%%%
In the previous subsection, identification reaches on graviton exchange mass
scales from azimuthal asymmetries allowed by transverse beam polarization are
found numerically equivalent to those obtained by longitudinally-polarized
beams. The point now is the distinction between different extra-dimensional
scenarios, in the specific chosen examples, the ADD and the RS scenarios, were
indirect signals of these mechanisms observed.

In the RS model, if $\sqrt s$ is far from the $Z$ and the $KK$ poles, the
imaginary part of the amplitude entering the $\sin{2\phi}$ term 
in the differential cross section with transversely-polarized beams
becomes vanishingly small. Conversely, the summation over
the essentially continuous spectrum of ADD gravitons can lead to a finite,
cut-off independent imaginary part, also sensitive
to the number of extra dimensions~\cite{Datta:2002tk}.
 Consequently, $f_G$ of
eq.~(\ref{f_g}) can acquire an imaginary part, strongly dependent on the number
of extra dimensions.

\begin{figure}[htb]
\begin{picture}(15,5)
\setlength{\unitlength}{1cm}
\put(4,0){\mbox{\includegraphics[width=5cm,height=8cm,angle=90]{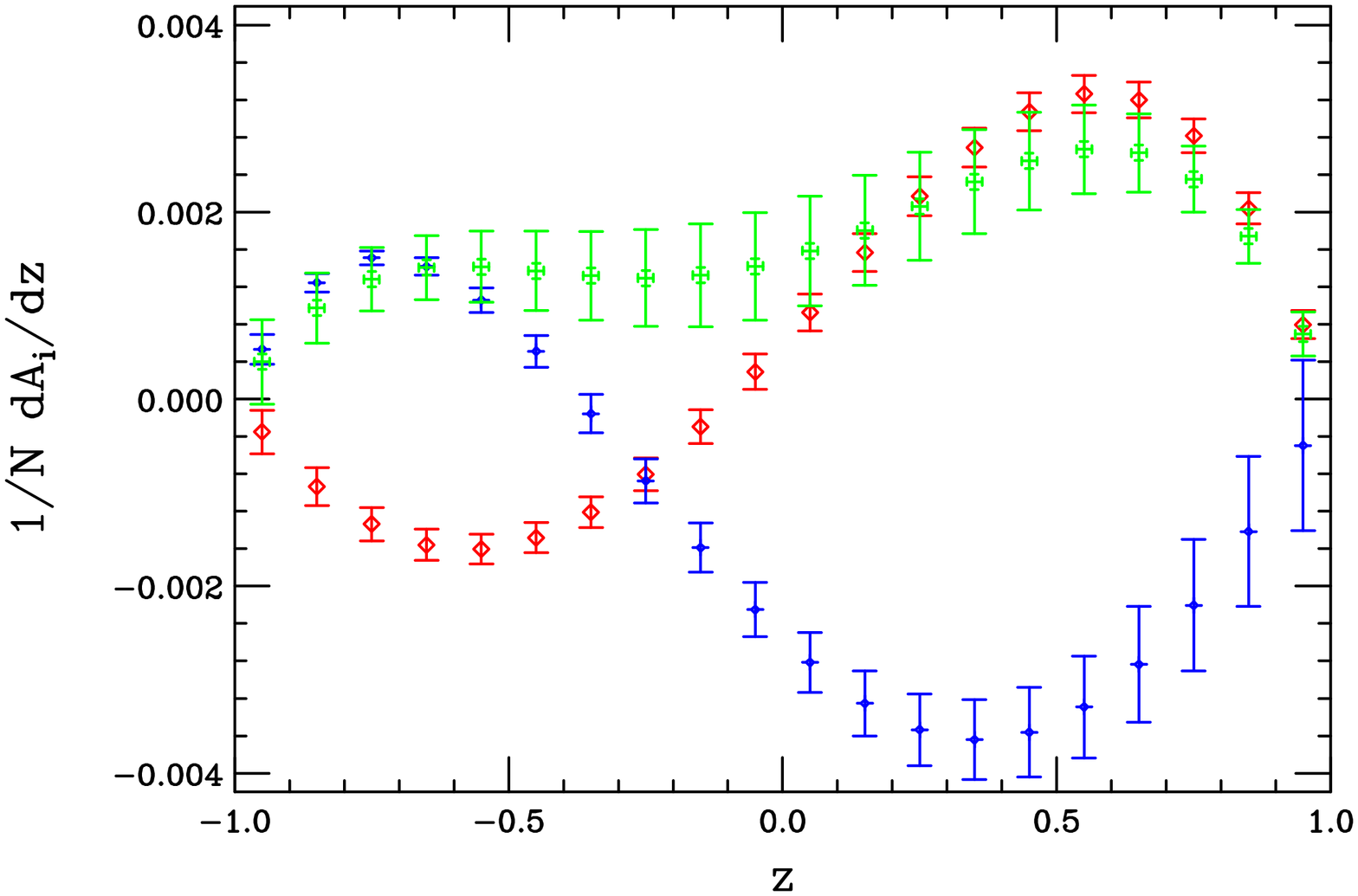}}}
\put(7.5,4){\small\color{Olive} $b\bar{b}$}
\put(9,1.8){\small\color{Blue} $\mu\bar{\mu}+\tau \bar{\tau}$}
\put(7,1.5){\small\color{Red} $c \bar{c}$}
\end{picture}
\vspace*{-.3cm}
\caption[Azimuthal distributions in fermion pair production in the ADD
model]{The $N^{-1} dA^T_i/dz$ distributions at $\sqrt{s}=
500~\text{GeV}$ assuming $M_H=M_D=1.5$ TeV and $d=3$ with an
integrated luminosity of 500~fb$^{-1}$. The plotted points from top to
bottom in the centre of the plot correspond, respectively, to $f=b,
\mu$ and $\tau$ combined together, and $c$~\cite{Rizzo:2002ww}.}
\label{fig:add-rs}
\end{figure}

With transverse polarization one can define a new asymmetry sensitive
to such an imaginary part, that exploits the
$\sin{2\phi}$ term:
\begin{equation}
\frac{1}{N}\frac{dA_i^T}{d\cos\theta}
=\frac{1}{\sigma}\,
\left[\int_+\frac{d\sigma}{d\cos\theta d\phi}
- \int_-\frac{d\sigma}{d\cos\theta d\phi}\right],
\label{ait}
\end{equation}
where now $\int_\pm$ indicate integrations over regions of positive and
negative values of $\sin{2\phi}$. It can be seen that the
$\cos{2\phi}$ terms proportional to the real part cancel for both 
the RS and ADD models. The
observable (\ref{ait}) vanishes identically for both the SM and the RS
scenario, thus a non-zero value unambiguously signals ADD graviton
exchange. Fig.~\ref{fig:add-rs} exemplifies the $\cos\theta$ behaviour
for selected values of the ADD mass
scales. The right panel of table~\ref{tab-ed2} reports the 5-$\sigma$ discovery
reach from the asymmetry (\ref{ait}), assuming $d=3$ for illustration
purposes, and indicates that the indirect manifestations of the ADD and RS
models could be distinguished from one another up to 
about $M_H\sim\left(2.5-3.0\right)\sqrt s$
\cite{Rizzo:2002ww}. This shows the essential role of transverse beam
polarization in this kind of analysis, which otherwise could not be performed.

%%%%%%%%%%%%%%%%%%%%%%%%%%%%%%%%%%%%%%%%%%%%%%%%%%%
\begin{table}[htb]
\begin{center}
\begin{minipage}{7cm}
\begin{center}
$M_H$ [TeV], $\sqrt{s}=0.5$~TeV, long. pol. \\
\begin{tabular}{|c||c|c|c|}
\hline
& \multicolumn{3}{|c|}{${\cal L}_\text{int}$ [fb$^{-1}$]}\\
$(P_{e^-},P_{e^+})$ & 100 & 300 & 500 \\ \hline\hline
$(0,0)$ & 2.3 & 2.6 & 2.9 \\ \hline
$(+80\%,0)$ & 2.5 & 2.8 & 3.05\\ \hline
$(+80\%,-60\%)$ & 2.45 & 3.0 & 3.25 \\ \hline
\end{tabular}
\end{center}
\end{minipage}
\begin{minipage}{8cm}
\begin{center}
$M_H$ [TeV], $(P^{\rm T}_{e^-},P^{\rm T}_{e^+})=(80\%,60\%)$ \\
\begin{tabular}{|c||c|c|c|c|}
\hline
& \multicolumn{4}{|c|}{${\cal L}_\text{int}$ [fb$^{-1}$]}\\
& 100 & 300 & 500 & 1000\\ \hline\hline
$\sqrt{s}=0.5$~TeV & 1.6 & 1.9 & 2.0 & 2.2 \\ \hline
$\sqrt{s}=0.8$~TeV & 2.4 & 2.6 & 2.8 & 3.1 \\ \hline
$\sqrt{s}=1.0$~TeV & 2.8 & 3.2 & 3.4 & 3.8 \\ \hline
\end{tabular}
\end{center}
\end{minipage}
\end{center}
\caption[5-$\sigma$ 
identification reach for the mass scale of extra dimensions]
{Left: 5-$\sigma$ identification reach 
on the mass scale $M_H$ vs.\ integrated luminosity
from the process $e^+e^-\to f{\bar f}$, with $f$ summed over $\mu,\tau,b,c$,
and for the energy 0.5~TeV \cite{Osland:2003fn} (see also
\cite{LC-TH-2001-007}, where the 95\% C.L. sensitivity for $M_H$ in $e^+e^-\to
\mu\bar{\mu}, c\bar{c},b\bar{b}$ has been simulated). Right: 5-$\sigma$
identification reach in $M_H$ vs.\ integrated luminosity using
$A^T_{\rm FB}$ for the process
$e^+e^-\to f\bar f$, with $f$ summed over $\mu,\tau,b,c$ and $t$
\cite{Rizzo:2002ww}.
\label{tab-ed1}.}
\end{table}
%%%%%%%%%%%%%%%%%%%%%%%%%%%%%%%%%%%%%%%%%%%%%%%%%%%

\begin{table}[htb]
\begin{minipage}{8cm}
\begin{tabular}{|c|c||c|}
\hline
 $E_{\rm CM}$ &Ident. reach&$95\%$ CL disc. reach\\
  ~[GeV]   & $M_H$ [TeV] & $M_H$  [TeV]  \\ \hline
\hline
500  & 5.4 & 10.2 \\ \hline
800  & 8.8 & 17.0 \\ \hline
1000  & 11.1 & 21.5 \\ \hline
1200  & 13.3 & 26.0 \\ \hline
1500  & 16.7 & 32.7  \\ \hline
\end{tabular}
\end{minipage}\hspace{1cm}
\begin{minipage}{8cm}
\begin{tabular}{|c||c|c|c|c|}
\hline
5-$\sigma$ disc. reach& \multicolumn{4}{|c|}{${\cal L}_\text{int}$ [fb$^{-1}$]}\\
$M_H$ [TeV] & 100 & 300 & 500 & 1000\\ \hline\hline
$\sqrt{s}=0.5$~TeV & 1.2 & 1.3 & 1.4 & 1.6 \\ \hline
$\sqrt{s}=0.8$~TeV & 1.8 & 2.0 & 2.2 & 2.4 \\ \hline
$\sqrt{s}=1.0$~TeV & 2.2 & 2.4 & 2.6 & 2.8 \\ \hline
\end{tabular}
\end{minipage}
\caption[Identification and discovery reach for mass scale in ADD model]
{Left: Identification reach and discovery reach $M_H$ in the ADD model 
by fitting the distribution
$N^{-1} dA^T/d\cos\theta \sim 1-\cos^2\theta$ for  
the final states $f=\mu,\tau$, $f=b$ and $f=c$ for different centre-of-mass 
energies.
Right:
5-$\sigma$ reach for the discovery of 
a non-zero value of the azimuthal asymmetry 
(\ref{ait}) vs.\ $\Lumint$ for $d=3$; 
$M_H=M_D$ is assumed throughout as is 
$(P_{e^-}^{\rm T},P_{e^+}^{\rm T})=(80\%,60\%)$~\cite{Rizzo:2002ww}.
\label{tab-ed2}}
\end{table}

{\bf \boldmath
Quantitative example: With transversely-polarized beams a new asymmetry
can be built, which is sensitive to imaginary parts of the graviton propagator.
Below the graviton poles no imaginary parts occur in the RS model, whereas in
the ADD model a cut-off independent imaginary part emerges. 
This new observable allows a distinction between the two models in LC studies
up to a scale $M_H\sim 3 \sqrt{s}$.}

%%%%%%%%%%%%%%%%%%%%%%%%%%%%%%%%%%%%%%%%%%%%%%%%%%%
\section{Other mechanisms of CP violation}
%%%%%%%%%%%%%%%%%%%%%%%%%%%%%%%%%%%%%%%%%%%%%%%%%%%

%%%%%%%%%%%%%%%%%%%%%%%%%%%%%%%%%%%%%%%%%%%%%%%%%%%
\subsection{Unconventional interactions in $t \bar{t}$ production 
\label{sec-trans-anant1}}
%%%%%%%%%%%%%%%%%%%%%%%%%%%%%%%%%%%%%%%%%%%%%%%%%%%

{\bf \boldmath
(Pseudo-) scalar or tensor interactions associated with a new physics
scale $\Lambda$ can lead to CP-odd observables from interference terms with
the virtual $\gamma$, $Z$ exchange in the annihilation channel.  Only with
both beams transversely polarized are these CP-violating effects measurable
without final-state spin analysis.}
\smallskip

The observation of 
CP violation in $e^+e^-$ collisions requires either the measurement of the polarization
of the final-state particles or the availability of polarized
beams. Transverse polarization of initial beams defines one more direction,
and can provide CP-odd asymmetries without the need for directly measuring
final-state polarizations. This may represent an advantage, e.g.\ as regards
the statistical significance of the signal.

Transverse polarization (TP) enables novel CP violation searches in
the inclusive process~\cite{Dass:1975mj} 
\begin{equation}
e^+e^-\to A+X.
\label{eq-transanant1}
\end{equation}
As regards the search for non-standard sources of CP violation, when the spin
of particle $A$ is {\it not} observed and $m_e$ 
is neglected, only (pseudo-)
scalar or tensor currents associated with a new-physics scale
$\Lambda$ can lead to CP-odd observables at the leading order 
in the new interaction if transversely-polarized beams are used.
They are due to the interference
 between these new currents and the $\gamma$ and $Z$ exchanges in
the $s$-channel, cf.\ also the matrix element squared in the general
expression, eq.~(\ref{eq_general}). Contributions from (pseudo-)
scalar or tensor currents lead to interference terms which are linear
in the transverse polarizations.  However, to construct CP-odd
observables both beams must be transversely polarized:
one must compare contributions with transversely-polarized electrons
with those resulting from transversely-polarized positrons 
(with the same polarization degree).
Vector-like four-fermion contact interactions can interfere with
the SM amplitudes, but for this kind of (helicity-conserving) non-standard
interactions no CP-odd distribution can arise even in the presence of
TP. Thus, the above-mentioned CP-odd observables have a great selective power
of novel (helicity-changing) interactions at the first order. With
longitudinal beam polarization there would be no interference of the SM with
scalar or tensor interactions. 

In the following the contributions to the
differential cross section due to (pseudo-) scalar and tensor 
contact-interactions at leading order in the interaction strengths for
the process $e^+e^-\to t\bar{t}$ \cite{Ananthanarayan:2003wi} are reviewed.
This allows to construct an effective up-down asymmetry and
a version of the same integrated over the polar angle.

An effective Lagrangian that can parametrize the above-mentioned non-standard
interactions in a general, model-independent way can be written as:
\begin{equation}\label{lag}
{\cal L}={\cal L}^{\rm S\!M}+
\frac{1}{{\mit\Lambda}^2}\sum_i(\:\alpha_i{\cal O}_i+{\rm h.c.}\:),
\end{equation}
where $\alpha_i$ are the coefficients which parametrize non-standard
interactions and are assumed to be of order unity, 
${\cal O}_i$ are the possible effective dimension-six operators, and
$\Lambda$ is the scale of new physics. 

After Fierz transformation the part of the Lagrangian containing the above
four-Fermi operators, and relevant to $e^+e^-\to t \bar t$, eq.~(\ref{lag}) can be rewritten
as
\begin{equation}\label{lag4f}
{\cal L}^{4F}
 =\sum_{i,j=L,R}\Bigl[\:S_{ij}(\bar{e}P_ie)(\bar{t}P_jt) 
 +V_{ij}(\bar{e}\gamma_{\mu}P_ie)(\bar{t}\gamma^{\mu}P_jt) 
 +T_{ij}
 (\bar{e}\frac{\sigma_{\mu\nu}}{\sqrt{2}}P_ie)
(\bar{t}\frac{\sigma^{\mu\nu}}{\sqrt{2}}P_jt)\:\Bigr] + \text{h.c.}, 
\end{equation}
where $P_{\rm L, R}$ are chirality projection operators and the coefficients 
must satisfy the relations
\begin{equation}
S_{\rm RR}=S^{*}_{\rm LL},\ \ \ S_{\rm LR}=S_{\rm RL}=0,\quad
V_{ij}=V^{*}_{ij},\quad
T_{\rm RR}=T^{*}_{\rm LL},\ \ \ T_{\rm LR}=T_{\rm RL}=0.
\end{equation}

The differential cross section depends on the combined couplings
\begin{equation}\label{S}
S\equiv S_{\rm RR} + \frac{2c_A^t c_V^e}{ c_V^t c_A^e} T_{\rm RR},
\end{equation}
where $c_V^{e,t}$, $c_A^{e,t}$ are the couplings of $Z$
to $e^+e^-$ and $t \overline t$. 
In (\ref{S}) the contribution of the tensor
term relative to the scalar term is suppressed by a factor
$2c_A^tc_V^e/c_V^tc_A^e \approx 0.36$. In what follows
only the combination $S$ is considered, not $S_{\rm RR}$ and $T_{\rm RR}$
separately.

One can construct a CP-odd asymmetry,  
up-down asymmetry, as\footnote{An
analogous up-down asymmetry which is available with transverse beam
polarization allows for the separation of light $u$ and $d$-type flavours
\cite{Olsen:1980rs}.}

\begin{equation}
\label{asym}
A(\theta)=
\dps{ \frac{ \dps{\int_0^\pi \frac{ d\sigma^{+-}}{d\Omega}} d\phi 
 - \int_{\pi}^{2\pi} \dps{\frac{ d\sigma^{+-}}{d\Omega}} d\phi
}{
 \dps{\int_{0}^{\pi} \frac{ d\sigma^{+-}}{d\Omega}} d\phi 
 + \int_{\pi}^{2\pi}\dps{ \frac{ d\sigma^{+-}}{d\Omega}} d\phi
}  },
\end{equation}
where the superscripts denote opposite transverse polarization of $e^-$, $e^+$
and $\phi$ is the azimuthal angle. One can see that $A(\theta)$ is
proportional to the imaginary part of $S$
\cite{Ananthanarayan:2003wi}, that appears in the differential
distribution as a factor of the $\sin\theta\sin\phi$ term.

Also a $\theta$-integrated version of the asymmetry can be defined,
\begin{equation}
\label{asymcutoff}
A(\theta_0)
=\frac{1}{\sigma^{+-}(\theta_0)}
\int_{-\cos\theta_0}^{\cos\theta_0}
\left[\int_0^\pi  - \int_{\pi}^{2\pi} \right]
\frac{d\sigma^{+-}}{d\Omega}d\cos\theta d\phi,
\end{equation}
where a cut-off angle $\theta_0$ has been introduced to be away from the
beam-pipe direction and $\sigma^{+-}(\theta_0)$ is the cross section
integrated with this cut-off.

In a numerical study, 
limits on $S$ can be put using the integrated
asymmetry $A(\theta_0)$ and optimized by tuning the value of $\theta_0$.  

\begin{figure}[htb]
\begin{center}
\begin{picture}(15,4.5)
\setlength{\unitlength}{1cm}
\put(-2,5){\mbox{\includegraphics[scale=0.29,angle=-90]{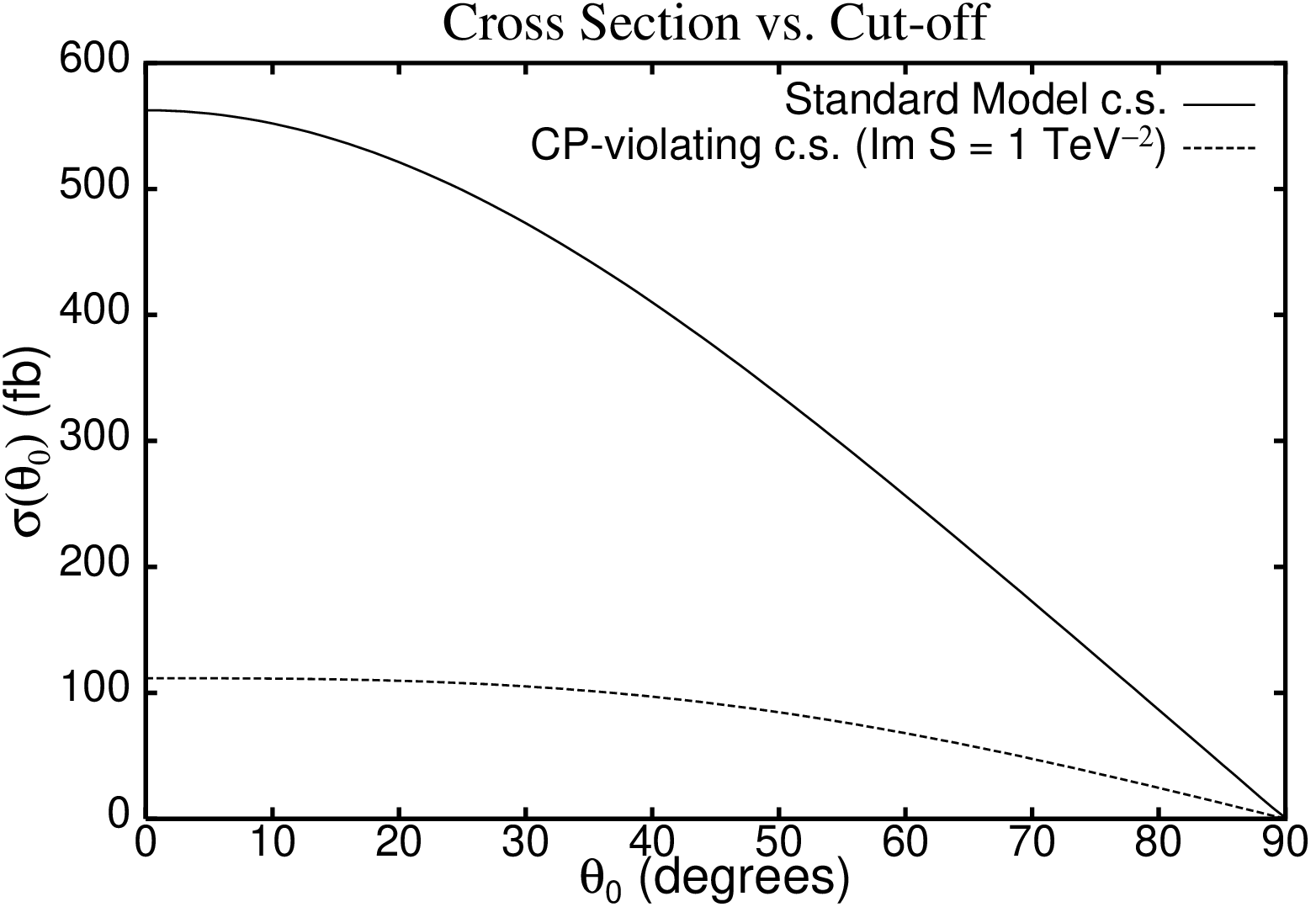}}}
\put(7,5){\mbox{\includegraphics[scale=0.29,angle=-90]{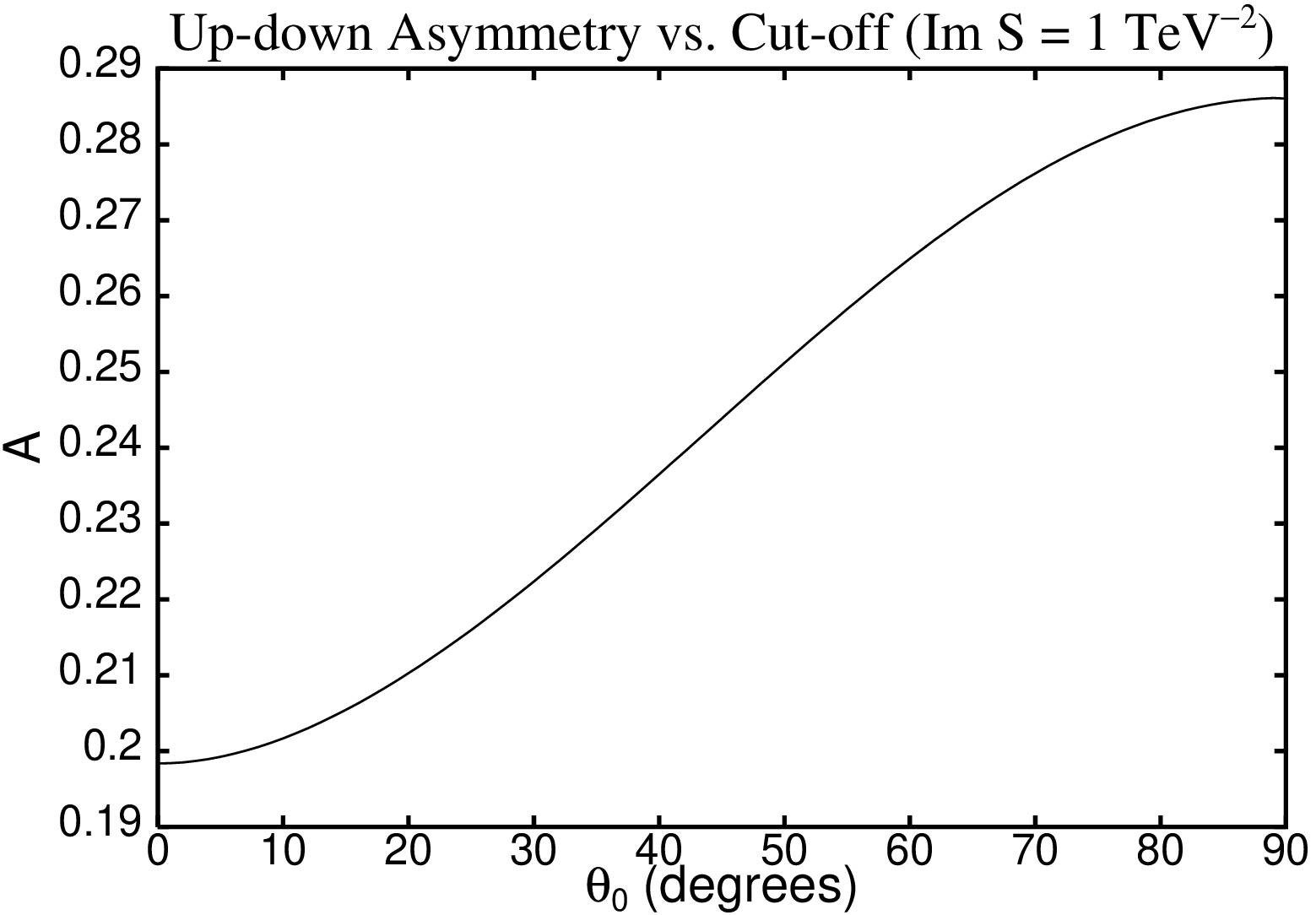}}}
\end{picture}
\end{center}
\vspace{-.5cm}
\caption[$S$-couplings in $t\bar{t}$:
Cross section and azimuthal asymmetry vs. cut-off angle]{
\label{fig-updownasy} Left: The SM cross section
(solid line) and the numerator
of the asymmetry $A(\theta_0)$ in eq. (\ref{asymcutoff}) (broken line) 
vs.\ $\theta_0$; 
Right:
The asymmetry $A(\theta_0)$ defined in eq.
(\ref{asymcutoff}) vs.\ $\theta_0$ for Im~$S=1$ 
TeV$^{-2}$~\cite{Ananthanarayan:2003wi}.}
\end{figure}

As can be seen from fig.~\ref{fig-updownasy}, the
value of $A(\theta_0)$ increases with $\theta_0$, because the SM
cross section in the denominator of eq.~(\ref{asymcutoff})
decreases with this 
cut-off faster than the numerator. Here, and in the subsequent figure, 
$\sqrt{s}=500$~GeV and 
$(P_{e^-}^{\rm T},P_{e^+}^{\rm T})=(100\%,100\%)$ are assumed.

\begin{figure}[htb]           
\begin{center}
\begin{picture}(15,4.2)
\put(4.,4.5){\psfig{file=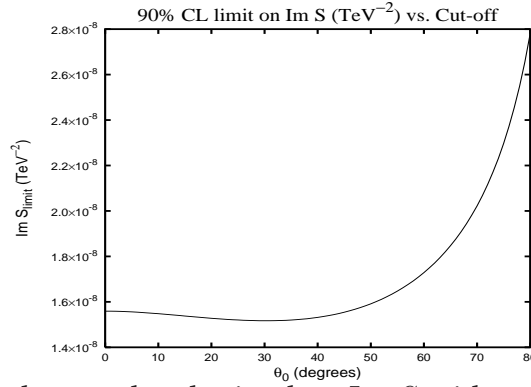,width=2.in,height=3.in,angle=-90}}
\end{picture}
\vspace{.1cm}
\caption[Imaginary part of the coupling $S$ vs. cut-off angle in 
$t\bar{t}$ production]{
The 90\% C.L.\ that can be obtained on 
Im $S$ with an integrated
luminosity of 500 fb$^{-1}$ vs.\ the cut-off angle
$\theta_0$~\cite{Ananthanarayan:2003wi}. \label{fig-cllevel}}
\end{center}
\end{figure}

Fig.~\ref{fig-cllevel} shows the 90\% confidence level (C.L.) limits that
could be placed on ${\rm Im}S$ for an integrated luminosity of $\Lumint
=500$~fb$^{-1}$.  This limit translates to a value of $\Lambda$ of the order
of 8 TeV, assuming that the coefficients $\alpha_i$ in (\ref{lag}) are of
order 1. The corresponding limit for $\sqrt{s}=800$~GeV with the same
integrated luminosity and complete TP is $\sim 9.5$ TeV.

Using the more realistic transverse polarization degrees of 80\% and 60\%,
the up-down asymmetry $A(\theta)$ or $A(\theta_0)$ 
gets multiplied by a factor $\frac{1}{2}(P_{e^-}^{\rm T} - P_{e^+}^{\rm T})$. 
For $(P^{\rm T}_{e^-},P^{\rm T}_{e^+})=(80\%,-60\%)$, this means a
reduction of the asymmetry by a factor of 0.7. Since the SM cross
section does not change, this also means that the limit on the
parameter Im $S$ goes up by a factor of $1/0.7\approx 1.4$, and
the limit on $\Lambda$ goes down by a factor of $\sqrt{0.7}\approx
0.84$, to about 6.7 TeV. If the positron beam is unpolarized,
however, the sensitivity decreases further.

In summary, TP can be used to study CP-violating asymmetries arising
from the interference of new-physics scalar and tensor
interactions with the SM $\gamma$- and $Z$-exchanges. 
These interference terms
cannot be seen with longitudinally-polarized or unpolarized beams.
Moreover, such an asymmetry would not be sensitive to new vector
and axial-vector interactions (as for example, from an extra $Z'$
neutral boson), or even electric or ``weak" dipole interactions of
heavy particles, since it vanishes if $m_e\sim 0$.
This manifests the selective power on new physics of this observable.
\smallskip

{\bf \boldmath
Quantitative example: With both beams transversely polarized, CP-odd
asymmetries can be constructed which are sensitive to new CP-violating
(pseudo-) scalar or tensor interactions.  The corresponding new physics
scale $\Lambda$ can be bounded at the 90\% confidence level, at 
about $7$-$10$~TeV,
with plausible assumptions on the centre-of-mass energy and the values of transverse beam polarizations.}

%%%%%%%%%%%%%%%%%%%%%%%%%%%%%%%%%%%%%%%%%%%%%%%%%%%
\subsection{Transversely-polarized beams for CP violation in $\gamma Z$ 
production \label{cpviol-trans}}
%%%%%%%%%%%%%%%%%%%%%%%%%%%%%%%%%%%%%%%%%%%%%%%%%%%
{\bf \boldmath
An anomalous CP-violating $\gamma \gamma Z$ vertex gives rise to
a novel asymmetry which is accessible with transversely-polarized beams in the
process $e^+e^-\to \gamma Z$.  This asymmetry, which is odd under
naive time reversal, is proportional to the real part of the $\gamma\gamma Z $
CP-violating coupling and is not accessible with 
longitudinally-polarized beams but could be measured with 
transversely-polarized beams.  This is in contrast to the simple 
forward-backward asymmetry
of the $\gamma$ (or $Z$) with unpolarized or
longitudinally-polarized beams, which is even under naive time reversal, hence
sensitive only to the imaginary part of the vertex.}
\smallskip

In $\gamma Z$ production a CP-violating contribution can arise if anomalous
CP-violating $\gamma \gamma Z$ and $\gamma ZZ$ couplings are present. The
contributions of the interferences of these anomalous couplings with the SM
contribution give rise to a polar-angle forward-backward asymmetry with
longitudinally-polarized beams \cite{Choudhury:1994nt}. New
combinations of polar and azimuthal asymmetries occur in the presence of
transversely-polarized beams
\cite{Ananthanarayan:2004eb,Ananthanarayan:2004xf}.  

Indeed, longitudinal beam polarization may play an important role in
improving the sensitivity to absorptive (or real) parts of
CP-violating anomalous couplings or form factors, which are measurable
even with unpolarized beams through the forward-backward asymmetry.
However, transverse polarization enables in addition measurements of
dispersive (or imaginary) parts of certain combinations of form
factors (or coupling constants) which are inaccessible with
longitudinally-polarized beams\cite{Ananthanarayan:2004xf}.

The role of transverse polarization in the context of CP violation has
previously been studied in section \ref{sec-trans-anant1} for the process
$e^+e^- \to t \bar{t}$.  Basically, since transverse beam polarization
provides an additional reference axis in addition to the $e^+e^-$ beam
direction, there is the possibility of studying the azimuthal distribution of
a single final-state particle instead of its polarization via the analysis of
its decay distribution.

In the process considered in section~\ref{sec-trans-anant1}, the SM
interaction occurs only via $s$-channel $\gamma$ and $Z$ exchanges, and
for $m_e=0$ 
CP-violating effects could only arise from the interference with chirality
changing (pseudo)scalar and tensor effective couplings. If, like in the
present case of $e^+ e^- \to \gamma Z$, $t$- and $u$-channel exchanges are
present, an additional dependence on the scattering (polar) angle $\theta$
exists and those considerations do not apply.
 Indeed, referring to eq.~(\ref{eq-transanant1}), if particle $A$ is
self-conjugate the forward-backward asymmetry
corresponding to $\theta \to \pi -\theta$ is CP-odd also in the absence of transverse
polarization \cite{Czyz:1988yt, Choudhury:1994nt, Rindani:1997qn}. 
Such forward-backward asymmetry would be even under `naive' time
reversal $T$, namely, the reversal of particle spins and momenta but not the
exchange of final with initial state. It can be shown that the CPT theorem implies this asymmetry
to be proportional to an absorptive part of the interfering amplitudes, hence
to the imaginary parts of the new-physics couplings (see, e.g.\ 
\cite{Rindani:1994ad,Ananthanarayan:2004eb}).

However, if there is transverse polarization, a T-odd but CP-even azimuthal
asymmetry can be combined with the T-even but CP-odd forward-backward
asymmetry to give an asymmetry which is both CP-odd and T-odd. In this
case, the CPT theorem would imply that such an asymmetry can measure the real parts of
the new-physics couplings that, as stated above, 
could not be measured without transverse 
polarization.

Turning now to $e^+e^-\to \gamma Z$, where both produced particles are
self-conjugate, at the tree level CP violation effects can arise if anomalous
CP-violating $\gamma\gamma Z$ and $\gamma ZZ$ couplings are present and
interfere with the SM amplitudes~\cite{Ananthanarayan:2004eb} 
{}\footnote{ For the analysis of the general set of form factors
see~\cite{Ananthanarayan:2004xf}.}.

The parametrization of
the anomalous vertex can be done in the most general form consistent with
Lorentz invariance, photon gauge invariance and chirality conservation, 
includes contact interactions as well as triple gauge boson couplings
\cite{lampe}, and can be written for $\gamma\gamma Z$ and $\gamma ZZ$ 
interactions as:
\begin{equation}
{\cal L} = 
   e \frac{\lambda_1}{ 2 m_Z^2} F_{\mu\nu}
    \left( \partial^\mu Z^\lambda \partial_\lambda Z^\nu
          - \partial^\nu Z^\lambda \partial_\lambda Z^\mu
      \right)
      +\frac{e}{16 c_W s_W} \frac{\lambda_2}{m_Z^2}
       F_{\mu\nu}F^{\nu \lambda}
       \left(\partial^\mu Z_\lambda + \partial_\lambda Z^\mu   \right),
      \label{lagrangian}
\end{equation}
where $c_W=\cos \theta_W$, $s_W=\sin \theta_W$ and $\theta_W$ is the weak
mixing angle. Notice that $\lambda_1$ and $\lambda_2$ do not arise in the SM
even at loop level, therefore their non-vanishing represents an unambiguous
signal of new physics. Only indirect limits are available from loop-induced
weak and electric fermion dipole moments, viz.\ $\vert\lambda_2\vert <
10^{-3}$ and similarly for $\lambda_1$\cite{Altarev:1992cf,Choudhury:1994nt}. 
The interest here, however, is in
model-independent direct limits on these constants.

The photon angular distribution contains interference terms dependent on $\Re
\lambda_2$, $\Im \lambda_1$, $\Im\lambda_2$ and proportional to the product of
electron and positron degrees of transverse polarization (parallel transverse
polarization directions of $e^+$ and $e^-$ are assumed).  In particular, their
dependence as $\cos\theta\cos 2 \phi$ and $\cos\theta\sin 2 \phi$, with $\phi$
the azimuthal angle, are of interest for the determinations of the
CP-violating couplings.

Define the following CP-odd asymmetries, that combine a
forward-backward asymmetry with an appropriate asymmetry in $\phi$, so as to
isolate the anomalous couplings:
\begin{align}
{\rm A}_1 &=
\frac{1}{\sigma_0}
\sum_{n=0}^3 (-1)^n
\int_{\pi n/ 2}^{\pi(n+1)/  2} d\phi
\left[
\int_{0}^{\cos \theta_0}  -
\int_{-\cos \theta_0}^{0} \right] d \cos\theta\, \frac{d \sigma}{d \Omega}, \\
{\rm A}_2& =
\frac{1}{\sigma_0}
\sum_{n=0}^3(-1)^n 
\int_{\pi (2 n-1)/4}^{\pi(2 n+1)/4} d\phi \,
\left[
\int_{0}^{\cos \theta_0} 
- \int_{-\cos \theta_0}^0 \right] d \cos\theta\,
\frac{d \sigma}{d \Omega} , \\
{\rm A}_3 &=
\frac{2}{\sigma_0}\biggl[
\int^{\cos \theta_0}_{0} 
-\int_{-\cos \theta_0}^{0} \biggr] d \cos\theta \,
\left[
\int_{0}^{\pi/4}  + \int_{3 \pi /4}^{5 \pi/4}+\int_{7 \pi /4}^{2 \pi}\right]  d\phi \,
\frac{d \sigma}{d \Omega},
\end{align}
where
$\sigma_0 \equiv \sigma_0(\theta_0)$ is the total cross section of the
process, integrated with a cut-off $\theta_0$ in the polar angle,
$\theta_0<\theta\le\pi-\theta_0$.

Explicitly, with $P^{\rm T}_{e^-}$ and $P^{\rm T}_{e^+}$ the degrees
of transverse polarization, one finds~\cite{Ananthanarayan:2004eb}:
\begin{align}
{\rm A}_1(\theta_0)
&=  - {\cal B}' \, g_A\,  
P_{e^-}^{\rm T} P_{e^+}^{\rm T}\, \Re \lambda_2, \\
{\rm A}_2(\theta_0)
&= {\cal B}'\,
P_{e^-}^{\rm T} P_{e^+}^{\rm T}\, [(g_V^2-g_A^2)\Im \lambda_1 - g_V \, \Im\lambda_2], \\
 {\rm A}_3(\theta_0)
 &= {\cal B}'\,
\left[ \frac{ \pi}{2} [  (g_V^2+g_A^2)\, \Im \lambda_1\! -\!g_V\, \Im\lambda_2]
+ P_{e^-}^{\rm T} P_{e^+}^{\rm T}
[(g_V^2-g_A^2)\Im \lambda_1\! -\! g_V \, \Im\lambda_2]\right].  
\end{align}
where 
\begin{equation}
{\cal B}' =
\frac{\alpha^2}{16 s_W^2 m_W^2}\,
\left(1-\frac{m_Z^2}{s}\right)^2
\frac{\cos^2\theta_0} {\sigma_0(\theta_0)}.
\end{equation}

It is seen that $A_1(\theta_0)$ is proportional to $\Re
\lambda_2$, while the other two asymmetries depend on both $\Im \lambda_1$ and
$ \Im\lambda_2$. These two couplings, therefore, can be constrained
independently from the simultaneous measurement of ${\rm A}_2$ and ${\rm
A}_3$. Notice that the products of electron and positron polarizations appear
in the expression of the asymmetries, therefore both $P_{e^-}^{\rm T}$ and
$P_{e^+}^{\rm T}$ must be non-zero to observe these CP-violating couplings.

The 90\% C.L.\ limit
that can be obtained with a linear collider with
$\sqrt{s} = 500$ GeV, $\Lumint = 500$ fb$^{-1}$, 
$(P^{\rm T}_{e^-},P^{\rm T}_{e^+}) = (80\%,60\%)$, 
making use of the asymmetries $A_i$, has been calculated 
in~\cite{Ananthanarayan:2004eb}. 
The limiting value $\lambda^{\rm lim}$ (i.e.\ the respective real or imaginary
part of the coupling) is related to the value $A$ of the asymmetry for unit
value of the coupling constant by
$\lambda^{\rm lim} = 1.64/(A\sqrt{N_{SM}})$, 
where $N_{SM}$ is the number of SM events.

$A_1$ depends on $\Re \lambda_2$ alone, and can therefore place an independent
limit on $\Re \lambda_2$. It should be emphasized that information on $\Re
\lambda_2$ cannot be obtained without transverse polarization. The limits are
summarized in table~\ref{tab_bartl}.  In the third and fourth columns of this
table, the constraints are obtained by assuming $\Im \lambda_1$ and $\Im
\lambda_2$ non-zero, one at a time.  In the last column, both $\Im \lambda_1$ and
$\Im \lambda_2$ have been taken simultaneously non-zero in ${\rm A}_2$ and
${\rm A}_3$, which determines an allowed region for those two couplings.

\begin{table}
\centering
\begin{tabular}{l|c|c|c|c}
\hline
Coupling & \multicolumn{3}{|c|} {Individual limit from}
 & Simultaneous limits \\
& $A_1$ & $A_2$ & $A_3$ &\\ 
\hline 
Re $\lambda_2$ & $6.2\times 10^{-3}$ & & & \\
Im $\lambda_1$ && $6.2\times 10^{-3}$ &$ 3.8\times 10^{-3}$ & $7.1\times
10^{-3}$ \\
Im $\lambda_2$ && $9.1\times 10^{-2}$ &$ 3.0\times 10^{-2}$ & $6.7\times
10^{-2}$ \\
\hline
\end{tabular}
\caption[Limits on CP-violating triple gauge couplings]{90\% CL limits on the couplings 
from asymmetries $A_i$ for a cut-off angle of $\theta_0=26^{\circ}$, 
which optimizes the sensitivity, $\sqrt{s} = 500$
GeV, and $\Lumint=500~\text{fb}^{-1}$. Assumed
transverse polarizations are 
$(P_{e^-}^{\rm T},P_{e^+}^{\rm T})= (80\%,60\%)$~\cite{Ananthanarayan:2004eb}. 
\label{tab_bartl}}
\end{table}

{\bf \boldmath\smallskip

Quantitative examples: Only with both beams transversely polarized is it
possible to put separate limits on the real as well as the imaginary part of possible
CP-violating anomalous $\gamma\gamma Z$ and $\gamma ZZ$
couplings.  The sensitivity is rather high, especially on $\Re
\lambda_2$, on which a direct reach of about $6.2\times 10^{-3}$ can
be achieved.}

%%%%%%%%%%%%%%%%%%%%%%%%%%%%%%%%%%%%%%%
\chapter{Summary of the physics case \label{chap-sum}}
%%%%%%%%%%%%%%%%%%%%%%%%%%%%%%%%%%%%%%%

In the previous chapters the benefits from
having the positron beams polarized at the linear collider, 
as well as the electron beam, have been
discussed. The
physics case for this option has been illustrated by explicitly analyzing
reference reactions in different physics scenarios. The results show that
positron polarization, combined with the clean experimental environment
provided by a linear collider, 
would allow one to improve dramatically the search
potential for new particles and the disentanglement of their dynamics.  
This would represent a crucial step towards 
the understanding of the
nature of fundamental interactions. Concurrently, the Standard Model and
its parameters could be scrutinized and determined 
with unparallelled precision. The
availability of positron polarization would allow significant
progress also in this respect.

In {\bf direct searches}, the physics potential of the  ILC is strongly 
improved if one can exploit simultaneously the independent polarizations
of both beams, particularly in the following regards, cf. also 
table~\ref{tab-sum}.
\begin{itemize}
\item 
The chiral structures of interactions in various processes can be 
identified independently and unambiguously. This provides the possibility 
of determining the quantum
numbers of the interacting particles and testing stringently model 
assumptions. Several of these tests are not possible with polarized electrons 
alone.
\item 
The larger number of available observables is crucial for
disentangling the new physics parameters in a largely model independent 
approach.
\item Transverse polarization of both beams enables the construction
of new CP-odd observables using products of particle momenta,
and further enlarges the number of observables available to constrain the 
new physics parameters.
\item 
The enhanced rates with suitable polarizations of the two beams
would allow for better accuracy in determining cross sections and asymmetries. 
This increase of the signal event rate
may even be indispensable, in some cases, for the observation of marginal 
signals of new physics.
\item A more efficient control of background processes can be obtained. 
The higher signal-to-background ratio may be crucial
for finding manifestations of particles related to new physics and 
determining their
properties. Important examples are the searches for signatures of 
massive gravitons, whose existence is foreseen by models with
extra dimensions, and of supersymmetric particles.
\end{itemize}

In {\bf indirect searches} for new physics, the clear advantages of
having both beams polarized simultaneously include the following.
\begin{itemize}
\item The enhancement of cross sections, and correspondingly of the rates, 
by effective use of 
the polarizations, leads to a reduction in the statistical uncertainties.
\item Significant increases in sensitivity to new physics
at high energy scales can be achieved, with the possibility
to elucidate the associated interactions.
\item The increase in the sensitivity to non-standard couplings due to the 
synergy of high energy, 
high luminosity and especially of
different possible initial polarization configurations 
will allow one to disentangle different kinds of interactions.
\item The left-right asymmetry, which can be crucial for distinguishing 
different models,
is often limited by systematic uncertainties. These can be 
reduced significantly when both beams are polarized.
\item There is an increase in the sensitivity to new interactions which 
are not of the current-current type,
such as those mediated by gravitons or (pseudo)scalar exchanges.
\end{itemize}

{\bf Transversely polarization} of both beams opens up new possibilities:
\begin{itemize}
\item This offers new observables to detect non-standard interactions,
including possible new sources of CP-violation.
\item It enhances the sensitivity to graviton interactions and enables 
one to draw a distinction between different scenarios of extra spatial 
dimensions, even far below the resonance production threshold.
\item It provides access to specific triple-gauge couplings which cannot 
be extracted with only longitudinally polarized beams.
\end{itemize}

The {\bf precision tests} of the Standard Model are unprecedented. In 
particular, the use of simultaneously polarized beams improves 
significantly the precision attainable in the following measurements.
\begin{itemize}
\item It improves the separation between the annihilation and 
scattering channels in $W^+W^-$ production, as required
for an optimal determination of triple-gauge-boson couplings.
\item 
High-luminosity GigaZ running at the $Z$~boson resonance or at the
$W^+W^-$ threshold with positron polarization allows for an improvement in 
the accuracy of the determination of $\sin^2\theta_W$ by an order of 
magnitude, through studies of the left-right asymmetry.
This will have far-reaching implications for consistency tests of the
electroweak theory, in particular in the Higgs sector.
\end{itemize}

The qualitative and quantitative consequences of having both beams
longitudinally polarized in comparison
with the $e^-$ polarization only 
for the reference reactions analysed in the
previous chapters are also listed in table~\ref{tab-sum}.  
These examples by no means exhaust the whole phenomenology of a polarized
positron beam. The quantitative improvement factors are relative to the
case with polarized electrons only. In most cases $|P_{e^-}|=80\%$ and
$|P_{e^+}|=60\%$ are used as conservative possibilities, but in some
examples also $|P_{e^-}|=90\%$ and $|P_{e^+}|=80\%$ or $40\%$ have been
considered.
  
It is apparent from the table that in some specific cases of non-standard
couplings factors of about an order of magnitude in the Figur-of-Merit, the
$S/\sqrt{B}$
ratio, could be gained when employing two polarized beams, compared with
the case of polarized electrons only.  Since in some new physics models
only very small rates are expected, such a large gain in the Figur-of-Merit
may be crucial to detect the signals.  In some examples, such
as testing the quantum numbers and properties of supersymmetric particles,
simultaneously polarized beams are definitely needed.

Indirect searches are a very important tool to find signatures of new
interactions characterized by higher mass scales far beyond the
kinematical reach of present, future and next-generation collider
experiments. Such new mass scales can be accessed by looking for
deviations of experimental observables from the Standard Model predictions,
and in principle the results depend sensitively on the particular model
adopted in order to analyse the data. Since the discovery reach scales
with the luminosity, it is often enhanced by about 20\% with both beams
polarized compared to the case with only polarized electrons. Furthermore,
the positron polarization option allows more observables to be measured,
reducing to a large extent the model dependence. The use of polarized
positrons also improves substantially the precision with which the
left-right asymmetry observables can be measured.

Specific advantages of transverse beam polarization are listed in
table~\ref{tab-sum}. For example, in indirect searches
for massive gravitons, one could even distinguish between different models
of the extra dimensions if both beams are transversely polarized.
Furthermore, transverse
polarization turns out to be particularly promising in the context of
searches of new CP-violating interactions, for example by exploiting
azimuthal asymmetries that may allow the detection of even marginal
manifestations of new sources of CP violation.

The physics studies are still ongoing, and
the listed examples should be understood solely as a contemporary status
report\footnote{
For further information, see {\tt
http://www.ippp.dur.ac.uk/$\sim$gudrid/power/ }.
}. Nevertheless, these examples already show that positron
polarization has excellent possibilities for enrichening considerably the 
physics output from the linear collider and that the full potential
of the ILC could be realized only with a polarized positron beam as
well.

{\small \begin{table}[htb]
\vspace*{-1.7cm}%\hspace*{-1cm}
\renewcommand{\arraystretch}{1.1}
\begin{tabular}{|l|ll|} \hline Case & Effects &
Gain \\ \hline\hline
{\bf SM}: &&\\
top threshold & Improvement of coupling measurement & factor 3 \\
$ t \bar{q}$ & Limits for FCN top couplings reduced & factor 1.8\\
CPV in $t \bar{t}$ &Azimuthal CP-odd asymmetries give  & $P_{e^-}^{\rm T} P_{e^+}^{\rm T}$ required\\
& access to S- and T-currents up to 10 TeV & \\[.2em]
$ W^+ W^-$ & Enhancement of $\frac{S}{B}$, $\frac{S}{\sqrt{B}}$ & up to a factor 2 \\
& TGC: error reduction of $\Delta \kappa_{\gamma}$, $\Delta \lambda_{\gamma}$, $\Delta \kappa_{Z}$,
$\Delta \lambda_{Z}$   & factor 1.8 \\ 
  & Specific TGC $\tilde{h}_{+}={\rm Im}(g_1^{\rm R}+\kappa^{\rm R})/\sqrt{2}$  
& $P_{e^-}^{\rm T} P_{e^+}^{\rm T}$ required \\[.1em]
CPV in  $\gamma Z$ & Anomalous TGC $\gamma\gamma Z$, 
$\gamma Z Z$ & $P_{e^-}^{\rm T} P_{e^+}^{\rm T}$ required\\[.2em]  
$HZ$ & Separation: $HZ\leftrightarrow H\bar{\nu}\nu$ & factor 4 with RL\\ 
& Suppression of $B=W^+\ell^-\nu$ & factor 1.7 \\[.2em] 
\hline
{\bf SUSY:} &&\\
$\tilde{e}^+ \tilde{e}^-$ & Test of quantum numbers $L$, $R$& $P_{e^+}$
required \\ & and measurement of $e^{\pm}$ Yukawa
couplings & \\ 
$ \tilde{\mu} \tilde{\mu}$ & Enhancement of $S/B$,
$B=WW$ & factor 5-7\\ & 
$\Rightarrow$ $m_{\tilde{\mu}_{L,R}}$ in the continuum & \\ 
$H A$, $m_A>500$~GeV &
Access to difficult parameter space & factor 1.6\\ $
\tilde{\chi}^+\tilde{\chi}^-$, $\tilde{\chi}^0 \tilde{\chi}^0$ &
Enhancement of $\frac{S}{B}$, $\frac{S}{\sqrt{B}}$ & factor 2--3\\ &
Separation between SUSY models, & \\ 
& 'model-independent' parameter
determination & \\
CPV in  $\tilde{\chi}^0_i \tilde{\chi}^0_j$ & 
Direct CP-odd observables & $P_{e^-}^{\rm T}P_{e^+}^{\rm T}$ required \\
RPV in $\tilde{\nu}_{\tau}\to \ell^+ \ell^-$ & Enhancement of
$S/B$, $S/\sqrt{B}$ & factor 10 with LL \\ & Test of spin quantum
number & \\[.2em]\hline
{\bf ED:} && \\ 
$ G \gamma$ & Enhancement of $S/B$, $B=\gamma \nu \bar{\nu}$, & factor 3\\ 
$e^+e^-\to f \bar{f}$ 
& Distinction between ADD and RS modes & $P_{e^-}^{\rm T} P_{e^+}^{\rm T}$ required\\[.2em]\hline
{\bf ${\bf Z'}$:} && \\  
$e^+e^-\to f \bar{f}$
 & Measurement of $Z'$ couplings & factor 1.5 \\[.2em]\hline
{\bf CI:} && \\
$ e^+ e-\to q \bar{q}$ & Model independent bounds & $P_{e^+}$ required
\\[.2em]\hline 
\multicolumn{3}{|l|}{\bf Precision measurements of the Standard Model at GigaZ:}\\[.1em]
$Z$-pole & Improvement of $\Delta\sin^2\theta_W$  & factor 5--10\\
& Constraints on CMSSM space & factor 5 \\
CPV in $Z\to b\bar{b}$ & Enhancement of sensitivity & factor 3 \\
\hline
\end{tabular}
\caption[Summary table of physics examples]{Some of the physics examples given in this report. 
The case of having both beams polarized is compared
with the case of using only polarized electrons; in many cases $(|P_{e^-}|,|P_{e^+}|)=(80\%,60\%)$
is compared to $(|P_{e^-}|,|P_{e^+}|)=(80\%,0\%)$, cf.\ see corresponding chapter;
B (S) denotes background (signal); CPV (RPV) means CP (R-parity) violation.\label{tab-sum}}
\end{table}
}

%%%%%%%%%%%%%%%%%%%%%%%

%%%%%%%%%%%%%%%%%%%%%%%%%
\chapter{Machine issues \label{machine}}
%%%%%%%%%%%%%%%%%%%%%%%%%
Polarized positrons are produced via pair production from 
circularly-polarized photons. The main issues associated with the development of
a polarized-positron source for a linear collider are: the generation of
polarized photons, the efficient capture of the positrons, and dealing
with the high heat loads (and accompanying radiation) due to the
required flux. This chapter discusses the photon-based schemes for
polarized-positron production being considered for the ILC. Status
updates of work towards the realization of viable polarized-positron
sources are also given. A report of work on polarized-electron sources
is included, followed by a discussion of the spin manipulation systems
required in the ILC. The chapter continues with a section about 
measuring the polarization with
conventional polarimetry. Emphasis is put on Compton polarimetry and designs for polarimetry 
up- as well as downstream of the IP 
are discussed. However, an overview of other methods such as M{\o}ller polarimetry is also given and
future design work is outlined.
The chapter concludes with a section about measuring the polarization from
physics processes in $W$ and $W$-pair production and from collider data
via application of the Blondel scheme.

%%%%%%%%%%%%%%%%%%%%%%%%%%
\section{General remarks}
%%%%%%%%%%%%%%%%%%%%%%%%%%

In this section, a short introduction to polarized positron production
is given along with brief descriptions of the possible sources of
polarization loss. A description of the polarized electron source for
the ILC is included for completeness.\\ \smallskip

The current parameter specifications for the ILC list polarized
electrons with $P_{e^-}\ge 80\%$ for the baseline machine whereas
positron polarization (with $P_{e^+}\ge 50\%$) is suggested as an
option for a later upgrade~\cite{scope}. If adequately planned in the
baseline, the upgrade to polarized positrons is relatively straight
forward and not prohibitively expensive in terms of either cost or
downtime required to implement the upgrade.  The 
two main methods for positron production under consideration for the ILC are
a photon-based source and a `conventional' source.  The photon-based
source uses multi-MeV photons and relatively thin targets (less than a
radiation length thick) to produce positrons. If the photons are
circularly polarized, the positrons (and electrons) are spin
polarized. This positron polarization can be preserved in the subsequent
capture, acceleration, damping, and transport to the collision
point(s). The conventional source uses a multi-GeV electron drive beam
in conjunction with thick, high-Z targets to produce 
positrons from the resultant electromagnetic cascade in the
target. The positrons produced by this method cannot be polarized.
These two schemes  ultimately
present very similar engineering challenges while at the same time
having distinct attributes and drawbacks. It is emphasized, however,
that only the photon-based schemes offer the promise of positron
polarization~\cite{Floettmann:2004vh}.

In addition to developing a source of polarized positrons (and electrons)
for a high energy linear collider, it is important to ensure that no significant
polarization is lost during the transport of the positron and electron beams
from
the source to the interaction region.  Transport elements downstream of
the sources which can contribute to a loss of polarization include the
initial acceleration structures, transport lines to the damping rings, the damping rings,
the main linacs,
and the high energy beam delivery systems. As discussed below, the largest
depolarization effect is expected to result from the collision of the two
beams at the interaction point(s). This effect is expected to
decreases the
polarization by about $0.25\%$, see section~\ref{pol-summary}. %%%%%%

%%%%%%%%%
\subsection{Polarized-positron source}
%%%%%%%%%% 
Circularly-polarized photons are required for the
generation of longitudinally-polarized positrons via $e^{\pm}$ pair
production in a thin target. The photons are in the energy range of a few MeV
up to about 100 MeV. Because the target is typically
a fraction of a radiation length thick, high strength materials,
such as titanium
alloys, can be considered as opposed to conventional targets, where
 high-Z, high-density materials are required to minimize the emittance
of the produced positrons. The two methods for generating the
polarized photons under consideration are: 
%\vspace{-.3cm} 
\begin{itemize}
\item a high-energy electron beam ( $\gsim$150 GeV) passing 
through a short period,
helical undulator. This scheme is discussed in greater detail in 
sections~\ref{und-e166},
\ref{und-dar}.  The E-166 experiment, which is currently running at SLAC,
is a demonstration of this undulator-based polarized positron production 
scheme.
The experiment will be finished in 2005.
\item Compton backscattering of laser light off a GeV energy-range
electron beam. This is discussed in detail in section~\ref{laser} and the
concept is being tested in an experiment which is currently running at KEK.
\end{itemize}
In both schemes a positron polarization of about $|P_{e^+}|\ge 60\%$
is expected at the ILC.

Both schemes would be
applicable for the ILC design \cite{scope}
 and also adaptable for a possible future
multi-TeV LC design~\cite{clic}.\\[.5em]

%%%%%%%%%%%%%%%%%%%%
\noindent {\bf Comparison with conventional (unpolarized) positron
sources}\\[.1em] Apart from the obvious advantage of generating
polarized positron beams, photon drive beam based positron sources
have various advantages and disadvantages as compared with
conventional sources using electron drive beams.  At this time both
the conventional and undulator-based positron source designs are
equally mature and can meet the ILC requirements. The big advantage of
the conventional source is that is completely decoupled from the rest
of the ILC whereas the undulator-based positron source needs a working
ILC main linac for operation. This leads to some advantages in
commissioning and uptime. With careful design and an additional simple
keep-alive positron source, these advantages can be mitigated, though
not completely eliminated. \\ The photon-based positron source has
three advantages over the conventional source~\cite{mikh04}.
\begin{itemize}
\item First the fact that the target is thinner allows for the use of  
lower-Z materials such as Ti-alloys that are stronger than the W-Re alloys used in the 
conventional sources. 
\item Secondly the photon energy can be chosen to be lower than the neutron 
photo-production cross-section and the activation of the target material may be much reduced. 
\item Thirdly the emittance of the produced positrons is less because the target is thin. This will
 have advantages for the design of the positron damping ring(s) and maybe reduce the 
 cost advantages of the conventional source.
\end{itemize}
 
Cost and availability issues associated with an undulator-based
positron source are discussed in~\cite{USLCTOS}.
In this study, a comparison is made between an
undulator-based system and a conventional (unpolarized) system.  Whereas the
detailed assumptions and quantitative results of \cite{USLCTOS} are
open to discussion, the general conclusions are reasonable: the cost
difference between an undulator source and an upgrade from a
conventional source is small ($\sim$ 1\%)
in comparison to the overall project cost; availability and schedule
are impacted due to the requirement of high energy electrons for
undulator-based positron production.  These issues need to be addressed
in future design studies.

%%%%%%%%%%%%%%%%
\subsection{Polarized electron source}
%%%%%%%%%%%%%%%%
The polarized electron source consists of a polarized high-power laser beam
and a high-voltage dc gun with a semiconductor photocathode.
This design was developed and successfully used at the SLC and is currently
in use at CEBAF. 
The SLC system can be adopted directly for the ILC using a
laser system with the appropriate time structure. While there is
discussion about the benefits of a low-emittance radio frequency (RF) 
gun for polarized-electron production, no R$\&$D is ongoing or planned
for such a gun. 
Recent work in the area of polarized
electron source design for the ILC is presented in
section~\ref{pol-elec} and it is expected to achieve
a degree of electron polarization of about $|P_{e^-}|\le 90\%$ at the ILC.

%%%%%%%%%%%%%%%%%%%%%%%%%%%%%%%%%%%%%%%
\subsection{Polarization preservation}
%%%%%%%%%%%%%%%%%%%%%%%%%%%%%%%%%%%%%%%
{\bf Low-energy beam capture and acceleration:}\\[.5em]
The transverse and longitudinal phase space of the positrons
coming out of the target is matched into the 6-D acceptance of the downstream system
 by a longitudinally varying solenoidal field.
The positrons are then accelerated up to the
damping-ring energy of 5 GeV.  Axial solenoid fields are used for the
transverse capture optics at the low energies up to about 250 MeV.
Thereafter, magnetic quadrupoles are used for focusing. The
longitudinal polarization is preserved while the transverse
component is washed out due to spin precession in the solenoidal magnetic
fields used for the initial capture. Because of the high incident beam power
on the target and constraints in the mechanical design of positron targets,
multiple target stations running in parallel may be required
for the ILC. The layouts of these target stations 
can be configured so that the polarization remains longitudinal after bunches
from the different targets are combined into the ILC beam. Space
charge effects are not expected to be a source of depolarization.
Because of the lower energy, space-charge forces are significantly
larger in the case of the polarized-electron source but 
even in this case no spin depolarization has been observed.  \\[.5em]

%%%%%%%%
\noindent {\bf Damping rings:} \\[.5em] 
Only polarization parallel or anti-parallel to the guide fields of
the damping ring is preserved. Thus the incoming longitudinal polarization 
must be rotated into 
the vertical prior to injection into the rings. This is done in the transport line
leading from the end of the 5 GeV linac to the damping ring, utilizing an
appropriate combination
of dipole and solenoid fields. The vertical spins can be rotated back to the
longitudinal direction with an analogous system in the
damping ring extraction line leading to the main linac. 
These spin manipulation systems are discussed in further 
detail in section~\ref{spinrotation}.

It is important to avoid spin-orbit coupling resonance depolarizing
effects by operating off the resonant 
energies. Depolarization due to 
the stochastic nature of photon emission can be estimated
by comparing the time that a beam spends in the damping ring to the
time constant for depolarization. This ratio is typically very small
and hence depolarization in the rings is expected to be vanishingly
small. Nevertheless it is important to revisit these issues as the ring designs
become available and a detailed simulation of
depolarization effects should be made. This is best carried out using
a Monte--Carlo simulation of the spin--orbit 
motion \cite{BarberHandbook,des-dr,clic-depol}. 
\\[.5em]

%%%%%%%%
\noindent{\bf Main linac:}\\[.5em] In a linac, the electric field and
the particle velocity are essentially parallel.  Then, according to
the Thomas-Bargmann-Michel-Telegdi (T-BMT) equation the electric field
will cause negligible spin precession. The effects of the
transverse field from the RF couplers and quadrupoles are also
negligible. There should be no loss of polarization in the main
linac.\\[.5em]
%%%%%%%%%

\noindent {\bf Beam delivery system:} \\[.5em] 
After acceleration in the main linacs, the
$e^{\pm}$ beams are brought to collision by the beam delivery
systems. These beam lines contain bend magnets. According to the T-BMT
equation, a beam deflection of $\delta \theta_b$ in a 
transverse magnetic field causes a spin rotation of $a \gamma \delta
\theta_b \approx E ({\mbox{GeV}})/0.441 ({\mbox{GeV}})\delta
\theta_b$. Depolarization associated with synchrotron radiation in the transport
system bends is expected to be negligible.
Care is needed to ensure that the polarization is
longitudinal (or perhaps transverse) as needed at the interaction point.  The
spin rotator systems between the damping rings and the main linacs
permit the setting of arbitrary polarization vector orientations at the
IP.\\[.5em]

%%%%%%%%%
\noindent{\bf Beam-beam interactions:} \\[.5em]
Loss of polarization can also occur as the electron and positron bunches collide.  
The two mechanisms responsible for this are 
spin rotation, according to the T-BMT equation, of the spins in one bunch due to the 
electric and magnetic fields in the oncoming bunch; and spin 
flip due to synchrotron radiation (the
Sokolov-Ternov effect). The combined effect of these two mechanisms has been studied 
analytically~\cite{mike6} as well as numerically~\cite{Thompson,CAIN}.
At 500 GeV center-of-mass energy for the nominal ILC parameters~\cite{scope},
the expected overall loss of polarization~\cite{Thompson} 
of either beam is expected to be 
$\Delta P/P \simeq 1\%$. However, the effective loss up to the point
of interaction (luminosity-weighted) is only $\simeq 0.25\%$:
see section~\ref{pol-summary} for more detail. 
Updates of these calculations using actual design parameters are
needed as the machine parameters become available.

%%%%%%%%%%%%%%%%%%%%%%%%%%%%%%%%%
\section{Positron polarization}

Polarized positrons are produced via pair production from 
circularly-polarized photons. The concepts under consideration for 
polarized-positron sources differ only in the manner in which the 
circularly-polarized photons are produced: either through the use of a helical
undulator or from Compton backscattering. The issues associated with
target engineering, optimization of the capture yield and positron
polarization are identical for both types of schemes.

This section begins with a discussion of undulator-based production of
polarized positrons. The basic concept is presented and followed by a
description of the SLAC E166 demonstration experiment. A summary of work at
Daresbury on the development of undulator prototypes concludes the
discussion of the undulator-based sources.  The design of a
laser-Compton based polarized positron source is described
in~\ref{laser}.  A discussion of a demonstration
experiment which has been developed at the KEK ATF is included.

%%%%%%%%%%%%%%%%%%%%%%%%%%%%%%%%%%%%%%%%%%%%%%%%%%%%%%%%%%%%%%%%%%%%%%%%
\subsection{Undulator-based polarized positron source \label{und-e166}}
%%%%%%%%%%%%%%%%%%%%%%%%%%%%%%%%%%%%%%%%%%%%%%%%%%%%%%%

\begin{figure}[htb]
\begin{picture}(10,7)
\setlength{\unitlength}{1cm}
\put(0,0){\mbox{\includegraphics[height=.25\textheight,width=.7\textheight]{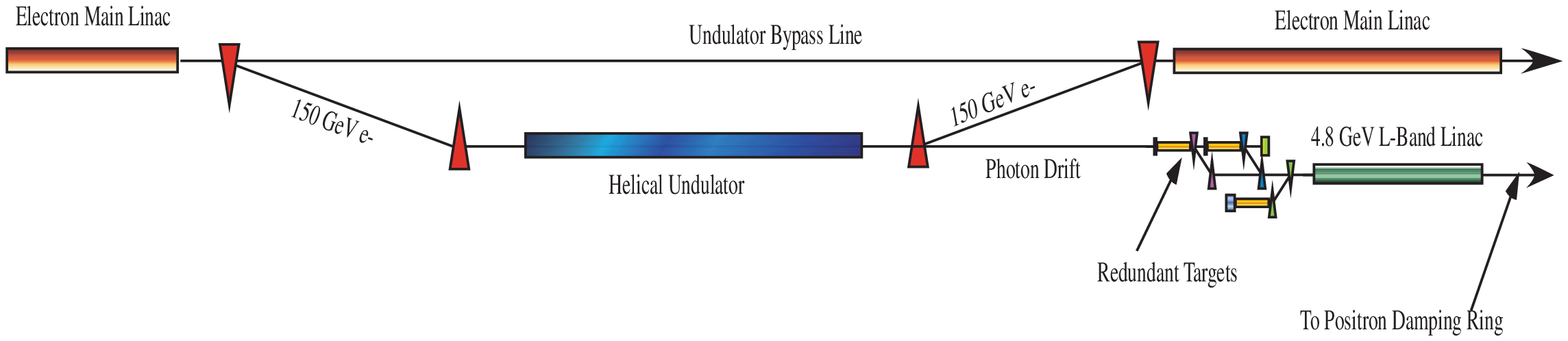}}}
\end{picture}\vspace{0cm}
\caption[Possible layout of the polarized positron source]{
A possible layout of the polarized positron source at the ILC.
The electrons could be extracted form the main linac with an energy 
of 150~GeV, as shown in the 
figure~\cite{USLCTOS}.
The undulator could also be placed at the end of the linac, where the 
electrons have an energy of $250$~GeV \cite{teslatdr01}.
\label{skizze-150}}
\end{figure}

A polarized positron source based on the radiation from an undulator
was first proposed by Balakin and Mikhailichenko~\cite{balakin79} in 1979
in the framework of the VLEPP project, several prototypes for short period helical super-conducting undulators have
been proposed~\cite{Mikhailichenko:2002wr}. 
This method requires a multi-hundred GeV electron beam
and in order to save on ILC cost, it is proposed to use the electron main linac to 
provide this beam.
Fig.~\ref{skizze-150} shows the layout of the
polarized positron source as developed for the USLCTOS cold reference
design~\cite{USLCTOS}. The electron beam is extracted from the electron main
linac at an energy of about 150 GeV and transported through $\sim 200$
m of the helical undulator. The electrons
are then injected back into the linac and are used for collisions.
The
period of the undulator is nominally 1 cm and the magnetic field on axis
has a strength of about B=1.1~T. 
Photons generated in the undulator have
an energy of the order of 10 MeV and are circularly polarized. These
photons are incident on a thin target (about 0.4 radiation lengths).
Within the target, the photons convert to electron-positron
pairs. The helicity of the initial state is conserved in the pair
creation process.  As described in detail in~\cite{Olsen}, the spin orientation of the
created positron is correlated to its energy relative to that of the 
initial, polarized photon. The polarization of the beam of
positrons that is collected from the target falls in the range of
40\%-70\%, and depends on the details of the initial photon polarization
spectrum, target material and thickness, and capture phase
space.  This scheme provides also a high intensity of polarized positrons.

The TESLA proposal has the undulator placed at the end of the linac
with consequentially higher energy electrons and photons~\cite{teslatdr01}. While the
details of the two designs vary due to the different available
electron energy (150 GeV in the case of the USLCTOS design and 250 GeV
in the case of TESLA design), see fig.~\ref{skizze-150}, either scenario (TESLA or USLCTOS) can be
used to produce a beam of polarized positrons for the ILC. Operational
aspects of the two different scenarios are somewhat different. The
USLCTOS design is based largely on the TESLA design with the 
difference that the undulator is placed along the electron linac rather
than at the end. This is done so that the electron drive beam energy does
not vary if the collision energy is changed. This design choice
requires that a linac bypass must be built which does not adversely
dilute the electron emittance. A principle concern in the design is
the effect of the limited energy passband of the bypass with regard to
machine protection in the event of a sudden loss of electron
energy. Design effort to mitigate this problem is ongoing.

The SLC positron source is a good starting point for extrapolation
to a positron source for the ILC. However,
the required ILC positron source flux is a factor of about 60 greater than that in the
SLC positron target system. The power of the incident drive beam and
the energy deposited in the target station and downstream systems is a
factor of about 10 greater than in the SLC design. In addition, the pulse
structure of the ILC beam is significantly different from that at
the SLC. The discrepancy between the factor of 60 in flux and the
factor of 10 in dissipated power is due to the expected increase in
the acceptance of the positron capture systems with respect to the SLC
design.  Both the TESLA Design Report and USLCTOS have descriptions of
plausible system designs. The design effort for the ILC positron
source is aimed at realizing the increase in positron capture
efficiency and handling the increase in power. This engineering effort
spans the entire range of component parameter space in the quest to
maximize the positron polarization while minimizing the incident beam
power. It should be noted that the design issues associated with both
the unpolarized conventional and the photon-based systems are essentially the same for
the systems downstream of the target: the matching optics, RF capture
sections, and damping ring acceptance.  Even the power absorption in
the conventional and photon-based target systems is very similar.
The work to develop a viable positron source includes the engineering
and testing of high strength target materials, development and testing
of magnetic capture optics, and prototyping of the RF structures used
for the initial acceleration and the target vault design
and associated remoted handling systems.

Some effort is still
required to meet the engineering challenges, whether or not positron
polarization is chosen for the ILC baseline.

%%%%%%%%%%%%%%%%%%%%%%%%%%%%%%%%%%%%%%
\subsubsection{Demonstration experiment E-166}
%%%%%%%%%%%%%%%%%%%%%%%%%%%%%%%%%%%%%%
A demonstration experiment
for undulator-based
polarized-positron production at the ILC has been approved at
SLAC~\cite{E166}. This experiment, E-166, is scheduled to run in 2005 in the SLAC Final
Focus Test Beam (FFTB) area.
The aim of the experiment is to test 
the fundamental process of generating circularly-polarized photons
in a helical undulator~\cite{blewett77} 
and the production of polarized positrons 
via pair production in a thin target with these photons.
In addition, the parameters of the experiment were chosen so that the
photon spectrum would be the same as that for the USLCTOS design and E166 would be a scaled
down demonstration of a polarized ILC positron source.

\begin{figure}
\setlength{\unitlength}{1cm}
\resizebox{.9\textwidth}{!}{%
\hspace{3cm}\includegraphics{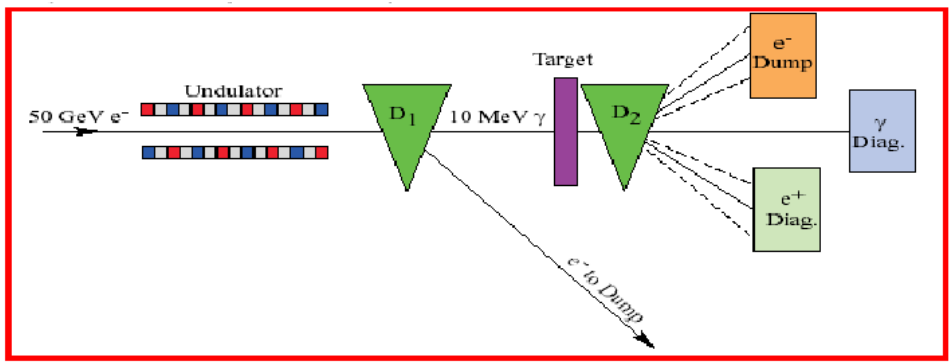}
}
\vspace{-.2cm}       
\caption[Conceptual layout of E166]{Conceptual layout of the 
experiment to demonstrate the
production of polarized positrons in the SLAC FFTB. The 50-GeV $e^-$
beam passes through an undulator, producing a beam of circularly
polarized photons of MeV energy. The electrons are deflected by
the $D_1$ magnet. The photons are converted to electrons and positrons
in a thin Ti target. The polarizations of the positrons and photons are
measured in polarimeters based on Compton scattering of electrons in
magnetised iron~\cite{E166}.}
\label{fig-skizze}       
\end{figure}

A schematic layout of E166 is shown in fig.~\ref{fig-skizze}. 
The FFTB electron beam passes through a 1-m long 
pulsed helical undulator to generate the circularly polarized photons.
Because the FFTB beam can have a maximum beam energy
of about 50 GeV, instead of the higher energies for the ILC design, the undulator
aperture has to be reduced to less than 1 mm in order to generate the same photon
spectrum (approximately 10 MeV cut-off for the first harmonic of the undulator
radiation).
The photons are converted to positrons (and electrons) in a thin, moveable target.
Titanium and tungsten targets, which are both candidates for
use in linear colliders, will be tested. The experiment will measure
the flux and polarization of the undulator photons, as well as the spectrum
and polarization of the positrons produced in the conversion target,
and compare the results of measurement with simulations. 
Thus the
proposed experiment directly tests the validity of
the simulation programs used for the physics of polarized pair
production in finite matter, in particular the effects of multiple
scattering on polarization.

Table~\ref{table-e166} provides a
comparison of the E-166 parameters with those of the unpolarized TESLA
TDR design and of the polarized design of the USLCTOS cold option. It is seen that 
E-166 produces
photons in the same energy range, uses similar target materials and
thicknesses, and produces positrons with the same polarization
characteristics as those in the two linear collider designs.

The achievable precision of 5-10\% of the proposed transmission polarimetry is sufficient. 
Good agreement between the experimental results and the simulations will lead to
greater confidence in the proposed designs of polarized-positron sources for the next
generation of linear colliders.

This experiment, however, will
not address detailed systems issues related to polarized-positron
production at a linear collider, such as capture efficiency, target thermal
hydrodynamics, radiation damage in the target, or an undulator
prototype suitable for use at the ILC; such issues are well within the
scope of R\&D of a linear collider project that chooses to implement a
polarized-positron source based on a helical undulator. 

%%%
%%%

\begin{table}
{\footnotesize
\hspace{3cm}\begin{tabular}{|lccc|}
\hline\noalign{\smallskip}
Parameter& TESLA & USLCTOS & E-166\\
\noalign{\smallskip}\hline\noalign{\smallskip}
Beam Energy, $E_e$ [GeV] & 150-250 & 150 & 50\\
$N_e$/bunch & $3\times 10^{10}$ & $2\times 10^{10}$ & $1\times 10^{10}$\\
$N_{bunch}$/pulse & 2820 & 2820 & 1\\
Pulses/s [Hz] & 5 & 5 & 30\\
Undulator Type & plan./helical & helical & helical\\
Und. Parameter, $K$ & 1 & 1 & 0.17\\
Und. Period, $\lambda_u$ [cm] & 1.4 & 1.0 & 0.24\\
Und. Length, $L$ [m] & 135 & 200 & 1\\
$1^{st}$ Harmon., $E_{c10}$ [MeV] & {9-25}&{11}&
{9.6}\\
Target Material & {Ti-alloy} & {Ti-alloy} &
{Ti-alloy, W}\\
Target Thickn. [rad. len.] & {0.4} & {0.4} &
{0.25-0.5}\\
Pos. Pol. [\%] & {0} & {59} &
{53}\\
\noalign{\smallskip}\hline
\end{tabular}
}
\caption[Polarized positron parameters]{Positron parameters for the
unpolarized TESLA baseline design, polarized USLCTOS design, and 
the E-166 experiment~\cite{E166}. \label{table-e166}}
\end{table}
%%%%%%%%%%%%%%%%%%%%%%%%%%%%%%%%%%%%%%

%%%%%%%%%%%%%%%%%%%%%%%%%%%%%%%%%%%%%%%%%%%%%%%%%%%%%%%%%
\subsection{Helical undulator development 
\label{und-dar}}
\begin{figure}[htb]
\begin{picture}(10,6)
\setlength{\unitlength}{1cm}
\put(1.5,-1){\mbox{\includegraphics[height=.32\textheight]{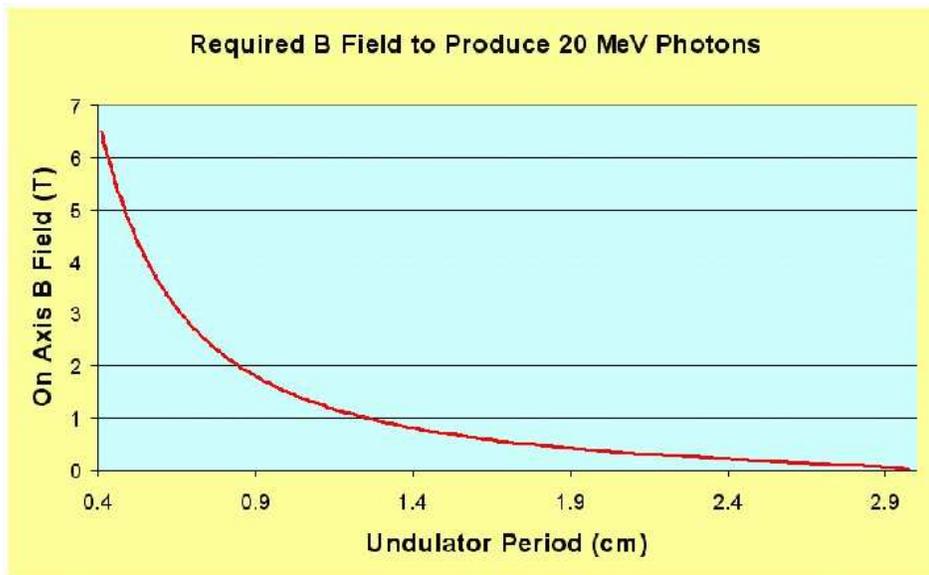}}}
\end{picture}\vspace{.8cm}
\caption[B-field vs. undulator period]{On-axis B field versus undulator 
period to produce circularly-polarized 20~MeV photons with a 250~GeV electron 
beam~\cite{scott04}.
\label{Figure2_duncan}}
\end{figure}

%%%%%%%%%%%%%%%%%%%%%%%%%%%%%%%%%%%%%%%%%%%%%%%%%%%%%%%%%
Two helical undulator prototypes for the ILC are being developed under the
guidance of the ASTeC group at Daresbury Laboratory~\cite{scott04}. 
The first device uses a bifilar helix of niobium-titanium
superconducting wire; the second device is based on 
neodymium-iron-boron permanent
magnet material.  Table~\ref{Table1_duncan} 
lists the parameters of the magnets. An undulator
period of 14 mm has been selected for the prototypes and fig.~\ref{Figure2_duncan} 
gives an indication
of the magnetic fields required to produce photons of approximately
the correct energy. The parameters
have been chosen to meet the requirements of the 
TESLA TDR, and can be scaled to the current ILC design.  
In table~\ref{Table1_duncan}, 
`Length' refers to the prototype and `Full
Module' is the length of an undulator section as envisaged for the
ILC. The full ILC undulator consists of a large number of modules to
make up the $\sim200$~m total required length.  Studies to date have
shown that both the superconducting and permanent magnet technologies
will meet the ILC undulator design criteria. To help in deciding which
design is better, short 10- to 20-period models have been developed. These
models will enable:
\vspace{-.3cm}
\begin{itemize}
\item the field quality and strength of each design to be checked.
\item assessment of the size and probability of magnet errors.
\item the evaluation of the effort required for full-sized module construction.
\item an improvement of cost estimates.
\end{itemize}

{\small
\hspace*{-1cm}
\begin{table}[htb]\hspace*{-1cm}
\begin{tabular}{|c|c|c|c|c|c|c|c|}
\hline
Type &Period & On-Axis Field & Magnet Aperture & Material & 
Current density & Length & Full Module\\ \hline
SC & 14 mm & 0.85 T & 4 mm & NbTi & 1000 $A/mm^2$ & 30 cm & 2-5 m\\ \hline
PPM & 14 mm & 0.83 T & 4 mm & NdFeB &(1.3 T) &  15 cm &  2-5 m \\ \hline
\end{tabular}
\caption[Design parameters of helical undulator prototypes]{Parameters
of the prototype of a 
super-conducting (SC) and permanent-magnet (PPM) helical 
undulator design parameters\cite{scott04}.
\label{Table1_duncan}}
\end{table}
}

It was decided that both prototype devices would have the same period
of 14~mm to facilitate the comparison. Modelling of the superconducting 
device indicates that an on-axis field greater than the specified value 
of 0.85~T is possible with the inclusion of iron.
The NdFeB undulator can also achieve a
greater field strength than the 0.83~T listed in table
\ref{Table1_duncan}.  Higher fields permit a reduction in the
period. However magnet inhomogeneities, magnetization vector
misalignments (in the case of the PPM device), and assembly errors
combine to reduce the field quality.  The dependence of the
field quality on these errors is one of the areas under investigation.

For the superconducting undulator, two wires
are wrapped in a double helix around the vacuum vessel~\cite{madey71}. When
current is passed through the wires the longitudinal components of
the magnetic field cancel leaving only the rotating (helical)
field required on axis. A 20 period (30 cm) long prototype 
has been made using NbTi wire. This device is
currently being tested.

\begin{figure}
\begin{picture}(10,3)
\setlength{\unitlength}{1cm}
\put(0,-1.4){\mbox{\includegraphics[height=.13\textheight]{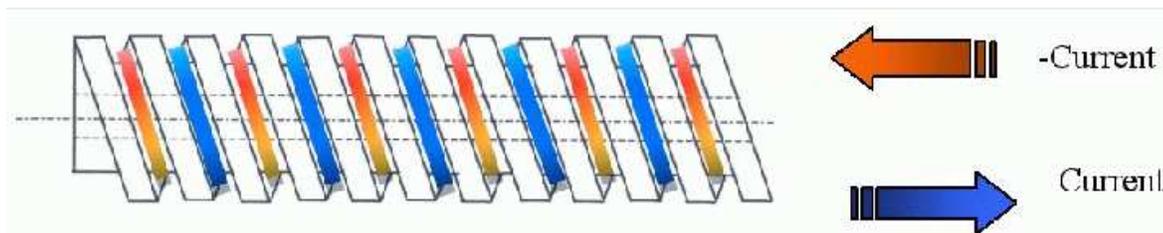}}
}
\end{picture}\vspace{.8cm}
\caption[Schematic of wires in helical undulator]{Schematic of wires wrapped in a helix 
around a former showing different current directions~\cite{scott04}. \label{Figure3_duncan}}
\end{figure}

\begin{figure}
\begin{picture}(10,5)
\setlength{\unitlength}{1cm}
\put(4.3,-1){\mbox{\includegraphics[height=.25\textheight]{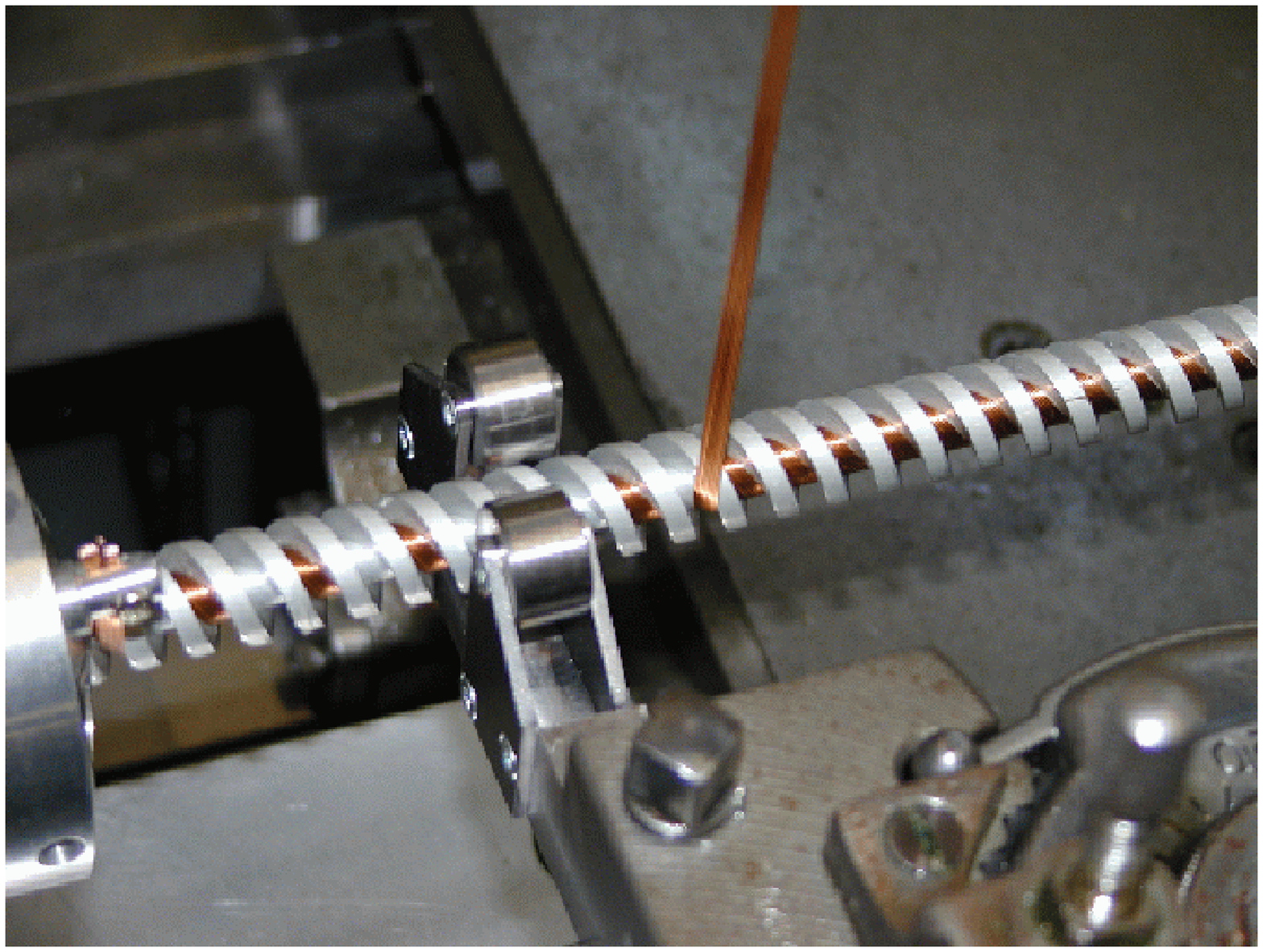}}
}
\end{picture}\vspace{.8cm}
\caption[Prototype of super-conducting undulator magnets]{ 
Undulator prototype with super-conducting (SC) 
wires being wound~\cite{scott04}.\label{Figure4_duncan}}
\end{figure}

In the case of the permanent magnet device, each period of the undulator is divided into
slices. Each slice comprises a ring of trapezoidally shaped 
permanent magnet material blocks whose magnetization vectors 
rotate through $720^{\circ}$ around the ring
to produce a transverse dipole field \cite{halbach80}. Adjacent rings are
rotated through $360^{\circ}$ over one undulator period to form a helical field pattern.
By changing the number of blocks in a ring and the number of rings in a period,
the field strength and quality is altered.
An $8\times 8$ prototype is being developed which uses 8 blocks to form a ring
with 8 rings to a period. The device is 10 periods long or about 15 cm,
including end pieces. This undulator will be ready for testing in the spring of 2005.

\begin{figure}
\begin{picture}(10,5)
\setlength{\unitlength}{1cm}
\put(1,-1){\mbox{\includegraphics[height=.20\textheight]{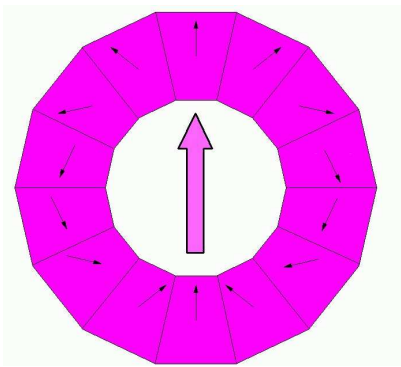}}
}
\put(9,-1){\mbox{\includegraphics[height=.20\textheight]{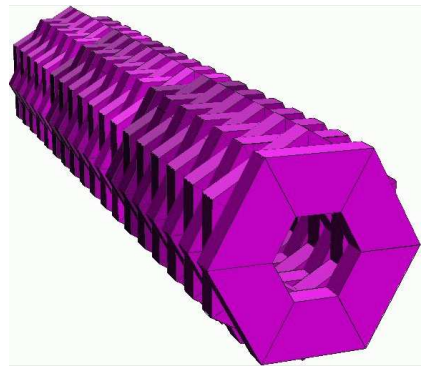}}
}
\end{picture}\vspace{.8cm}
\caption[Undulator design with permanent-magnet design]{Dipole
 field created by many permanent-magnets (PM) blocks 
arranged in a ring. Many rings
are stacked together and rotated to create the helical field~\cite{scott04}.
\label{Figure5_duncan}}
\end{figure}
%%%%%%%%%%%%%%%%%%%%%%%%%%%%%%%%%%%%%%%%%%%%%%%%%%%%%%%%%%%%
\subsection{Laser-Compton based polarized positron source
\label{laser}}
%%%%%%%%%%%%%%%%%%%%%%%%%%%%%%%%%%%%%%%%%%%%%%%%%%%%%%%%%%%%

Polarized positrons can be also created by a laser-based scheme, where
circularly polarized gamma-photons are obtained by backscattering
laser light off an electron beam.  These circularly polarized gammas are then 
used to generate longitudinally-polarized positrons via pair production in a thin radiator
in the same manner as in the undulator-based case.
One of the advantages of this scheme is that because the required electron beam has a 
few GeV and can be generated in a stand-alone linac,
the positron source is independent of the
electron main linac in contrast to the case with the undulator-based source.
The main technical challenge in this scheme is to realize the total laser light intensity
needed for an ILC-capable positron source.
Three possible ways are under consideration to realize this 
scheme~\cite{Omori2004}.

\begin{figure}[htb]
\begin{picture}(10,6)
\setlength{\unitlength}{1cm}
\put(1.5,-.5){\mbox{\includegraphics[height=.25\textheight]{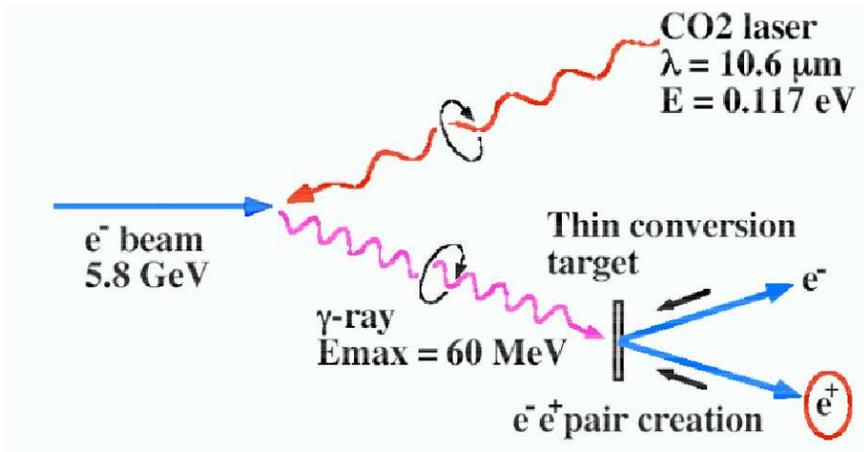}}}
\end{picture}\vspace{.2cm}
\caption[Laser-based scheme]{Collision of a 5.8~GeV electron beam and CO$_2$
laser light creates circularly-polarized gamma-photons with maximum energy of 60~MeV.
Polarized positrons are then generated via pair production in a 
target~\cite{OmoriNIM}. \label{Fig.A}}
\end{figure}

\begin{itemize}
\item The first uses CO$_2$ lasers. Collision of a 5.8~GeV electron beam and CO$_2$
laser light creates photons with a maximum energy of 60~MeV, see fig.~\ref{Fig.A}, whose
energies are suitable for positron production.  A
multi-bunch pulse CO$_2$ laser needs to be developed for this purpose.  The possibility
currently considered is to create 100~laser bunches with an inter-bunch spacing of 2.8~nsec. 
A 5.8~GeV electron beam 
must then have the same bunch time spacing to get collisions of all bunches.
The above design is similar to the design for a warm LC~\cite{OmoriNIM}.

To meet ILC requirements, about 30~lasers are used to obtain 3000~bunches.
Since the ILC operates at 5~Hz, one has 200~msec
between pulses. 100~msec are used for the generation of the 
3000 bunches and 100~msec are used for damping.
Fast kickers are necessary to handle such bunches when one injects (extracts)
into (from) a damping ring.

\item A second technique employs FELs of CO$_2$ wave length for the 
light source~\cite{hiramatsu2003}.  
By using an FEL, one can
produce 3000~bunches of light in one pulse. The electron beam
also has 3000~bunches to get collisions with all photon bunches.
The bunch spacing of both beams should be identical for that purpose.  
The bunch spacing can be chosen in a very wide range when one employs FELs as light
sources: for example one can choose a bunch time spacing of 2.8~nsec, 20~nsec,
or 300~nsec .

\item Lastly, a new idea requires accumulation of positrons in a ring.  Since the ILC
operates at 5~Hz, one has 100~msec to accumulate positrons and get the
required population. Then the number of positrons created in one
collision can be drastically reduced. Thus the collision repetition
rate should be very high (more than MHz) in order to get the required
population.  To achieve this, one proposal is to use pulse stacking by
inserting a cavity of laser light in an electron
ring~\cite{Moenig-private}.  Another proposal to realize 
high-repetition collisions is to use a high-repetition FEL based on ERL
(energy recovery linac) technology developed for light
sources~\cite{Pitthan-private}.
\end{itemize}
Work is still necessary to develop 
these schemes to the point where they can be considered as 
viable options for polarized positron production.

%%%%
%%%%%%%%%%%%%%%%%%%%%%%%%%%%%%%%%%%%%%%%%%%%%%%%%%%%%%%%
\subsubsection{Laser-Compton demonstration experiment}
\begin{figure}[htb]
\begin{picture}(10,6)
\setlength{\unitlength}{1cm}
\put(2,-2.5){\mbox{\includegraphics[height=.35\textheight]{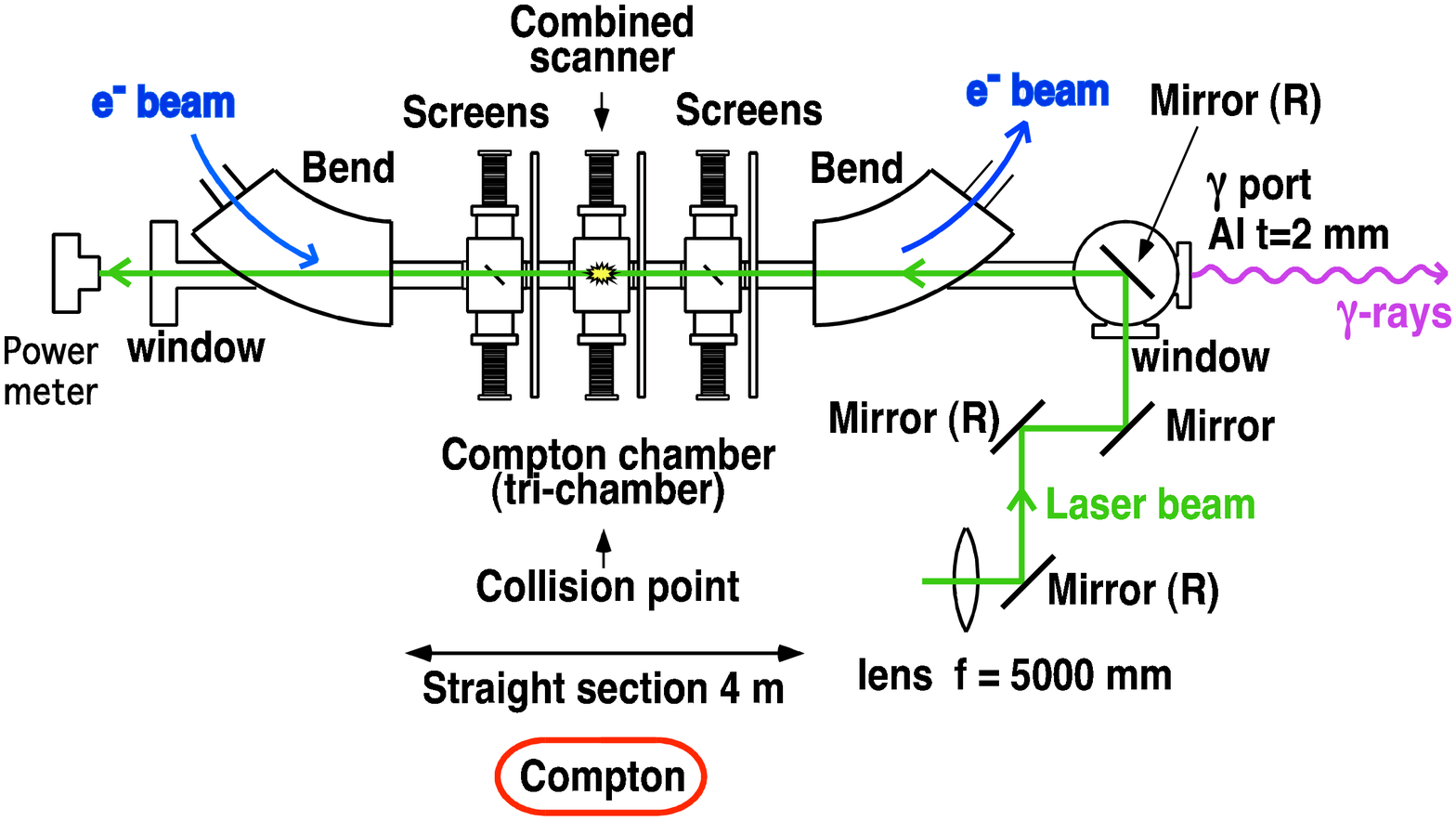}}}
\end{picture}\vspace{.2cm}
\caption[Compton chamber at the experiment at KEK-ATF]{Compton chamber: 
with screens at the collision point as well as  upstream and downstream of the collision point,
the positions and angles of the laser and electron beams are adjusted so that accurate head-on 
collisions can be realized. The spot sizes of the electron and laser beams are measured
with a wire and a knife-edge scanner.
\label{Fig.B}}
\end{figure}

\begin{figure}[htb]
\begin{picture}(10,7)
\setlength{\unitlength}{1cm}
\put(0,-.5){\mbox{\includegraphics[height=.35\textheight]{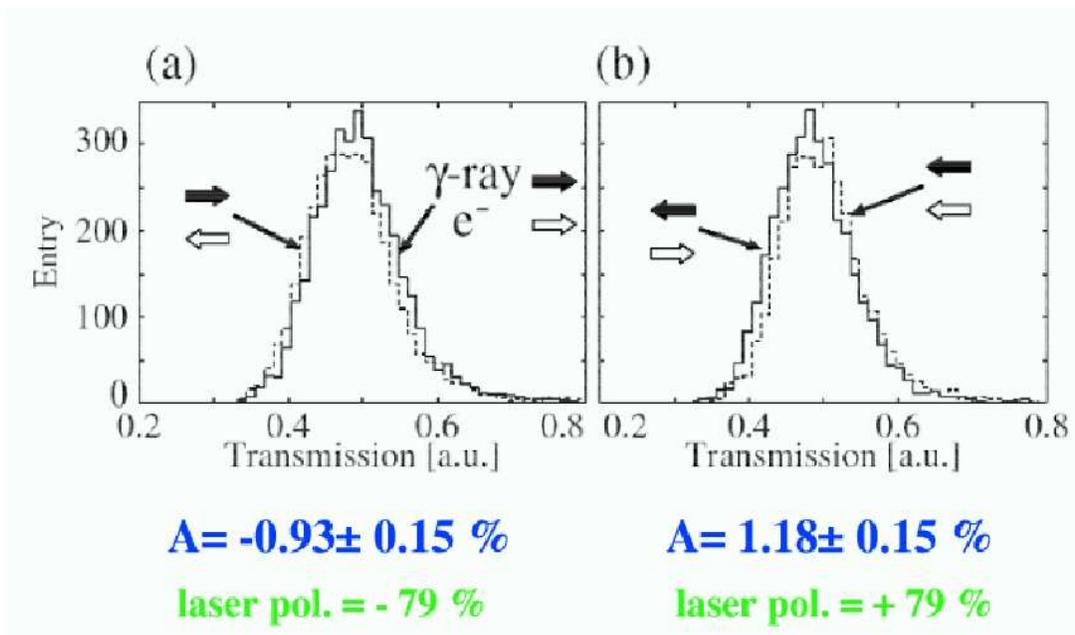}}}
\end{picture}\vspace{.2cm}
\caption[Transmission asymmetry]{Measured transmission asymmetry of the 
photons~\cite{FukudaPRL}.
This asymmetry is proportional to the polarization of the photons and corresponds to 
a photon polarization of about $90\%$. The photons are measured above a threshold of 21~MeV.
\label{Fig.C}}
\end{figure}

%%%%%%%%%%%%%%%%%%%%%%%%%%%%%%%%%%%%%%%%%%%%%%%%%%%%%%%%
An experiment is ongoing at the KEK-ATF to make a proof-of-principle 
demonstration of laser-based schemes, and to develop the polarimetry of short pulses
of photons and positrons.  In this experiment, a 1.28~GeV electron beam from the
ATF and the 2nd harmonic of a YAG laser are used to produce polarized photons
with a maximum energy of 56~MeV~\cite{Sakai:2002xp}. The collision point is located in the
extraction line of the ATF.  At the collision point a specially designed
chamber, the Compton chamber, is installed; fig.~\ref{Fig.B}.  
%%%
%%%
This chamber has
screens at the collision point, and also downstream and upstream of the
collision point.  By using these screens, the positions and angles of
the laser and electron beams are adjusted in order to realize accurate 
head-on collisions. The Compton
chamber also has a wire scanner and a knife-edge scanner to measure the spot
sizes of the electron and laser beams at the collision point. The polarization
of the produced photons is measured by a transmission method.  In this
method the intensities of transmitted photons for the parallel and the
anti-parallel cases are measured downstream of magnetized iron.
Here the parallel (anti-parallel) case means that the spins of the photons
and the electrons in the magnetized iron are parallel (anti-parallel)
to each other. Since no event identification is necessary in this
transmission method, it is suitable for measuring the polarization of a
high-intensity beam in which a very large number of photons in the beam
interact in a very short time (~30 psec). The iron pole in an
electromagnet is used as the magnetized iron. The polarity of the magnet is flipped to
measure the transmission asymmetry in the two cases. The transmission asymmetry
 is proportional to the polarization of the photons.
Fig~\ref{Fig.C} shows the measured transmission asymmetry 
of the photons~\cite{FukudaPRL}. 
This asymmetry corresponds to a photon polarization of about $90\%$. 
Photons are measured above a threshold of 21~MeV.\\[1em]
%%%

\begin{figure}[htb]
\begin{picture}(10,9)
\setlength{\unitlength}{1cm}
\put(0,0){\mbox{\includegraphics[height=.35\textheight]{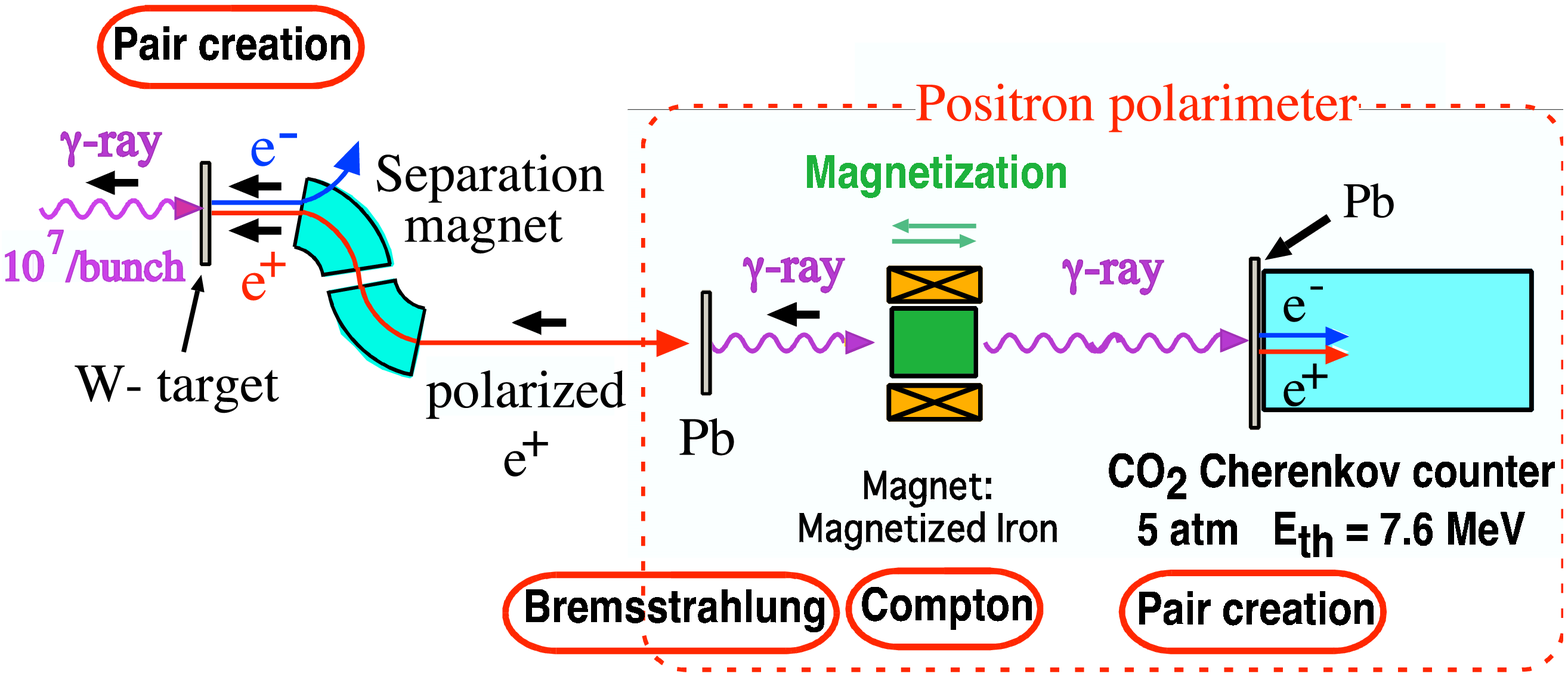}}}
\end{picture}\vspace{-2.2cm}
\caption[Apparatus for polarized positron production at KEK-ATF]{Production of 
polarized positrons
in a thin tungsten target. Dipole magnets separate the polarized electrons and positrons. 
The positron polarization is measured with the Compton-transmission method. 
\label{Fig.D}}
\end{figure}

Fig.~\ref{Fig.D} shows the apparatus for producing polarized positrons from
polarized photons and for measuring the polarization of these positrons. 
%%%
%%%
In this
design, $\sim 10^7$ gamma-photons are produced in a bunch.   
Polarized electrons and positrons are produced by pair creation in
a thin tungsten target.  A pair of 
dipole magnets separates positrons and electrons. The dipole
magnets select only the high-energy part of the positron spectrum (about 25 -- 45~MeV).
The number of positrons after momentum selection is $\sim 10^4$ per bunch.
One uses again a Compton-transmission method to measure the positron polarization, fig.~\ref{Fig.D}. 
The expected value of the transmission asymmetry is
about $0.7~\%$.  The experiment is in progress.

%%%%%%%%%%%%%%%%%%%%%%%%%%%%%%%%%%%%%%%%%%%%%%%
\section{Polarized electrons\label{pol-elec}}
%%%%%%%%%%%%%%%%%%%%%%%%%%%%%%%%%%%%%%%%%%%%%%%

\begin{figure}[htb]
\begin{picture}(10,11)
\setlength{\unitlength}{1cm}
\put(3,0){\mbox{\includegraphics[height=.5\textheight,width=0.48\textheight]{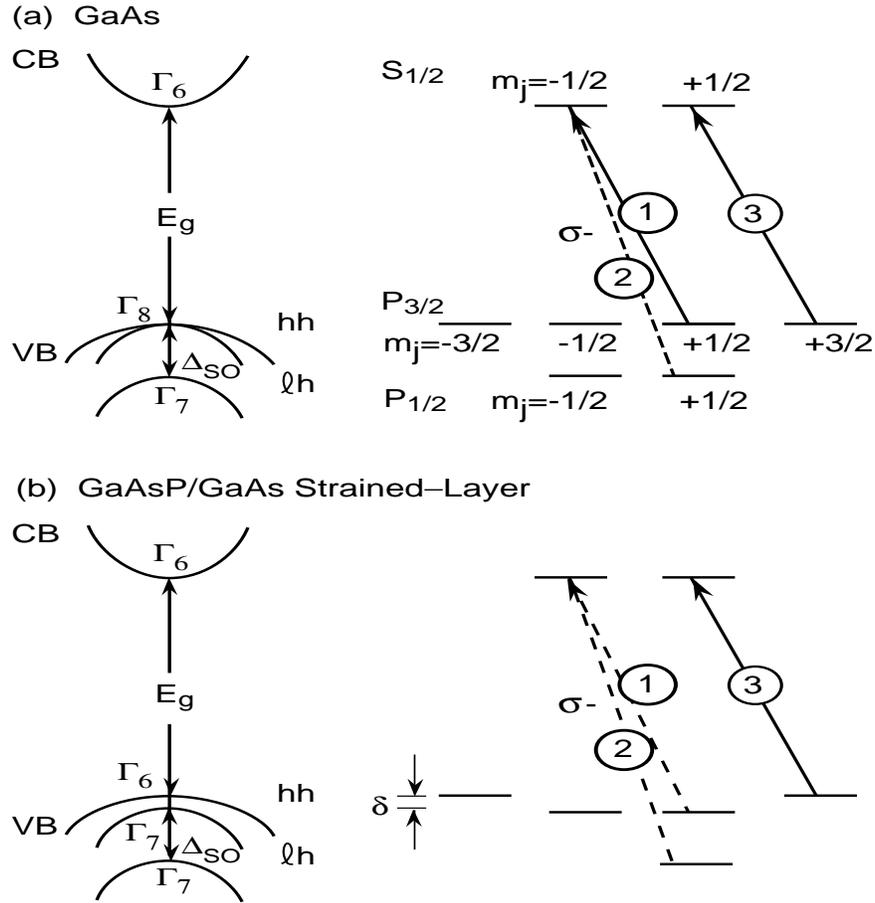}}}
\end{picture}\vspace{0cm}
\caption[Band gap diagram and transition probabilities
for photocathodes]{Band gap diagram 
and transition probabilities at the
$\Gamma$ point for (a) GaAs and (b) strained-layer GaAsP/GaAs.
$\Delta_{\rm SO}$ is the spin-orbit splitting, $\delta$ is the hh-lh
(heavy hole-light hole) splitting due to strain.  A non-zero $\delta$
can also be produced with a superlattice structure.  At room
temperature, ${\rm E}_g=1.424$~eV, $\Delta_{\rm SO}=0.340$~eV. On the right,
relative transition probabilities between s and p orbitals for
excitation by $\sigma^-$ light are shown. Transitions at the band gap
minimum are shown as solid lines.
\label{fig_elpol}}
\end{figure}

Highly-polarized electrons are obtained from GaAs by directing a
circularly-polarized laser beam on to a thin (typically 100 nm)
p-doped epilayer of strained GaAs~\cite{nakanishi-A}, of strained
InGaAs~\cite{nakanishi-B}, and of a short period 
superlattice~\cite{nakanishi-CD}. Only a small fraction of the photons are
absorbed in the epilayer. Electrons with zero momentum are promoted
from the highest level of the 
valence band (VB) to the lowest level of the conduction band (CB) upon
the absorption of laser photons, see fig.~\ref{fig_elpol}. 
A biaxial compressive strain produced
by a lattice mismatch with the substrate or by the quantum confinement
associated with short-period superlattice (SL) structures breaks the
degeneracy of the heavy-hole (hh) and light-hole (lh) bands at the
highest level of the VB~\cite{Prepost}. A hh-lh separation of
$\sim$80 MeV is readily achieved, which in carefully grown structures,
is sufficient to allow the selection of electrons from the hh band
only. With the laser wavelength tuned to the band gap, angular
momentum selection rules result in the production of CB electrons in
one spin state exclusively. A thin epilayer is chosen to minimize
strain relaxation. As the CB electrons diffuse to the surface, they
undergo some depolarization, primarily by interaction with holes. This
effect can be considerably reduced by decreasing the dopant density~\cite{nakanishi-E}
(everywhere except the last few nanometers near the surface, which
is often called {\em gradient doping method}~\cite{nakanishi-FG}). 
Near the surface, the energy levels for
p-doped GaAs bend downwards. Most of the electrons reaching the
surface are confined to this band-bending region (BBR) for a finite
time until they are emitted to vacuum or lose sufficient energy for them to be
trapped in surface states. Although the BBR is depleted of holes, the
confined but still mobile electrons in the BBR lose energy by
scattering from optical phonons, as a result of which the amplitude
and phase of the spin precession vector is continually reoriented,
leading to a significant depolarization. The probability for electrons
to escape to vacuum can be as high as 20\% if the surface is properly
activated with caesium and an oxide to create a negative electron
affinity.

%%%

The SLC demonstrated that electron beams with
polarization at the IP approaching 80\% can be generated reliably for
a period of years. The SLC beam was produced using a 100-nm thick
GaAsP/GaAs strained-layer photocathode. By decreasing the dopant
density in the bulk, the polarization of this type of cathode was
slightly improved for the initial run of a parity violation experiment
E158 at SLAC. Higher polarization is an ongoing R\&D
aim. Several laboratories have reported electron beams with
polarization of 90\% or even higher. 
The problem is that such high
values are universally achieved only with a cathode surface having a
relatively low quantum efficiency (QE), defined as the number of
photoemitted electrons per incident photon. Improvements in
polarization while maintaining a high QE have recently been achieved with
strained GaAsP/GaAs superlattice (SL) structures~\cite{nakanishi-H}. 
Each layer of the
superlattice (typically 4 nm) is considerably thinner than the
critical thickness ($\sim$10 nm) for the onset of strain relaxation,
while the transport efficiency for electrons in the conduction band
can still be high. In addition the effective band gap for such
superlattices is larger than for GaAs alone, which improves the QE. Today,
100-nm thick, gradient-doped GaAsP/GaAs superlattice photocathodes
routinely yield at least 85\% polarization at low energy with a
maximized QE of $\sim$1\%. This type of cathode was successfully used
during the summer 2003 dedicated run at SLAC of E158-III, for which
the polarization at high energy, measured by a M\o ller polarimeter,
was (89 $\pm$~4)\% except for short periods following refreshment of the
cathode QE (accomplished every few days by adding a small amount of Cs
to the surface). This experience points to the possibility of a
constant 90\% polarization if a technique to control the minimum
surface barrier can be developed. Recent data for 2 SL photocathodes
are compared with that for the best of the strain-layer cathodes in
table~\ref{table1}. In the table, $\lambda_0$ is the wavelength
corresponding to the maximum polarization, $P_{e,max}$. The absolute
accuracy of the polarimeters is on the order of $\pm 5\%$.

The ILC beam is a train of 2820 bunches spaced 337 ns apart at a
repetition rate of 5 Hz. The electron source should be able to
generate at least twice the charge required at the interaction point
(IP). Thus the required charge at the source will be $\sim$6.4 nC per
bunch, $\sim$20 $\mu$C per train.  A dc-biased photocathode gun as
used for the SLC is expected to be used. The bunching systems necessary
for the ILC electrons are very similar to those used at the SLC
notwithstanding the requirement of the ILC bunch format.

\begin{table}[htb]
\begin{center}
{\footnotesize
\begin{tabular}{|cccrrrl|}
\hline
&&&&&&\\
Cathode & Growth & Polarimeter & $P_{e,max}$ & $\lambda_0 $ & $QE_{max}(\lambda_0)$ & Ref. 
\\[1ex]
Structure & Method & {} &{} & (nm) & {} & {} \\[1ex]
\hline
&&&&&&\\
GaAsP/GaAs strained SL: &&&&&&\\ 
Nagoya &MOCVD &Mott &$\geq$ 0.90 &775 &0.004 &\cite{watanabe} \\[1ex]
SLAC &MBE & CTS Mott &0.86 &783 &0.012 &\cite{maruyama1} \\[1ex]
{} &{} &M\o ller E158-III &0.89 &780 &0.008 & \cite{mike-neu}\\[1ex]
GaAsP/GaAs strained-layer: &&&&&&\\ 
SLAC &MOCVD &CTS Mott &0.82 &805 &0.001 &\cite{maruyama2} \\[1ex]
{} &{} &M\o ller E158-I &0.85 &800 &0.004 &\cite{anthony} \\[1ex]
\hline
\end{tabular}
}
\caption{Comparison of 2 strained superlattice and 1 strained-layer photocathodes.
\label{table1} }
\end{center}
\end{table}

%%%%%%%%%%%%%%%%%%%%%%%%%%%%%%%%%
\section{Spin manipulation systems \label{spinrotation}}
%%%%%%%%%%%%%%%%%%%%%%%%%%%%%%%%%

\begin{figure}[htb]
\begin{picture}(10,9)
\setlength{\unitlength}{1cm}
\put(.5,0){\mbox{\includegraphics[height=.35\textheight,width=0.63\textheight]{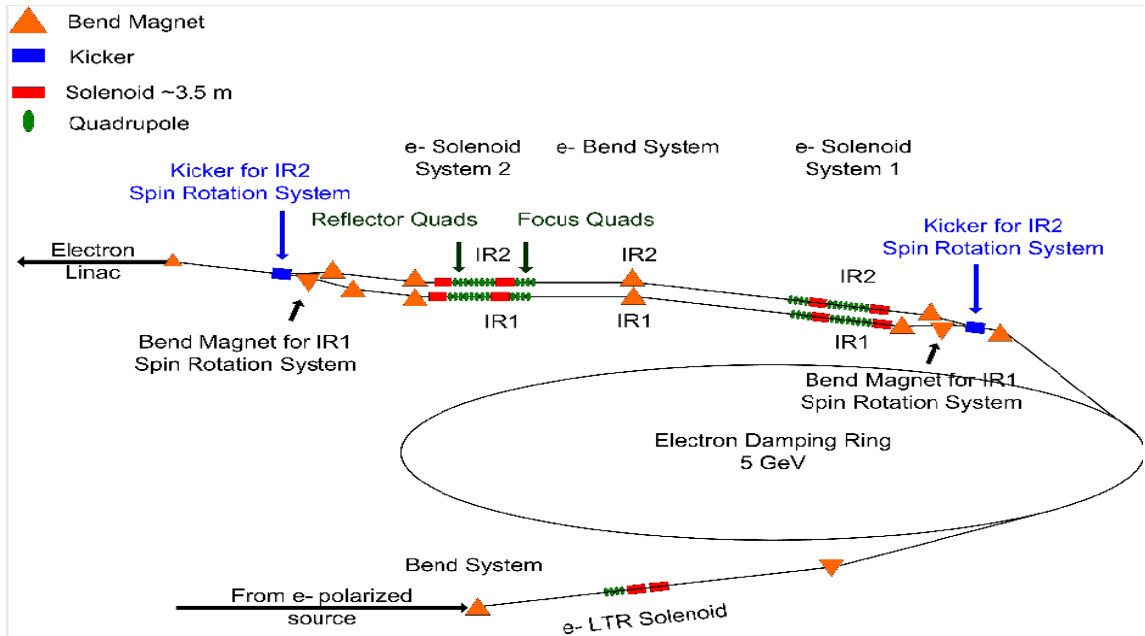}}}
\end{picture}\vspace{0cm}
\caption[Layout of the electron damping ring for serving two
IR's]{Layout of the electron damping ring system showing the parallel
spin rotation beam lines for IR1 and IR2. A pair of kicker magnets is
turned on between pulse trains to deflect the beam to the spin
rotation solenoids for IR2~\cite{Moffeit-twoir}.
\label{Fig_emdr}}
\end{figure}

\begin{figure}[htb]
\begin{picture}(10,9)
\setlength{\unitlength}{1cm}
\put(0,0){\mbox{\includegraphics[height=.35\textheight,width=0.67\textheight]{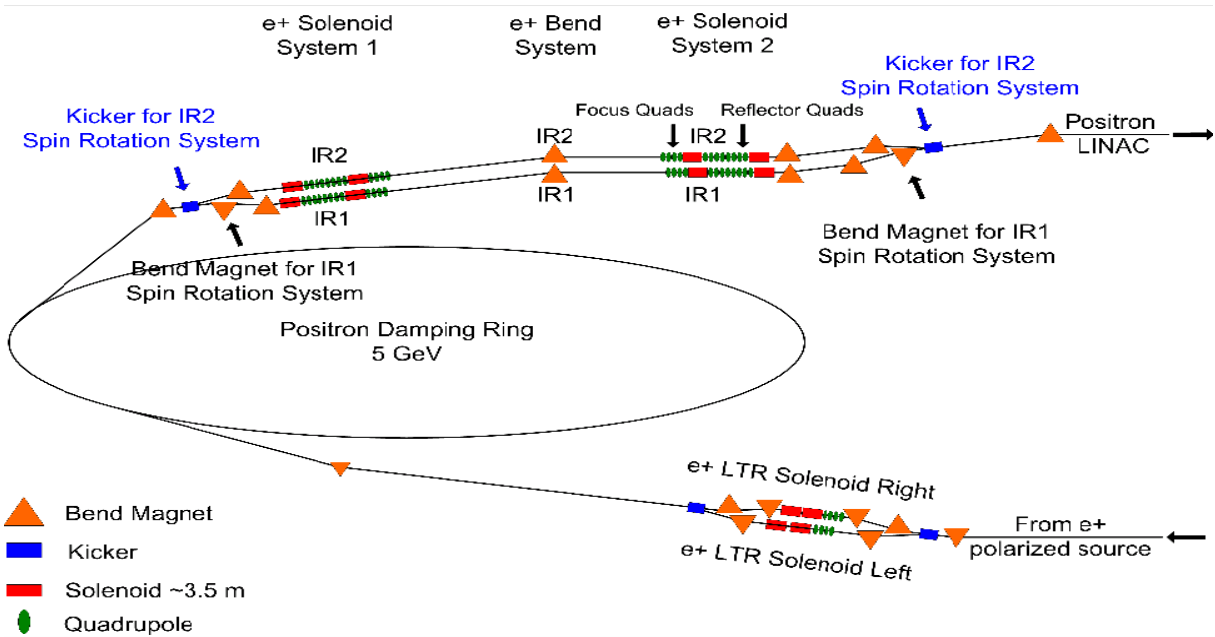}}}
\end{picture}\vspace{0cm}
\caption[Layout of the positron damping ring for serving two
IR's]{Layout of the positron damping ring system showing the parallel
spin rotation beam lines for randomly selecting the positron polarization
direction. A pair of kicker magnets is turned on between pulse trains
to deflect the beam to the spin rotation solenoids with negative
B field~\cite{Moffeit-twoir}. This is obviously not necessary with electrons whose
polarization is switched at the source.
\label{Fig_epdr}}
\end{figure}

A spin rotation scheme for the ILC that allows the polarization 
vector of the electron and positron beams to be tuned independently
for two interaction regions (IR) is described in~\cite{Moffeit-twoir}.
The correct polarization direction for a particular IR can be selected by
directing the beam into one of two parallel spin rotation beam lines
located between the damping ring and the linac. With fast
kicker magnets, it is possible to rapidly switch between these
parallel beam lines, so that polarized beams can be delivered to two
IRs on a pulse-train by pulse-train basis. 
Fig.~\ref{Fig_emdr} shows the layout of the electron damping ring system and fig.~\ref{Fig_epdr}
shows the layout for the positron damping ring system.
A similar scheme can be
employed in the low-energy transport (LTR) to the positron damping ring, to
allow rapid helicity switching for polarized positrons. In that 
case the axial solenoid fields
must have equal but opposite directions in the two lines. A pair
of kicker magnets is turned on between pulse trains to deflect the beam to the spin rotation
solenoids with negative B-field.

%%%
%%%
%%%
%%%

%%%
\begin{figure}[htb]
\begin{picture}(10,5)
\setlength{\unitlength}{1cm}
\put(1,0){\mbox{\includegraphics[height=.2\textheight,width=0.6\textheight]{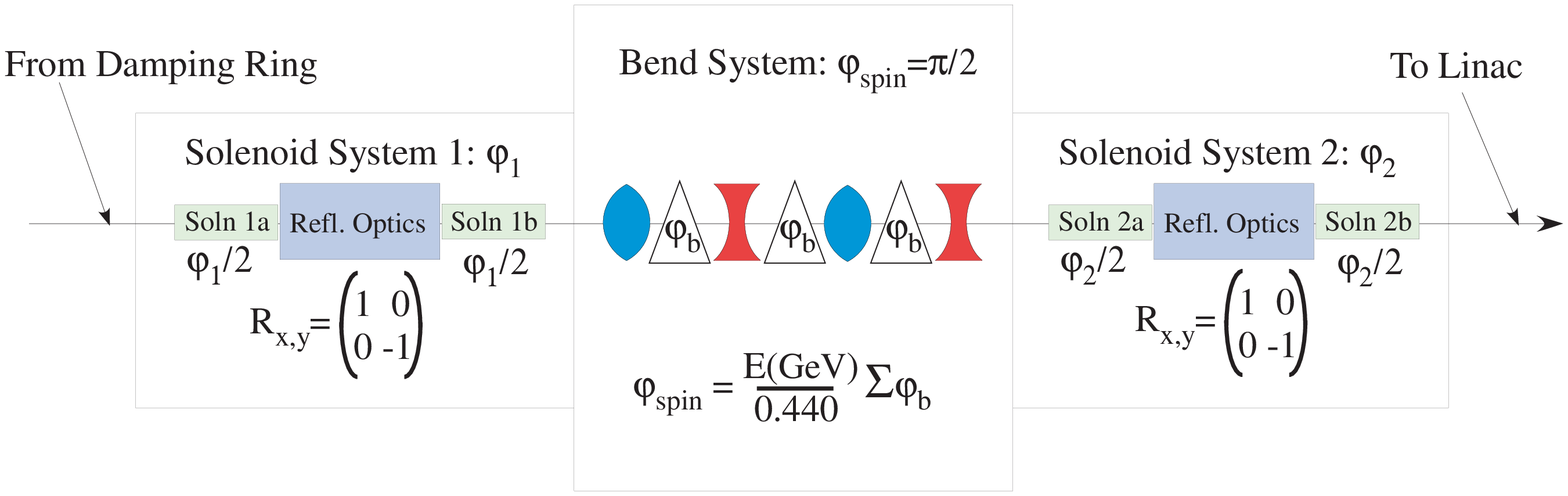}}
}
\end{picture}
\caption[Spin rotation system for transverse polarisation]{Spin rotation system to 
rotate a vertical spin coming from the damping ring to an arbitrary spin 
orientation. See text for explanation. \label{fig_trans-john} }
\end{figure}

%%%%%%%

The system
shown in fig.~\ref{fig_trans-john} forms the basic spin rotation building block in
each of the IR1/IR2 ring-to-linac transport lines.

If properly designed, such a spin rotation system after the damping ring
will allow the spin of the beams to be set to any arbitrary
orientation by the time they reach the IP. The system described in
\cite{refPE} accomplishes the desired spin manipulation while
minimizing emittance dilution of the flat
beams. Fig.~\ref{fig_trans-john} illustrates the basic features of the
system which consists of two solenoidal rotation systems with a bend
rotation system in between.  Starting with an initial vertical spin
vector coming out of the damping ring, the final spin vector at the
exit of the full system is $\vec{S}$~\cite{refPE},

\begin{equation}
\vec{S}=\stackrel{\leftrightarrow}{\Omega}_{\rm tot}\cdot 
\left( \begin{array}{r} 0 \\ \pm 1 \\ 0 \end{array}
\right)=\left( \begin{array}{c} \mp \sin\phi_1 \cos\phi_2 \\
\pm \cos\phi_1 \cos\phi_2\\ \pm \sin\phi_2 \end{array} \right),
\label{eq_trans-john}
\end{equation}
where $\phi_1$ and $\phi_2$ are the spin rotations due to the 
first and second solenoidal rotation systems. From 
eq.~(\ref{eq_trans-john}), it is seen that any arbitrary 
orientation 
of the spin vector is possible through a suitable choice of $\phi_1$ and $\phi_2$.

Each solenoid system is made up of an initial solenoid followed by an
optical transport section which gives a unity transformation in the
horizontal plane and a $-1$ transformation in the vertical plane, an
optical reflection. The optical section is followed by a second
solenoid. Each solenoid is set to provide a spin rotation of
half the total required for the particular solenoidal system,
$\phi_i$. The optical reflection transport is required to avoid
transverse betatron coupling of the beam by cancelling the rotation due
to the solenoids.

The bend section has a constant spin rotation contribution of $\pi/2$ in the 
horizontal-longitudinal plane; the spin rotates about the longitudinal axis in the solenoids. 
Upstream of the damping rings, emittance dilution is not of concern in
the longitudinal to vertical spin transformation. This system consists of
a simple bend followed by a solenoid of suitable strength.

%%%%%%%%%%%%%%%%%%%%%%%%%%%%%%%%%%%%%%%%%%%%%%%%%%%%
\section{Polarimetry at the ILC \label{pol-summary}}
%%%%%%%%%%%%%%%%%%%%%%%%%%%%%%%%%%%%%%%%%%%%%%%%%%%%%

The primary polarimeter measurement at the ILC will be performed by a
Compton polarimeter. An accuracy of $\left(\Delta
P_{e^-}/P_{e^-}\right)=0.25\%$ should be achievable \cite{Rowson:2000qn}. The
polarimeters can be located upstream or downstream of the Interaction
Point (IP) and it is desirable to implement polarimeters at both
locations.  The downstream polarimeter may require a large crossing
angle at the IP; polarimetry is one of many issues to be considered
in a decision on the IR crossing angles.
 
Compton polarimetry is chosen as the primary polarimetry technique
for several reasons:
\begin{itemize}
\item The physics of the scattering process is well
understood in QED, with radiative corrections less than 0.1\%
\cite{mike7};
\item Detector backgrounds are easy to measure and correct for
by using laser off pulses;
\item Polarimetry data can be taken parasitic to physics data;
\item The Compton scattering rate is high and small
statistical errors can be achieved in a short amount of time
(sub-1\% precision in one minute is feasible);
\item The laser helicity can be selected on a pulse-by-pulse
basis;
\item The laser polarization is readily determined with 0.1\%
accuracy.
\end{itemize}

%%%%%%%%%%%%%%%%%%%%%%%%%%%%%%%%%%%%%%%%%%%%%
\subsubsection{M\o ller and Bhabha polarimetry}

A M\o ller or Bhabha scattering polarimeter can be considered 
as a cross check of the principle high-energy Compton polarimetry,
or at various intermediate stages of the acceleration process. This
well-known method has been applied in numerous
experiments~\cite{ref4,ref5,ref6} with energies ranging from MeVs to
multi-GeVs. M\o ller type polarimeters have in particular been used for
many years in conjunction with fixed-target experiments at
SLAC~\cite{ref7,ref8,ref9,ref10,ref11,ref12} and elsewhere~\cite{ref13,ref14,ref15,ref16}.
                                                                                
The application of this technique at linear collider beams appears to be
generally inferior, at least in its conventional incarnations, in
comparison with laser-based Compton polarimetry. 

M\o ller polarimetry at a linear collider has currently still
some disadvantages with respect to Compton polarimetry:
\begin{enumerate}                                                                                
\item low target polarization: only about $8\%$ electron polarization is
possible with ferromagnetic materials (to be compared with $100\%$ laser
polarization in Compton polarimetry) and it is not easy to calibrate this
important quantity with great accuracy;
\item  limited precision: only one single M\o ller polarimeter with rather
exceptional properties has reported a precision
$\Delta P/P=0.5\%$~\cite{ref16,ref17}. The typical performance of more
conventional devices has been only at the level of
2-5\% or worse~\cite{ref7,ref8,ref9,ref10,ref11,ref12,ref13,ref14,ref15};
\item  invasive: insertion of the target is invasive to the beam, i.e.
such a polarimeter can only operate in the dumpline of the beam, or only
intermittently, if used upstream of the collider interaction point.
\end{enumerate}
             
Additional issues that have to be considered are beam heating of the
target and multiple event topologies~\cite{ref18}.
                                                                                
Since all of the major drawbacks of the conventional M\o ller technique
originate in the ferromagnetic foil target, one may wonder if not a
radically different target with nearly $100\%$ electron polarization is
possible, which furthermore should be thin enough to preserve the
pristine quality of the beam.
                                                                                
A polarized gas target with similar properties has in fact been
in operation since several years in the HERA storage ring at the HERMES
experiment~\cite{ref19}, and M\o ller and Bhabha polarimetry has routinely
been employed to diagnose the performance of the spin-polarized target
cell~\cite{ref20}. However, since the average $e^+/e^-$ beam currents of HERA
are 1000$\times$ larger than those at the linear collider, it is doubtful if
something like the HERMES gas target could be developed to be suitable for
polarimetry at the ILC.\\ 
                                                                                
One could summarize 
that dedicated M\o ller/Bhabha beam polarimetry based on conventional target
technology is generally inferior to laser-based Compton polarimetry, but
may be useful in limited precision applications at intermediate energies.           
With an advanced target design, 
an ultimate
precision $\Delta P/P=0.5\%$ may be achievable and useful as an occasional
crosscheck of the principle Compton results, but the general operational
drawbacks remain.
                                                                                
%%%%%%%%%%%%%%%%%%%%%%%%%%%%%%%%%%%%%%%%%%%%%%%%%%%%%%%%%%%%%%%
\subsection{Compton scattering basics}
%%%%%%%%%%%%%%%%%%%%%%%%%%%%%%%%%%%%%%%%%%%%%%%%%%%%%%%%%%%%%%%

One defines $E_0$ and $\omega_0$ to be the incident energies of the
electron and photon, and $E$ and $\omega$ to be the scattered energies
of the electron and photon. The dimensionless {\it x}, {\it y} and
{\it r} scattering parameters are defined by:
\begin{eqnarray}
x & = & \frac{4E_0 \omega_0}{m^2} \cos^2(\theta_0 /2) \simeq
\frac{4E_0 \omega_0}{m^2} \\ y & = & 1-\frac{E}{E_0} =
\frac{\omega}{E_0} \\ r & = & \frac{y}{x(1-y)}
\end{eqnarray}                                                            
where {\it m} is the mass of the electron and $\theta_0$ is the
crossing angle between the electron beam and the laser beam.  For
polarimeters with small crossing angles at the Compton IP,
$\cos^2(\theta_0 /2) \simeq 1$.

The spin-dependent differential Compton cross section is given by:
\begin{eqnarray}
\left(\frac{d\sigma}{dy}\right)_{Compton} & = &
\left(\frac{d\sigma}{dy}\right)_{unpol} \left[1+P \lambda A_z
(x,y)\right] \\ \left(\frac{d\sigma}{dy}\right)_{unpol} & = &
\frac{0.499 {\rm barn}}{x} \left[\frac{1}{1-y}+1-y-4r(1-r)\right] \\
A_z (x,y) & = & rx(1-2r)(2-y)
\end{eqnarray}
where $P$ is the longitudinal polarization of the electron and
$\lambda$ is the circular polarization of the laser photon.  The
Compton asymmetry analyzing power, $A_z (x,y)$, is maximal at the
kinematic endpoint, corresponding to 180$^\circ$ backscattering in the
center-of-mass frame, with
\begin{eqnarray} 
E_{\rm min} = E_0 \frac{1}{1+x}
\end{eqnarray}
For a 250 GeV electron beam colliding with a 532-nm laser, the
Compton-scattered electrons have their kinematic endpoint at $E_{\rm
min} = 25.1$ GeV with an analyzing power $A_z = 98 \%$. Figure 1 shows
the resulting $J_z$=3/2 and $J_z$ =1/2 Compton cross sections and
analyzing power.
\begin{figure}
\begin{center}
\epsfig{file=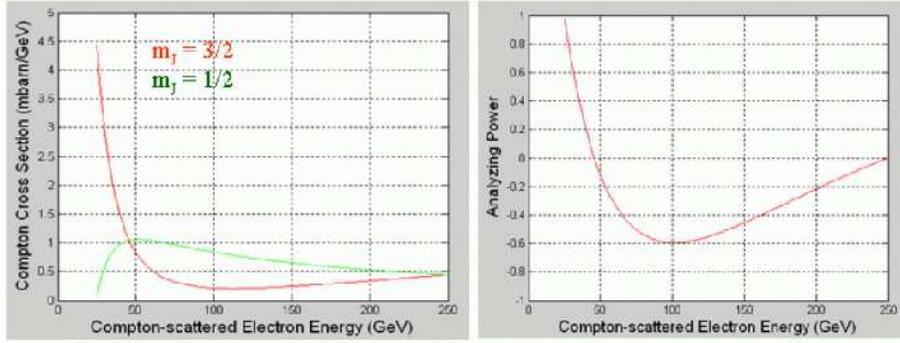,width=12cm}
\end{center}\vspace{-.8cm}
\caption[Compton cross section]{Compton cross section for scattering of 532 nm photons with a
250 GeV electron beam. The $J_z$=3/2 ($J_z$=1/2) cross section for
electron and photon spins aligned (anti-aligned) is shown in 
red (light green).}
\label{figure_Compton_kin}
\end{figure}

%%%%%%%%%%%%%%%%%%%%%%%%%%%%%%%%%%%%%%%%%%
\subsection{Upstream Compton polarimeter}
%%%%%%%%%%%%%%%%%%%%%%%%%%%%%%%%%%%%%%%%%%

The Compton Polarimeter upstream of the collider IP requires the
Compton IP be just upstream of the energy collimation region so that
off-energy Compton electrons are not lost in the Interaction Region.
An additional requirement is that the beam trajectory at the Compton
IP be parallel to the trajectory at the collider IP.  A polarimeter
satisfying these requirements was included in the TESLA design and we
provide a brief description here~\cite{polnote}.

\begin{figure}[hbt]
\vspace{5mm}
\centerline{\epsfig{file=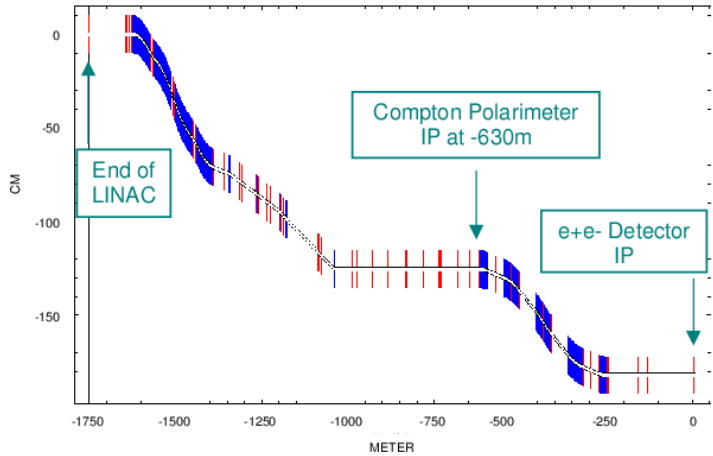, width=10cm, angle=0}}
\vspace{-.5cm}
\caption{Upstream Compton polarimeter in the TESLA design.
\label{pol-fig1}}
\end{figure}

The Compton IP is located 630 meters upstream of the collider IP, see Fig.~\ref{pol-fig1}.
The Compton laser is pulsed with a pattern that matches the pulse and
bunch structure of the ILC. In this way it is possible to achieve high
luminosity, typically six orders of magnitude higher than with
continuous lasers of comparable average power.

Compton electron detection in the multi-event (integrating) mode is
the principal detection method. Event rates and statistical errors
have been calculated for different beam energies and laser
parameters~\cite{polnote}.  Table~\ref{table:par250} gives typical
polarimeter parameters for $\sqrt{s}=500$~GeV. The performance is
similar for other energy regimes. For much higher or lower beam
energies, it will be advantageous to change the wavelength of the
laser or to adapt a four magnet chicane around the Compton IP comparable
to the one of the downstream polarimeter, 
in order to retain maximum coverage of the electron detector hodoscope.

\begin{table}[t]
\centerline{\begin{tabular}{|l c c c|}\hline
                      &      & {\bf Upstream}	& {\bf Downstream} \\
			&	{\bf $e^+/e^-~beam$} & {\bf laser~beam} & {\bf laser~beam}   \\
\hline
Energy                     & 250~GeV            & 2.3~eV 	& 2.3~eV  \\
Charge or energy/bunch     & $2 \cdot 10^{10}$  & $35~\mu J$ & 100 mJ \\
Bunches/sec                & 14100              & 14100    &  5       \\
Bunch length $\sigma_t$    & 1.3~ps             & 10~ps    &  1 ns    \\
Average current(power)     & $45~\mu A$         & 0.5~W    &  0.5~W   \\
$\sigma_x \cdot \sigma_y~(\mu m)$ & $10 \cdot 1$ upstream   & $50 \cdot 50$ & $100 \cdot 100$  \\ 
   &  $30 \cdot 60$ downstream	&	&	\\
\hline
			&		&	{\bf Upstream}	& {\bf Downstream} \\
			&		&	{\bf polarimeter} & {\bf polarimeter} \\
Beam crossing angle     &	&     10~mrad	 	& 11.5 mrad         \\
Luminosity & & $1.5 \cdot 10^{32}cm^{-2}s^{-1}$ & $5 \cdot 10^{30}cm^{-2}s^{-1}$ \\
Event rate at 25-GeV Endpoint 	& & 300,000/GeV/sec & 10,000/GeV/sec   \\
$\Delta P / P$ stat. error & & $<1\%$ / sec & $<1\%$ / min     \\
$\Delta P / P$ syst. error     & & $0.25\%$ & $0.25\%$   \\
\hline
\end{tabular}}
\caption[Compton polarimeter parameters]{\label{table:par250}Compton
polarimeter parameters at 250~GeV.}
\end{table}

%%%%%%%%%%%%%%%%%%%%%%%%%%%%%%%%%%%%%%%%%%%%%
\subsection{Downstream Compton polarimeter}
%%%%%%%%%%%%%%%%%%%%%%%%%%%%%%%%%%%%%%%%%%%%%

The design criteria for an extraction line polarimeter may prevent
a downstream polarimeter if there is no crossing angle at the IP.  The
20-mrad crossing angle design considered for NLC, however, permitted a
downstream polarimeter and we give a brief description for it here
with beam parameters appropriate for the ILC design.

\begin{figure}
\hspace{-.5cm}
\epsfig{file=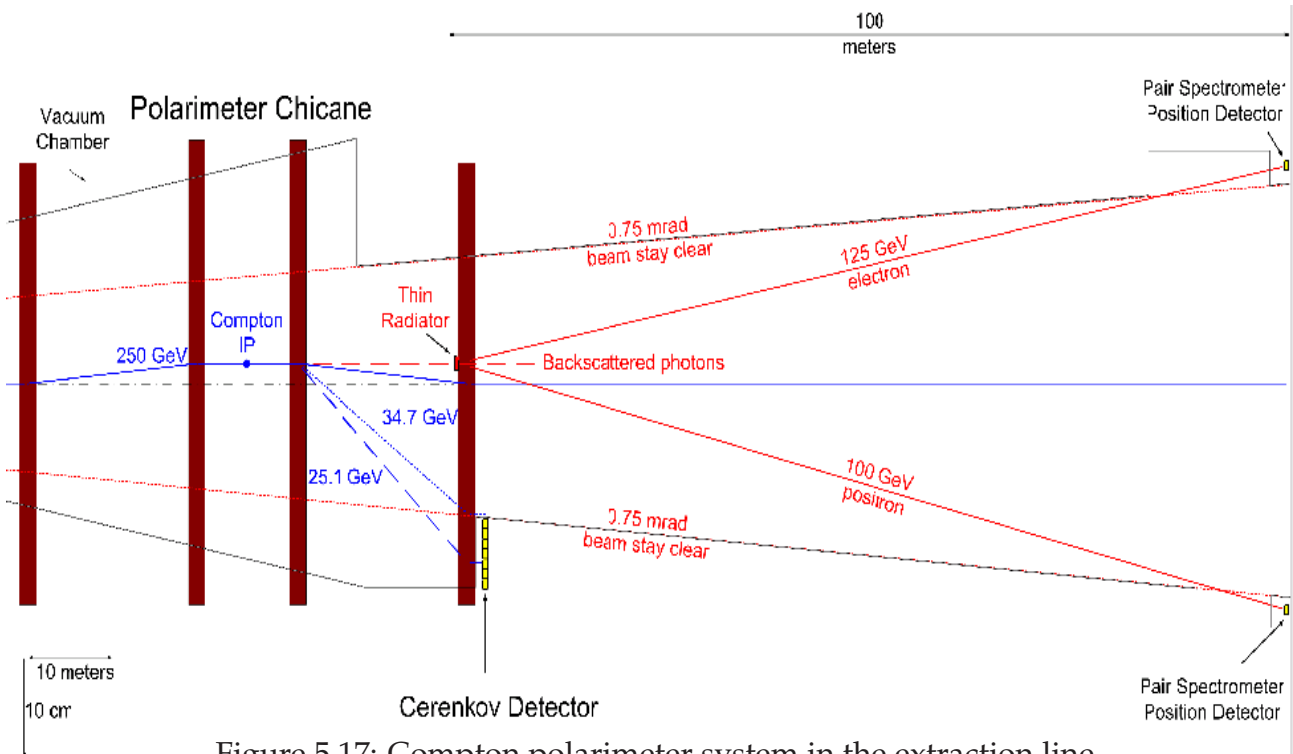,height=10cm,width=17.1cm}
\vspace{-1.3cm}
\caption{Compton polarimeter system in the extraction line.}
\label{fig:polarimeter}
\end{figure}

The Compton IP is located approximately 60 meters downstream from the
collider IP (see Figure~\ref{fig:polarimeter}).\cite{woods} The
Compton IP is at a secondary focus in the middle of a chicane with 20
mm dispersion, but with no net bend angle with respect to the $e^+e^-$
collision IP. Beam losses in the extraction line are acceptable, both
for machine protection and for ILC detector backgrounds.

The Compton laser has a low repetition rate of 5-10 Hz, but has high
power to give good signal-to-background ratio in the more difficult
downstream environment with disrupted primary electrons and
beamstrahlung photons present.  The Compton laser pulses collide with
a small fraction of the ~14 kHz electron bunch rate, but the timing of
the Compton laser pulses can be varied so as to sample systematically 
all electron bunches in the train and determine any variation of the
beam polarization along the train.

\begin{figure}
\begin{center}
\epsfig{file=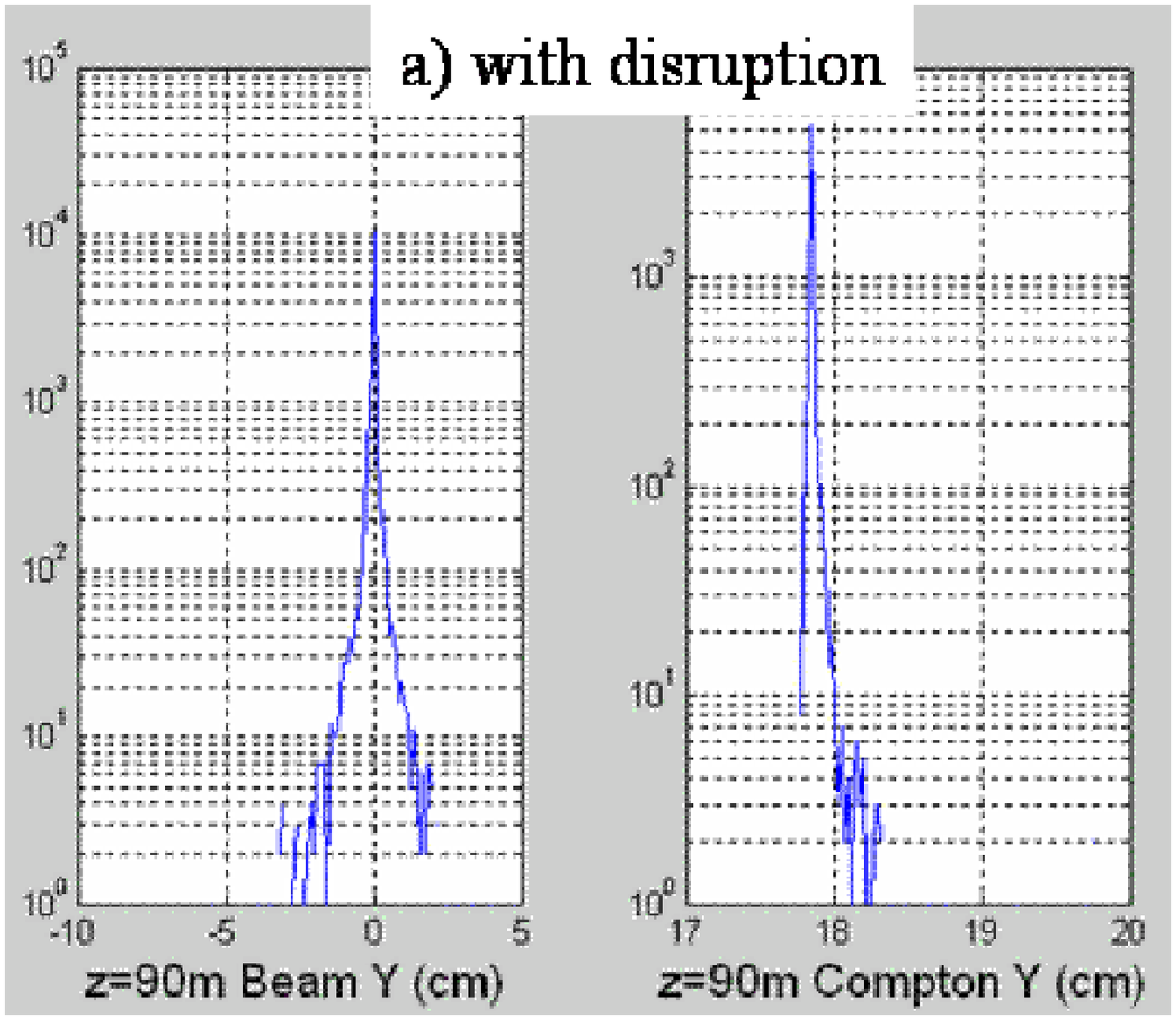,width=5.5cm}
\hspace{1cm}
\epsfig{file=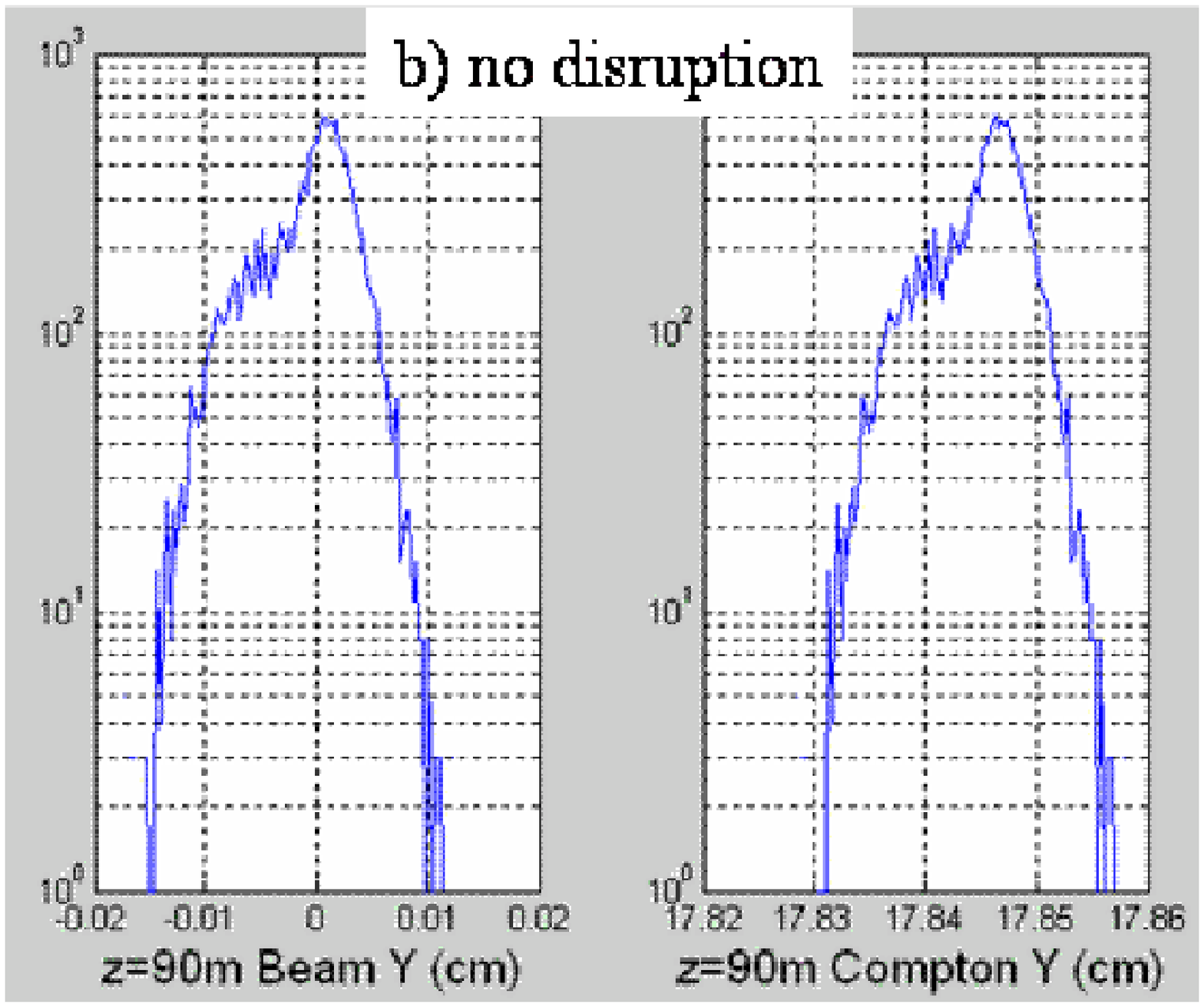,width=5.7cm}
\end{center}\vspace{-.6cm}
\caption[Vertical distributions and Compton-edge]{Vertical 
distributions of disrupted beam and Compton-edge
electrons at the Compton detector plane a) with $e^+e^-$ collisions
and b) without collisions.}
\label{fig:z90-ypos}
\end{figure}

The primary Compton detector is envisioned to be a segmented electron
detector operating in the multi-electron (integrating) mode, sampling
the electron flux at energies near the Compton kinematic edge at 25.1
GeV (for 250 GeV primary electron energy and 532 nm laser photons).
The Compton electron detector must discern between the Compton edge
electrons and low energy disrupted primary electrons, and it must be
located outside a 0.75-mrad cone from the IP that contains the
intense flux of beamstrahlung photons.  Figure~\ref{fig:z90-ypos}
shows the y distribution of 25.1-GeV Compton-scattered electrons at
the detector located downstream of the polarimeter chicane magnets. 
The Compton-edge electrons peak at $\approx18$ cm, well separated
from the tails of the disrupted electron beam.  For polarimeter
operation at electron beam energies other than 250 GeV, the chicane
dipoles should retain the same B field.  This changes the dispersion
at the Compton IP, but since the Compton edge endpoint energy does not
change quickly with incident electron beam energy, the location of
Compton-edge electrons at the Compton detector plane will only shift
slightly.  The parameters for the downstream Compton polarimeter are
summarized in Table~\ref{table:par250}. In that context one should note that
due to the strong fields 
significantly higher depolarization effects are expected at the possible 
multi-TeV LC design\cite{clic-depol}, so that a downstream polarimeter might 
be indispensable in that case.

%%%%%%%%%%%%%%%%%%%%%%%%%%%%%%%%%%%%%%%%%%%%%%%%%%%%%%%%%%%%%%%
\subsection{Luminosity-weighted polarization}
%%%%%%%%%%%%%%%%%%%%%%%%%%%%%%%%%%%%%%%%%%%%%%%%%%%%%%%%%%%%%%%
The luminosity-weighted beam polarization will differ from the
measured polarization due to disruption and radiation in the beam-beam
collision process. There are also effects from  polarization spread and
spin transport.

The spin motion of a deflected electron or positron beam in a transverse
magnetic field follows from the familiar T-BMT expression
\begin{equation}
\theta^{spin} \;=\; \gamma\: \frac{g-2}{2}\: \theta^{orbit}
\;=\; \frac{E_0}{0.44065~GeV}\: \theta^{orbit}
\label{eq:spinmotion} \\
\end{equation}
\noindent
where $\theta^{orbit}$ and $\theta^{spin}$ are the orbit deflection and spin
precession angles, $E_0$ is the beam energy, $\gamma=E_0/m$, and
$(g-2)/2$ is the famous g-factor anomaly of the magnetic moment of
the electron.  

The difference between the luminosity-weighted beam
polarization and the polarimeter measurement is written as $dP=P_z^{lum-wt} -
P_z^{CIP}$.  
To minimize $dP$, it is required that the beam direction at
the IP of the polarimeter (Compton IP) be aligned with the collision axis at the $e^+e^-$ IP
to within $50\mu rad$.  We consider estimates of several contributions
to $dP$, assuming full longitudinal spin alignment at the polarimeter
IP.  The use of both an upstream and a downstream polarimeter will assist
in achieving the desired small $dP$ and estimating the associated systematic
error~\cite{moffeit2}.

The contributions to $dP$ for the downstream polarimeter
are considered first, using estimates in the NLC design study~\cite{moffeit2}. Orbit
misalignments between the polarimeter IP and the collision IP are
expected to be below 50 $\mu$rad, which would give $dP=-0.04\%$.
Imperfect compensation for steering effects due to the angle between
the beam trajectory and the detector solenoid is expected to give an
additional trajectory alignment uncertainty of 30 $\mu$rad, and gives
a contribution $dP = -0.01\%$.  The effect of Sokolov-Ternov spin
flips is expected to contribute $dP=+0.3\%$ ($dP=-0.1\%$) for
downstream polarimeter measurements with (without) collisions.  The
angular divergence of the incoming beam is expected to contribute
$dP=-0.03\%$, while the angular divergence of the outgoing beam is
expected to contribute $dP=+0.1\%$ ($dP=-0.25\%$) for downstream
polarimeter measurements with (without) collisions.  Lastly, effects
from chromatic aberrations, which were important at SLC, are expected to
be negligible.  Adding all these contributions together, we expect a
total difference $dP=+0.32\%$ ($dP=-0.43\%$) for downstream
polarimeter measurements with (without) collisions. 

A similar compilation of these effects for an upstream polarimeter
leads to $dP=-0.42\%$~\cite{moffeit2}. The systematic errors associated
with $dP$ should be substantially smaller than the size of $dP$, and
having independent information from 3 polarimeter measurements
(upstream, downstream with and without collisions) will be important
in minimizing this.

%%%%%%%%%%%%%%%%%%%%%%%%%%%%%%%%%%%%
\subsection{Transverse Polarimetry \label{subsec:trans}}
%%%%%%%%%%%%%%%%%%%%%%%%%%%%%%%%%%%%
The spin rotation system with two spin rotators following the Damping
 Ring can achieve arbitrary spin orientation in the linac or at the IP
 as described in section~\ref{spinrotation}.  
Eq.~\ref{eq_trans-john} describes the spin vector
 at the beginning of the main linac.  A unitary spin transport matrix,
 $R$, then describes spin transport from that location to the IP or
 polarimeters.  To determine $R$ using the longitudinal Compton
 polarimeter one performs a 3-state measurement 
 by choosing appropriate spin rotator settings to align the electron
 spin along the $x$, $y$, or $z$ axis at the start of the
 main linac.  The $z$-component of the spin transport matrix can
 be measured with the Compton polarimeter, which measures the
 longitudinal electron polarization,
\begin{equation}
P_z^C = R_{zx} \cdot P_x^L + R_{zy} \cdot P_y^L + R_{zz} \cdot P_z^L.
\label{eq_transverse}
\end{equation}  
The three $P_z^C$ measurements for each $x$, $y$, or $z$ 
spin orientations at the start of the main linac determines the
spin rotation matrix elements $R_{zx}$, $R_{zy}$, and $R_{zz}$.  This is
sufficient to determine the full rotation matrix, which is described
by three Euler angles.  The matrix $R$ can be inverted to determine the
required spin rotator settings for the desired spin orientation at the
IP.
                                                                                
When running transverse polarization at the IP, the longitudinal
Compton polarimeter can monitor that the longitudinal beam
polarization stays close to zero. Calibrations of the beam
polarization can be done periodically (perhaps every 1-2 days,
depending on stability and required precision) by changing the spin
rotator settings to achieve longitudinal polarization.  The ILC beam
polarization is likely to be fairly stable and well understood and
this method should achieve a precision of about 1$\%$.  Another
possibility is to use the transverse asymmetries in $\mu^+\mu^-$-pair
production.

%%%%%%%%%%%%%%%%%%%%%%%%%%%%%%%%%%%%%%%%%%%%%%%%%%%%%%%%%%%%%%%
\subsection{Future design work}
%%%%%%%%%%%%%%%%%%%%%%%%%%%%%%%%%%%%%%%%%%%%%%%%%%%%%%%%%%%%%%%

The beam delivery systems for each of 2 IRs at the ILC will be updated
in the coming year.  The same is true for the extraction lines from
the collider IPs to the beam dumps.  Design choices for the crossing
angles of the IRs and linacs will impact polarimeter designs, as will
other design choices such as the optics and locations for the upstream
energy collimation.  Polarimeter design studies need to be an integral
part of the beam optics design studies.

One wants to find optics solutions where the beam trajectories are
parallel at the upstream and downstream polarimeter Compton IPs and
the collider IP.  For the upstream polarimeter we need adequate energy
collimation downstream of the Compton IP to mitigate against
backgrounds from off-energy Compton-scattered electrons.  For the
downstream polarimeter, we need to evaluate its feasibility and
expected performance for different IR crossing angle geometries.
Extraction line optics for a downstream polarimeter should also
accommodate an energy spectrometer.

\begin{figure}
\begin{center}
\epsfig{file=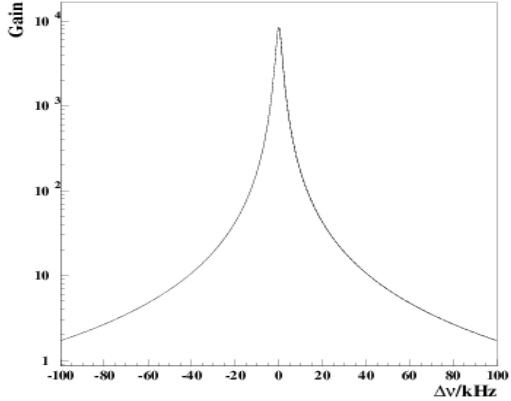,width=7cm,height=5.5cm}
\end{center}
\vspace{-.8cm}
\caption[Fabry-Perot cavity]{Gain of a 2 m long cavity as a function of the difference
between the laser and cavity resonance frequencies. the values of the
cavity mirror reflection and transmission coefficients are those of the HERA
cavity\cite{Zomer}.}
\label{figure_zomer}
\end{figure}

The designs of the laser systems of the
Compton polarimeters will be re-evaluated for optimizing
the pulse energy, pulse length and pulse repetition rate, while
achieving a simple and robust laser system.  This will include
revisiting design criteria for the measurement time needed to achieve
$0.1\%$ statistical precision and the time scale for monitoring
dependence of polarization along the 1-ms train.  
It is also planned to
evaluate a recent proposal to implement a Fabry-Perot cavity for the
laser beam around the Compton IP \cite{Zomer}. A commercial laser with
$\approx 10$ nJ/pulse and 3-5 MHz pulse frequency could be used
together with a high finesse cavity to reach in excess of $\approx 1$
J/pulse inside the cavity.  In order to amplify laser beam power with
a Fabry-Perot cavity one must keep such a cavity at resonance,
i.e. match the cavity length and laser beam frequency very
precisely. The cavity gain factor is shown in
Figure~\ref{figure_zomer}.  A fast feedback system would be required
to ensure these conditions.  Such a device is used at the CEBAF
accelerator \cite{Jefferson} and at HERA.

%%%%%%%%%%%%%%%%%%%%%%%%%%%%%%%%%%%%%%%%%%%%%%%%%%%%%%%%%%
\section{Polarization measurements with collider data \label{sect:colldata}}
%%%%%%%%%%%%%%%%%%%%%%%%%%%%%%%%%%%%%%%%%%%%%%%%%%%%%%%%%%

Polarization measurements with polarimeters are limited to a total precision
around 0.25\%. In addition polarimeters measure either the polarization of the
incoming beam that has not been depolarized by the beam-beam interaction or
the one of the outgoing beam which has been depolarized three to four time as
much as the interacting particles.
On the other hand there are several processes at a linear collider whose
polarization structure is known and which might be used to measure the
polarization directly from data. The large luminosity of the linear collider
offers the possibility to reach a precision much better then the polarimeters.

One example is the $\sin^2 \theta_{{\rm eff}}$ measurement with
the Blondel scheme at GigaZ (see sec. \ref{sec:gigaz})
where the relevant observables can be extracted directly from the data without
the use of polarimeters.
One has, however, to take into account that all methods using annihilation
data involve some physics assumptions that have to be considered in the
framework of the model in which the data are analysed.
The data driven methods also cannot replace completely the polarimeters.
The data methods need a large luminosity to get to a precise result while
polarimeters are completely systematics limited and statistics is no
problem. In any case polarimeters are thus needed for a fast machine
tuning. In addition there are some assumptions in the data-methods that have
to be verified or corrected with the polarimeters.
In all cases the data methods need the assumption that the absolute values of
the polarizations of the left and right handed states are the same.
If electron and positron polarization is available the effective
polarizations, explained in sec. \ref{subsec:2} and the formulae to obtain
the polarization involve linear and quadratic terms of the polarizations. For
these reasons any correlations between the two beam polarizations need to be
known from beam-beam simulations and polarimeters.

%%%%%%%%%%%%%%%%%%%%%%%%%%%%%%%%%%%%%%%%%%%%%%%%%%%%%%%%%%
\subsection{Measurements with electron polarization only}
%%%%%%%%%%%%%%%%%%%%%%%%%%%%%%%%%%%%%%%%%%%%%%%%%%%%%%%%%%

\begin{figure}[htb]
\begin{center}
\includegraphics[width=0.5\linewidth,bb=22 10 467 515]{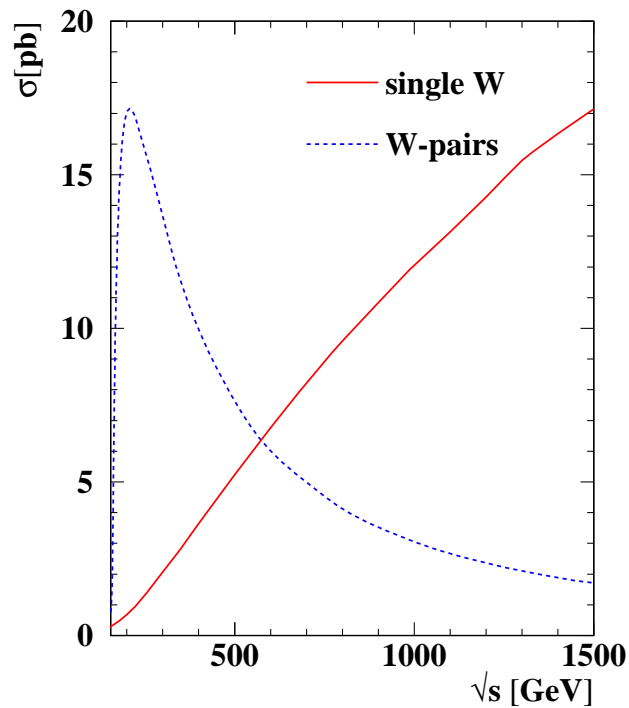}
\end{center}\vspace{-.7cm}
\caption[$W$ and $W$-pair production]{Cross section for single $W$ and $W$-pair production at a LC}
\label{fig:wcross}
\end{figure}

If only electron polarization is available not only the Lorenz structure of
the used process is needed but the exact helicity structure needs to be known.
The only process fulfilling this requirement is the V-A structure of the
$W$-fermion couplings. This coupling can be utilised in two processes at a
linear collider, single $W$ production and $W$-pair production.
As can be seen from figure \ref{fig:wcross} both processes have a cross
section of several pb so that a few million events are expected.

\begin{figure}[htb]
\begin{center}
\includegraphics[width=0.5\linewidth,bb=16 10 547 384]{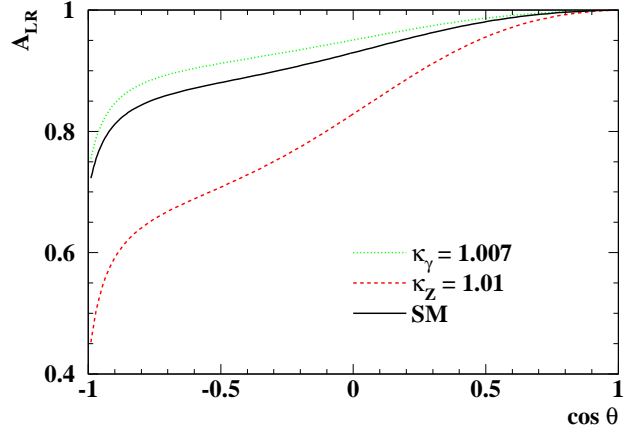}
\end{center}\vspace{-.7cm}
\caption[Left-right asymmetry of $WW$ production]{Left-right 
asymmetry of $W$-pair production as a function of
  the polar angle in the Standard Model and with anomalous triple-gauge couplings.}
\label{fig:wwalr}
\end{figure}

$W$-pair production proceeds through the Feynman diagrams shown in 
fig.~\ref{fig:wwfeyn}. In general the process is a complicated mixture of the
neutrino t-channel exchange, only determined by the $W$-fermion couplings, and
the $\gamma$ and Z s-channel exchange that also involve (anomalous) gauge
couplings. However, as shown in fig.~\ref{fig:wwalr}, the forward pole is
completely determined by the neutrino exchange and insensitive to the anomalous
couplings. For this reason it is possible to extract the polarization and the
triple gauge couplings \cite{ref:gounaris}
simultaneously from the $W$-pair data sample. The
expected precision is $\Delta P_{e^-}/P_{e^-} = 0.1\%$ for a luminosity of 
$500 \, \fbi$ at $\sqrt{s} = 340 \GeV$. The correlation with the anomalous
gauge couplings is negligible and the only assumption involved is that no
right-handed $W$-fermion couplings appear.
Experimental details of the analysis can be found in \cite{lcnote}.

If electron and positron polarization are available, both can be measured
simultaneously from the $W$-pair sample.
With equal luminosity at all four helicity
combinations and $P_{e^-}=80\%,\,P_{e^+}=60\%$
one gets $\Delta P_{e^-}/ P_{e^-} \approx 0.1\%$ and $\Delta P_{e^+} / P_{e^+}
\approx 0.2\%$ and negligible correlations between the polarizations
and between the polarization and the couplings.
If only 10\% of the luminosity is spent on the equal helicities the
polarization errors increase by roughly a factor two with $-50\%$ correlation.

Single $W$ production is dominated by the Feynman diagram shown in fig.~\ref{fig:swfeyn}. 
Since this process involves the V-A coupling of the $W$ to
fermions a $W^-$ can only be produced from a left-handed electron and a $W^+$
from a right-handed positron. Measuring the $W$ charge, the polarization can thus
be measured for electrons and positrons separately. 
The outgoing electron usually disappears in the beampipe so that the $W$ charge
has to be reconstructed from the $W$ decay products.
This means that only leptonic $W$-decays can be used for the analysis for which 
 no detailed simulation study exists yet. 
The experimental
signature is a single lepton in the detector which can be measured with high
efficiency and small background.
Because of the usually small $W$ energy also the interference with $W$-pair 
production should be very small.
Assuming  $\sqrt{s} = 500 \GeV$, ${\cal L} = 1 \abi$
and 100\% efficiency for $W^- \rightarrow e^-,\mu^-$ an error of 
$\Delta P_{e^-}/ P_{e^-} \sim 0.15\%$ is expected.

%%%%%%%%%%%%%%%%%%%%%%%%%%%%%%%%%%%%%%%%%%%%%%%%%%%%%%%%%%
\subsection{The Blondel scheme \label{sect:blondel}}
If a process $e^+ e^- \rightarrow \bar{f}f$ is mediated by pure s-channel
vector-particle exchange the cross section for the different
polarization states with electron and positron polarization available
can be written as
\begin{equation}
\sigma \, = \, \sigma_u \left[ 1 - P_{e^+} P_{e^-} + \ALR (P_{e^+} - P_{e^-}) \right],
\label{eq:sigmaff}
\end{equation}
where $P_{e^+}$ and $P_{e^-}$ are the longitudinal polarizations of the positrons 
and electrons measured in the direction of the particle's velocity, $\sigma_u$
denotes the unpolarized cross section and $\ALR$ the left-right asymmetry.
If the signs of the two polarizations can be switched independently four cross
sections can be measured to determine the four unknowns. From these cross sections
the polarizations can be obtained, if $\ALR \ne 0$:
\[
{P}_{e^\pm} \, = \, \sqrt{\frac{
( \sigma_{+-}+\sigma_{-+}-\sigma_{++}-\sigma_{--})
(\mp \sigma_{-+} \pm \sigma_{+-}-\sigma_{++}+\sigma_{--})}{
( \sigma_{-+}+\sigma_{+-}+\sigma_{++}+\sigma_{--})
(\mp \sigma_{-+} \pm \sigma_{+-}+\sigma_{++}-\sigma_{--})}}
\]
where in $\sigma_{ij}$ $i$ denotes the sign of the positron and $j$
the sign of the electron polarization. 
A drawback of this method is that some luminosity needs to be spent with the same 
helicities
for both beams which is not very interesting for most physics processes.

To measure the polarization with this scheme two processes have
been considered, 
\begin{itemize}
\item $\ee \rightarrow \bar{f}f$ with $\sqrt{s'} \approx \sqrt{s}$;
\item radiative return events 
$(\ee \rightarrow \rm{Z}\gamma \rightarrow \bar{f}f \gamma)$.
\end{itemize}
The cross section and left right asymmetry for the two processes at 
$\sqrt{s}=350$ and $500\GeV$ are given in table \ref{tab:cross}.
Both cross sections scale approximately with $1/s$.
\begin{table}[htbp]
\begin{center}
\begin{tabular}[c]{|c|c|c|c|c|}
\hline
$\sqrt{s}$ & $\sigma_{RR}$ & $\ALR(RR)$ & $\sigma_{HE}$ & $\ALR(HE)$ \\
\hline
$340\GeV$ & $17 \,\pb$ & 0.19 & $5 \,\pb$ & 0.50 \\
$500\GeV$ & $7 \,\pb$ & 0.19 & $2 \,\pb$ & 0.50 \\
\hline
\end{tabular}
\end{center}\vspace{-.5cm}
\caption[Left-right asymmetry and radiative return]{Cross section and asymmetry 
for high-energy and radiative-return $\bar{f}f$ events.}
\label{tab:cross}
\end{table}
The high-energy events can be measured with high efficiency and almost
no background. However the analysis relies on the assumption of
s-channel vector-exchange, so for analyses like the search for
R-parity-violating
sneutrinos the results cannot be used.

On the contrary radiative return events contain on-shell Z-decays
which are well understood from LEP'1 and SLD. In about 90\% of the
events the high-energy photon is lost in the beampipe. These events
can be reconstructed kinematically and most backgrounds can be
rejected. However, at ILC energies the cross section for the fusion
process $\ee \rightarrow \rm{Z}\ee$ is of the same order as the
signal. In those events one electron has almost the beam energy and
stays at low angle while the other is extremely soft and also often
lost in the beampipe resulting in a $\sim 30 \%$ background of Zee
events in the radiative return sample.
The only way Zee events can be rejected is to require a photon
above $7^\circ$ where photons and electrons can be separated by the
tracking detectors. Applying some additional event selection cuts on
the hadronic mass and the balance of the event, about 9\% of the
radiative return events are accepted with only a small Zee background.
However in these events the slow electron is seen in the detector, so
that they can easily be rejected by vetoing on an isolated electron.

Assuming $|P_{e^-}|=80\%, \, |P_{e^+}|=60\%$, 
an integrated luminosity of $500 \, \fbi$ at $\sqrt{s}=340 \GeV$
and 50\% or 10\% of the luminosity spent with both beam polarizations
with the same sign table \ref{tab:polresbl} shows the obtainable errors
on the two polarizations and their correlation. Due to the scaling of
the cross sections the errors are about a factor $\sqrt{2}$ larger at 
$500 \GeV$.
It should be noted that the relative errors scale approximately with
the product of the polarizations.
\begin{table}[htbp]
  \begin{center}
    \begin{tabular}[c]{|c|c|c|c|c|c|c|}
      \hline & \mth{${\cal L}_{\pm\pm} /{\cal L} = 0.5$} & \mth{${\cal
      L}_{\pm\pm} /{\cal L} = 0.1$} \\ & HE & rr & $WW$ & HE & rr & $WW$\\
      \hline $\Delta P_{e^-}/P_{e^-}\, [\%]$ & $\pd0.10$ & $\pd0.51$ &
      $\pd0.07$ &$\pd0.21$ & $\pd1.11$ &$\pd0.11$ \\ $\Delta
      P_{e^+}/P_{e^+}\, [\%]$ & $\pd0.12$ & $\pd0.53$ & $\pd0.11$
      &$\pd0.15$ & $\pd1.13$ &$\pd0.21$ \\ corr & $ -0.49$ & $ -0.91$
      & $ 0 $ &$ -0.56$ & $ -0.93$ &$ -0.52$ \\ \hline
    \end{tabular}
    \caption[Relative polarization with Blondel scheme]{Relative polarization error 
using the Blondel scheme for 
      $\sqrt{s}=340 \GeV, \, {\cal L}=500 \, \fbi, \, |P_{e^-}|=80\%, \, |P_{e^+}|=60\%$
      (HE = High energy events, rr = radiative return, $WW$ =  $W$-pair production).
      }
    \label{tab:polresbl}
  \end{center}
\end{table}

Radiative corrections to the form of eq.~(\ref{eq:sigmaff}) have
been checked with the KK Monte Carlo~\cite{kkmc}. For the high-energy
events and for the radiative-return events with a seen photon they are
negligible. For the radiative return events where the photon is lost
in the beampipe, which are not used in this analysis, the corrections
are at the percent level.

Because of the high losses in the selection of the radiative-return 
events, the errors in the single polarizations seem rather
large. However the large negative correlation reduces the error
substantially for the effective polarizations needed in the
analysis. Table \ref{tab:effpol} compares the errors on the effective
polarizations for the setups shown in table \ref{tab:polresbl} and for
polarimeter measurements assuming 0 or 50\% correlation between the
two polarimeters.

The effective polarizations considered are:
\begin{itemize}
\item $P_{\rm eff} = \frac{|P_{e^-}|+|P_{e^+}|}{1+|P_{e^-} P_{e^+}|}$,
relevant for $\ALR$ with s-channel vector exchange;
\item $|P_{e^-} P_{e^+}|$, relevant for the cross section suppression/enhancement
  with s-channel vector exchange;
\item $|P_{e^-}|+|P_{e^+}| -|P_{e^-} P_{e^+}|$, relevant for the cross section
  suppression/enhancement for t-channel $W$-pair production.
\end{itemize}
Due to the high anti-correlation, even the results from the radiative-return 
analysis with one tenth of the luminosity at the low cross
sections are competitive with Compton polarimetry. 

\begin{table}[htbp]
  \begin{center}
    \begin{tabular}[c]{|c|c|c|c|c|c|c|c|c|c|}
      \hline
      & value & \multicolumn {8}{|c|}{Rel. error [\%]} \\
      \cline{3-10}
      &
      & \mth{${\cal L}_{\pm\pm} /{\cal L} = 0.5$} 
      & \mth{${\cal L}_{\pm\pm} /{\cal L} = 0.1$} 
      & \mch{Polarimeter}\\
                                          &        & HE     & rr     & $WW$    & HE     & rr     &
$WW$ & $\rho \! = \! 0$ & $\rho\! =\! 0.5$  \\
      \hline
      $(|P_{e^-}|\!\!+\!|P_{e^+}|\!)/(1\!+\!|P_{e^-} P_{e^+}|\!)$ & $0.95$ & $0.02$ & $0.08$ & $0.02$& $0.05$ & $0.17$ & $0.02$& $0.13$ & $0.16$ \\
      $|P_{e^-} P_{e^+}|$                         & $0.48$ & $0.11$ & $0.22$ & $0.13$& $0.18$ & $0.42$ & $0.18$& $0.71$ & $0.87$ \\
      $|P_{e^-}|\!+\!|P_{e^+}| \!-\!|P_{e^-} P_{e^+}|$      & $0.92$ & $0.03$ & $0.12$ & $0.03$& $0.06$ & $0.25$ & $0.03$& $0.19$ & $0.21$ \\
      \hline
    \end{tabular}
    \caption[Relative error on the effective polarization]{Relative error 
on the effective polarizations for the
      discussed setups and
      $\sqrt{s}=340 \GeV, \, {\cal L}=500 \, \fbi, \, |P_{e^-}|=80\%, \, |P_{e^+}|=60\%$.
      For the polarimeter a total error of $0.5\%$ has been assumed.
      (HE = High energy events, rr = radiative return, $WW$ =  $W$-pair production).
      }
    \label{tab:effpol} 
  \end{center}
\end{table}

Additional information can be obtained from Bhabha 
scattering~\cite{Cuypers:1996it}.
Elastic electron-electron (M\o ller) and positron-electron (Bhabha)
scattering processes have large and
well-known~\cite{ref1,ref2,ref3,Gideon-bhabha} spin asymmetries which
can be exploited for polarimetry. However, in contrast to weak
interaction processes (such as single-W production) it is necessary
that both colliding particles are polarized. If both beams in the
collider are polarized, it is therefore possible to determine the
product of the beam polarizations relatively quickly from the copious
Bhabha events observed in the principal collider detector,
even without dedicated beam polarimeters. In this context it is
worthwhile to mention that the spin asymmetries for 
Bhabha scattering reach their maximum value of $7/9$\footnote{The
maximum value of the M\o ller and Bhabha asymmetry is 7/9 for the case
of longitudinal beam polarizations, which is the case of principal
interest. For transverse beam polarization, the maximum value is only
$\pm 1/9$ for parallel beam spins, with the sign depending on the
azimuth angle between the spin and the scattering plane.  The
asymmetry vanishes for transverse beam polarizations with orthogonal
spin directions. These properties are useful for
transverse spin experiments, cf. section~\ref{subsec:trans}.} for symmetric pairs at
$\theta_{cm}=90\deg$ which guarantees excellent coverage in the
collider detector.
 
A detailed study including Z-exchange remains to be done, however the pure
QED term constraints the two polarisations along the small axis of the error
ellipse from the Blondel scheme, so that the final gain is probably not very
large.

%%%%%%%%%%%%%%%%%%%%%%%%%%%%%%%%%%%
\subsection{Experimental aspects}
%%%%%%%%%%%%%%%%%%%%%%%%%%%%%%%%%%%

Although the methods presented here measure the luminosity-weighted
polarization directly from the annihilation data, some experimental assumptions
are involved.
In all cases it is assumed that the absolute values of the polarization of the
left- and right-handed state are the same and possible corrections have to be
obtained from polarimeters. If the polarization is written as 
${P} = \pm \langle |{ P}| \rangle  + \delta {\ P}$ The shift
in the measured polarization using $W$s in the case of electron polarization
only is given by $\Delta { P}/P = \delta { P}$

Using the Blondel scheme with polarized electrons and positrons, the
corresponding errors are 
\begin{eqnarray*}
  \Delta P_{e^-} & = & \pd 1.0 \delta P_{e^-} + 0.6 \delta P_{e^+}\\
  \Delta P_{e^+} & = & -0.5 \delta P_{e^-} -0.7 \delta P_{e^+}
\end{eqnarray*}
for the high-energy events and
\begin{eqnarray*}
  \Delta P_{e^-} & = & \pd  2.4 \delta P_{e^-} + 2.1 \delta P_{e^+}\\
  \Delta P_{e^+} & = & -1.7 \delta P_{e^-} -1.7 \delta P_{e^+}
\end{eqnarray*}
for the radiative-return sample.

The corresponding corrections have to be obtained from polarimeters. This is
possible in a Compton polarimeter where the laser polarization can be flipped
easily. To assure that the electron-laser luminosity does not depend on the
laser polarization, or to correct for such effects, one should have a
multichannel polarimeter with a large lever arm in the analysing power.

If electron and positron polarization is available, in the formulae for the
effective polarizations and for the Blondel scheme products of the two
polarizations appear so that one has to understand the correlations between
the electron and positron polarization.
In principle there can be a correlation due to the depolarization in the
bunch. Studies with CAIN \cite{cain}, however, indicate that these
correlations are small. Another source of correlation can come from time
dependences or spatial correlations due to the beam delivery system. If half
of the luminosity is taken with a polarization 5\% higher and the other half
5\% lower than average, the polarizations obtained with the Blondel scheme are
off by around 0.25\%, affecting the effective polarization by the same amount.
Measuring the polarization with polarimeters would only result in a 0.16\%
error in the effective polarization.

Time correlations have to be tracked with polarimeters. Spatial correlations
due to the beam-delivery system have to be obtained from simulations and
should be minimised already in the design.

%%%%%%%%%%%%%%%%%%%%%%%
%%%%%%%%%%%%%%%%%%%%%%%%%%%%%%
\chapter{Concluding Remarks}
%%%%%%%%%%%%%%%%%%%%%%%%%%%%%%
Strong theoretical arguments and perhaps some experimental evidence
indicate that the Standard Model of elementary particles is not the final
theory of everything, and a variety of possible extensions have been
proposed. Many of these envisage new building blocks and new interactions
near the Fermi scale, and perhaps at higher mass scales. The choice among
such scenarios, and hence the direction of new physics beyond the Standard
Model, is of paramount importance. As the first machine to probe directly
the TeV scale, the LHC will surely provide
new discoveries and  valuable information in this
regard. However, it is generally agreed that the clean and precise
environment of $e^+e^-$ collisions at the ILC is ideally suited to the search
for new physics and for determining precisely the underlying structure of
the new interactions, whatever direction the LHC results may favour.

As we have demonstrated here, polarized beams will be very powerful tools
to help reach these goals. The physics examples presented here have shown
that having both beams polarized simultaneously will provide 
high flexibility and a very
efficient means for the disentanglement of non-standard effects 
in various new physics scenarios, and their positive identification, as
well as having the capacity to observe surprises in precision tests of the
Standard Model.  Having two polarized beams available is crucial for
determining the properties and the quantum numbers of new particles, and
to test fundamental model assumptions, as we have shown in the specific
example of supersymmetry. 

The larger number of observables accessible with two polarized beams
provide better tools for revealing the structure of the underlying
physics and determining new physics parameters in model-independent
analyses. New signals may become accessible by maximizing the analyzing
power using suitable beam polarizations combined with high luminosity. In
many cases, as we have shown, double beam polarization enables better
statistics to be obtained and the dominant systematical errors in indirect
searches to be reduced. This can give access to physics scales that may be
far beyond the direct kinematical reach of accelerators. 

To exploit fully high-precision tests of the Standard Model at GigaZ, 
both beams must be polarized. To make full use of the extremely high
statistics, the beam polarizations must be known with high precision,
which cannot be provided by conventional polarimetry methods.  However,
the required precision can be achieved in the Blondel scheme, where both
electron and positron polarizations are needed.

Further, with both beams polarized, one has the possibility to exploit
transversely-polarized beams for physics studies. This option provides new
and efficient observables for the detection of possible sources of CP
violation. Additionally, it becomes easier to observe the effects of
massive gravitons and to distinguish between different models with extra
spatial dimensions, far below the threshold for the spin-2 excitations.
One can test specific triple-gauge-boson couplings which are not
accessible otherwise. The power to identify new physics with polarized
beams would represent a step forward of the utmost importance for our
understanding of fundamental interactions.

An overview of the physics benefits coming from positron polarization
has been given in this report, also in comparison with the case of
only polarized electrons. An overview of the machine design for
polarized beams at the ILC has also been given. SLAC experiments
established that reliable high-energy electron beams with a
polarization of $85$-$90$\% can be provided for an LC.  The standard
source for polarized electrons is a DC gun with a strained
superlattice photocathode.  To obtain polarized positrons at an LC,
two methods are discussed:
\begin{itemize}
\item[a)] Circularly-polarized photons from a
helical undulator can be used to generate lon\-gi\-tu\-dinally-polarized
positrons in a target via $e^{\pm}$ pair production; 
\item[b)]
Circularly-polarized photons can also be obtained by backscattering of
laser light off the electron beam.
\end{itemize}
A demonstration experiment for an
undulator-based polarized positron source, E-166 at SLAC, is currently
running. Prototypes of both superconducting and
permanent magnets for helical undulators for the ILC with lengths 
of 10 or 20 periods
are already under construction at the Daresbury and Rutherford
Laboratories. Meanwhile, the concept of the laser-based positron source
is being tested in an experiment at KEK.

As already emphasized, precise physics analyses at an LC require
accurate beam polarization measurements. The primary polarimetry
measurement at the ILC will be performed with a Compton polarimeter,
with an expected accuracy $\Delta P_{e^-}/P_{e^-}=0.25\%$.  The
polarimeters can be located upstream or downstream of the Interaction
Point (IP), and it would be desirable to implement polarimeters at
both locations. A downstream polarimeter may require a large crossing
angle at the IP. 

Since polarimeters measure either the polarization of the incoming beam
before it has been depolarized by the beam-beam interaction, or that of
the outgoing beam after depolarization in the interaction
region, it is desirable to measure the polarization also directly from the
data via processes with precisely known polarization structures, such as
$WW$ production. Alternatively, in the GigaZ option, one may apply the
Blondel scheme, for which polarized positrons are needed. Methods for
measuring the polarization using annihilation data involve physics
assumptions that have to be considered in the framework of the model in
which the data are analysed. The data-driven methods therefore cannot
replace completely the polarimeters but provide an independent and
complementary measurement.  The data methods offer the possibility to reach a
precision even better than the polarimeters, if the underlying physics is
well understood.

{\it In summary}, it has been demonstrated that having simultaneously
polarized $e^-$ and $e^+$ beams is 
a very effective tool for
direct as well as indirect searches for new physics.  This option provides
ideal preparation even for unexpected new physics. Polarized positrons are
necessary for several specific physics issues, and enrich the
physics potential considerably. 
Techniques and engineering design for a polarized positron source
are becoming well advanced. Therefore, including a polarized positron source
in the ILC baseline design can be considered.

%%%%%%%%%%%%%%%%%%%%%%%
%%%%%%%%%%%%%%%%%%%%%%%%%%%%%%%%%%%%%%%%%%%%%%%%%%%%%%%%%%%%%%%
\appendix
\chapter{Tools for simulation studies: MC event generators}
\setcounter{equation}{0}
\renewcommand{\thesection}{A}

%%%%%%%%%%%%%%%%%%%%%%%%%%%%%%%%%%%%%%%%%%%%%%%%%%%%%%%%%%%%%%%
The use of numerical programs based on Monte Carlo (MC) techniques
has become essential in performing any detailed experimental analysis
in collider physics.
In this Appendix the so-called event generators with
inclusion of beam polarizations  are briefly described. These
programs must be interfaced to both detector simulations and beam energy
spectra to give a complete picture of the actual physical process.\\[.3em]
%%%%%%%%%%%%%%%%%%%%%%%%
Most MC event generators fall into one of two classes:
\begin{itemize}
\item[{\bf a)}] {\bf general-purpose (or multi-purpose) event generators}\\
which aim to
perform the full simulation of the event starting with the
initial-state collider beams, proceeding through 
the parton-shower (wherein coloured particles are perturbatively evolved from 
the hard scale of the collision to an infrared cut-off) and hadronization
 (wherein partons left after the said perturbative evolution
are formed into the observed mesons and baryons) stages;
\item [{\bf b)}] {\bf parton-level event generators}\\
which typically
perform the hard scattering part of the simulation only, perhaps
including decays, and rely on one of the general-purpose generators
for the rest of the simulation.
\end{itemize}
During the LEP-era the experiments relied on the general-purpose event
generators for the description of hadronic final states together with more
accurate parton-level programs interfaced to the former ones for specific
processes, e.g., two- and four-fermion production.  At a future linear
collider, as one wishes to study final states with higher multiplicities, e.\
g.\ six or even eight particles, this combined approach will become more
important as these final states cannot be described by the general-purpose
event generators.
 
%%%%%%%%%%%%%%%%%%%%%%%%%%%%%%%%%%%%%%%%%%%%%%%
\subsection*{General-purpose event generators}
%%%%%%%%%%%%%%%%%%%%%%%%%%%%%%%%%%%%%%%%%%%%%%%
\begin{sloppypar}
General-purpose event generators include
HERWIG~\cite{Corcella:2000bw}, ISAJET~\cite{Baer:1999sp} and
PYTHIA~\cite{Sjostrand:2000wi}, which use different phenomenological models
and approximations. The major differences between these programs are in the
approximations used in the parton shower evolution and the hadronisation
stage.  There are also major differences between the generators in the
treatment of spin correlation and polarization effects. Both ISAJET and HERWIG
include longitudinal polarization effects in both SM and supersymmetric (SUSY)
production processes, while PYTHIA includes both longitudinal and transverse
polarizations in many processes.
\end{sloppypar}  

Another important difference is in the treatment of the subsequent decay of
any heavy particle produced in the hard process. While HERWIG includes the
full correlations in any subsequent decays using the method described
in~\cite{Richardson:2001df}, both ISAJET and PYTHIA only include these effects
in some processes, such as $W$-pair production. Including transverse
polarization also in the HERWIG production stage is certainly possible.
 
The only program currently available in C++ which is capable of generating
physics results is SHERPA (based on the APACIC++~\cite{Kuhn:2000dk} parton shower).  Work
is however underway to rewrite both PYTHIA~\cite{Bertini:2000uh} and
HERWIG~\cite{Gieseke:2002sg} in C++. Given the new design and structure of
these programs, the treatment of both spin correlation and polarization effects
should be much better than in the current FORTRAN programs. HERWIG++
should include full polarization and correlation effects using the method
of \cite{Richardson:2001df}.
 
%%%%%%%%%%%%%%%%%%%%%%%%%%%%%%%%%%%%%%%%%%%%
\subsection*{Parton-level event generators}
%%%%%%%%%%%%%%%%%%%%%%%%%%%%%%%%%%%%%%%%%%%%
There are a large number of programs available which calculate some set of
hard processes, and are interfaced to one of the general-purpose generators,
most often PYTHIA.  Many of the two- \cite{Kobel:2000aw} and four-fermion
\cite{Grunewald:2000ju} generators were used by the LEP collaborations (see
report of the LEP-II MC workshop for their detailed discussion).  Some
programs are written for six-fermion processes, e.g.,
ref.\ LUSIFER~\cite{Dittmaier:2002ap}, SIXFAP \cite{Montagna:1997dc}, EETT6F
\cite{Kolodziej:2001xe} and SIXPHACT \cite{Accomando:1998ju}. These generators
usually include state-of-the-arts calculations of the generated processes.
Many of these
codes use helicity-amplitude techniques to calculate the matrix elements and
therefore either already include polarization effects or could easily be
modified to do so.
     
There are a number of codes available, which are capable of calculating and
integrating the matrix elements for large numbers of final-state particles
automatically:\\
-- {\tt AMEGIC++} \cite{Krauss:2001iv} makes use of helicity
amplitudes and is part of SHERPA.\\
-- {\tt COMPHEP} \cite{Pukhov:1999gg} uses
the traditional trace techniques to evaluate the matrix elements; therefore it
is at present not suitable to investigate polarization/spin effects.  However,
the conversion to the use of helicity amplitudes techniques is currently
planned.\\
-- {\tt GRACE}~\cite{Ishikawa:1993qr} (with the packages {\tt
BASES} and {\tt SPRING}) uses the calculation of matrix elements via helicity
amplitude techniques.\\
-- {\tt HELAC/PHEGAS} calculates cross
sections\cite{Kanaki:2000ey,Papadopoulos:2000tt}.\\
-- {\tt MADGRAPH/MADEVENT}
\cite{Maltoni:2002qb} uses helicity amplitude techniques for the matrix
elements to calculate cross sections (based on the {\tt HELAS}
\cite{Murayama:1992gi} subroutines).\\
-- {\tt WHIZARD} \cite{Ohl:2000pr} is
an integration package which can use either {\tt COMPHEP, MADGRAPH} or {\tt
O'MEGA}\footnote{{\tt O'MEGA} uses the approach of \cite{Caravaglios:1995cd}
to evaluate the matrix elements but does not yet include any QCD
processes.}~\cite{Moretti:2001zz} to calculate the matrix elements.
Transverse beam polarization is included.

The implementation of polarization and correlation effects differs between
these programs. In general, apart from {\tt COMPHEP} (as noted), these
programs are all based on helicity amplitude techniques at some point in the
calculation and therefore the inclusion of both transverse and longitudinal
beam polarization is possible even where it is not currently implemented.
%%%%%%%%%%%%%%%%%%%
\subsection*{Event generators for supersymmetry}
%%%%%%%%%%%%%%%%%%%
 
Polarization and spin correlation effects are particularly important in
studying SUSY scenarios, in order to measure the fundamental parameters of the
underlying model.
  
HERWIG, PYTHIA and ISAJET all include longitudinal polarization effects in
SUSY production processes. There is also a parton-level program
SUSYGEN~\cite{Ghodbane:1999va}, interfaced to PYTHIA, which includes these
effects.
 
These programs also differ in the inclusion of the correlations in the
subsequent decays of the particles. While SUSYGEN includes these correlations
using helicity amplitude techniques and HERWIG uses the method
of~\cite{Richardson:2001df}, these effects are generally not included in
either PYTHIA or ISAJET.
 
Among the parton-level programs, at present only {\tt COMPHEP} and
{\tt GRACE} and {\tt AMEGIC++} include SUSY processes, limitedly to
the MSSM (although extensions to non-minimal SUSY models are now
planned within {\tt COMPHEP}).  {\tt WHIZARD} also supports the MSSM,
optionally with anomalous couplings, but model extensions or
completely different models can be added.  Finally, {\tt MADGRAPH} can
be extended to add the additional SUSY interactions that are needed,
as (most of) the prototype subroutines for the latter already exist in
the HELAS library.
 
%%%%%%%%%%%%%%%%%%%%%%%%%%%%%%%%%%%%%%%%%%%%%%%%%%%

%%%%%%%%%%%%%%%%%%%%%%%%%%%%%%
\newpage
\subsection*{Acknowledgements}
%%%%%%%%%%%%%%%%%%%%%
The work was partially supported by the Commission of the European 
Communities under the 6th Framework Programme `Structuring the European
Research Area', contract number RIDS-011899.
This work is also supported
by the 'Fonds zur F\"orderung der wissenschaftlichen Forschung' (FWF)
of Austria, projects No. P16592-N02 and by the
European Community's Human Potential Programme under contract
HPRN-CT-2000-00149, by the 'Deutsche Forschungsgemeinschaft'
(DFG) under contract Fr 1064/5-1, by the Polish State Committee
for Scientific Research Grant No 2 P03B 040 24
and by the DOE project with the Contract No. DE-AC02-76SF00515.
S.Hes. is supported by G\"oran Gustafsson Foundation.
GMP would like to thank F.~Richard, D.~Schulte, T.~Teubner, 
P.~Uwer and N. Walker for many helpful discussions. 

%%%%%%%%%%%%%%%%%%%%%%%%%%%%%%%%%%%%%%%%%%%%%%%%%%%%%%%%%%%%%%

%%%%%%%%%%%%%%%%%%%%%%%%%%%%%%%%%%%%%%%%%%%%%%%%%%%%%%%%%%%%%%

\end{document}